\documentclass[preprint,trackchanges]{aastex62}

\usepackage{amsmath}

\hypersetup{linkcolor=cyan,citecolor=magenta,filecolor=yellow,urlcolor=blue}

\received{December 21, 2017}
\accepted{March 29, 2018}
\published{May 17, 2018 }

\shorttitle{Constraining the Type of Central Engine of GRBs with {\it Swift} Data}
\shortauthors{Li et al.}

\begin{document}
\title{Constraining the Type of Central Engine of GRBs with {\it Swift} Data}

\correspondingauthor{Liang Li}
\email{liang.li@fysik.su.se, xfwu@pmo.ac.cn, leiwh@hust.edu.cn, dzg@nju.edu.cn}

\author[0000-0002-0786-7307]{Liang Li}
\affil{Purple Mountain Observatory, Chinese Academy of Sciences, Nanjing 210008, China}
\affiliation{Department of Physics, Stockholm University, AlbaNova, SE-106 91 Stockholm, Sweden}
\affiliation{Department of Physics, KTH Royal Institute of Technology, SE-106 91 Stockholm, Sweden}

\author{Xue-Feng Wu}
\affiliation{Purple Mountain Observatory, Chinese Academy of Sciences, Nanjing 210008, China}

\author{Wei-Hua Lei}
\affiliation{School of Physics, Huazhong University of Science and Technology, Wuhan 430074, China}

\author{Zi-Gao Dai}
\affiliation{School of Astronomy and Space Science, Nanjing University, Nanjing, Jiangsu 210093, China}

\author{En-Wei Liang}
\affiliation{GXU-NAOC Center for Astrophysics and Space Sciences, Department of Physics, Guangxi University, Nanning 530004, China}

\author{Felix Ryde}
\affiliation{Department of Physics, KTH Royal Institute of Technology, SE-106 91 Stockholm, Sweden}

\begin{abstract}

The central engine of gamma-ray bursts (GRBs) is poorly constrained. There exist two main candidates: a fast-rotating black hole and a rapidly spinning magnetar.
Furthermore, X-ray plateaus are widely accepted to be the energy injection into the external shock.
In this paper, we systematically analyze the \emph{Swift}/XRT light curves of 101 GRBs having plateau phases and known redshifts (before 2017 May).
Since a maximum energy budget ($\sim2\times10^{52}$ erg) exists for magnetars but not for black holes, this provides a good clue to identifying the type of GRB central engine.
We calculate the isotropic kinetic energy $E_{\rm K,iso}$ and the isotropic X-ray energy release $E_{\rm X,iso}$ for individual GRBs.
We identify three categories based on how likely a black hole harbors a central engine: 'Gold' (9 out of 101; both $E_{\rm X,iso}$ and $E_{\rm K,iso}$ exceed the energy budget), 'Silver' (69 out of 101; $E_{\rm X,iso}$ less than the limit but $E_{\rm K,iso}$ is greater than the limit), and 'Bronze' (23 out of 101, the energies are not above the limit).
We then derive and test the black hole parameters with the Blandford-Znajek mechanism, and find that the observations of the black hole candidate ('Gold'+'Silver') samples are consistent with the expectations of the black hole model.
Furthermore, we also test the magnetar candidate ('Bronze') sample with the magnetar model, and find that the magnetar surface magnetic field ($B_{p}$) and initial spin period ($P_{0}$) fall into reasonable ranges.
Our analysis indicates that if the magnetar wind is isotropic, a magnetar central engine is possible for 20\% of the analyzed GRBs.
For most GRBs, a black hole is most likely operating.

\end{abstract}

\section{Introduction}\label{sec:intro}

One of the most fundamental, yet unanswered, questions in gamma-ray burst (GRB) physics is the nature of the central engine. Observationally, they have to fulfill several requirements. First, they should be able to release a huge amount of energy ($\sim10^{49}-10^{55}$ erg). Second, indirect estimates of the jet Lorentz factor $\sim$100, together with the compactness problem, require the central engine to be made up of baryons as much as possible. Finally, the rapid variability hints towards an intermittent central engine \citep[e.g.,][]{1996IAUS..165..467F,2001AJ....121.2879B,2010ApJ...725.2209L}.

There are two main progenitor models for the central engine. On the one hand, the magnetar model involves a rapidly spinning and highly magnetized neutron star. In this scenario, the rotational energy of the neutron star powers the GRB \citep[e.g.,][]{1992Natur.357..472U, 1994MNRAS.270..480T,1998A&A...333L..87D, 2000ApJ...537..810W, 2001ApJ...552L..35Z,2008MNRAS.385.1455M, 2011MNRAS.413.2031M, 2012MNRAS.419.1537B, Margutti:2013dg, 2014ApJ...785...74L,Lu:2015bt}. The limited energy budget provided by the magnetar model might, however, be at odds with observed GRB energies \citep{2010ApJ...711..641C}.
On the other hand, a newly formed black hole (hereinafter BH) surrounded by an accretion disk \citep[e.g.,][]{1991AcA....41..157P,1992ApJ...395L..83N,1999ApJ...524..262M} can produce bipolar jets through the Blandford-Znajek (BZ) mechanism \citep{1977MNRAS.179..433B} or through the production and polar annihilation of neutrino-antineutrino pairs \citep{2017arXiv170505516L}.
The rapid rotation of the central engine required by GRB models might, however, prevent the collapse to a BH \citep{2007ApJ...664..416B,2008ApJ...673L..43D}.
Distinguishing between a BH and a neutron star as the central engine is of the utmost importance as it would help to better constrain the progenitor of GRBs and their emission mechanisms.

The long time activity and decay of the central engine can leave footprints in the X-ray and optical afterglows and can thus help distinguish between the proposed progenitors. For example, the optical rebrightening of GRB 970228 was interpreted to be the result of energy injection by a millisecond pulsar \citep{1998A&A...333L..87D}. In addition, \cite{2001ApJ...552L..35Z} showed that a shallow decay followed by a normal decay phase is naturally obtained for the energy ejection by a millisecond magnetar.
The widely accepted scenario for the X-ray plateaus (also called the shallow decay) is the energy injection, either matter dominated or Poynting-flux (electron-positron wind) dominated, into the external shock. The simplest way to deal with the energy injection process is by assuming that the central engine is a millisecond magnetar, and that the injected energy is carried out by a Poynting-flux/electron-positron wind extracting the spin energy of the magnetar through the magnetic dipole radiation (MDR).
Of particular importance is the interpretation of the X-ray plateau as energy injection into the external forward shock, which is an evidence for extended central engine activity. The energy can be provided either in terms of Poynting flux from the central magnetar or by the deposition of kinetic energy into the forward shock by a stratified jet.
Both models are successful in explaining the normal decay phase at the end of the plateau.
However, the plateau is followed by a sharp flux drop as steep as $F \sim t^{-9}$. Such a sharp decay is very difficult to explain using the forward shock model, since it requires the radiation to be released from a small radius. In this case, the plateau has to be of internal origin (as opposed to an external one, when it can be explained by energy injection). Such plateaus are easily explained in the magnetar framework as residual rotational and magnetic energy that can be extracted after the main extraction event. However, no model for the emission mechanism exists to date.
The spin-down luminosity of the energy injection from a central engine would evolve with time as (\citealt{2001ApJ...552L..35Z}; the magnetar injection corresponds to {\it q}=0 for $t<\tau$ and {\it q}=2 for $t>\tau$)
\begin{eqnarray}
\label{eq:q} L(t)\simeq L_{0} \left(\frac{t}{t_{0}}\right)^{-q}=\left\{\begin{array}{ll}
L_{0}, & , t\ll\tau\\
L_{0}(t/\tau)^{-2}, & , t\gg\tau \\
\end{array} \right.
\end{eqnarray}
Here, {\it q} is the luminosity injection index. For BHs, they can also inject energy, but {\it q} will depend on density profiles of the star \citep[e.g.,][]{2008MNRAS.388.1729K, 2008Sci...321..376K}.

An additional criterion that could be used to differentiate between the magnetar and the BH model is the total energy emitted during the bursts. Indeed, heuristic arguments lead to a maximum energy that can be released by a magnetar, corresponding to the maximum rotational energy of $2\times10^{52}$ erg (\citealt{2014ApJ...785...74L}). Such a limit does not exist for BHs.
The emitted radiation energy will only be some fraction of the rotational energy depending on the radiative efficiency (see, e.g., \citealt{2017MNRAS.472.3058B}).
More importantly, such a limit requires an assumption to be made on whether or not the energy emitted from a magnetar is isotropic. The reason is that the inferred total energy is given by integrating the observed flux over all solid angles. The angular distribution of the flux must therefore be assumed.
For a millisecond magnetar, the energy is carried away by a Poynting-flux/electron-positron wind extracting the spin energy through the MDR.  The wind expected to be launched is nearly isotropic wind (e.g. \citealt{1992Natur.357..472U}). For instance, for the pulsar wind nebular of Crab, the MDR of a neutron star is nearly isotropic.
Moreover, in the case of a neutron star merger as the origin of the magnetar central engine in short GRBs, the total energy is also expected to be emitted quasi-spherically (e.g. \citealt{2014MNRAS.439.3916M, 2006MNRAS.372L..19F}). However, if the magnetar is born at the center of a collapsar, the progenitor star could cause a collimation of the emitted energy, both during the prompt phase (\citealt{2007MNRAS.380.1541B, 2009MNRAS.396.2038B}) and the afterglow phase (\citealt{2014ApJ...785...74L}). However, \cite{2014MNRAS.443...67M} argue for the magnetar wind to be close to isotropic within a collapsar based on the energetics of the associated supernovae (SNe). Only a small fraction of the total energy is then channeled into the GRB jet, while most of the total energy is responsible for the SN explosion.
Therefore, the isotropic injection has a low efficiency.
However, the energy injection from a fast-rotating BH through the BZ mechanism is naturally considered to be collimated \citep{1977MNRAS.179..433B}.
The collimated injection has a high efficiency, since a large fraction of the energy can be injected into the GRB jet.

\textcolor{orange}{In this paper, we assume that the magnetar wind is quasi-spherical during the energy injection phase of the X-ray afterglow.} We analyze the plateau afterglow of GRBs in order to investigate the plausibility of having a magnetar as a central engine. We extensively search the \emph{Swift}/XRT data for GRBs whose light curves have a plateau phase and calculate the X-ray and kinetic energy released during this phase. \textcolor{blue}{We then investigate whether or not the data are consistent with the BH central engine model.}

The paper is organized as follows.
The XRT data reduction and the sample selection are presented in \S 2.
The physical parameters of the GRBs and the hypothetical BH/magnetar parameters are derived in \S 3.
The statistical analyses of the external shock model parameters are presented in \S 4.
The discussions and conclusions are presented in \S 5.
Throughout this paper, a concordance cosmology with parameters $H_{0}= 71$ ${\rm km}$ s$^{-1}$ ${\rm Mpc^{-1}}$, $\Omega_{M}=0.30$, and $\Omega_{\Lambda}=0.70$ is adopted.

\section{Data Analysis\label{Sec:data}}

We restricted our study to GRBs observed by \emph{Swift}/XRT from 2004 December to 2017 May, which show a transition from a plateau to either a normal decay phase (for the case of canonical light curves) or a very steep decay (for the case of "internal plateaus").
We first identify such bursts by visual inspection of the X-ray light curves. To derive the properties of the GRBs in the rest frame, we only select the bursts with known redshifts.

\subsection{Sample Definition}

Two possible forms of light curves might confuse our sample selections performed through visual inspection. First, we might take an intrinsic single power-law (SPL) decay as a broken power-law decay. Second, we might also take a normal decay followed by a postjet break phase as a plateau followed by a normal decay phase. 

Therefore, to characterize a plateau, we perform a temporal fit to the X-ray light curves within a time interval ($t_{s}$, $t_{e}$) for each burst.
Here, $t_{s}$ is the time when the plateau begins and $t_{e}$ when the segment after the plateau break ends.
We employ either a smoothly broken power law (BKPL),
\begin{equation}
F =  F_0 \, [(t/t_{\rm b})^{\alpha_1\omega}+(t/t_{\rm b})^{\alpha_2\omega}]^{-1/\omega},
\label{eq:bkpl}
\end{equation}
or an SPL\textbf{\footnote{Note that our basic model is the BKPL model; in order to exclude an intrinsic SPL decay cases, we also apply the SPL model for some cases.}} to fit the light curves,
\begin{equation}
 F = F_0\, t^{-\alpha}.
\end{equation}
In the above equations, $\alpha$, $\alpha_1$, and $\alpha_2$ are the temporal slopes; $t_{\rm b}$ is the break time; $F_{\rm b} = F_0\, 2^{-1/\omega}$ is the flux of the break time; and $\omega$ describes the sharpness of the break\footnote{The smaller the $\omega$, the smoother the break; we often fixed it to be 3.}.
We perform the fits to the data with a nonlinear, least-squares fitting using the Levenberg-Marquardt algorithm, and an IDL routine called mpfitfun.pro\footnote{\url{http://purl.com/net/mpfit}} \citep{2009ASPC..411..251M}.

To exclude an intrinsic SPL decay, we first apply the above two models. We then choose the best model according to the following two principles.
(i) We compare the $\chi^2_{\rm r}$ value for both models and choose the model having the $\chi^2_{\rm r}$ closest to 1\footnote{We excluded GRB 050416A, GRB 080928, GRB 081118, and GRB 101219B from our samples since we confirmed that SPL is the best-fit model. We retained GRB 060418, GRB 061021, GRB 070529, GRB 081221, and GRB 131030A in our samples since we confirmed that BKPL is the best-fit model.}.
For example, the value of the $\chi^2_{\rm r}$ of GRB 070529 for the BKPL model fitting is given 29/30 (close to 1), and for the SPL model fitting it is 68/33 (much larger than 1). We then adopt the BKPL as the best model for GRB 070529.
(ii) In case both models have similar $\chi^2_{\rm r}$ values (close to 1), we always chose the simplest model (SPL).
For instance, GRB 101219B has $\chi^2_{\rm r}$ 16/18 (close to 1) for the BKPL model and 21/21 (close to 1) for the SPL model. In this case, the SPL model is the best model for our choice.
We then remove those cases that have the SPL model as the best fit\footnote{We excluded GRB 080210, GRB 130514A, and GRB 140430A from our samples.}.

To properly select only plateau phases and remove the normal decay followed by a jet-break phase, we only keep the temporal decays that are too fast to be explained by an external shock. For this, we use the closure relations to obtain an estimate of what should be the spectral index $\alpha(\beta)$ in the case of different spectral regimes (ISM/wind),
\begin{eqnarray}
\label{n} \alpha^{s}_{1}<\alpha(\beta_{1})\equiv \left\{\begin{array}{ll}
\frac{3\beta_{1}-1}{2}, &  \nu>max{(\nu_{m},\nu_{c})}   \textrm{(ISM and wind)}\\
\frac{3\beta_{1}}{2},   &  \nu_{m}<\nu<\nu_{c}  \textrm{(ISM)}\\
\frac{3\beta_{1}+1}{2}, &  \nu_{m}<\nu<\nu_{c} \textrm{(wind)}\\
\end{array} \right.
\end{eqnarray}
Here, $\alpha_{1}$ is the temporal decay index before the break time, $\Gamma_{1}$ is the photon spectral index, $\beta_{1}$ is the spectral index ($\beta=\Gamma-1$), $\nu_{m}$ is the minimum frequency, and $\nu_{c}$ is the cooling frequency.
Comparing the theoretical spectral indexes to the measured ones, we only keep the bursts that satisfy\footnote{With this selection, the bursts removed from the study are GRB 081208, GRB 081203A, GRB 080413B, GRB 090313, GRB 111123A, GRB 130427B, and GRB 140213A.} $\alpha^{s}_{1}<\alpha(\beta)$.

\begin{figure}
\gridline{\fig{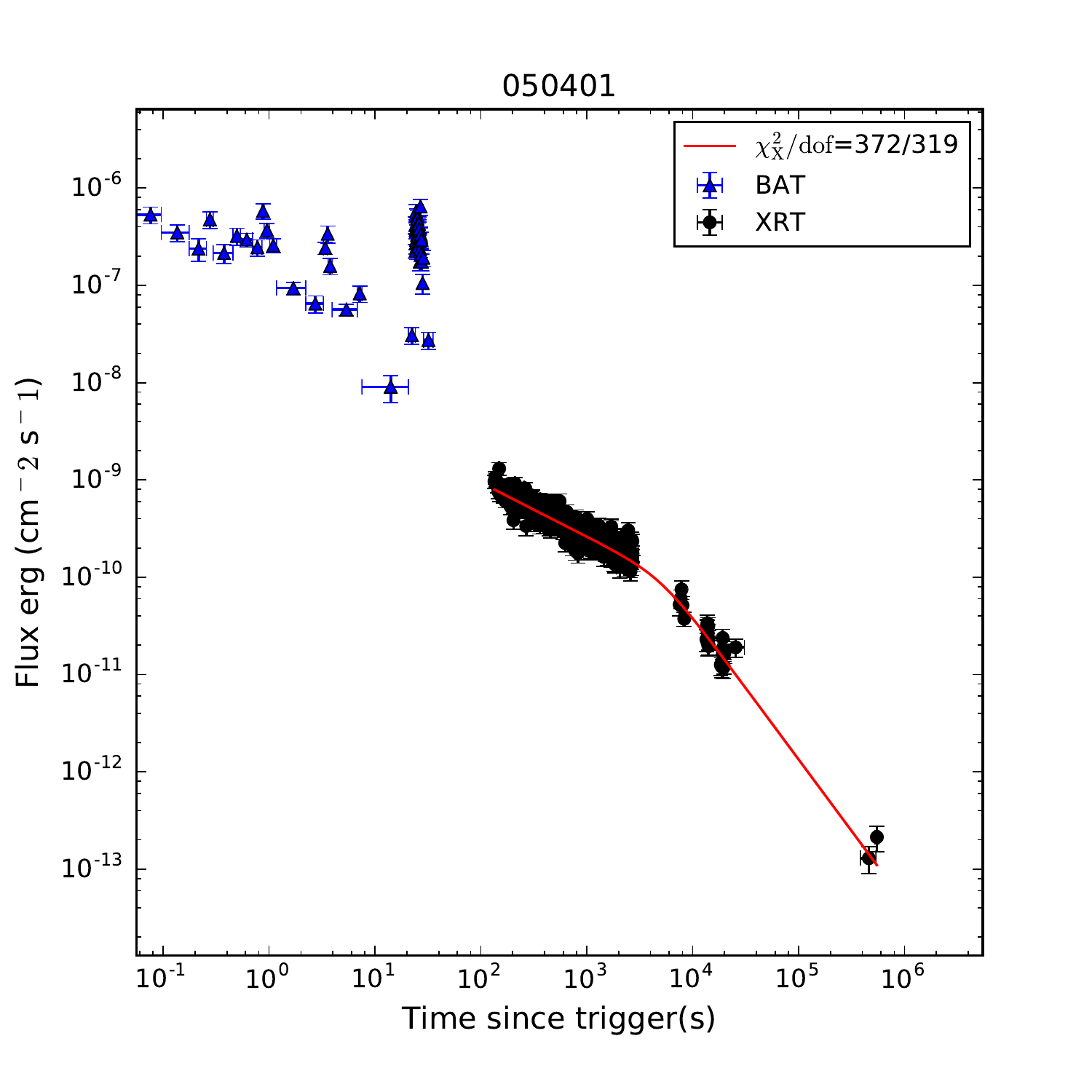}{0.28\textwidth}{}
          \fig{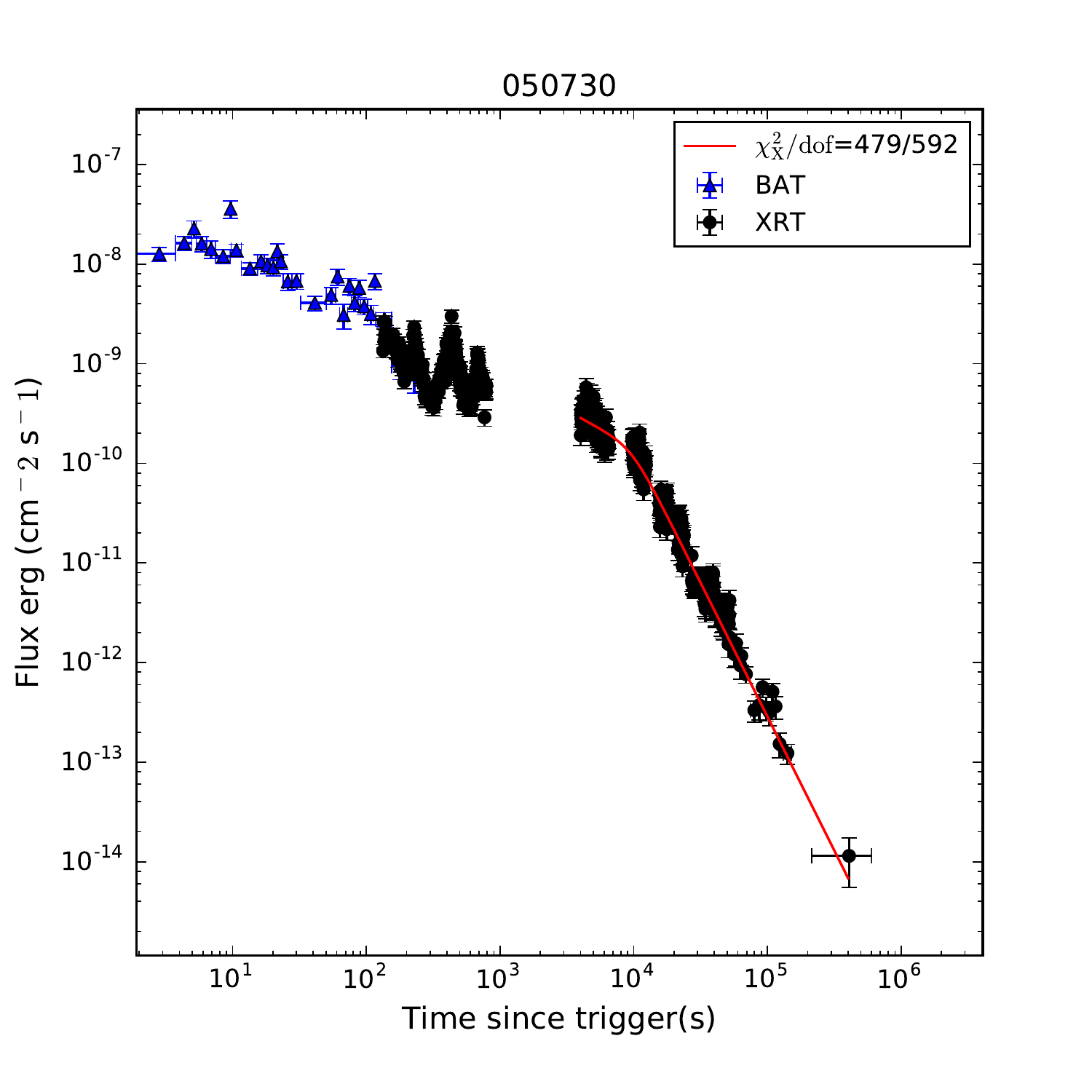}{0.28\textwidth}{}
          \fig{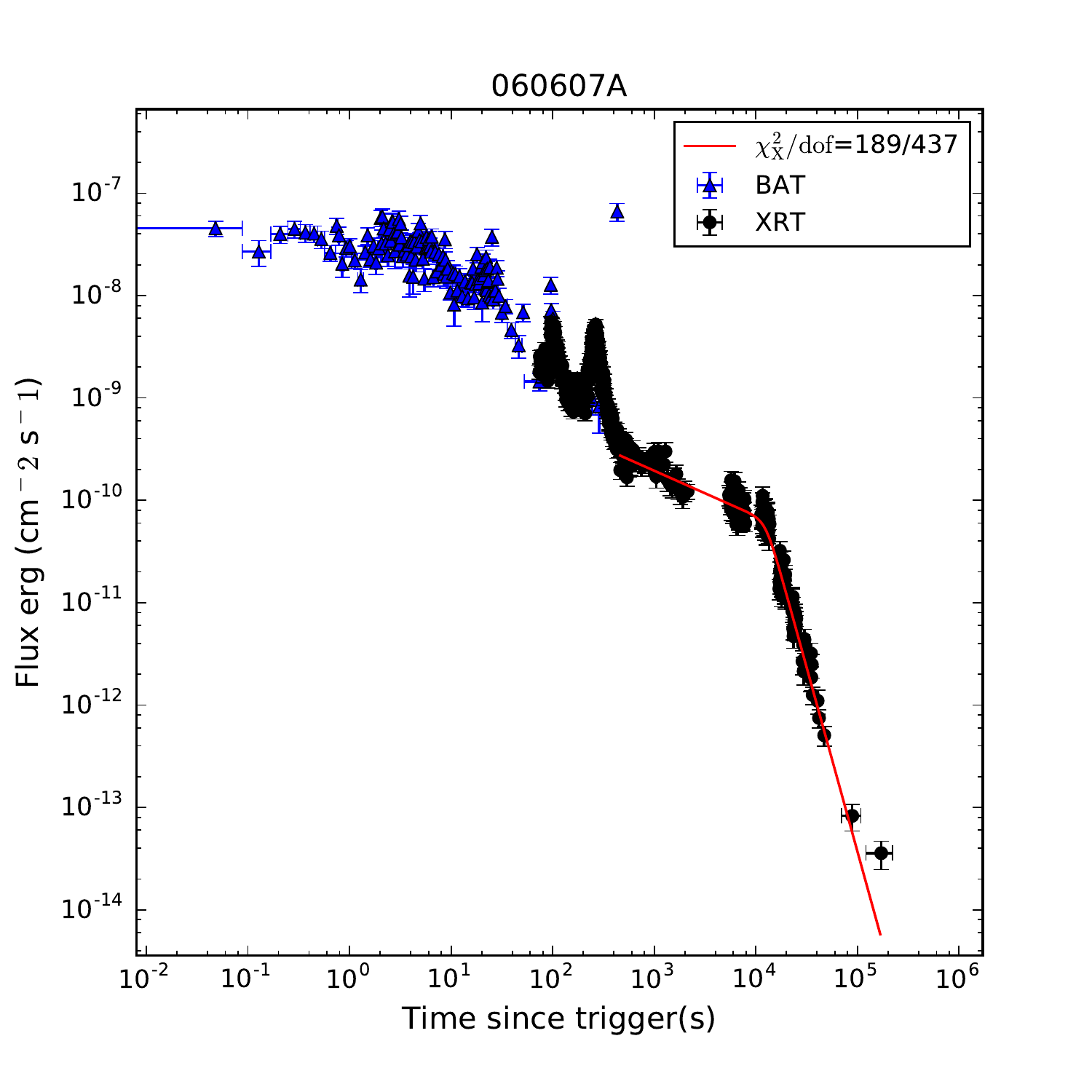}{0.28\textwidth}{}
          }
\gridline{\fig{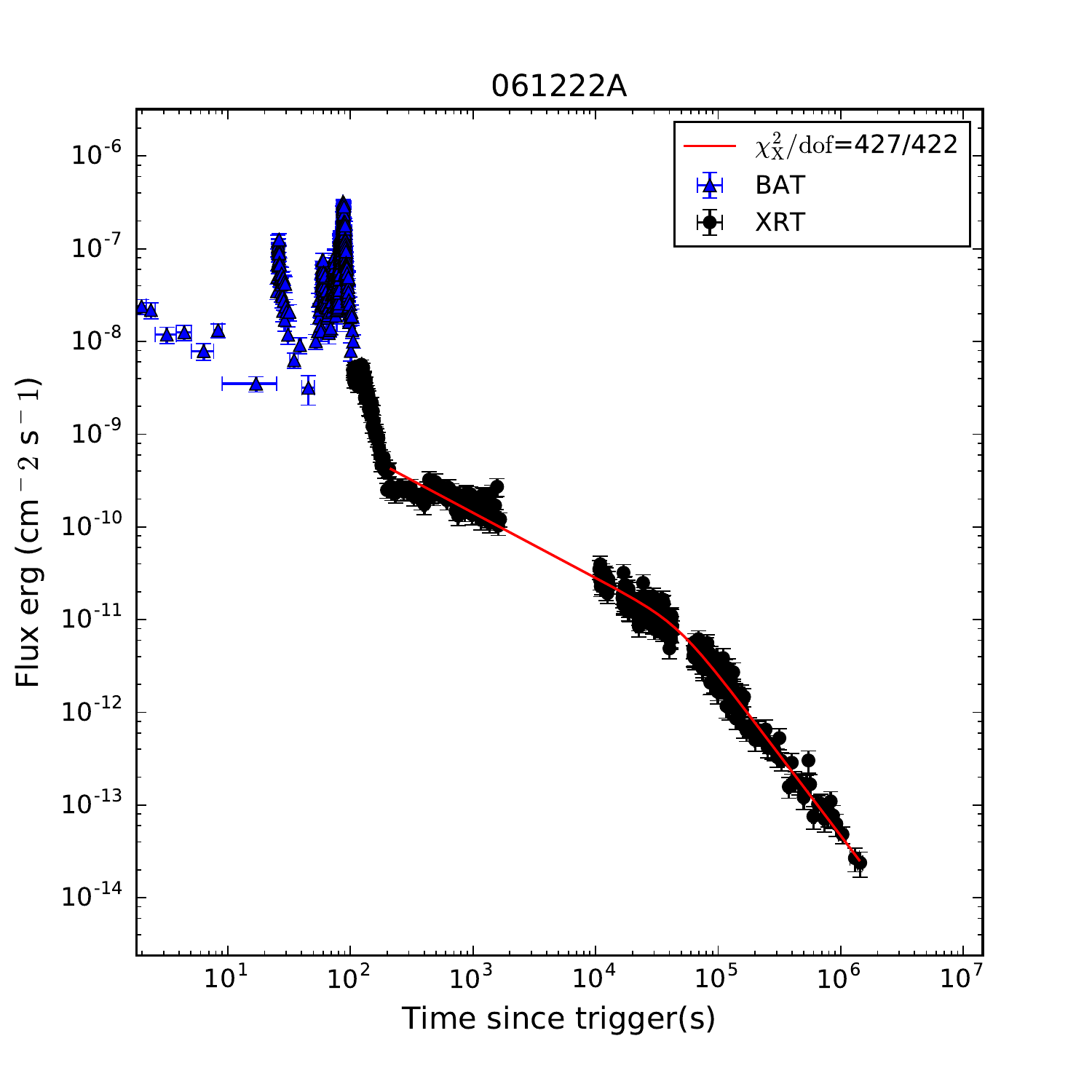}{0.28\textwidth}{}
          \fig{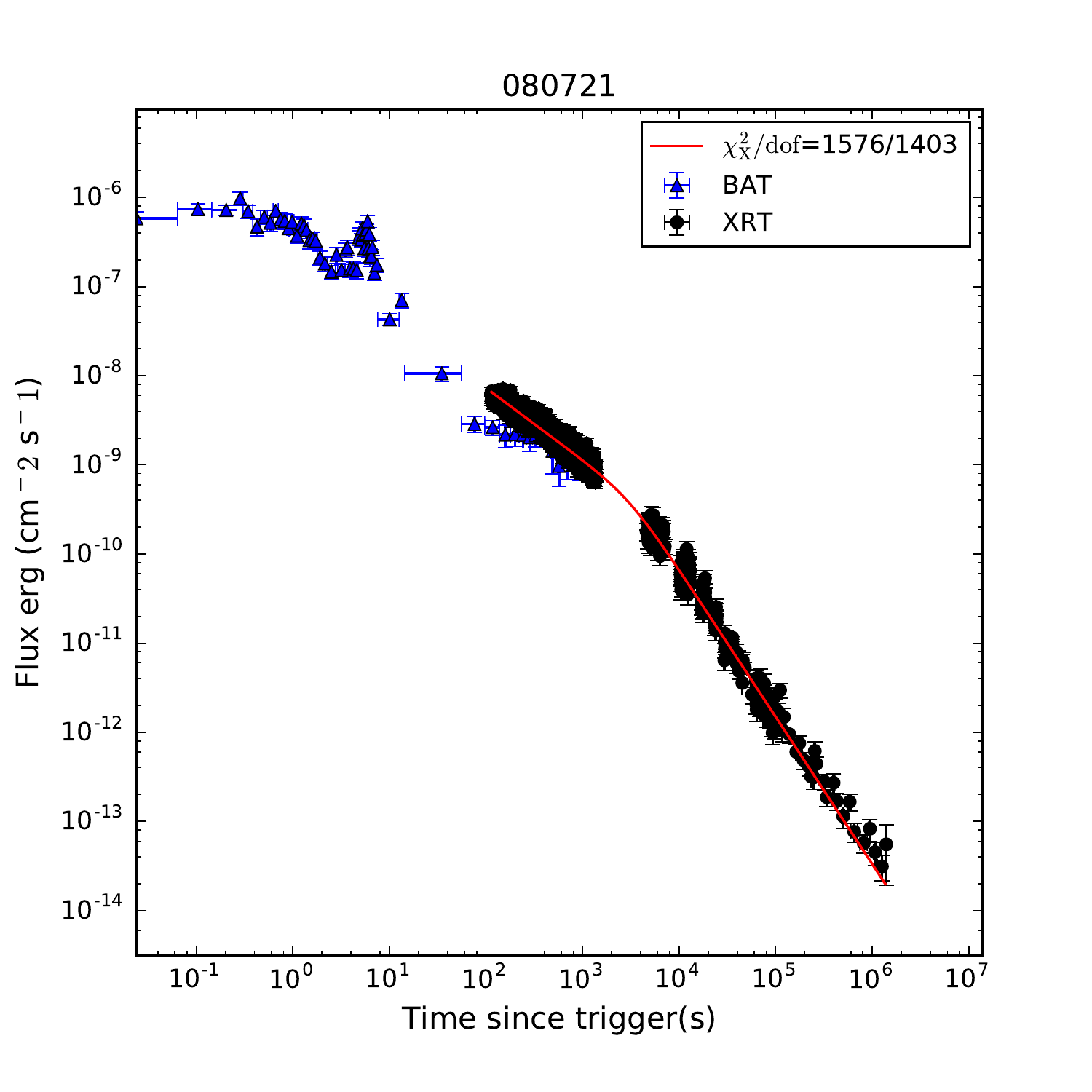}{0.28\textwidth}{}
          \fig{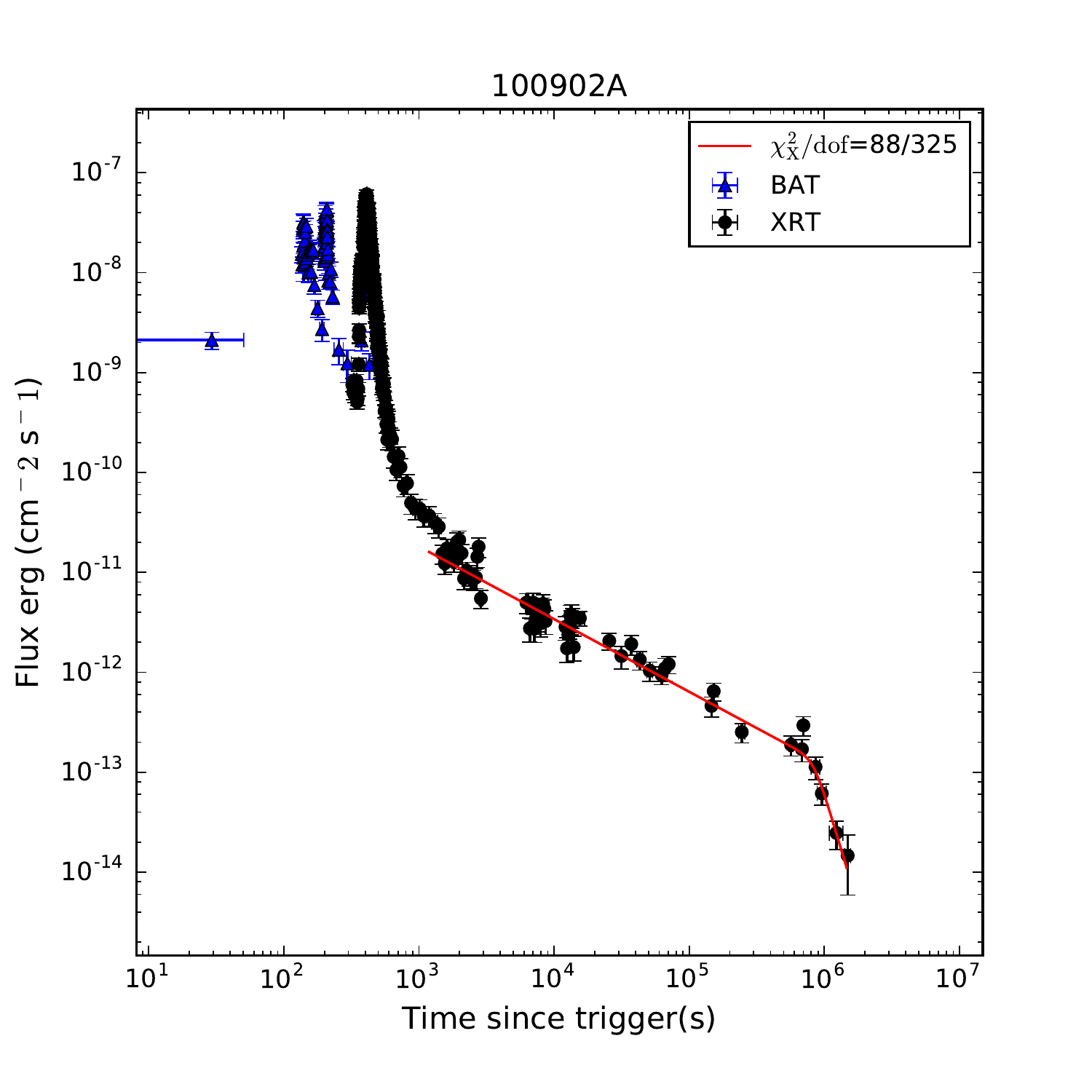}{0.28\textwidth}{}
          }
\gridline{\fig{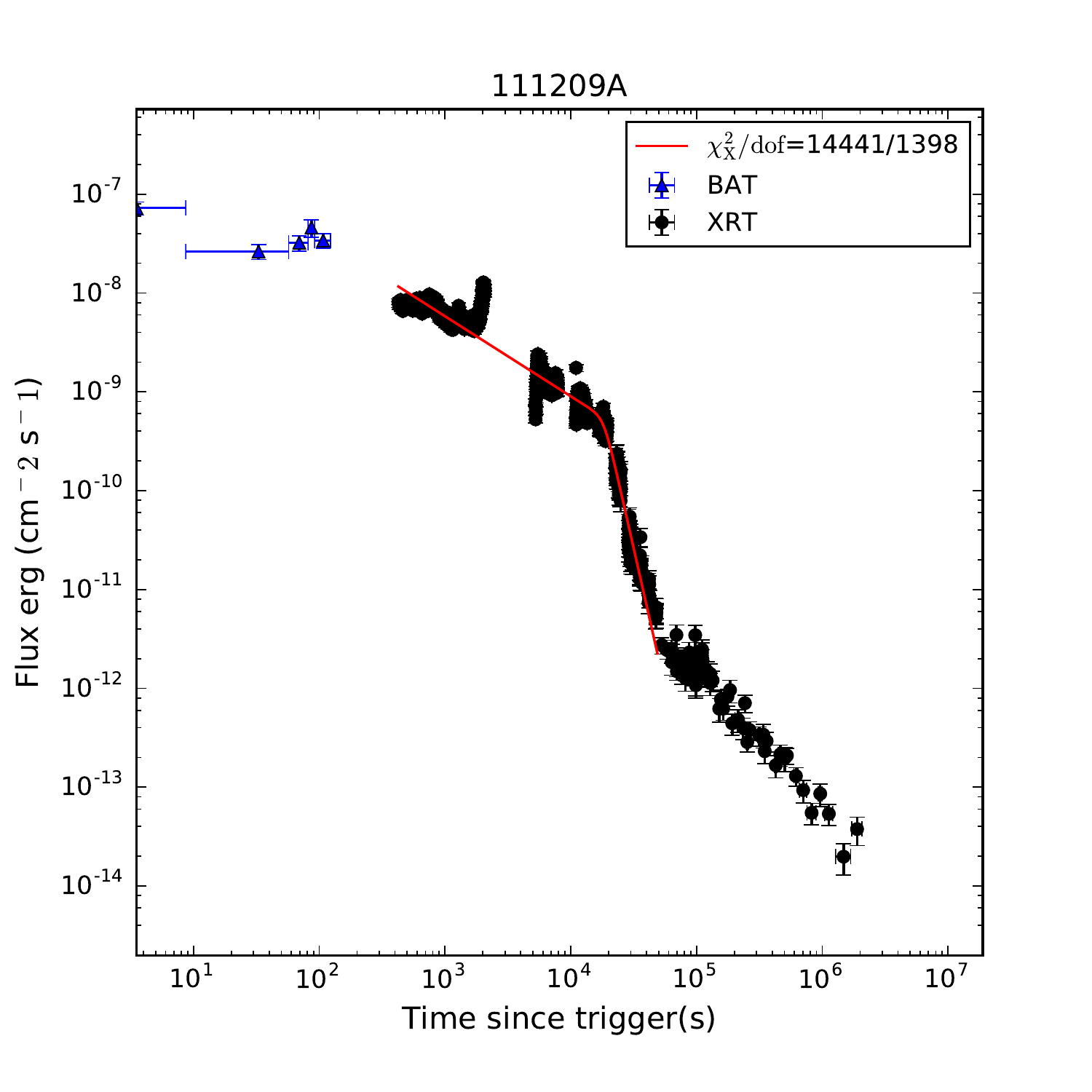}{0.28\textwidth}{}
          \fig{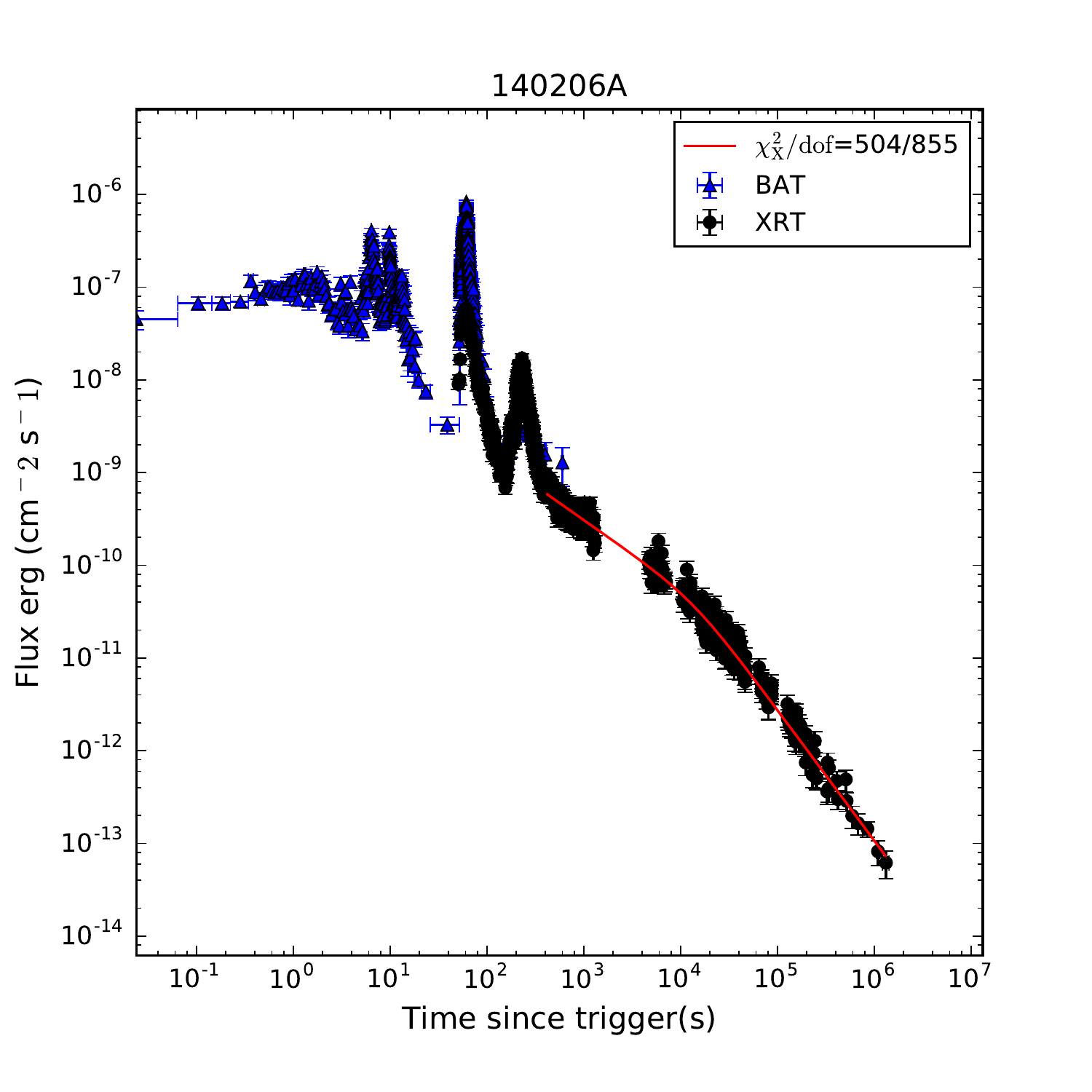}{0.28\textwidth}{}
          \fig{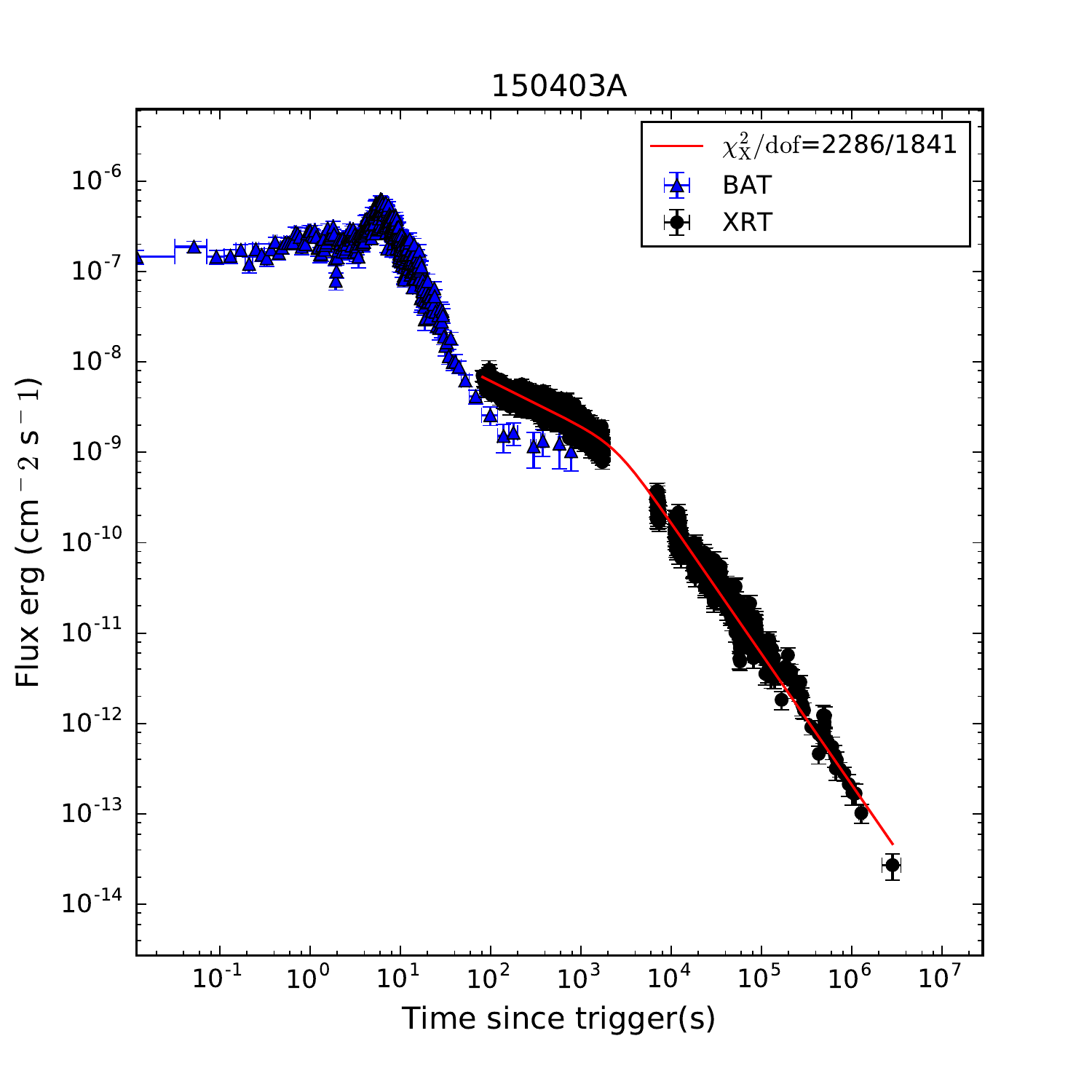}{0.28\textwidth}{}
          }
\caption{Best-fitting light curves of the X-ray plateau for the Gold sample.\label{Gold}}
\end{figure}

\begin{figure}
\gridline{\fig{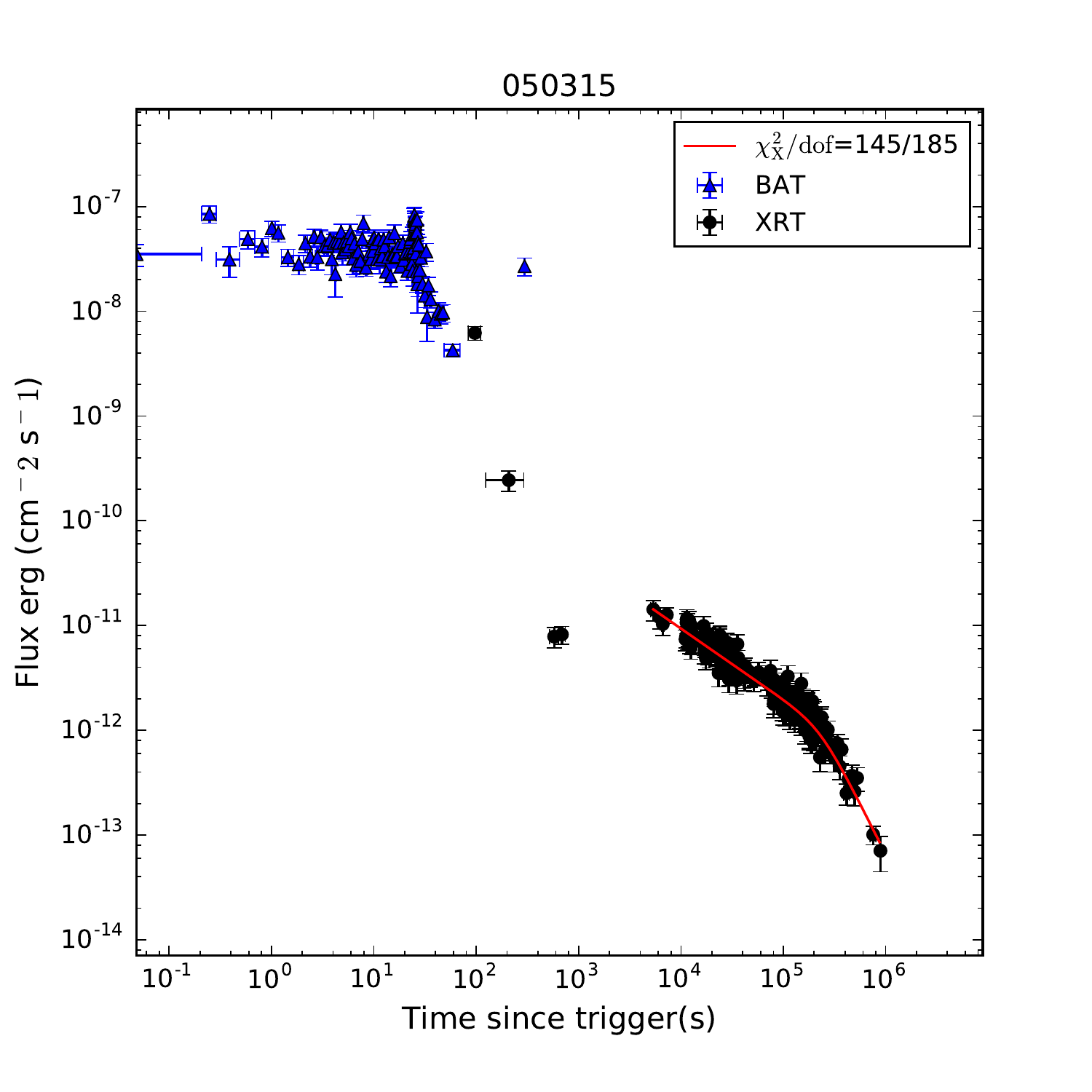}{0.28\textwidth}{}
          \fig{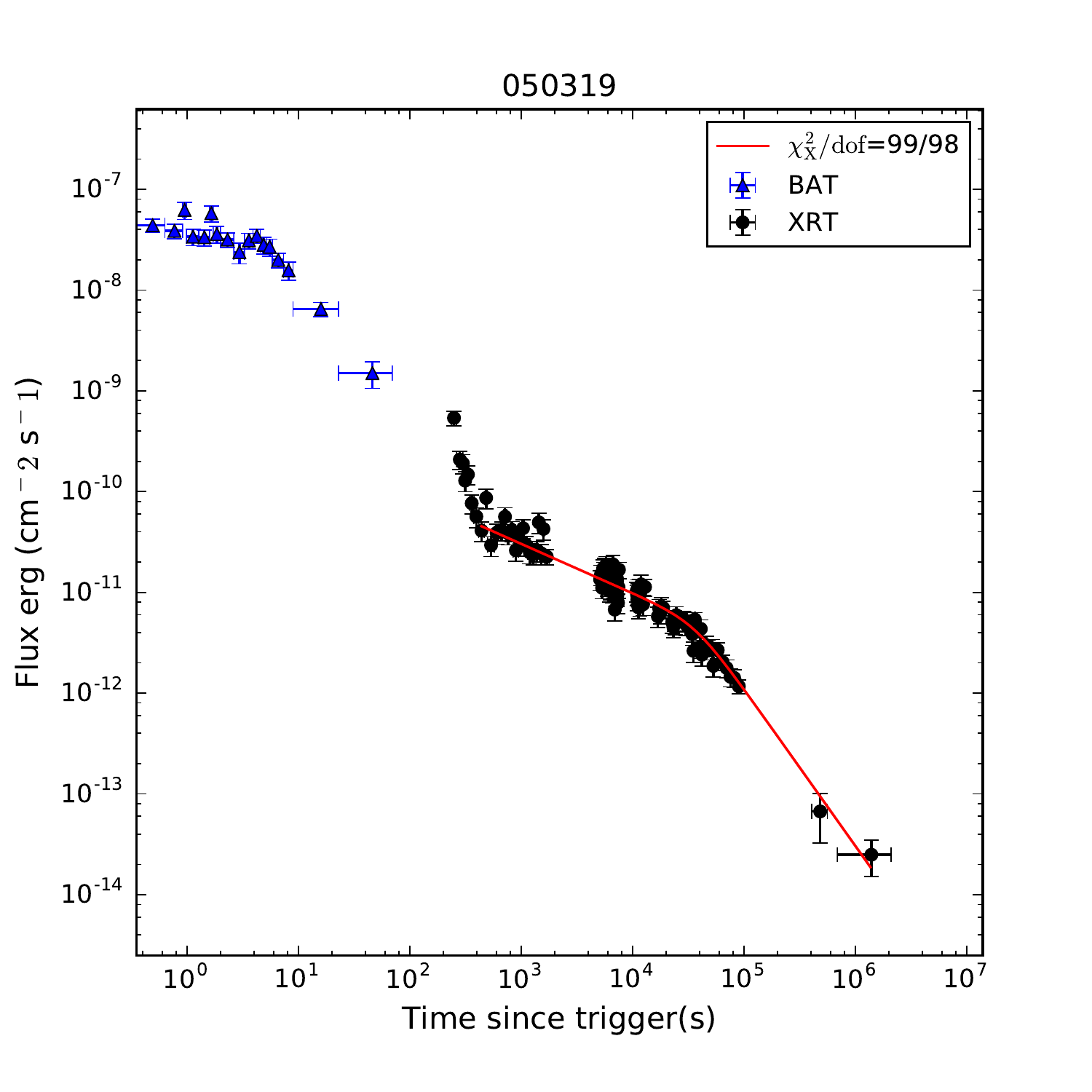}{0.28\textwidth}{}
          \fig{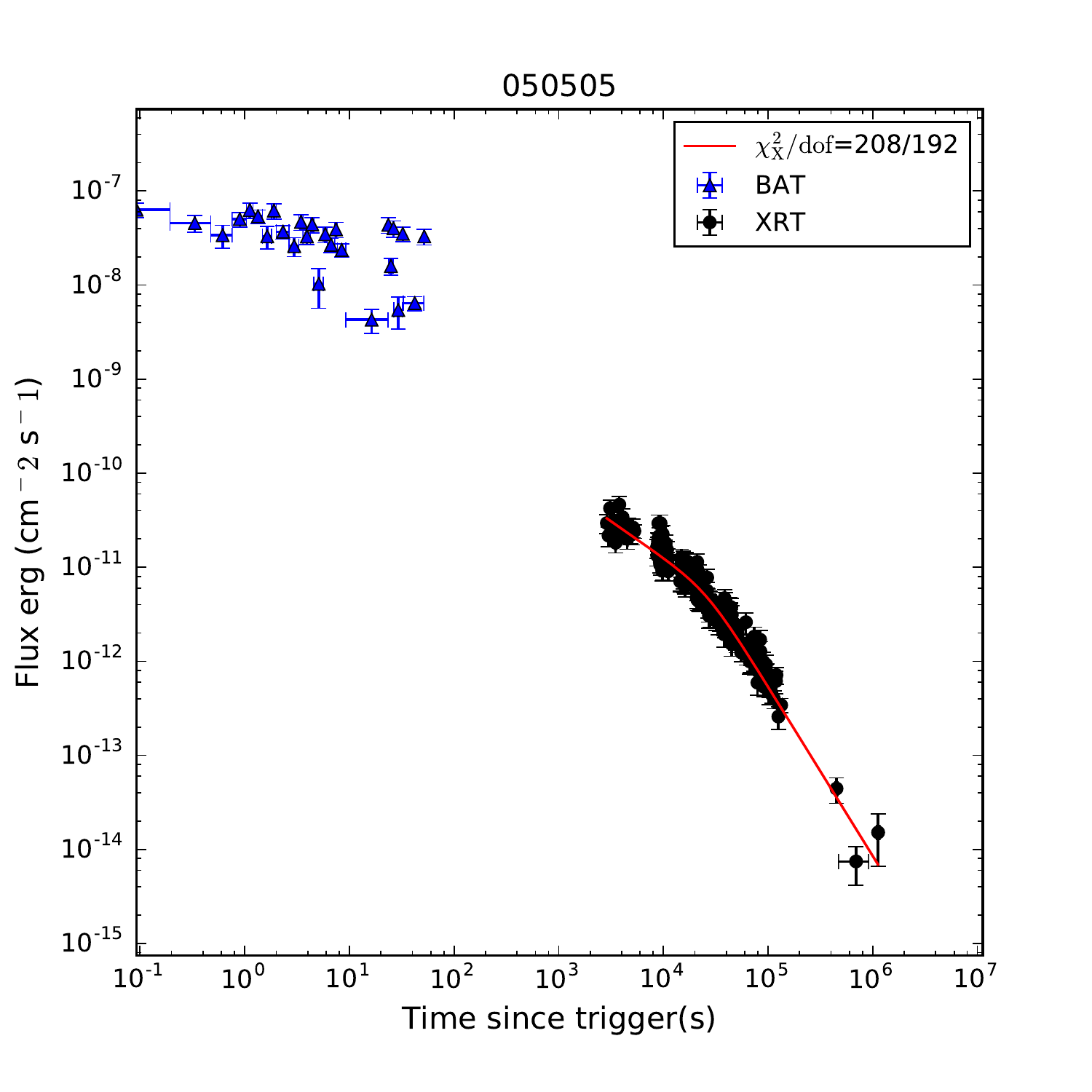}{0.28\textwidth}{}
          }
\gridline{\fig{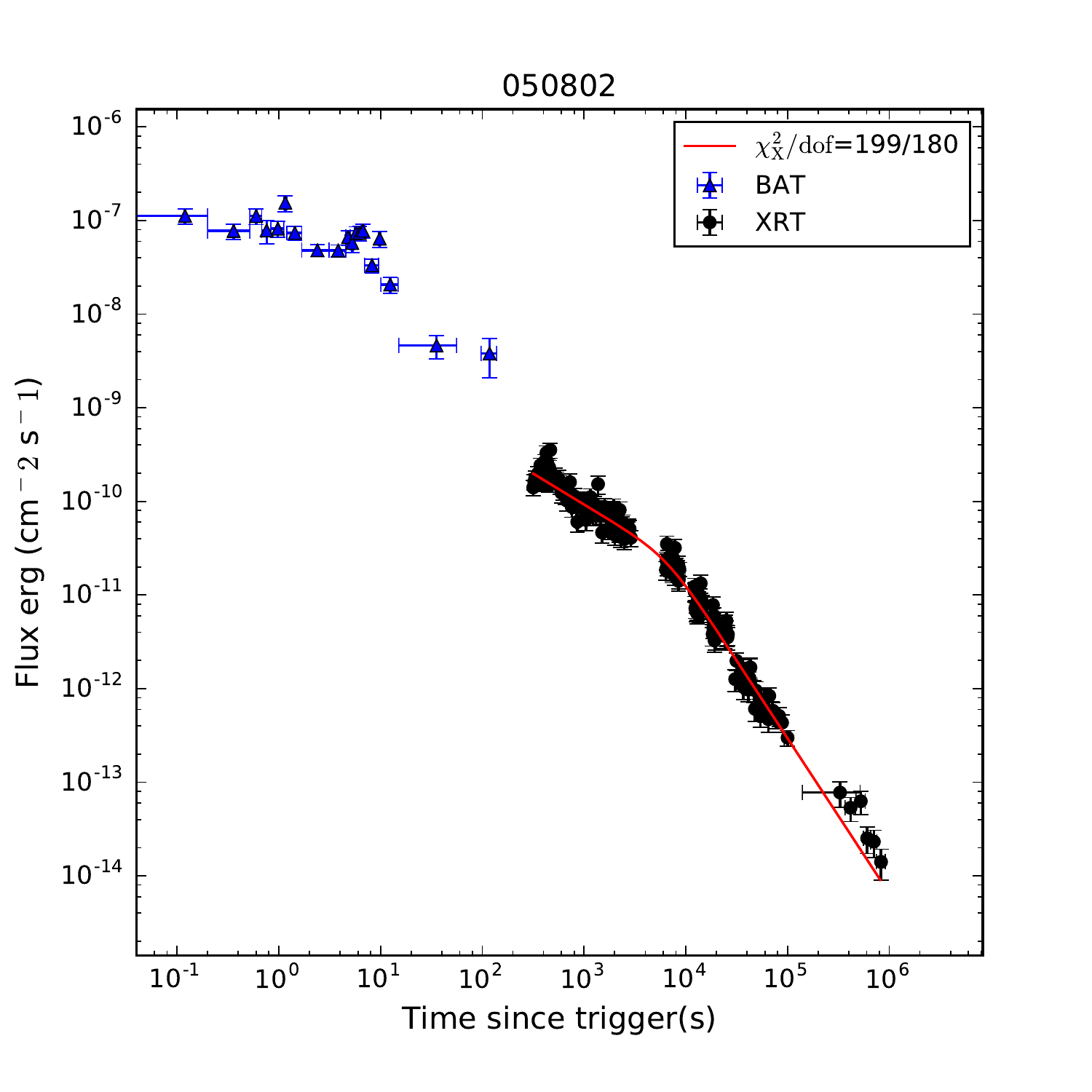}{0.28\textwidth}{}
          \fig{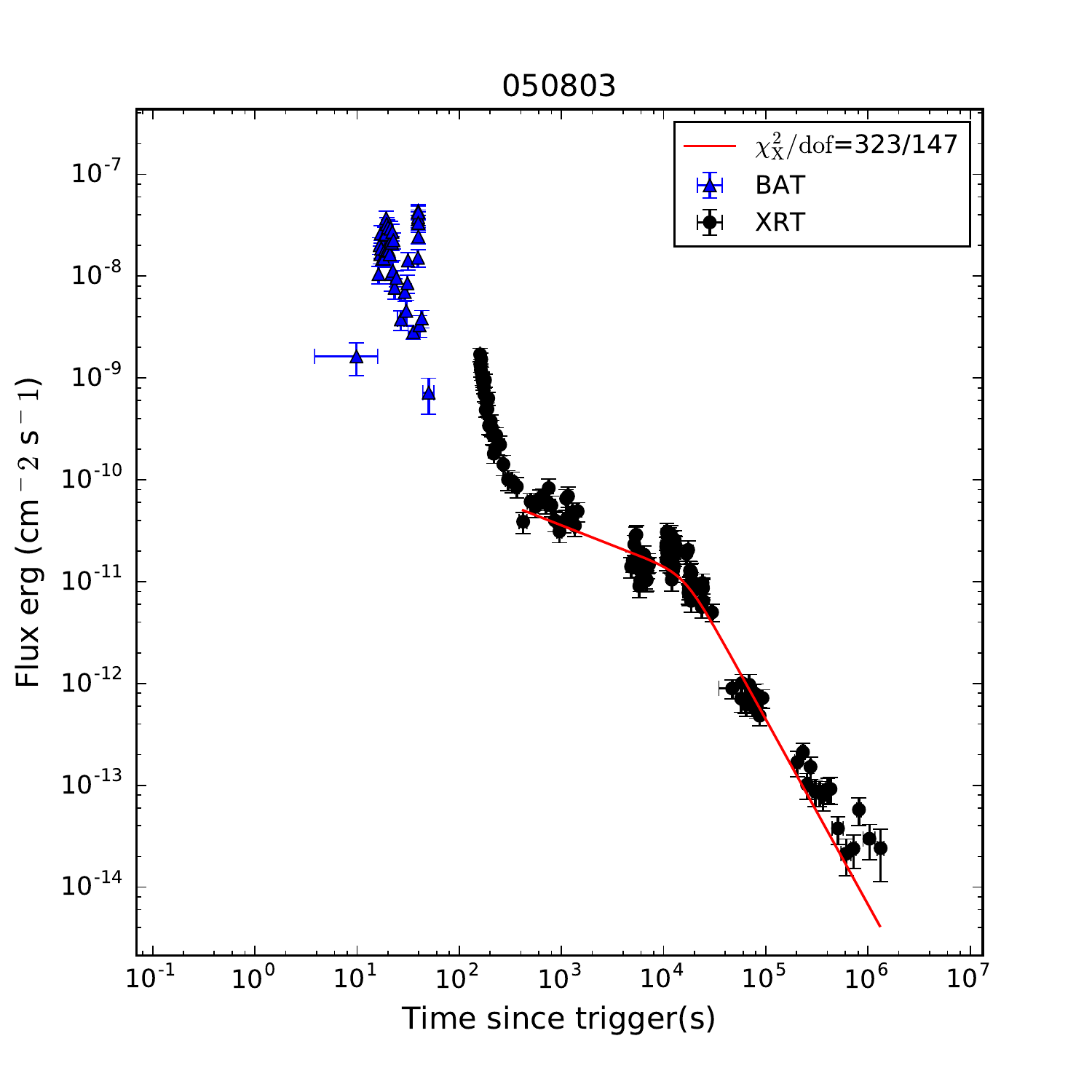}{0.28\textwidth}{}
          \fig{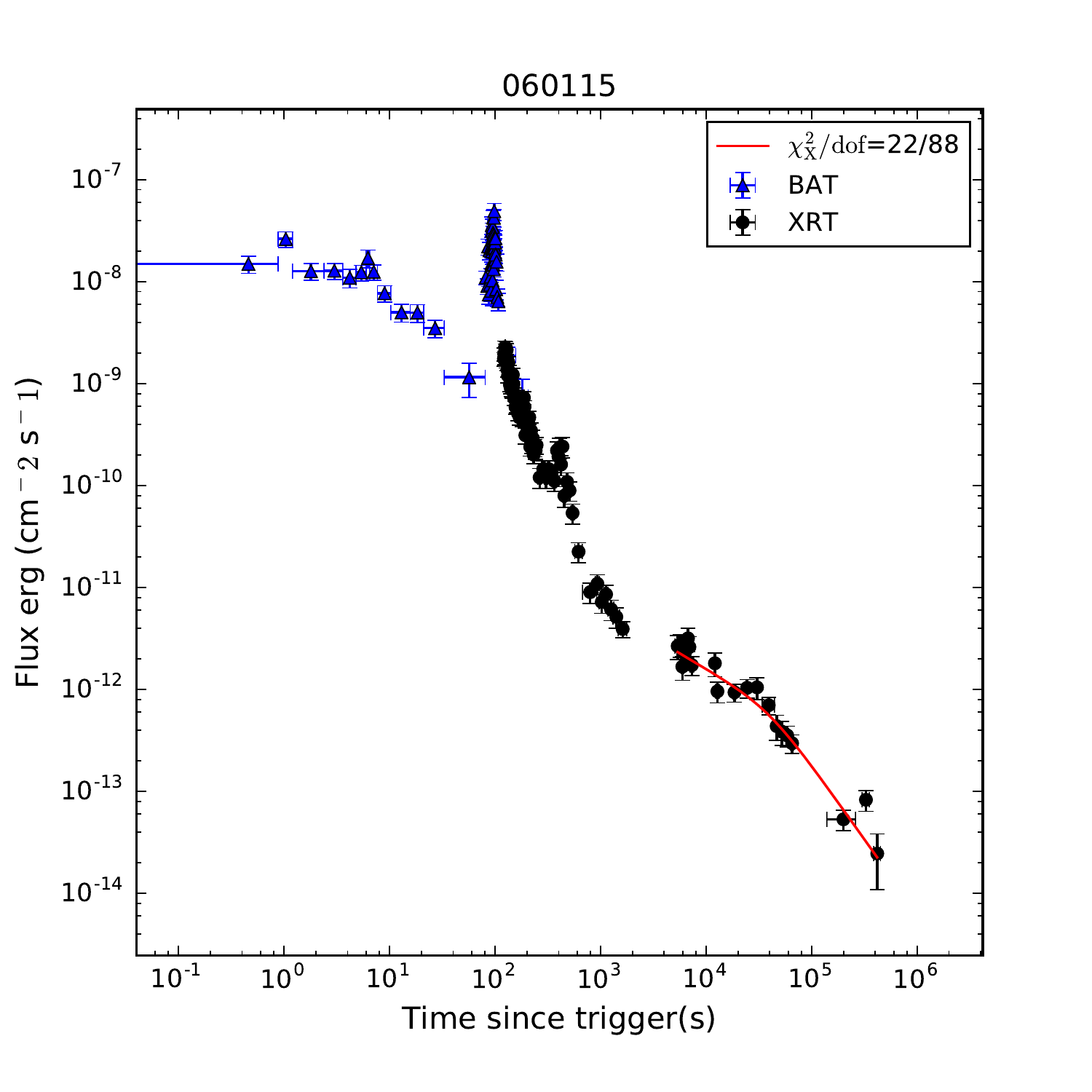}{0.28\textwidth}{}
          }
\gridline{\fig{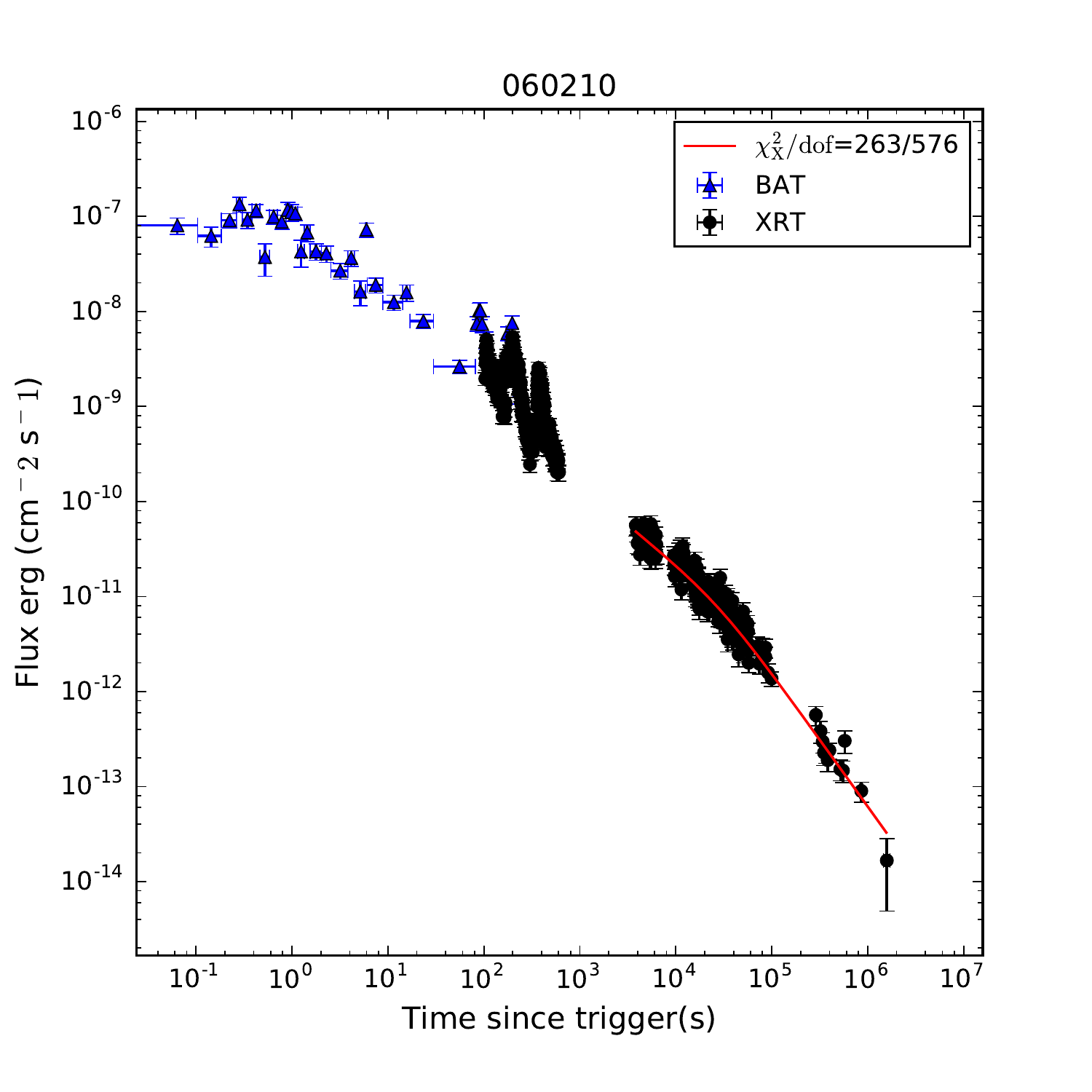}{0.28\textwidth}{}
          \fig{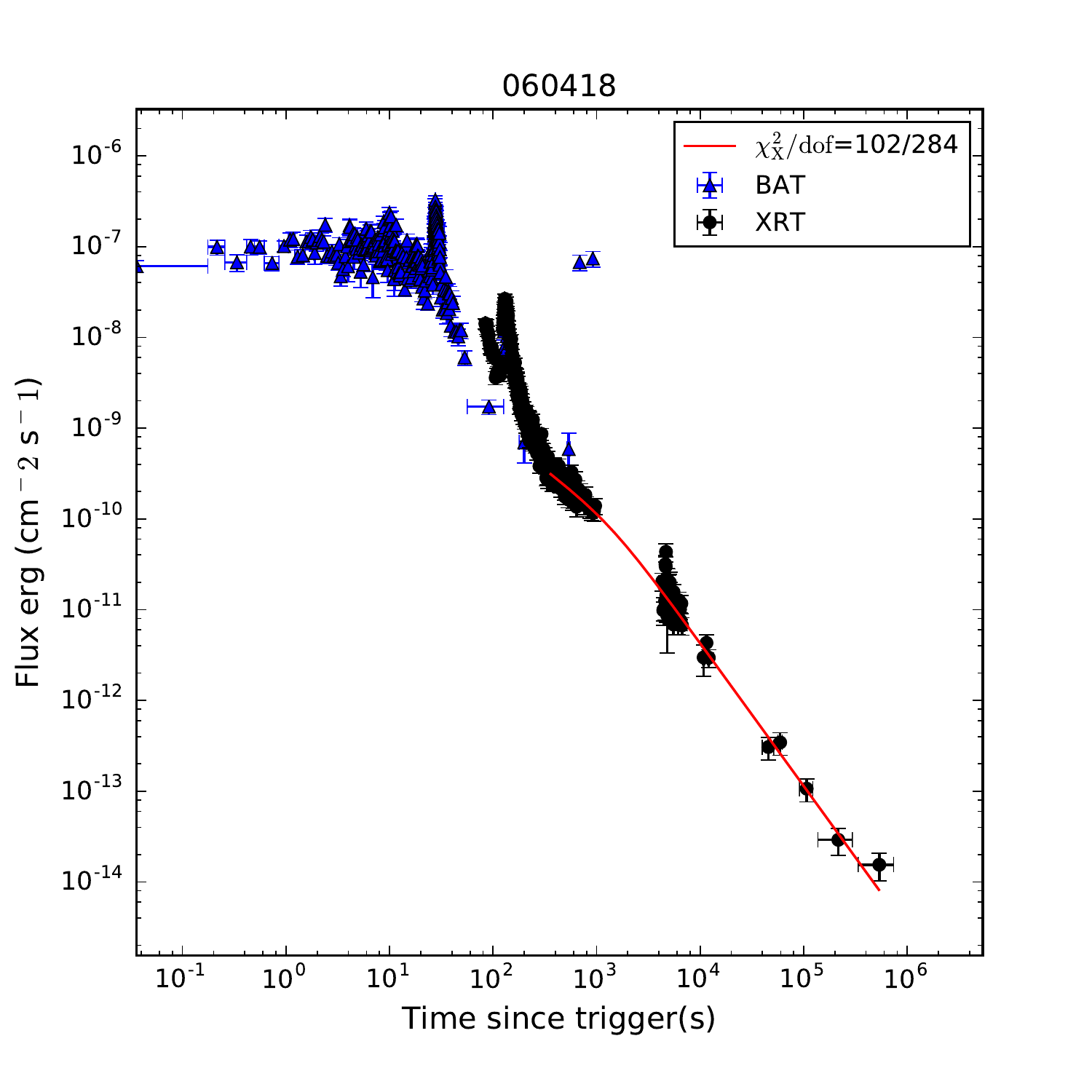}{0.28\textwidth}{}
          \fig{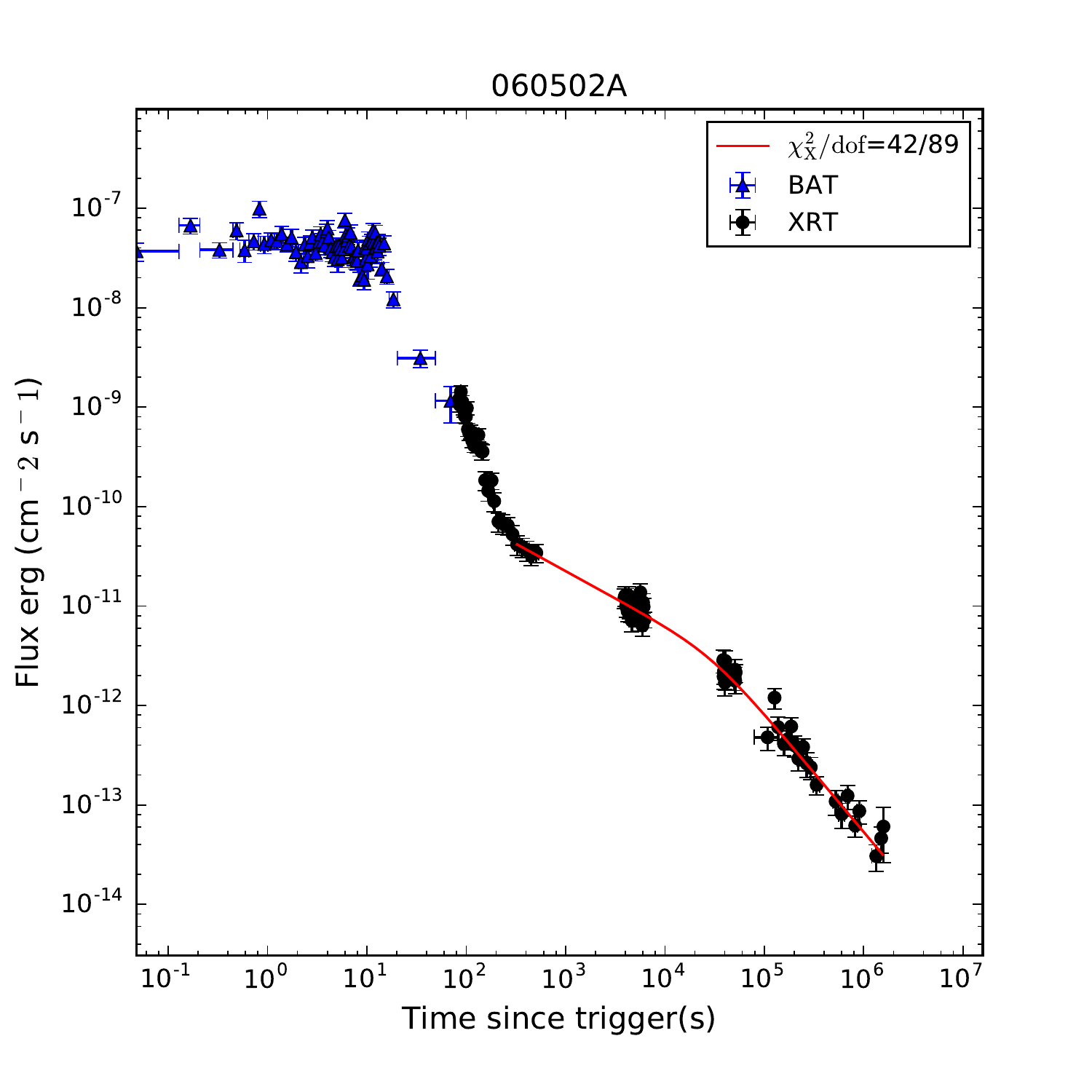}{0.28\textwidth}{}
          }
\gridline{\fig{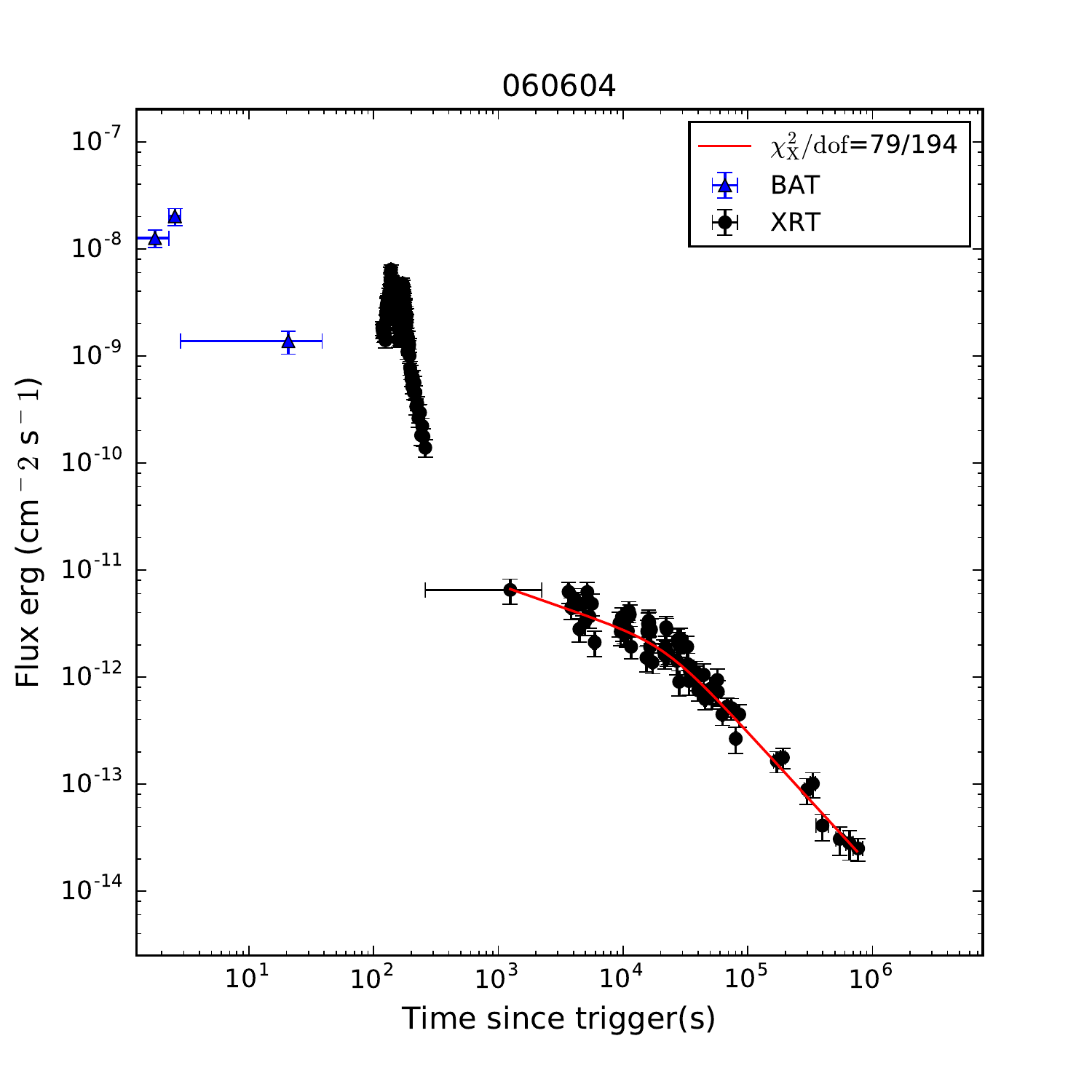}{0.28\textwidth}{}
          \fig{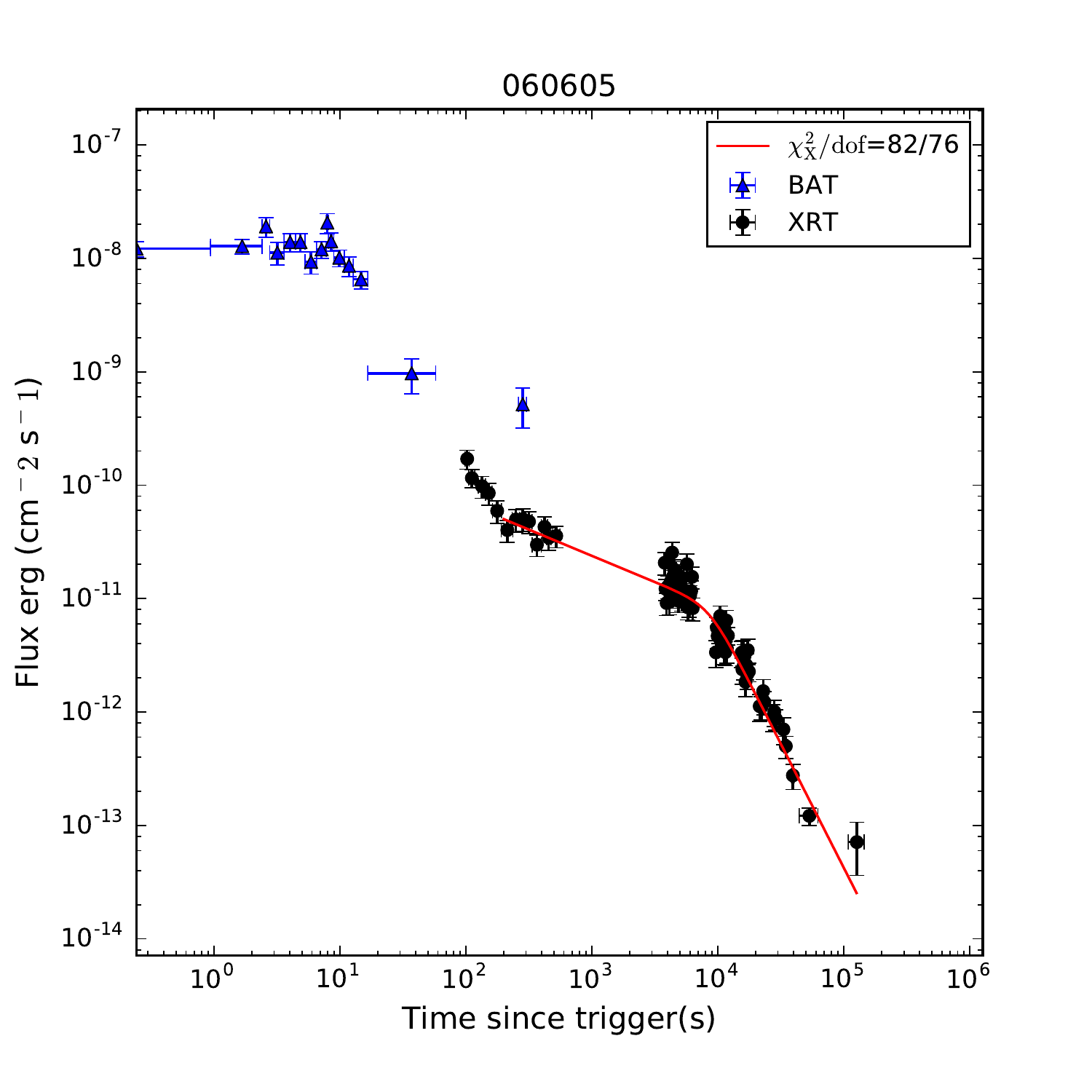}{0.28\textwidth}{}
          \fig{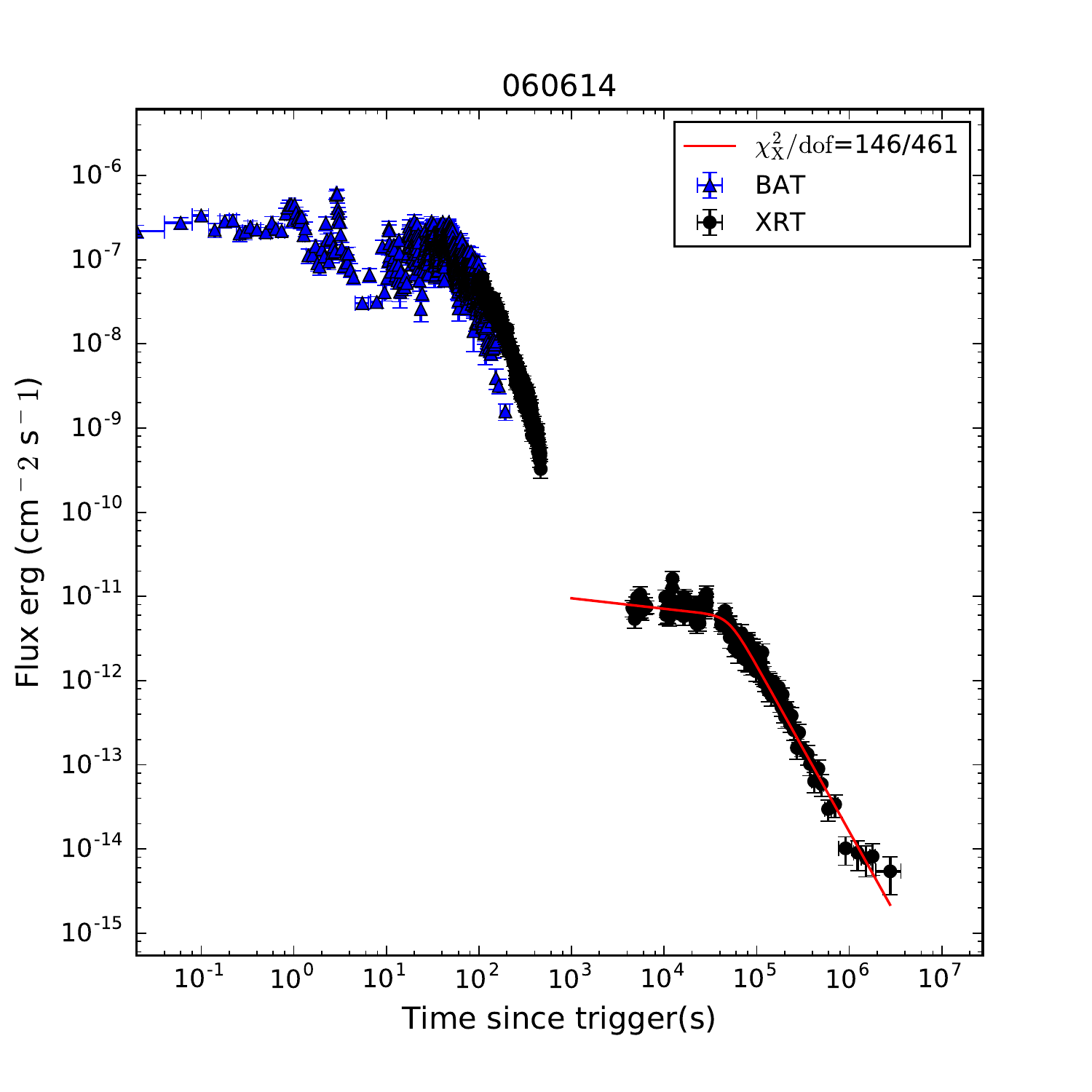}{0.28\textwidth}{}
          }
\caption{Best-fitting light curves of the X-ray plateau for the Silver sample.} \label{Silver}
\end{figure}

\begin{figure}
\gridline{\fig{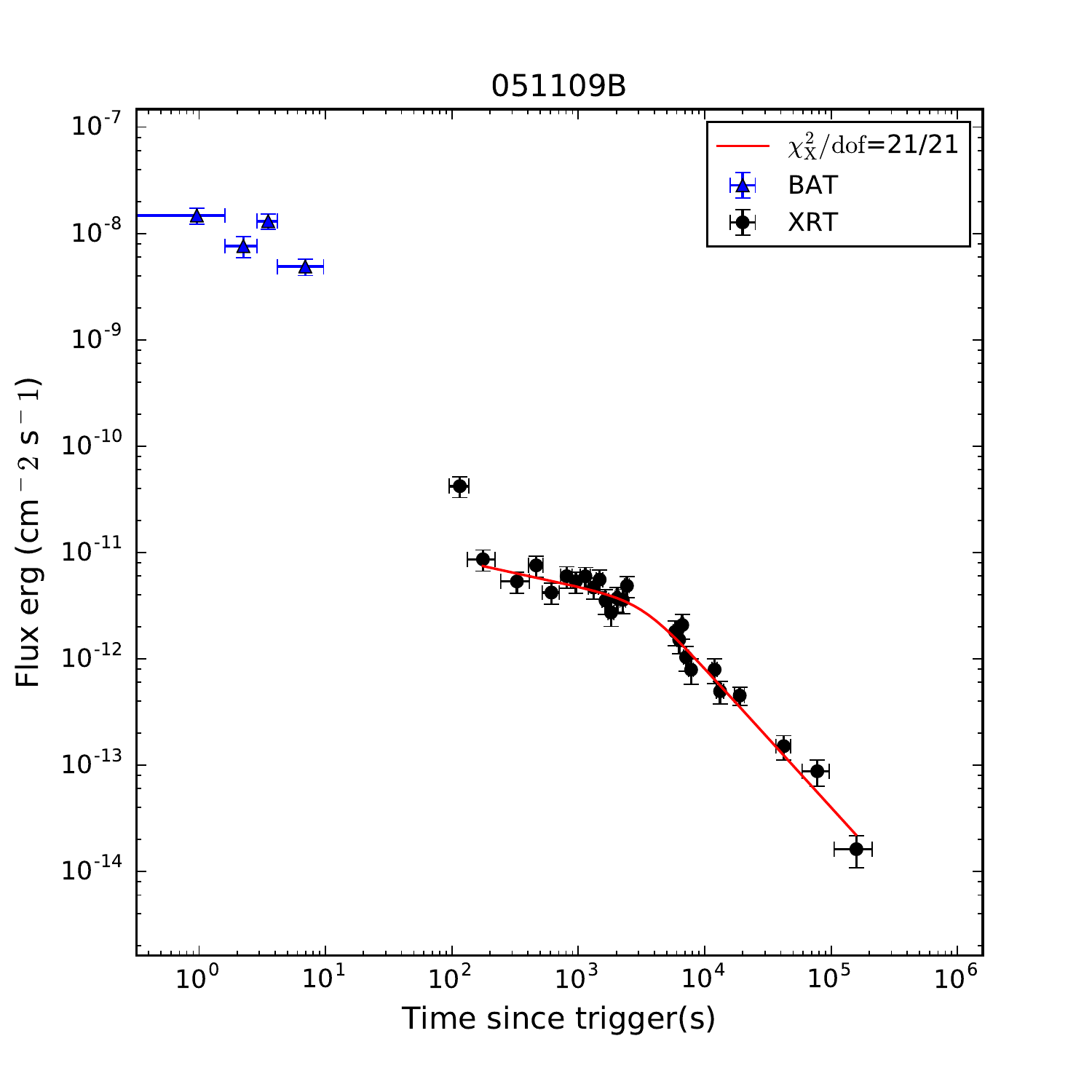}{0.28\textwidth}{}
          \fig{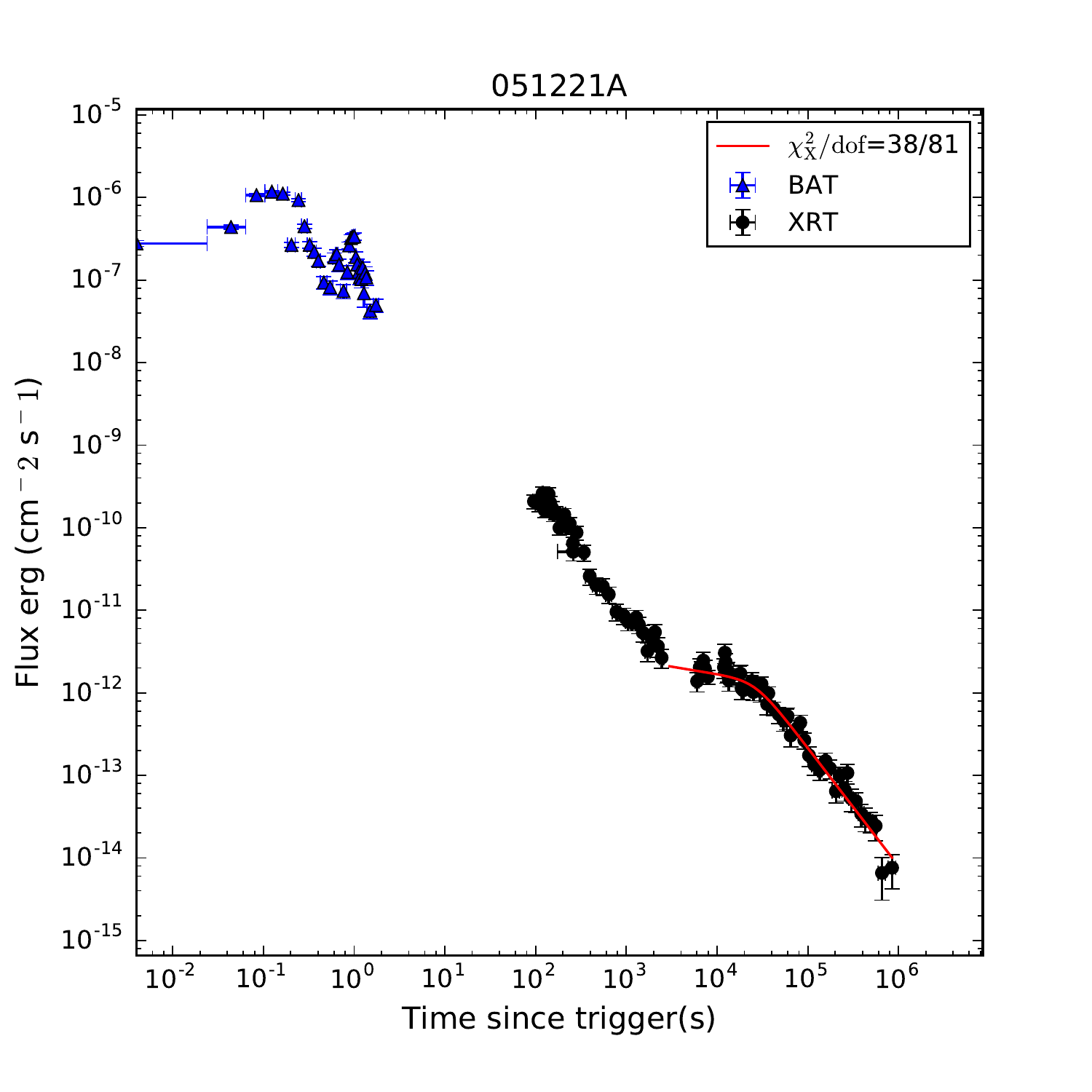}{0.28\textwidth}{}
          \fig{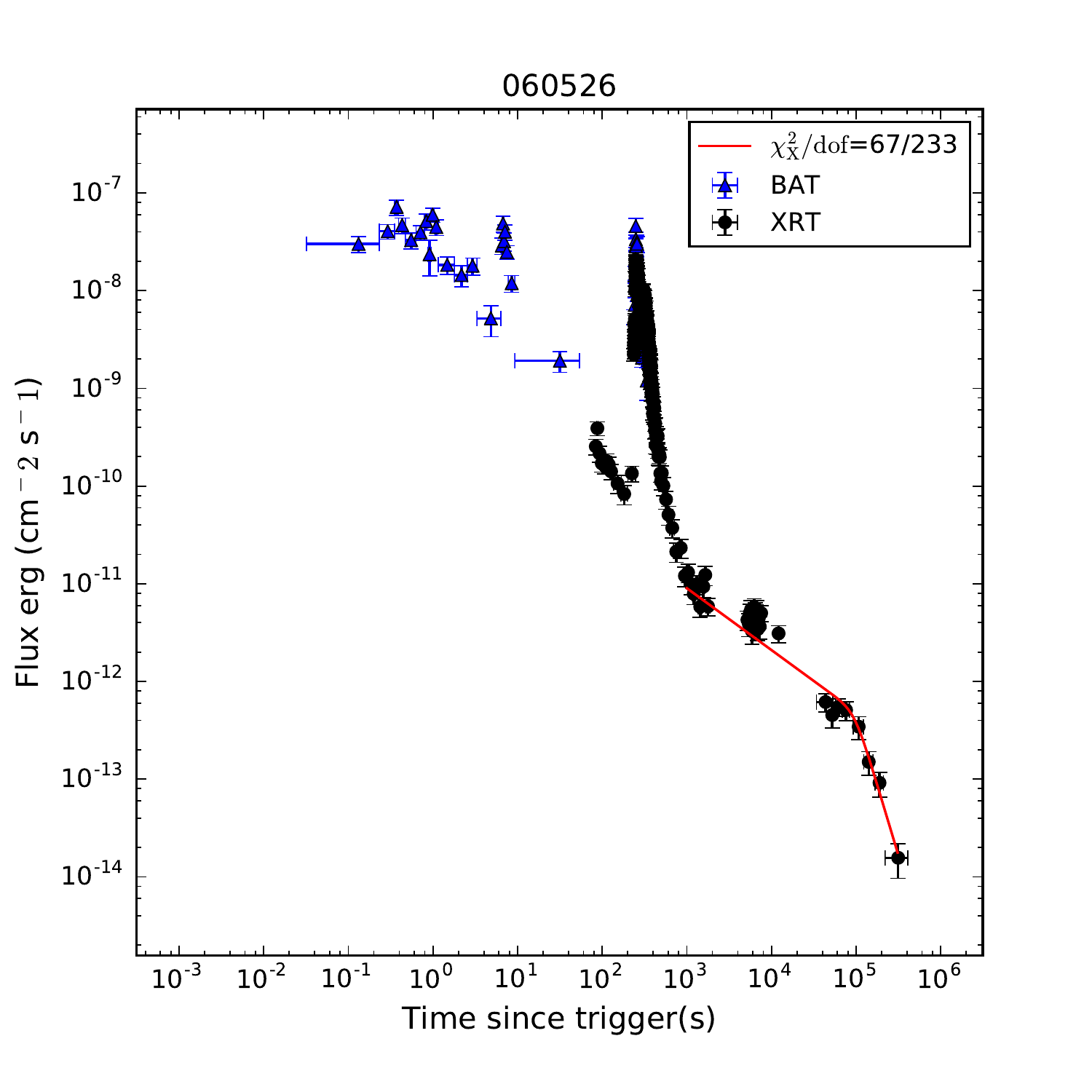}{0.28\textwidth}{}
          }
\gridline{\fig{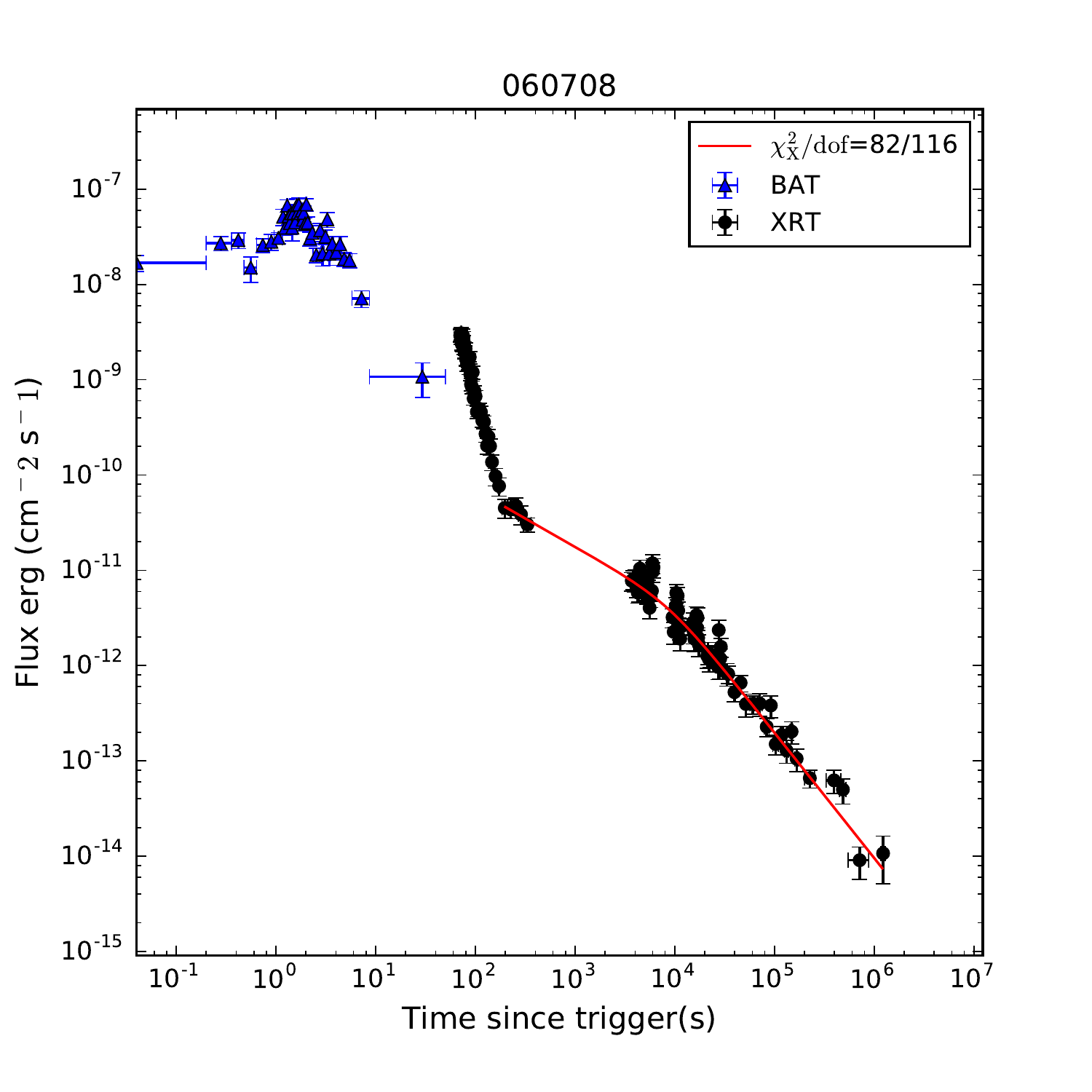}{0.28\textwidth}{}
          \fig{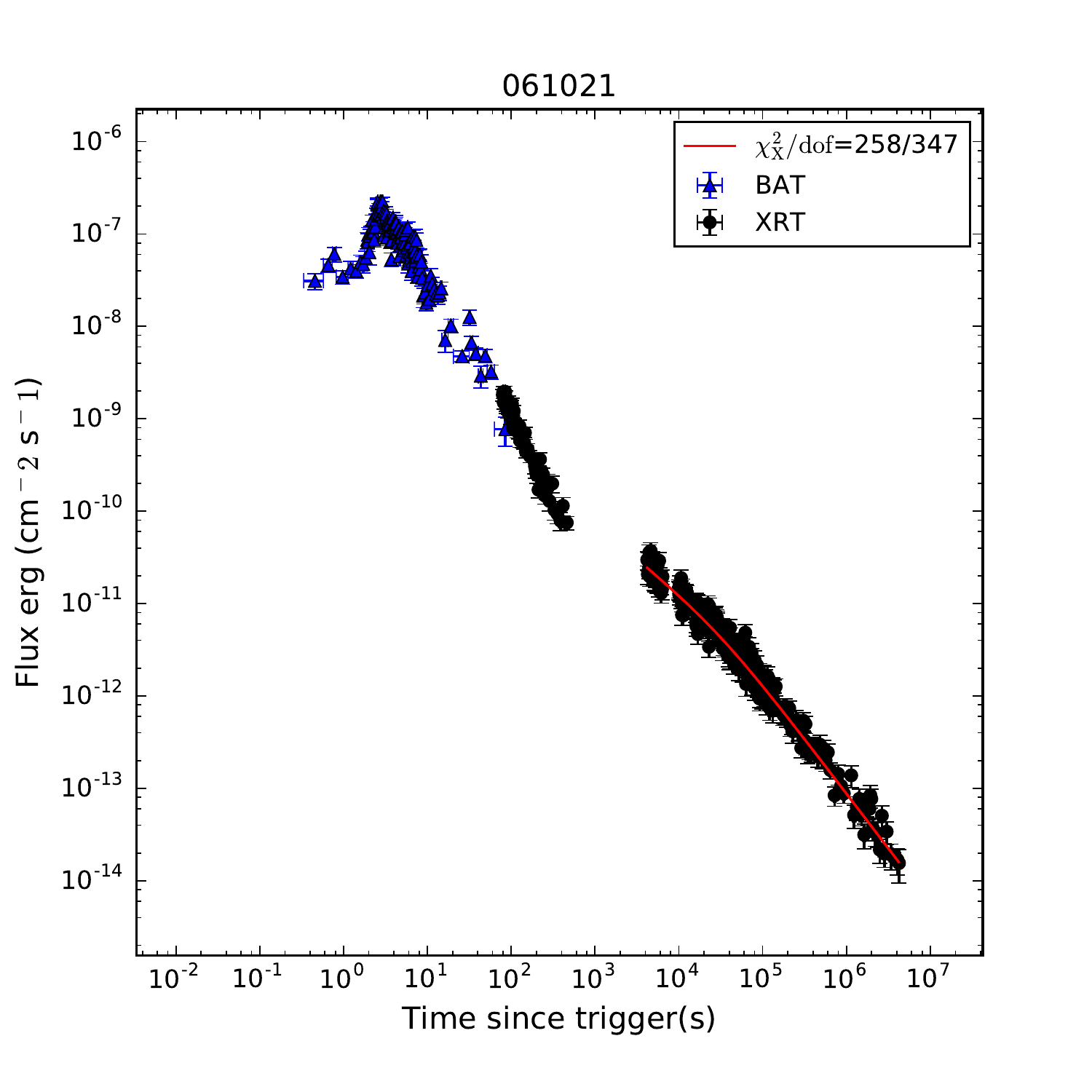}{0.28\textwidth}{}
          \fig{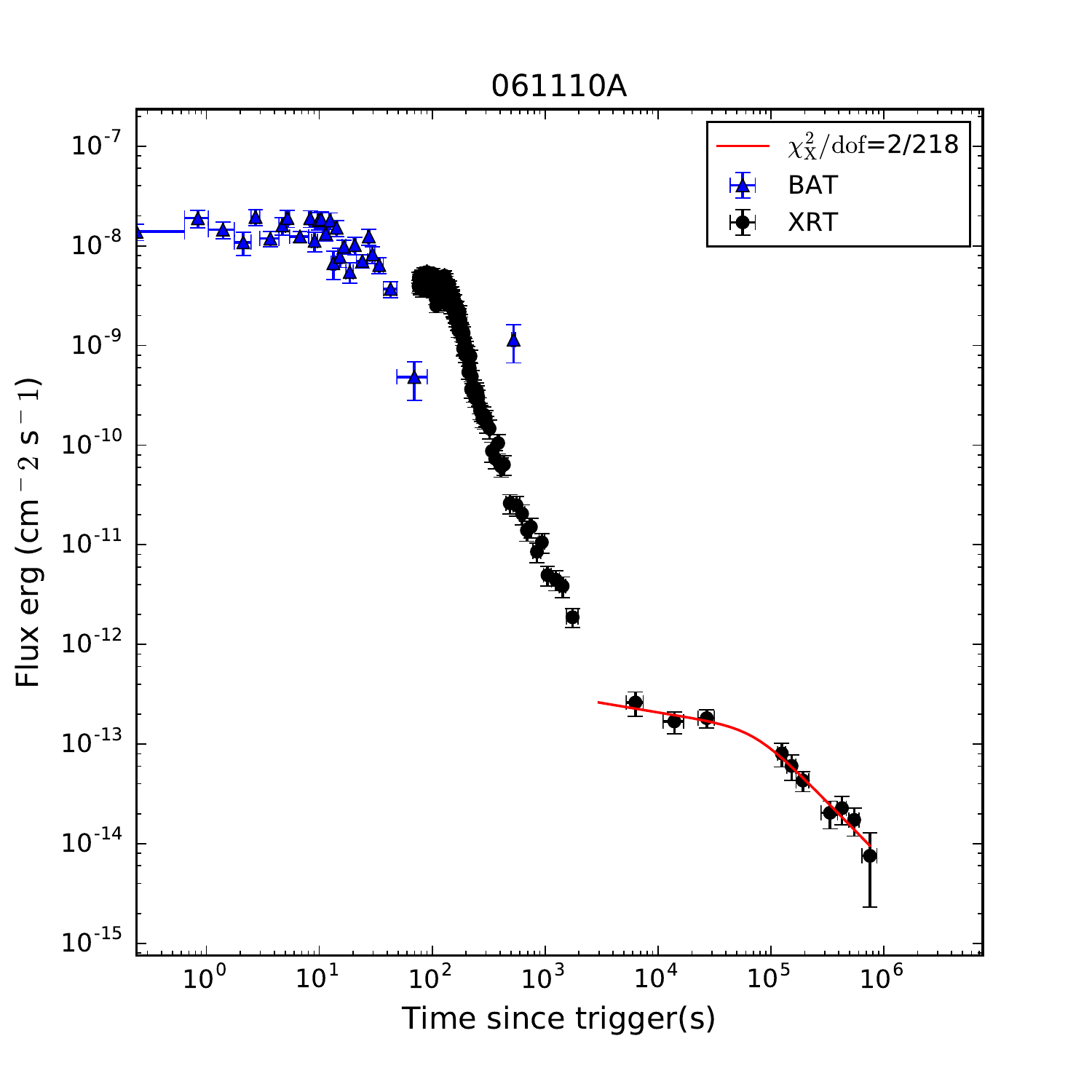}{0.28\textwidth}{}
          }
\gridline{\fig{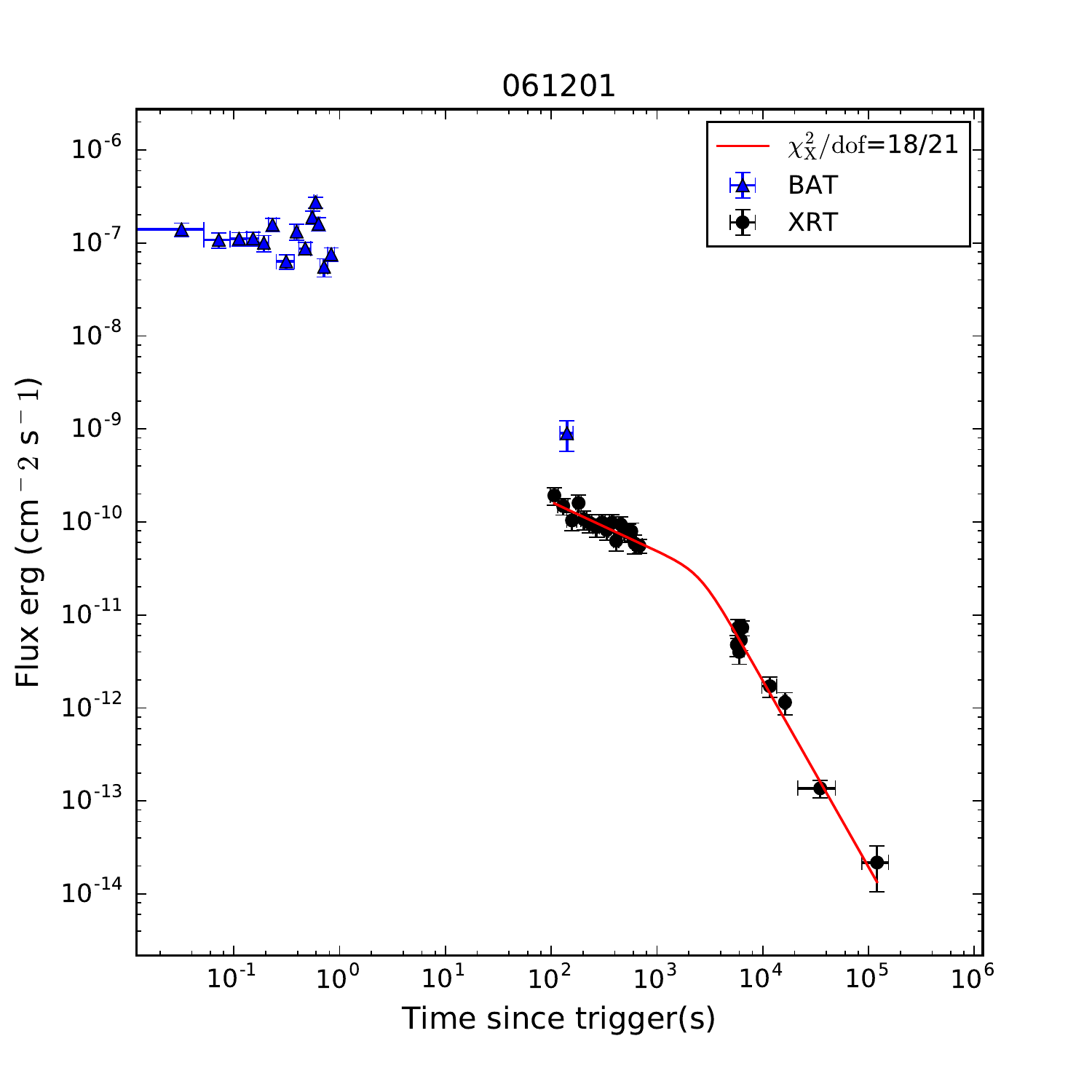}{0.28\textwidth}{}
          \fig{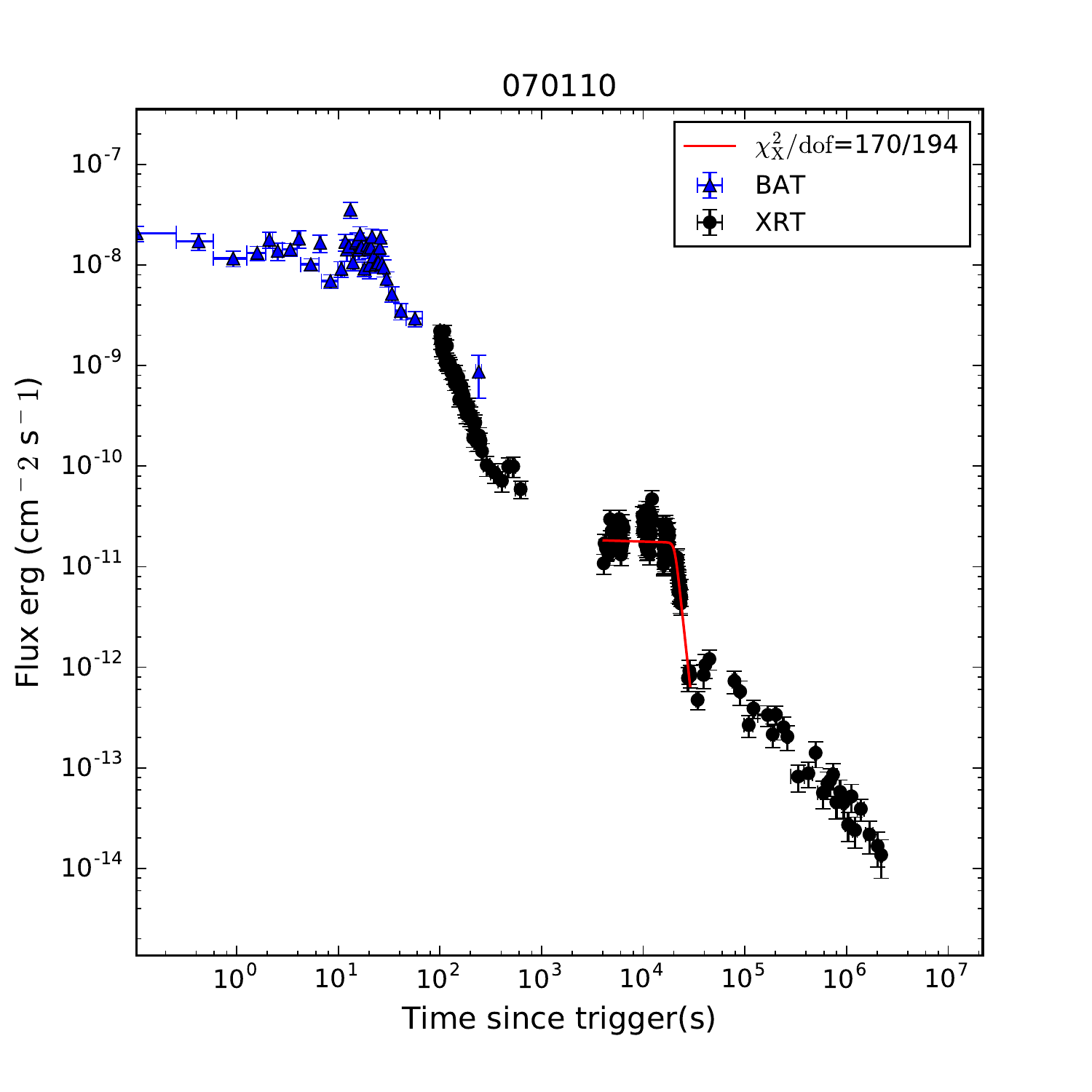}{0.28\textwidth}{}
          \fig{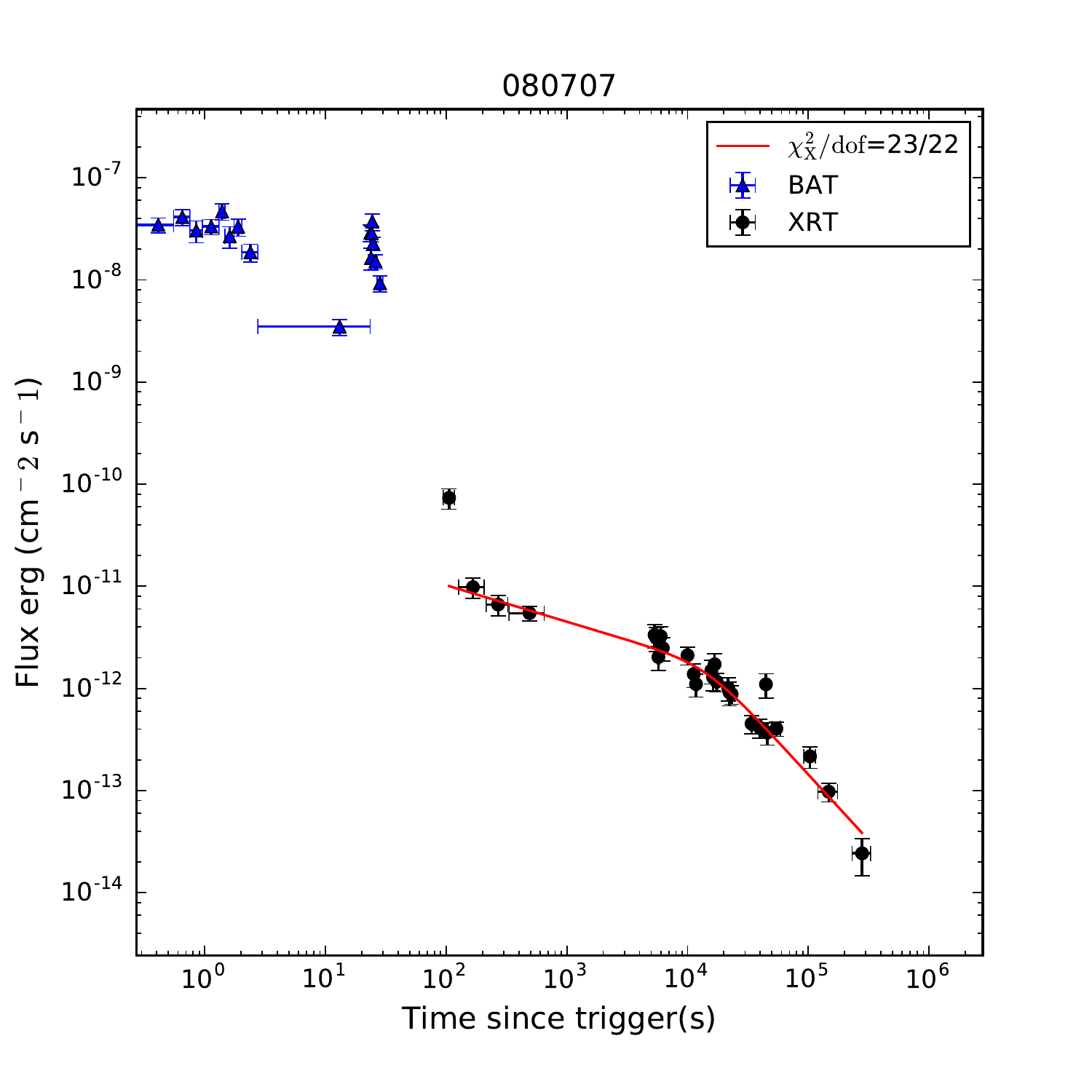}{0.28\textwidth}{}
          }
\gridline{\fig{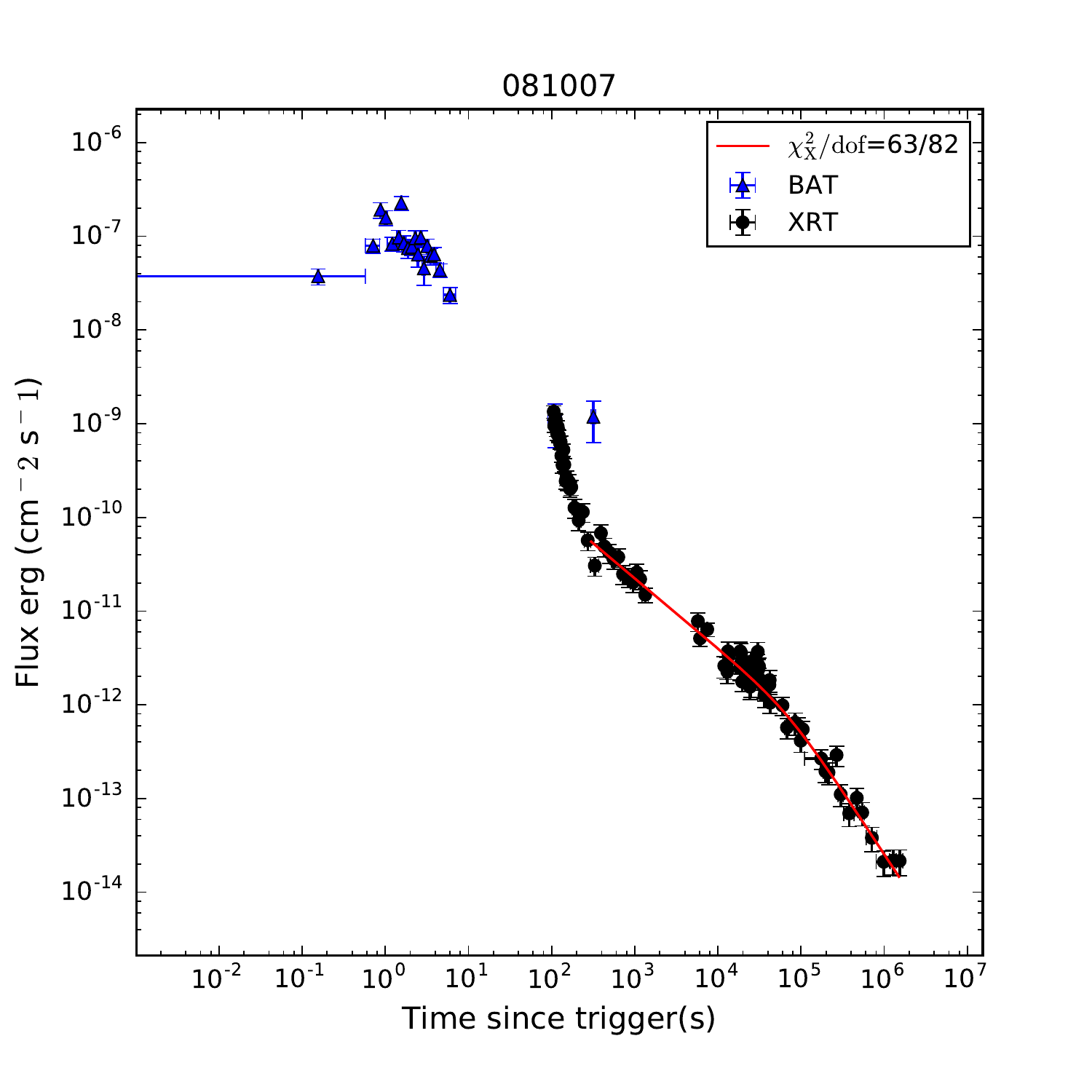}{0.28\textwidth}{}
          \fig{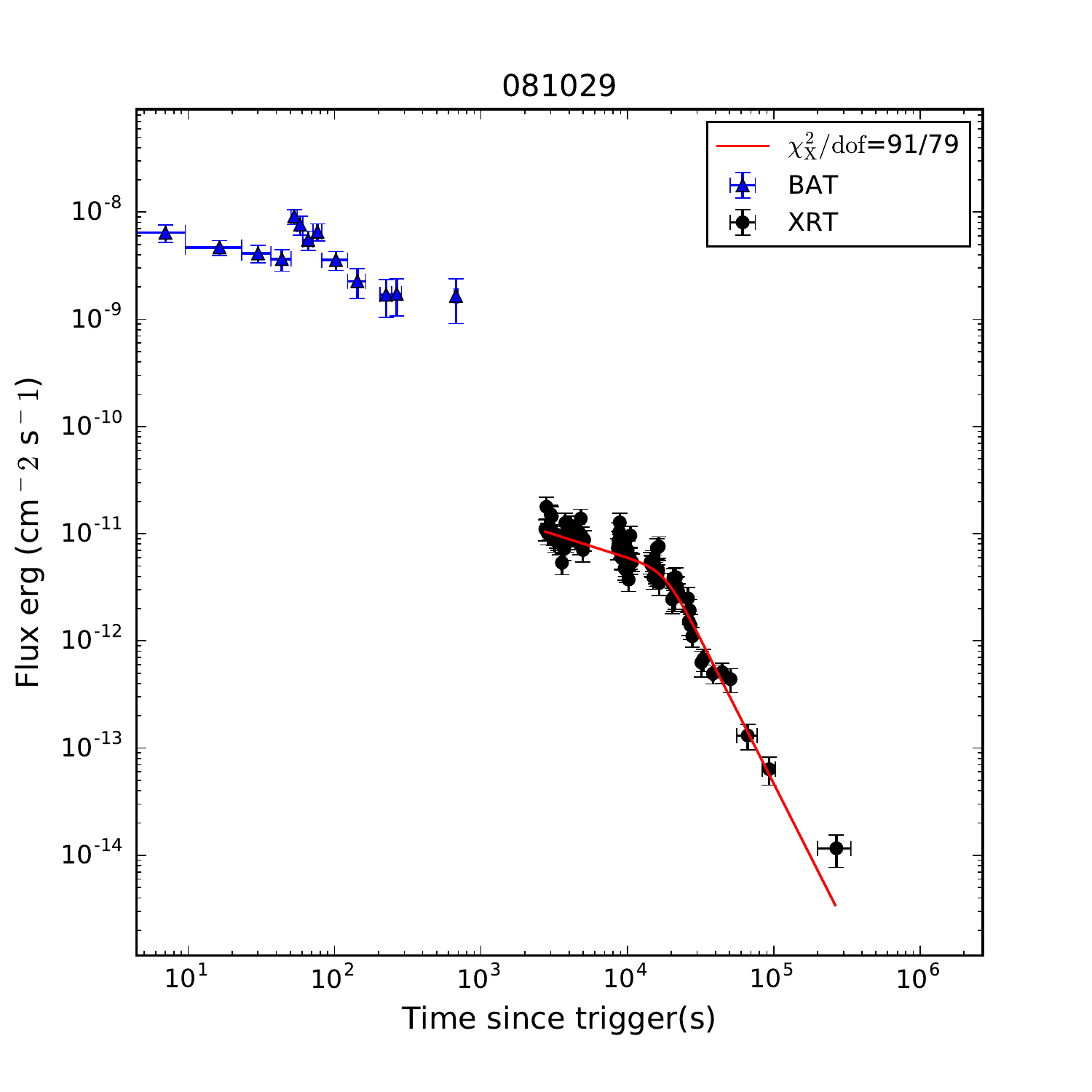}{0.28\textwidth}{}
          \fig{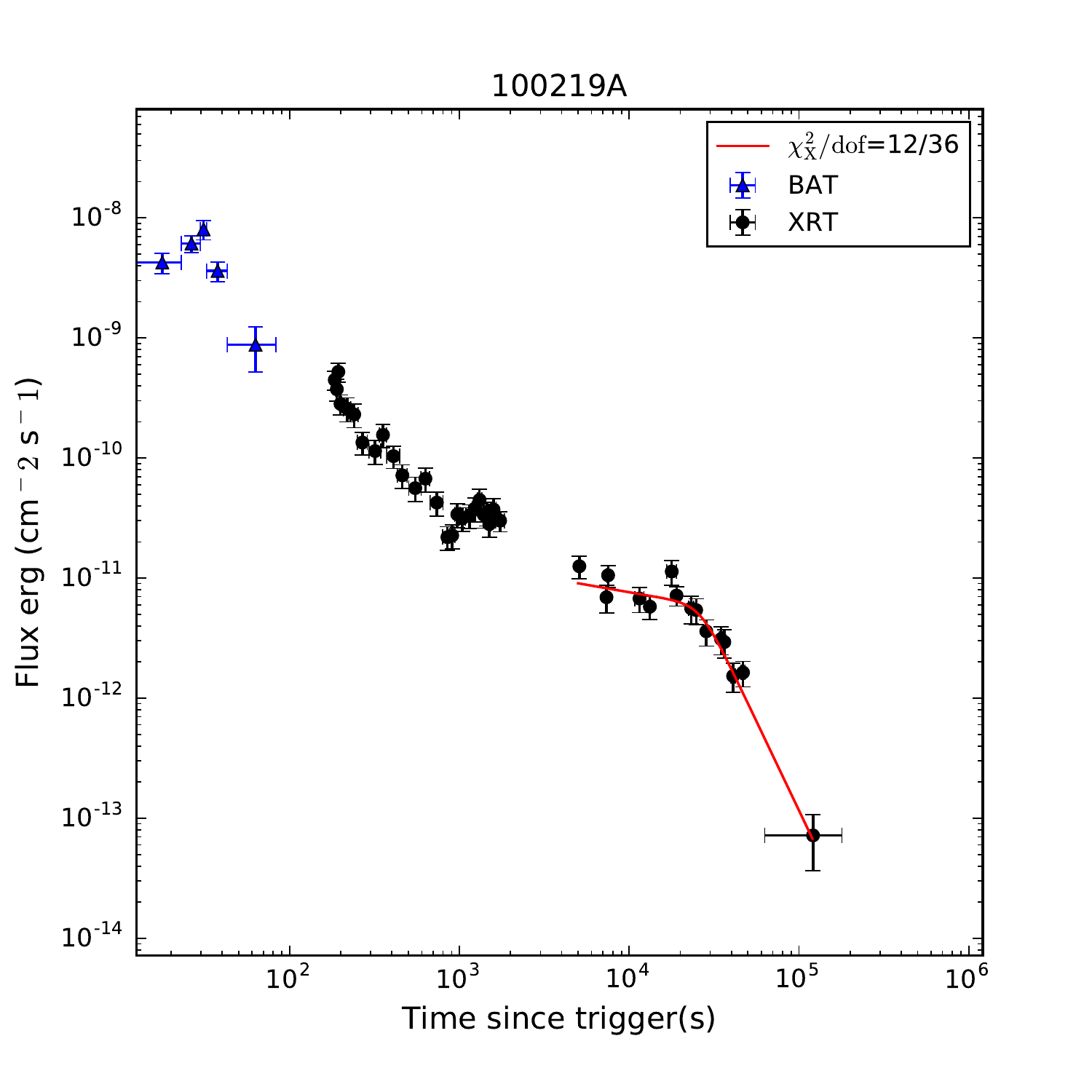}{0.28\textwidth}{}
          }
\caption{Best-fitting light curves of the X-ray plateau for the Bronze sample.} \label{Bronze}
\end{figure}

A total of 101 GRBs are selected. These are shown in Figs. \ref{Gold}, \ref{Silver} and \ref{Bronze}.

\subsection{Type of Plateau}

\begin{figure*}
\centering
\includegraphics[angle=0, scale=0.8]{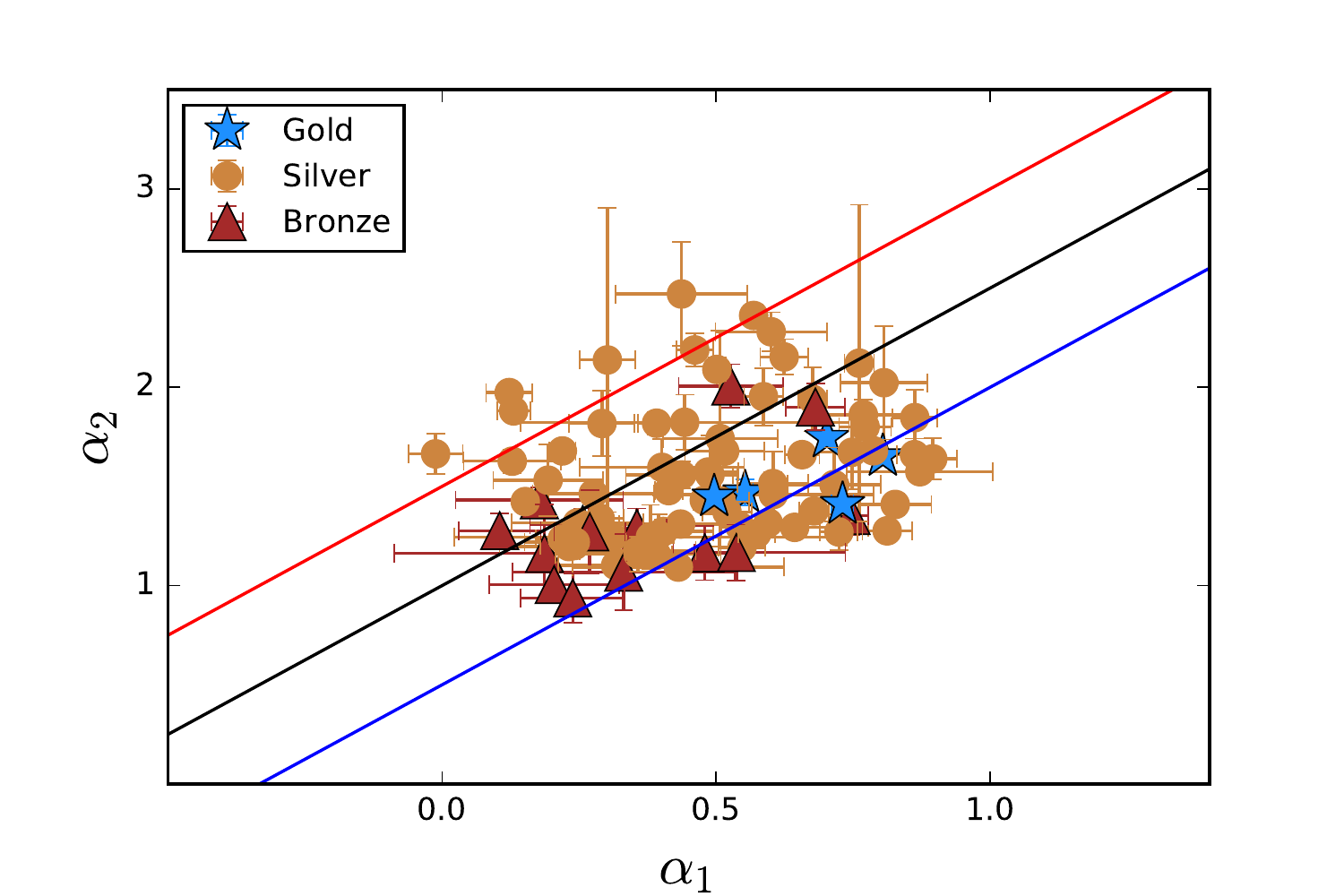}
\caption{The temporal decay indices $\alpha_{2}$ versus $\alpha_{1}$ for our samples (excluded the internal plateau bursts). The three solid lines indicate the closure relations of three specific external shock models invoking energy injection with the parameter $q$=0, as is expected in the BH/magnetar central engine models. The red, black and blue solid line are the model:
$\alpha_{2}=(3\alpha_{1}+3)/2, (\nu_{m}<\nu_{x}<\nu_{c}$, ISM);
$\alpha_{2}=(3\alpha_{1}+1)/2, (\nu_{m}<\nu_{x}<\nu_{c}$, wind);
$\alpha_{2}=(3\alpha_{1}+2)/2, (\nu_{m}<\nu_{x}<\nu_{c}$, ISM and wind).}\label{alpha1alpha2}
\end{figure*}

\begin{figure*}
\includegraphics[angle=0,scale=0.11]{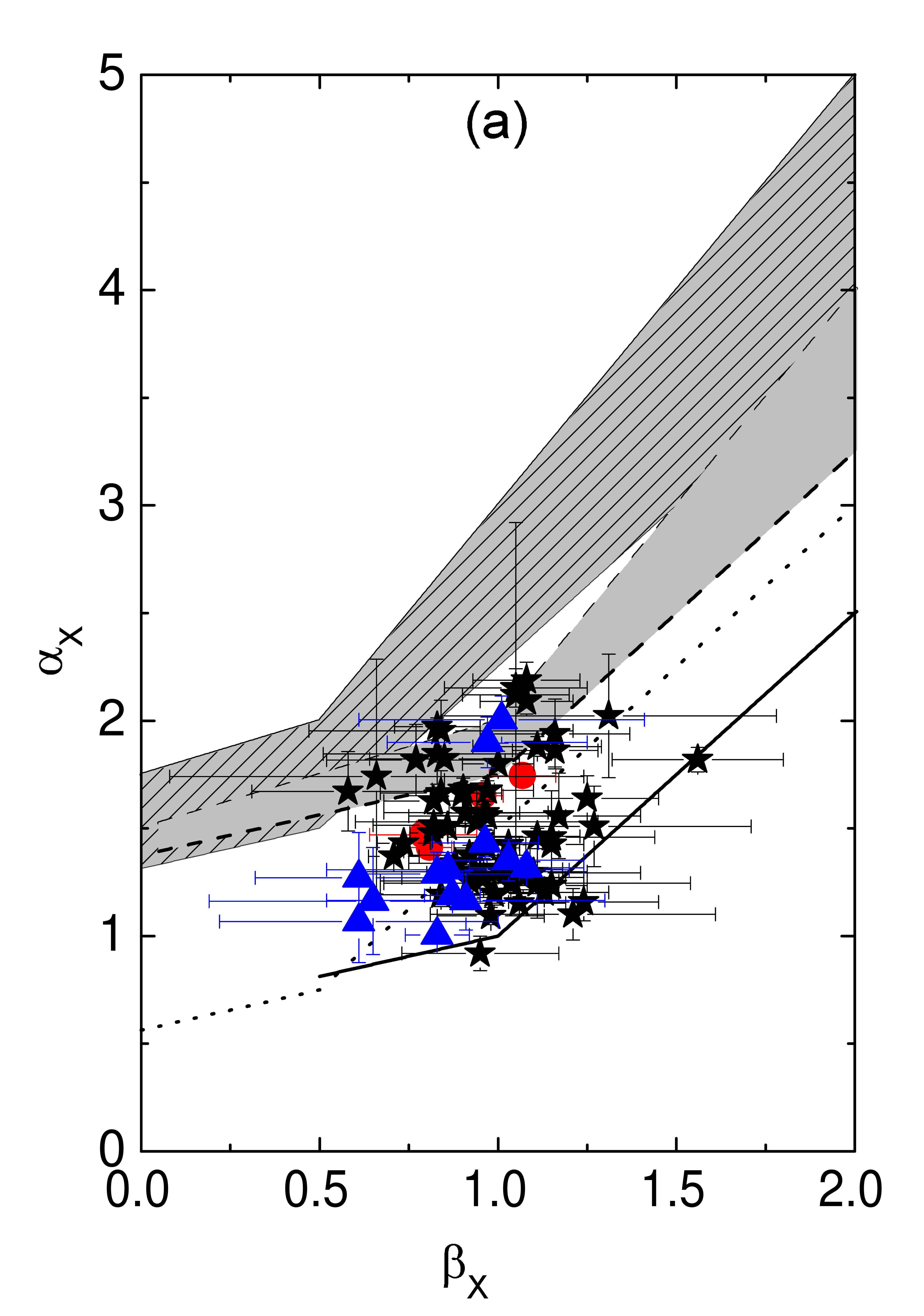}
\includegraphics[angle=0,scale=0.11]{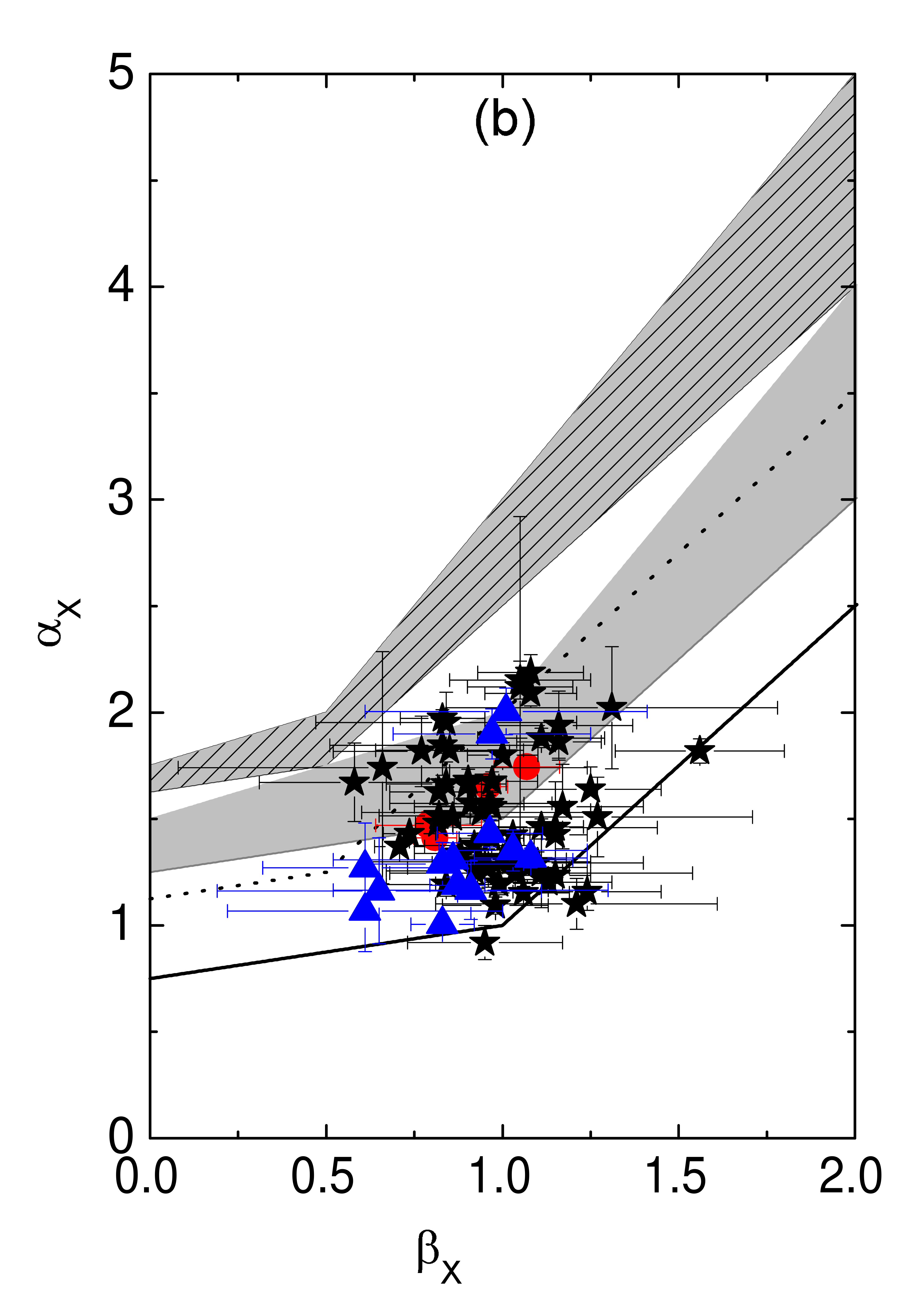}
\caption{The temporal decay index $\alpha$ against spectral index $\beta$ along with the closure relations of the external shock models for our samples. (a) The case of the ISM model: the solid line (pre-jet break) and the shaded region (post-jet break) are for the spectral regime I $(\nu_{x}> max(\nu_{m}, \nu_{c})$, while the dashed line (pre- jet break) and hatched region (post jet break) are for the spectral regime II $(\nu_{m}<\nu_{x}<\nu_{c})$; solid (black) dots and solid (red) dots are for regime I and II, respectively. (b) The case of the wind medium case. Same conventions, except that triangles (blue) denote the spectral regime II. The red-dot (Gold), black-star (Silver) and blue-triangle (Bronze) points represent the different samples, respectively.}\label{alphabeta}
\end{figure*}

In order to identify the type of the plateau (internal or external), we compare the temporal ($\alpha_{1}$ and $\alpha_{2}$) and spectral ($\Gamma_{1}$ and $\Gamma_{2}$) indices\footnote{Throughout the paper, the convention $F_{\nu}\propto \nu^{-\beta}t^{-\alpha}$ is adopted.} before and after the plateau break, and combine them through the closure relations. Since the external plateau is purely dynamical, no change in the spectral index is expected. As such, we require $\Gamma_{1}$=$\Gamma_{2}$ (within errors), which is associated with the equality between the measured decay index and that obtained by closure equations from $\Gamma_{1}$ and $\Gamma_{2}$.
The prebreak slope $\alpha_{1}$ (corresponding to {\it q}=0) is given by
\begin{eqnarray}
\label{n} \alpha_{1}= \left\{ \begin{array}{ll}
\frac{2\alpha_{2}-3}{3}, & {\rm \nu_{m}<\nu_{}<\nu_{c} \qquad (ISM)} \\
\frac{2\alpha_{2}-1}{3}, & {\rm \nu_{m}<\nu_{}<\nu_{c} \qquad (Wind)} \\
\frac{2\alpha_{2}-2}{3},& {\rm \nu_{}>\nu_{c} \qquad (ISM/Wind)}
\end{array} \right.
\label{eq:alpha1alpha2}
\end{eqnarray}
while the postbreak slope $\alpha_{2}$ (corresponding to $q=1$; \citealt{2001ApJ...552L..35Z}), according to Zhang et al. (2006), is
\begin{eqnarray}
\label{n} \alpha_{2}= \left\{ \begin{array}{ll}
\frac{3\beta_{2}}{2}=\frac{3(p-1)}{4}, & {\rm \nu_{m}<\nu_{}<\nu_{c} \quad (ISM)} \\
\frac{3\beta_{2}+1}{2}=\frac{3p-1}{4}, & {\rm \nu_{m}<\nu_{}<\nu_{c} \quad (Wind)} \\
\frac{3\beta_{2}-1}{2}=\frac{3p-2}{4}, & {\rm \nu_{}>\nu_{c} \qquad (ISM/Wind)}
\end{array} \right.
\label{eq:closure}
\end{eqnarray}
where $\beta_{2}$ is the spectral index during the normal decay phase, and $p$ is the electron's spectral distribution index. These criteria gave 89 bursts in this sample.
Fig.\ref{alpha1alpha2} shows all GRBs' external plateaus in the $\alpha_{1}-\alpha_{2}$ plane, with three theoretically favored lines (Equation \ref{eq:alpha1alpha2}) plotted.
Fig.\ref{alphabeta} shows the $\beta-\alpha$ plane based on either the ISM or the wind model.
The GRBs satisfying the "closure relations" are identified as our sample GRBs with colored data points.

The internal plateau \citep{2007ApJ...665..599T, 2010MNRAS.402..705L,2010MNRAS.409..531R, 2013MNRAS.430.1061R} cases are selected based on $\Gamma_{1}$, different $\Gamma_{2}$, and $\alpha_{2}>2+\beta$. This sample is made up of only 12 bursts. The bursts that cannot be placed in either of these two categories are discarded from further analysis\footnote{This is the case for GRB 050730, GRB 060526, GRB 061021, GRB 060607A, GRB 070110, GRB 081029, GRB 100219A, GRB 100302A, GRB 100902A, GRB 111209A, GRB 120521C, GRB 130408A.}.

\subsection{X-Ray Plateau and Afterglow Kinetic Energies}

We derive the X-ray isotropic energy released ($E_{\rm X,iso}$) during the plateau phase by integrating the fluence from $t_{\rm s}$  to $t_{\rm b}$,
\begin{equation}
E_{\rm X,iso}=\frac{4\pi k D_{\rm L}^{2}S_{\rm X}}{1+z},
\end{equation}
where $k=(1+z)^{\beta-1}$, $\beta$ is the spectral index, $S_{\rm X}$ is the X-ray fluence\footnote{To obtain more precise X-ray fluence, we corrected them from absorbed to unabsorbed by multiplying the ratio of the observed and unabsorbed flux, which are collected from the \emph{Swift} archive.} integrated over the plateau phase, $D_{L}$ is the luminosity distance, and $z$ is the redshift.

Another important parameter is the isotropic kinetic energy ($E_{\rm K,iso}$), which is measured from the afterglow flux during the normal decay.
Here, $E_{\rm K,iso}$ is calculated following the method discussed in \cite{2007ApJ...655..989Z}.
Since $E_{\rm K,iso}$ depends on the afterglow models (we note it here again briefly for completeness), it is important to identify the afterglow spectral regimes. We first use the observed quantity ($\alpha$,$\beta$) in the normal decay phase to constrain/determine the afterglow spectral regimes of each individual burst (\citealt{2015ApJ...805...13L}). We find that 18 of the 101 GRBs belong to spectral regime I ($\nu_{\rm x}>max(\nu_{m},\nu_{c})$), and 83 out of the 101 GRBs belong to spectral regime II ($\nu_{m}<\nu_{\rm x}<\nu_{c}$).
Here, $\nu_{x}$ is the frequency of the X-ray band. In addition, 42 out of the 83 bursts satisfy the ISM model, and 31 out of 83 satisfy the wind model.

We consider the relations obtained by \cite{2007ApJ...655..989Z} to compute the kinetic energy of the external plateau sample using the normal decay segment. Since the normal decay phase is always observed at late times ($t\sim 10^{4}$ s) in our samples, we consider the slow cooling scenario ($\nu_{m}<\nu_{\rm x}<\nu_{c}$; see also \citealt{2015MNRAS.454.1073B}). In the case ($\nu_{\rm x}>max(\nu_{m},\nu_{c})$), the same express expression can be applied to both the wind-like and the constant interstellar medium since the kinetic energy does not depend on the medium profile in this spectral regime.
\begin{equation}
\begin{split}
E_{\rm K,iso,52}=\left[\frac{\nu F_{\nu}(\nu=10^{18}Hz)}{5.2\times10^{-14} \rm erg s^{-1}cm^{-2}}\right]^{4/(p+2)}\\
\times D_{28}^{8/(p+2)}(1+z)^{-1}t_{d}^{(3p-2)/(p+2)}\\
\times (1+Y)^{4/(p+2)}f_{p}^{-4/(p+2)}\varepsilon_{B,-2}^{(2-p)/(p+2)}\\
\times \varepsilon_{e,-1}^{4(1-p)/(p+2)}\nu_{18}^{2(p-3)/(p+1)},
\end{split}
\end{equation}
where $\nu F_{\nu}$ ($\nu=10^{18})$ Hz is the energy flux\footnote{Note that since $\nu F_{\nu}$ is not an observed quantity, and it can be converted to the observed flux through integral frequencies over the X-ray energy range (0.3-10 keV).
We note that for a few cases, the photon index $\Gamma_{\rm X,2}=2.0$ during the normal decay phase, we adopt $\Gamma_{\rm X,2}\simeq 2.01$ (within the errors) make a approximate treatment in order to avoid the $1-\beta=0$ ($\beta=\Gamma-1)$ for the calculations.} at frequency $10^{18}$ Hz in units of ${\rm erg s^{-1} cm^{-2}}$, $t_{\rm d}$ is the time in the observer frame in units of days, $D_{28}=D/10^{28}$, is the luminosity distance in units\footnote{The convention $Q=10^{x}Q_{x}$ is adopted in cgs units for all other parameters throughout the paper.} of $10^{28}$ cm, {\it n} is the number density in the constant ambient medium, $A_{*}$ is the parameter of the stellar wind, {\it Y} is the Compton parameter, and $\epsilon_{e}$ and $\epsilon_{B}$ are the energy fraction assigned to the electron and magnetic field, respectively. $f_{p}$ is a function of the power-law electron distribution {\it p} index\footnote{$f_{p}$ is derived from the "external plateau" sample with $p>$2. Only one case, GRB 100302A, has p$<$2, and it was discarded.} \citep{2007ApJ...655..989Z},
\begin{equation}
f_{p}=6.73 \left(\frac{p-2}{p-1}\right)^{p-1}\left(3.3\times10^{-6}\right)^{(p-2.3)/2}.
\end{equation}

We adopted standard values of the microphysical parameters of the shock derived from observations \citep[e.g.,][]{2002ApJ...571..779P,2003ApJ...597..459Y}, namely $n$=1, $Y$=1, and $(\epsilon_{e},\epsilon_{B})= (0.1, 0.01)$.

When $\nu_{m}<\nu<\nu_{c}$, the kinetic energy depends on the type of external medium.
For a wind medium, one has
\begin{equation}
\begin{split}
E_{\rm K,iso,52}=\left[\frac{\nu F_{\nu}(\nu=10^{18}Hz)}{7.4\times10^{-14} \rm erg s^{-1}cm^{-2}}\right]^{4/(p+1)}\\
\times D_{28}^{8/(p+1)}(1+z)^{-(p+5)/(p+1)}t_{d}^{(3p+1)/(p+1)}\\
\times f_{p}^{-4/(p+1)}\varepsilon_{B,-2}^{-1}\varepsilon_{e,-1}^{\frac{4(1+p)}{(p+1)}}\\
\times A_{*,-1}^{-4/(p+1)}\nu_{18}^{2(p-3)/(p+1)},
\end{split}
\end{equation}
while for a constant ISM, one has
\begin{equation}
\begin{split}
E_{\rm K,iso,52}=\left[\frac{\nu F_{\nu}(\nu=10^{18}Hz)}{6.5\times10^{-14} \rm erg s^{-1}cm^{-2}}\right]^{4/(p+3)}\\
\times D_{28}^{8/(p+3)}(1+z)^{-1}t_{d}^{3(p-1)/(p+3)}\\
\times f_{p}^{-4/(p+3)}\varepsilon_{B,-2}^{-\frac{(p+1)}{(p+3)}}\varepsilon_{e,-1}^{\frac{4(1-p)}{(p+3)}}\\
\times n^{-2/(p+3)}\nu_{18}^{2(p-3)/(p+3)}.
\end{split}
\end{equation}

After deriving $E_{\rm K,iso}$ at the break time $t_{\rm b}$, we next estimate the upper limit of $E_{\rm K,iso}(t_{0})$ by $E_{\rm K,iso}(t_{\rm b})$. If $t_{s}>t_{0}$, the upper limit of $E_{\rm K,iso}(t_{0})$ can be estimated by
\begin{equation}
E_{\rm K,iso}(t_{0})<E_{\rm K,iso}(t_{b})/\left(\frac{t_{\rm b}}{t_{\rm s}}\right)^{(1-q)}.
\end{equation}

The error on $E_{\rm K,iso} (t_{\rm b})$ is derived by  bootstrapping. For other parameters with simple mathematical expressions, the errors are calculated by error propagation.

\begin{figure*}
\centering
\includegraphics[angle=0,scale=0.70]{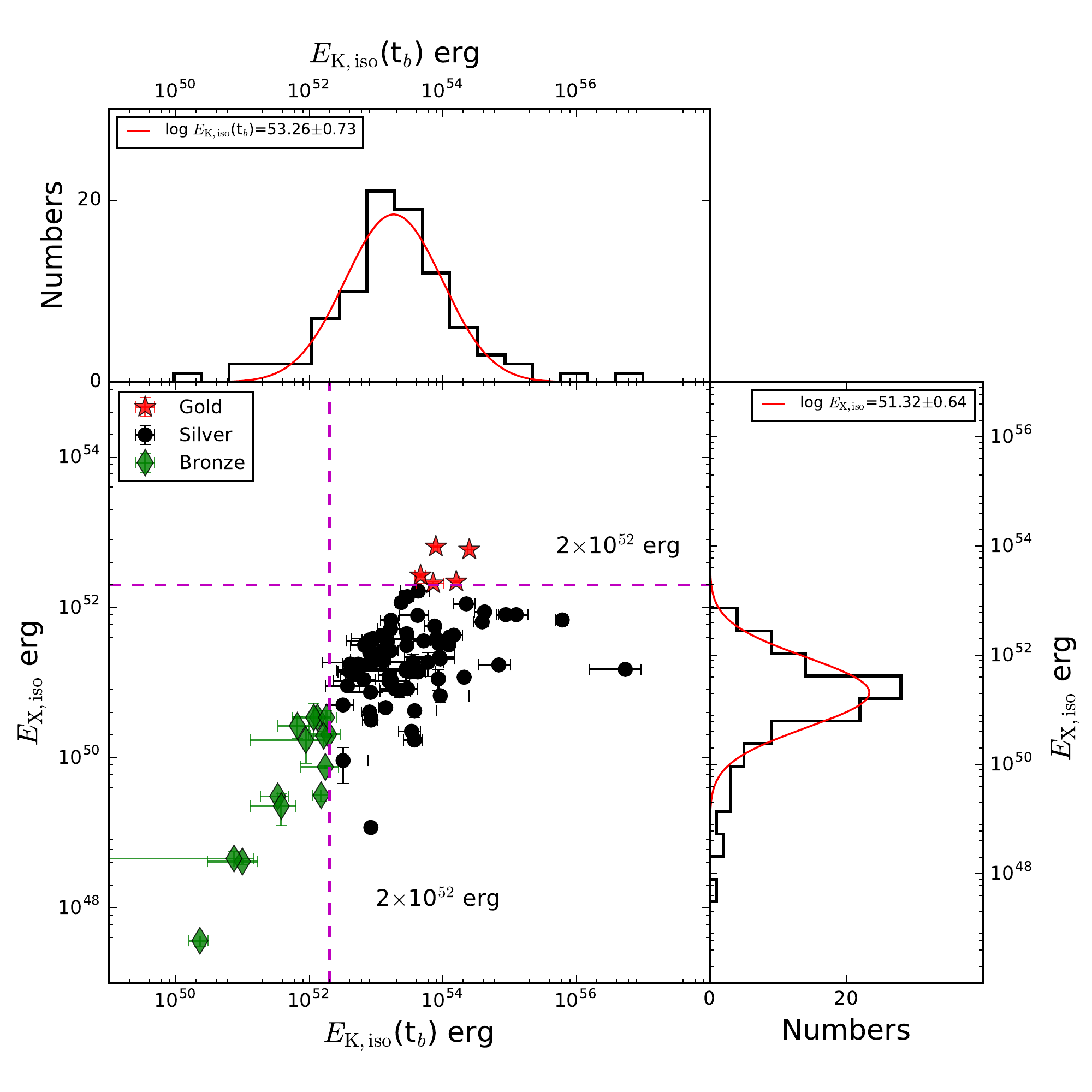}
\caption{The $E_{\rm X,iso}$ compared against the $E_{\rm K,iso}(t_{b})$ with their distributions (black solid-lines) and the best Gaussian fits (red solid-lines). Different colors of scatter data points represent different sub-samples (Gold, Silver and Bronze). Two dash-lines represent $E_{\rm X,iso}$ and $E_{\rm K,iso}(t_{b})$ equal 2$\times$10$^{52}$ erg, respectively.}\label{Groups}
\end{figure*}

After determining the bursts with an external plateau, we further split them into three groups. The classification is made depending on whether the X-ray isotropic energy $E_{\rm X,iso}$ and or the kinetic isotropic energy $E_{\rm K,iso}$ is larger than $2\times10^{52}{\rm erg}$, corresponding to the maximum energy released by a magnetar. This is illustrated in Fig.\ref{Groups}.

\begin{itemize}
\item Gold Sample: bursts for which the $E_{\rm X,iso}$ of the plateau phase is greater than $2\times10^{52}\rm erg$.
The sample is made up of nine GRBs and the power-law fits to the light curves are shown in Fig.\ref{Gold}. This corresponds to 9\% of all the bursts.

\item Silver Sample: bursts for which the $E_{\rm X,iso}$
 of the plateau phase is less than $2\times10^{52} \rm erg$, but  $E_{\rm K,iso}$ is greater than $2\times10^{52} \rm erg$. It is also consistent with the theoretical prediction of the BH central engine model.
We note that because the computation of $E_{\rm K,iso}$ relies on assumed values for the microphysical parameters, it is not possible to completely rule out the magnetar progenitor for the bursts. This sample is made up of 69 bursts, and their light curves are presented in Fig.\ref{Silver}. This corresponds to 68\% of all the bursts.

\item Bronze Sample: bursts for which neither $E_{\rm X,iso}$ nor $E_{\rm K,iso}$ are above $2\times10^{52}$ erg. For the 23 bursts in this sample, it is not possible to completely rule out the BH progenitor. Fig.\ref{Bronze} shows the fitted light curves.
This corresponds to 23\% of all the bursts.
\end{itemize}

Under the assumption that a magnetar emits its energy isotropically, this classification indicates that 77\% of the bursts have a BH central engine; see further discussion in \S 5.1.

\subsection{$\gamma$-Ray Energetics}

The $k$-corrected isotropic energy released in the gamma-ray band ($E_{\gamma,\rm iso}$) is retrieved from the published paper.
For bursts that did not have published values, we derived them from the observed fluence ($S_{\gamma}$) in the detector band and $k$-correct to the rest-frame (1-10$^{4}$) keV using the spectral parameters.

\begin{figure*}
\centering
\includegraphics[angle=0,scale=0.80]{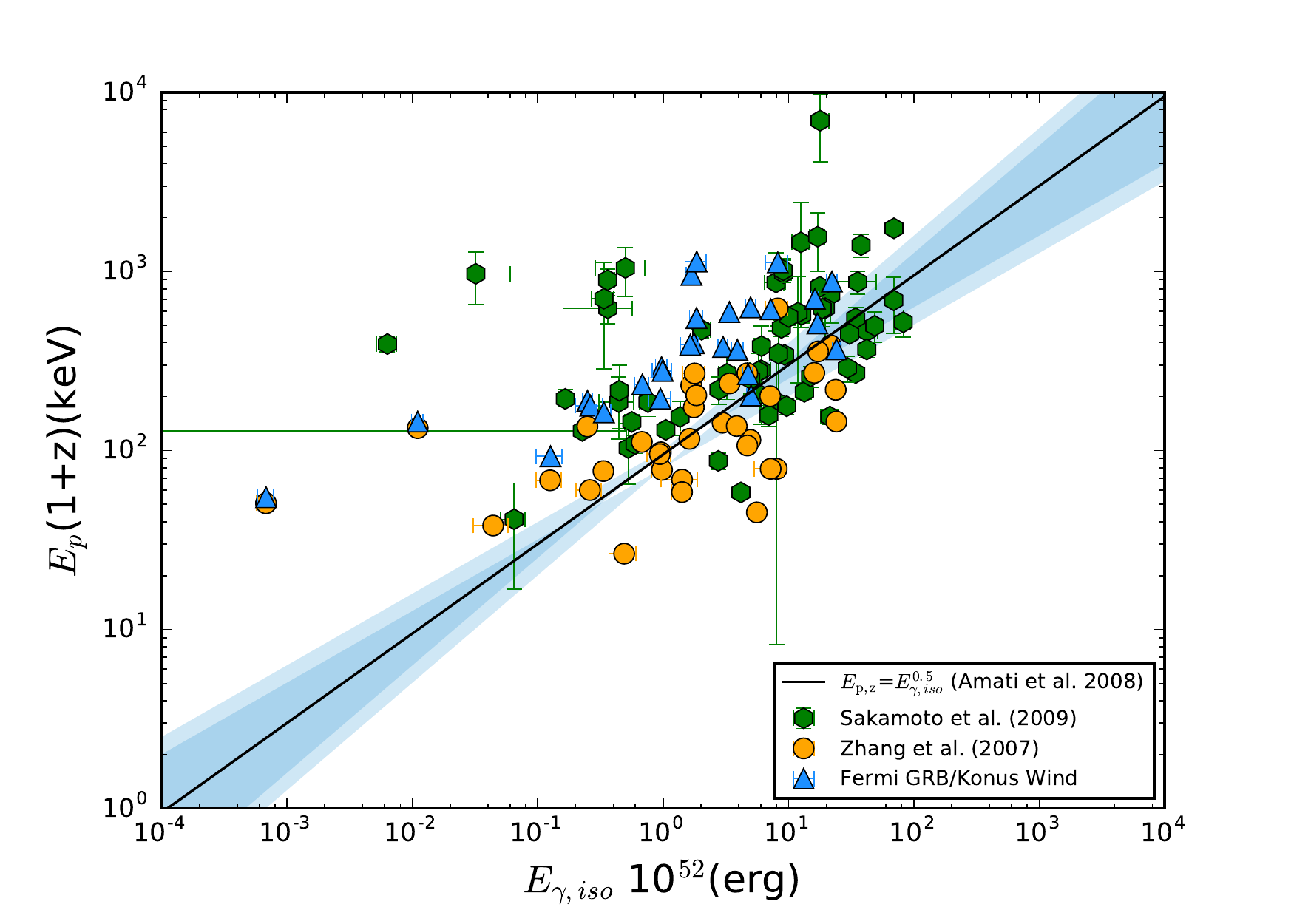}
\caption{$E_{\rm p,z}$ compared against the $E_{\gamma,\rm iso}$. The yellow data points are derived from \citep{2007ApJ...655L..25Z}, the green data points derived from \citep{2009ApJ...693..922S}, and the blue data points derived from {\it Fermi}/GBM or Konus/Wind (collected from the published papers or the GCN Circulars Archive). The solid line and the shaded region represent the empirical Amati relation.}\label{EpEiso}
\end{figure*}

We obtained the fluence and spectral parameters (including the $\alpha$ and $\beta$ spectral indexes and the peak energy $E_{\rm p}$) of the Band function from published papers \citep[e.g.,][]{2016MNRAS.455.1027D} or the GCN Circulars Archive\footnote{\url{https://gcn.gsfc.nasa.gov}}.
First, if a burst was detected by \emph{Swift}/BAT and also by {\it Fermi}/GBM and/or Konus/Wind, we considered the spectral parameters from either the \emph{Fermi}/GBM or Konus/Wind instrument.
Second, if a burst was only detected by the \emph{Swift}/BAT instrument and fit with a power-law or a cutoff power-law model, we proceeded with the following steps:
(1) for the cutoff power-law model, the fit parameters include $\alpha$ and $E_{\rm p}$ \citep{2011ApJS..195....2S}, so we adopt the typical value of $\beta$=-2.3.
(2) for the SPL model, we derive $E_{\rm p}$ from the empirical correlation $\log E_{\rm p}=(2.76\pm 0.07)-(3.61\pm0.26) \log \Gamma^{\rm BAT}$ \citep{2007ApJ...655L..25Z}, where $\Gamma$ is the photon index of the \emph{Swift}/BAT band, which was found from the \emph{Swift} data archive, and adopted the typical $\beta$ value of -2.3.
(3). We finally test the results against the Amati relation \citep{2002A&A...390...81A}, as shown in Fig.\ref{EpEiso}. Our results are consistent with the Amati relation, which indicates that our method is reasonable.

We derive the $k$-correction \citep{2001AJ....121.2879B} factor in the rest-frame 1-10$^{4}$ keV band, and derive the isotropic energy $E_{\gamma, \rm iso}$ as
\begin{equation}
E_{\gamma, iso}=\frac{4\pi k D_{L}^{2}S_{(\gamma)}}{1+z}.
\end{equation}
The isotropic luminosity at the break ($L_{\rm X}$) can be derived as
\begin{equation}
L_{\rm X, iso}=4\pi k D_{L}^{2}F_{\rm X,b},
\end{equation}
where $F_{\rm X,b}$ is the flux at the break time ($t_{\rm b}$), which was obtained from the light curve fit.

The luminosity at the break time ($L_{\rm X, iso}$) and the isotropic release energy of the plateau phase ($E_{\rm X,iso}$), together with the time intervals ($t_{s}$, $t_{e}$) for the light curve fitting, the temporal indices of the plateau phase ($\alpha_{1}$) and the normal decay or steeper decay phase ($\alpha_{2}$), the break time ($t_{\rm b}$), the X-ray fluence for the plateau phase ($S_{\rm X}$), which is the integrated observed flux during the time interval ($t_{s}$, $t_{b}$), and the photon indices\footnote{The photon indices of the plateau phase were taken from the table published in \cite{2010ApJ...722L.215D}, and the rest of the values were computed from the {\it Swift} BA+XRT repository: \url{http://www.swift.ac.uk/burst_analyser/}} for the shallow decay/plateau phase ($\Gamma_{1}$) and a normal or steeper decay phase ($\Gamma_{2}$), are obtained from the \emph{Swift} archive,\footnote{We applied the same photon model for the both pre- and postbreak phases, either from the PC model to the PC model, or from the WT model to the WT model, since the data for the late time observed is usually a WT model.}, which are reported in Table 1.
In addition, the table also gives the spectral indices of the plateau phase and of the following segment, both obtained from either the \emph{Swift} archive or from the closure relation.

\subsection{Jet-opening angle}\label{Beaming}

In this section, we study the beaming correction with three methods and compare their results. The beaming correction is defined as
\begin{equation}
f_{\rm b}=1-cos\theta_{j}\simeq(1/2)\theta_{j}^{2}.
\end{equation}
It is used to correct the $\gamma$-ray and kinetic energies.
We obtain the geometrically corrected gamma-ray energy through $E_{\gamma}=f_{\rm b} E_{\gamma, \rm iso}$， and $E_{\rm K}=f_{\rm b} E_{\rm K,iso}$.  The latter  quantity is used for collimated outflows from a BH central engine (See \S 3.2).

The jet-opening angle can be derived from the observational data \citep{Rhoads:1999hp,Sari:1999ge,Frail:2001hd},
\begin{equation}
\theta_{j}=
0.124 \left(\frac{t_{j}}{1\rm day} \right)^{3/8} \left(\frac{1+z}{2}\right)^{-3/8} \left(\frac{E_{\rm k,iso}}{10^{52}\rm erg}\right)^{-1/8} \left(\frac{n}{1\rm cm^{-3}}\right)^{1/8},
\label{eq:thetaj}
\end{equation}
where $\theta_{j}$ is the jet-break time, $E_{\rm K,iso}$ is the isotropic kinetic energy, and {\it n} is the medium density.
If the light curve shows a jet break, we measure $\theta_{j}$ using Equation (\ref{eq:thetaj}). In our sample, 18 GRBs show a jet break. For the rest of the GRBs without a jet break, we use the observed time of the last data point as the jet-break time ($t_{j}$) to estimate $\theta_{j}$, and get an upper limit.

Another method to estimate the jet-opening angle is by empirical relations \citep{2012cosp...39.1354N}, $\theta_{j}\approx5.0/\Gamma_{0}$, where $\Gamma_{0}$ is the initial Lorentz factor, also derived using the empirical relation as
\begin{equation}
\Gamma_{0}=27 E_{\gamma,\rm iso,52}^{0.26},
\end{equation}
where $E_{\gamma,\rm iso, 52}$ is the isotropic energy, in units of 10$^{52}$ erg.

\begin{figure*}
\includegraphics[angle=0,scale=0.60]{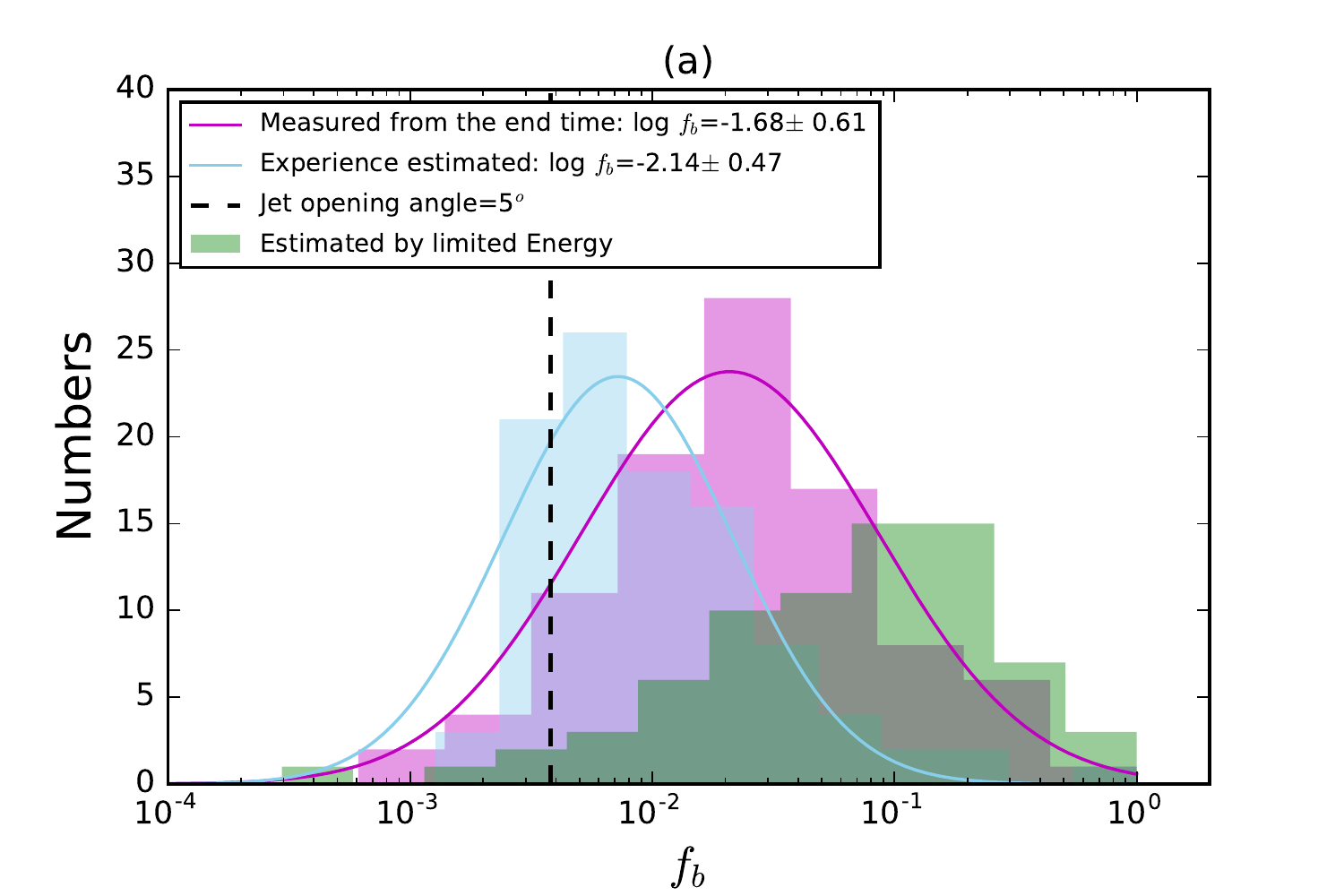}
\includegraphics[angle=0,scale=0.60]{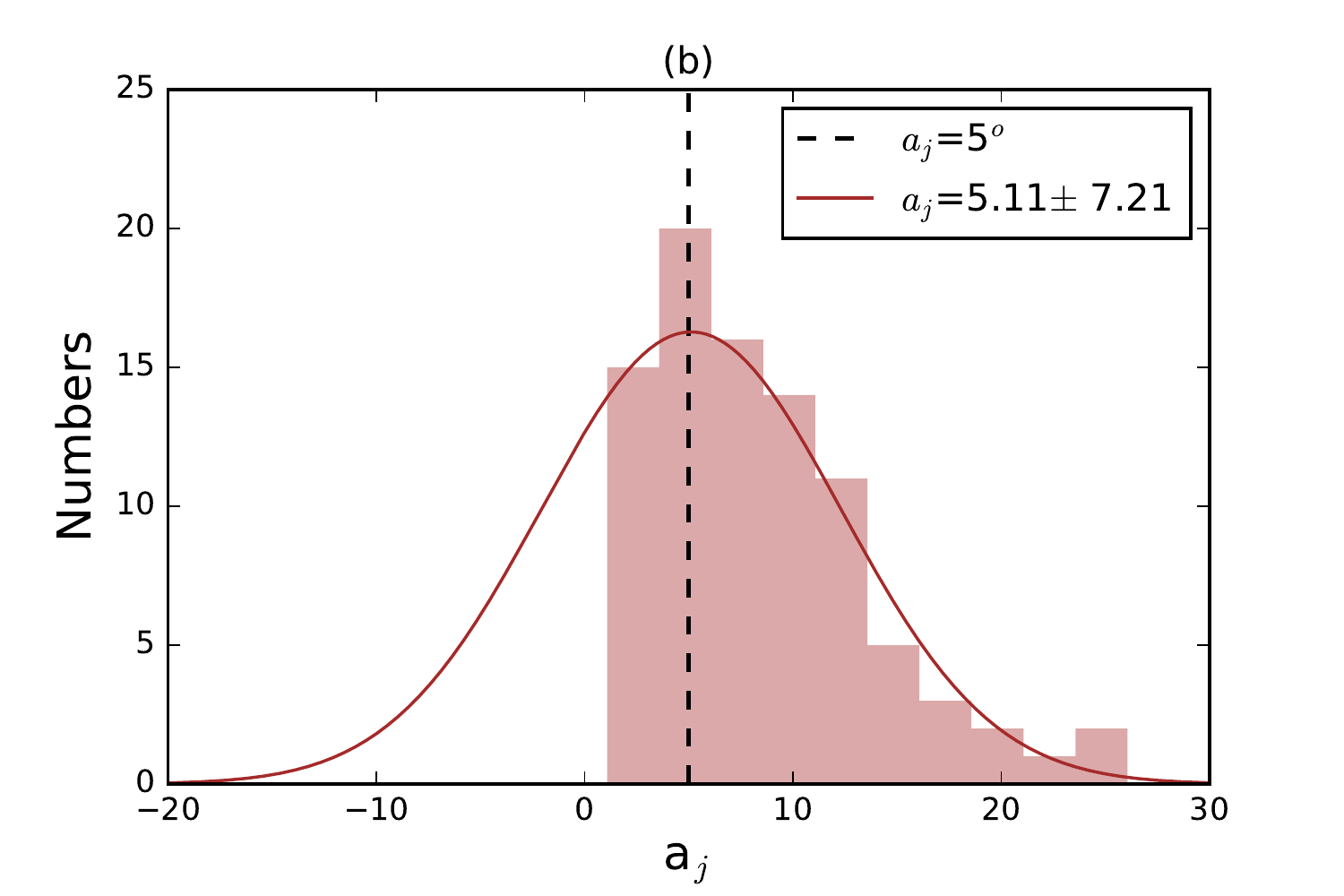}
\caption{Left panel: the distribution of the beaming correction factor $f_{\rm b}$ for different methods and their best Gaussian fits. The green histogram shows the distribution of $f_{\rm b}$ for individual bursts to be consistent with that of a magnetar. Right panel: the distribution of $a_{j}$ and the best Gaussian fit, which were derived from an initial assumption $\theta_{j,0}\approx\theta_{j}$ (see the text).}\label{angle}
\end{figure*}

Fig.\ref{angle} shows the distributions of the beaming factors $f_{\rm b}$.
We compare the distributions of these two methods.
These distributions can be fitted with Gaussian functions; we have log$f_{\rm b}=-1.68\pm0.61$ from using the late data point method and log$f^{'}_{\rm b}=-2.14\pm0.47$ from the empirical relation method. We find that the $f_{\rm b}$ values from the first method is greater than those from the second method, and the $f_{\rm b}$ of $\theta_{j}=5^{o}$ lies between the two distributions.
In \cite{2012cosp...39.1354N}, a coefficient 5 was adopted for the estimation.
In turn, if it is initially defined as $\theta_{j}\equiv a_{j}/\Gamma_{0}$, we then can estimate $a_{j}$. Fig.\ref{angle}b shows the distribution of $a_{j}$, with the best Gaussian fit $a_{j}= 5.11 \pm 7.21$. This result is consistent with the finding in \cite{2012cosp...39.1354N}.
The results, which include $\theta_{j}$, $f_{\rm b}$, $a_{j}$, $\theta^{'}_{j}$, and $f^{'}_{b}$, are presented in Table 4.

We finally adopted the following criterion to calculate the beaming factor: if a burst has an observed light curve jet break, we then use it to estimate $\theta_{j}$; otherwise, we adopt $\theta_{j}=5^{o}$.

\subsection{$\gamma$-Ray Radiative efficiency}\label{Rfficiency}
Another point we want to investigate is whether the $\gamma$-ray radiative efficiency is different among the samples.
The radiative efficiency is defined as \citep{2004ApJ...613..477L}
\begin{equation}\label{Eq:efficiency}
\eta_{\gamma}=\frac{E_{\gamma,\rm iso}}{E_{\gamma,\rm iso}+E_{\rm K,iso}}=\frac{E_{\gamma}}{E_{\gamma}+E_{\rm K}}.
\end{equation}
Since the kinetic energy ($E_{\rm K,iso}$) increases with time during the plateau phase but remains the same during the normal decay phase, so the radiation efficiency $\eta_{\gamma}$ thus depends on which $E_{\rm K}$ epoch is adopted.
Here we focus on the radiation efficiency at the end of the shallow decay phase, $\eta_{\gamma}(t_{\rm b})$.
The physical meaning for this is the efficiency of converting the spin-down energy of a BH/magnetar to $\gamma-$ray radiation.

The X-ray efficiency can also be defined as the conversion of the BH/magnetar spin-down energy to radiation, i.e.
\begin{equation}
\eta_{\rm X}=\frac{L_{\rm X,iso}}{L_{\rm X,iso}+L_{\rm K,iso}}=\frac{L_{\rm X}}{L_{\rm X}+L_{\rm K}}.
\end{equation}

\begin{figure*}
\includegraphics[angle=0,scale=0.60]{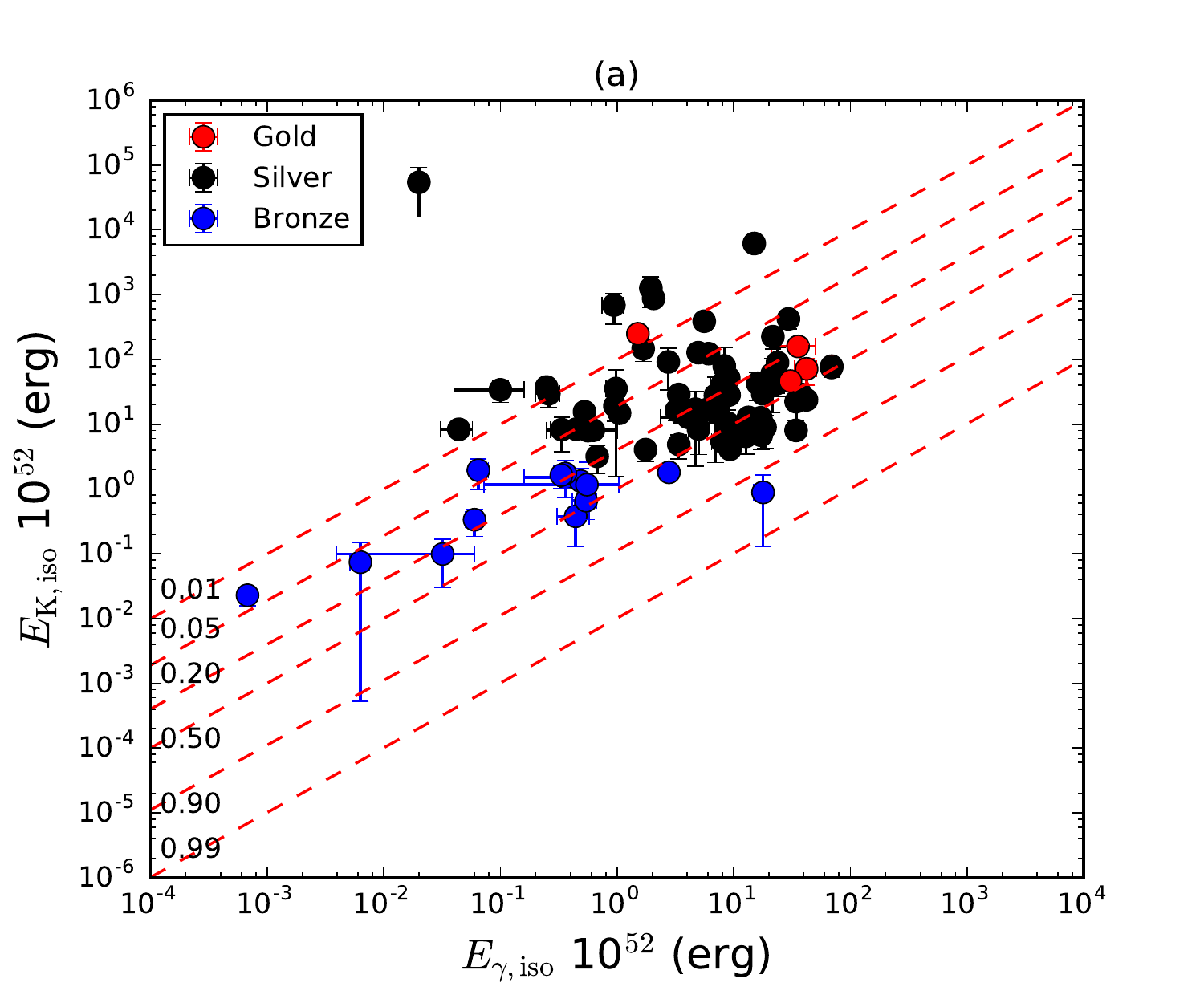}
\includegraphics[angle=0,scale=0.60]{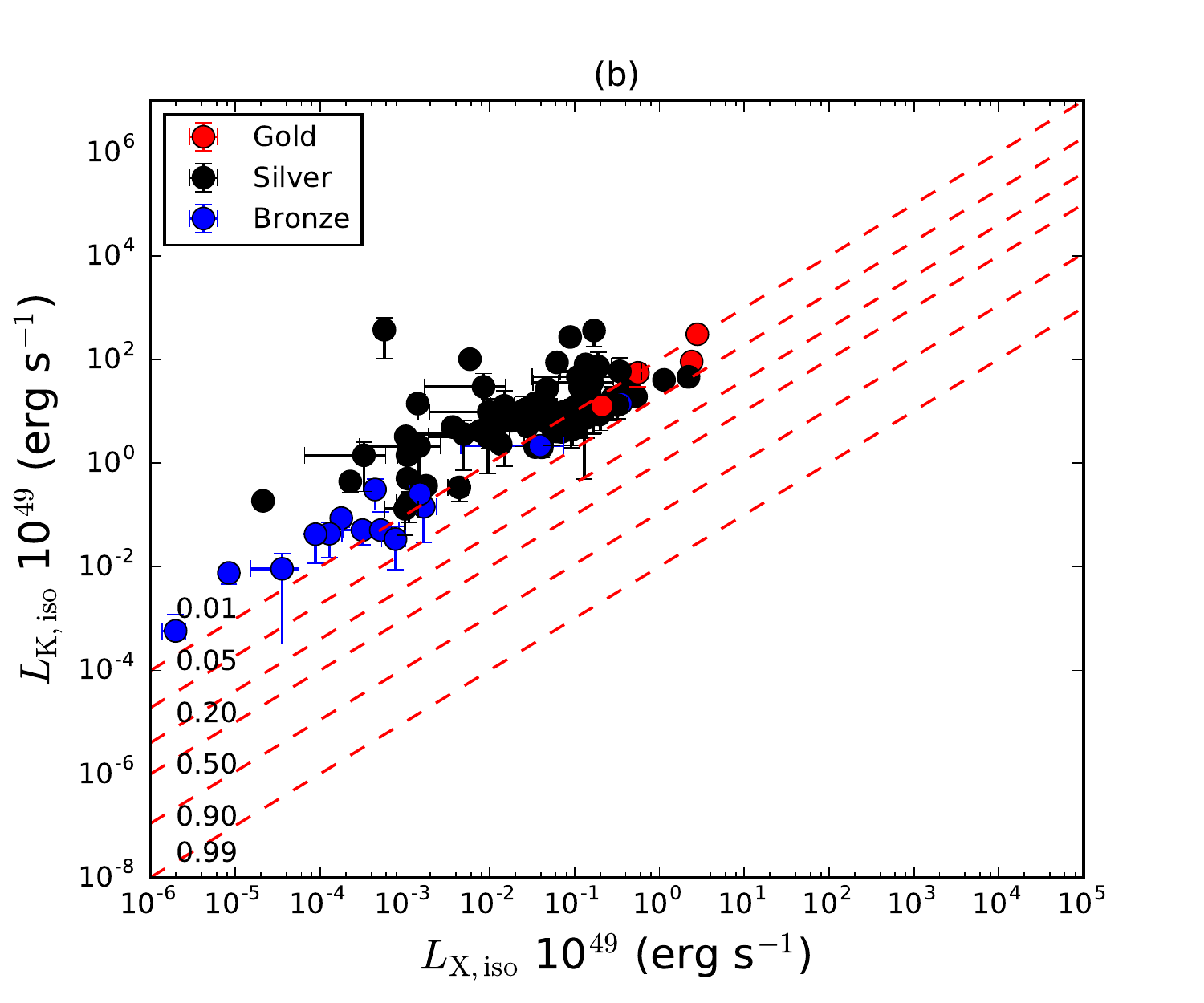}
\caption{(a) $E_{\gamma,\rm iso}-E_{\rm K,iso}$ scatter plot for all of the GRBs (excluding internal plateaus). Slanted dashed lines mark the constant $\gamma-$ ray efficiency ($\eta_{\gamma}$) lines. $E_{\rm K,iso}$ is calculated at $t_{\rm b}$; (b) $L_{\rm X,iso}-L_{\rm K,iso}$ scatter plot for the BH/magnetar samples (excluding internal plateaus). The constant X-ray efficiency $\eta_{\rm X}$ lines are overplotted. }\label{efficiency}
\end{figure*}

\begin{figure*}
\includegraphics[angle=0,scale=0.60]{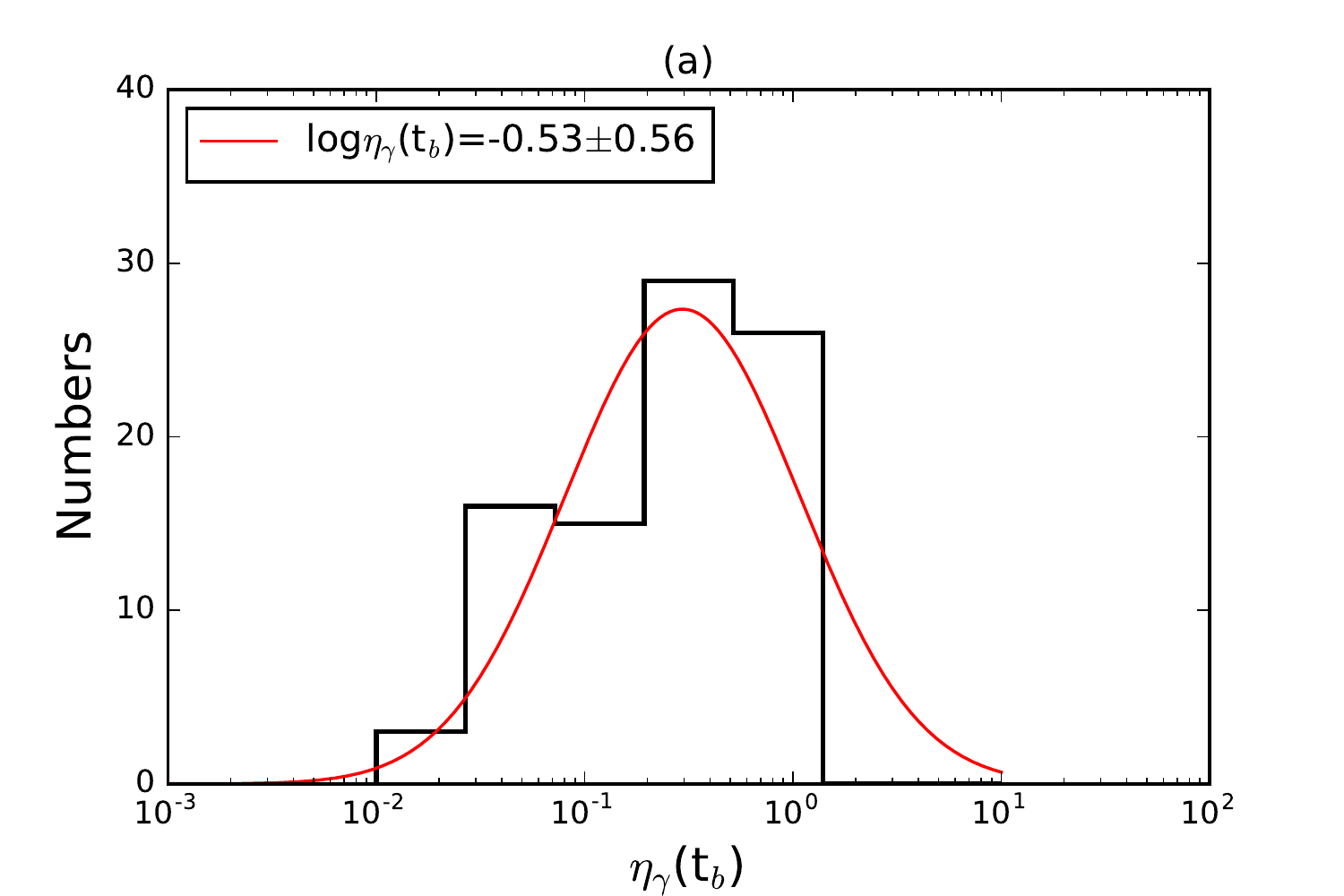}
\includegraphics[angle=0,scale=0.60]{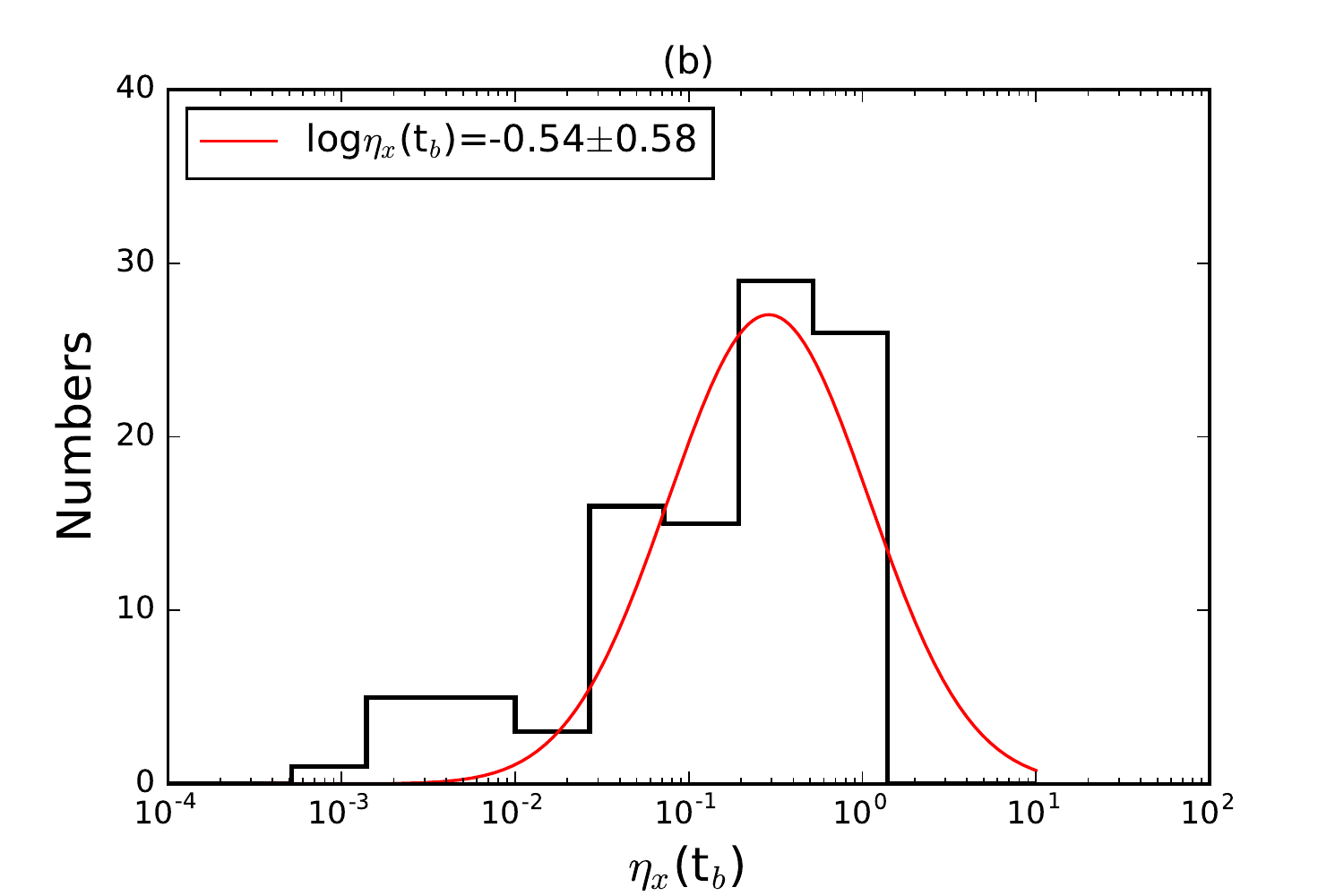}
\caption{Histograms of $\eta_{\gamma}(t_{\rm b})$ and $\eta_{\rm X}(t_{\rm b})$ of our samples (black lines) and their best Gaussian fits (red lines).}\label{efficiencyDis}
\end{figure*}

In Fig.\ref{efficiency}, we compare $E_{\gamma,\rm iso}$ and $E_{\rm K,iso}$ (Fig.\ref{efficiency}a) and $L_{\rm X,iso}$ and $L_{\rm K,iso}$ (Fig.\ref{efficiency}b) for different samples.
The same distribution is also shown in Figs. \ref{efficiencyDis}a and \ref{efficiencyDis}b. The typical value is log $\eta_{\gamma}$($t_{\rm b}$)= -0.53$\pm$0.56 for $\eta_{\gamma}$ and is log $\eta_{x}$($t_{\rm b}$)=-0.54$\pm$0.58 for $\eta_{x}$. The radiative efficiencies in the $\gamma-$ and X-ray energy bands, in general, present the same distribution.

\section{Central Engine Properties \label{Models}}

In this section, we derive the properties of the central engine based on our observations, assuming two models.

\subsection{BH Central Engine}\label{sec:BH}

One of the leading models for the GRB central engine is a stellar-mass BH surrounded by a hyper-accreting disk \citep[e.g.,][]{1999ApJ...518..356P,2001ApJ...557..949N,2002apa..book.....F,2008bhad.book.....K,2013LRR....16....1A,2013ApJ...765..125L,2014SSRv..183...21B,2014ARA&A..52..529Y}, with a typical accretion rate of 0.01-1 $M_\sun s^{-1}$. There are two main energy reservoirs that provide the jet power, namely the accretion energy in the disk that is carried by neutrinos and antineutrinos, which annihilate each other and power a bipolar outflow \citep[e.g.,][]{2002ApJ...579..706D,2002ApJ...577..311K,2006ApJ...643L..87G,2007ApJ...657..383C,2007ApJ...664.1011J,2007ApJ...661.1025L, 2008ChJAA...8..404L,2009ApJ...700.1970L}, and the spin energy of the BH, which can be tapped through magnetic fields via the BZ \cite{1977MNRAS.179..433B} mechanism \citep[see also][]{Lee:390303,Li:2000hq,2013ApJ...765..125L}.

As suggested by \cite{2013ApJ...765..125L}, the jet might be dominated by the BZ power especially at late times \citep{2005MNRAS.364L..42F,2011ApJ...726...90Z}. The rotational energy of a BH with angular momentum $J_\bullet$ is a fraction of the BH mass $M_\bullet$ (or dimensionless mass $m_{\bullet}=M_{\bullet}/M_\odot$),
\begin{equation}\label{Erot}
E_{\rm rot}=f_{\rm rot}(a_\bullet)\frac{M_{\bullet}}{M_\sun} c^{2} {\rm erg}= 1.8 \times 10^{54} f_{\rm rot}(a_\bullet) \frac{M_\bullet}{M_\sun} {\rm erg},
\end{equation}
where $a_\bullet = J_\bullet c/(GM_\bullet^2)$ is the BH spin parameter and
\begin{equation}
f_{\rm rot}(a_\bullet)=1-\sqrt{(1+\sqrt{1-a_\bullet^2})/2 }.
\end{equation}

We then connect the kinetic energy of the jet to the spin energy of the BH through the relation
\begin{equation}\label{Beam}
\eta E_{\rm rot}=f_{\rm b} E_{\rm K,iso},
\end{equation}
where $\eta$ is the efficiency of converting the spin energy to the kinetic energy of the jet through the BZ mechanism, and $f_{\rm b}$ is the beaming factor (see \S \ref{Beaming} for a detailed discussion).
For a maximally rotating BH, about 31\% of the rotational energy can be taken out to power a GRB \citep{Lee:390303}. This efficiency is not sensitive to the BH spin $a_\bullet$.
Therefore, we take $\eta \simeq 0.3$ in the following.

From Equations (\ref{Erot}) and (\ref{Beam}), one can constrain the BH spin parameter once we know the BH mass $M_\bullet$, the jet-opening angle $\theta_{\rm j}$, and the kinetic energy of the jet $E_{\rm K,iso}$. Following \cite{2015MNRAS.446.1213K}, the typical value for the BH mass in a long GRB is $6 M_\sun$. In this paper, we adopt $M_\bullet =3 M_\sun$ for short GRBs\footnote{Our samples include five short GRBs, and four of them belong to the Bronze sample.} and $6 M_\sun$ for long GRBs.

For $a_\bullet >0.1$, the bimodal distribution of the index $q$ can be explained by the spin evolution of the BH. The BH evolves with time during a GRB, since it would be spun up by accretion and spun down by the BZ mechanism \citep{2017ApJ...849...47L}.
Due to and the different initial parameter sets for different GRBs, the evolution of the characteristics varies from burst to burst. Therefore, we have $q>0$ for the spin-down evolution and $q<0$ for the spin-up case.

For $a_\bullet <0.1$, the BH spin does not necessarily need to be very low. Instead, the value we obtained based on Eq.(\ref{Beam}) should be the change in the BH spin $\Delta a_\bullet$ during a GRB. The very small value of $\Delta a_\bullet$ means that the BH spin does not evolve. This is possible when the BH spin is at a critical value where either the spin-up due to accretion is comparable to the spin-down due to the BZ mechanism or the accretion rate is quite low. A clear evidence comes from the distribution of $q$ when $a_\bullet <0.1$, for which $q$ is close to zero.

\begin{figure*}
\centering
\includegraphics[angle=0,scale=0.40]{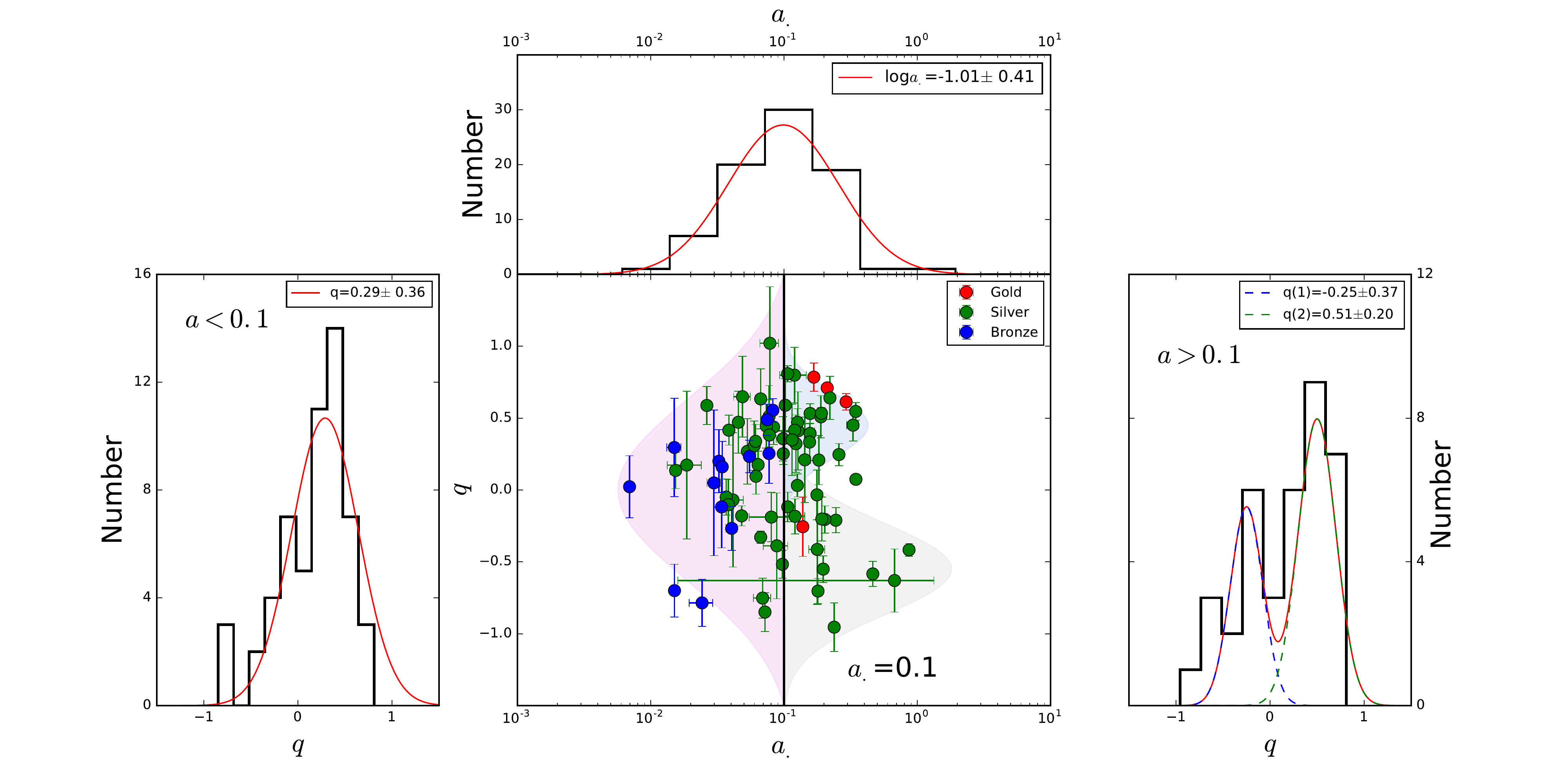}
\caption{Top panel: the distribution of the parameter $a_\bullet$ of the black hole with its the best Gaussian fit;
bottom panel: the distribution of the luminosity injection index {\it q} for $a<0.1$ (left) and $a_\bullet>0.1$ (right); the luminosity injection index $q$ compared against the parameter $a_\bullet$ (middle). The different dot colors represent the different subsamples, the black perpendicular line represents $a_\bullet$=0.1, and the different color of the shaded area describe $q$ for different distributions of different $a_\bullet$ values.
}\label{BH}
\end{figure*}

The correlation of $a_\bullet-q$ for our different samples is presented in Fig.\ref{BH}.
We find that the Gold sample bursts cluster into the region where $a_\bullet>0.1$. The results are consistent with the prediction of the BH central engine model.
Here we note that since the Bronze sample bursts could still have a BH central engine, we also calculate $a_\bullet$ for them. Interestingly, we found all Bronze sample bursts have  $a_\bullet<$ 0.1.

The BH parameters ($a_\bullet$), the duration of the prompt emission $T_{\rm 90}$, the isotropic energy release in the gamma-ray band $E_{\gamma,\rm iso}$, the isotropic kinetic energy $E_{\rm K,iso}$, the electron spectral index $p$, the luminosity inject index $q$, and a redshift $z$ (searched from the published papers or the GCN Circulars Archive) are reported in Table 2 for the Gold and the Silver samples.

For a magnetized BH-accretion disk system, there is another magnetic mechanism: the magnetic coupling effect between the BH and the disk through closed magnetic field lines \citep{1999Sci...284..115V,2000ApJ...534L.197L,2002MNRAS.335..655W,2009ApJ...700.1970L,2010A&A...509A..55J}. Similar to the BZ mechanism, the magnetic coupling effect also extracts rotational energy from the spinning BH. Only if the BH spin is initially small would the magnetic coupling would act as an additional spin-up process. A similar discussion of this aspect was made by \cite{2012ApJ...759...58D} within the context of the magnetar central engine model. In more general cases, the magnetic coupling effect would not significantly affect the BH spin evolution \citep{2009ApJ...700.1970L}.

\subsection{Magnetar Central Engine}\label{sec:magnetar}

A rapidly spinning magnetar model invokes a rapidly spinning, strongly magnetized neutron star called a "magnetar". 
There are two critical magnetar parameters for a magnetar undergoing bipolar spin-down: the initial spin period $P_{0}$ and the surface polar cap magnetic field $B_{p}$, which can be derived using the characteristic luminosity $L_{0}$ and the spin-down time scale $\tau$. We derive the magnetar parameters ($P_{0}$ and $B_{\rm P}$) using Eqs. (\ref{eq:L0})-(\ref{eq:LK}), assuming isotropic emission from the magnetar.

For a magnetar, the characteristic spin-down luminosity is:
\begin{equation}
L_{0}=1.0\times10^{49} \rm ergs^{-1} (B_{p,15}^{2}P_{0,-3}^{-4}R_{6}^{6}).
\label{eq:L0}
\end{equation}
where $R$ is the stellar radius\footnote{We assume the typical value for $R_{6}$=1, $M=M_{1.4}$ and $I_{45}\sim2$.}.
In Eguation (\ref{eq:q}), $t$ is a measure of time in the rest frame, and $\tau$ is the initial spin-down timescale, which is defined by
\begin{equation}
\tau=\frac{P}{2\dot{P}}_{0}=2.05\times10^{3}s(I_{45}B_{p,15}^{-2}P_{0,-3}^{2}R_{6}^{-6}),
\label{eq:spintime}
\end{equation}
where $\dot{P}$ is the rate of period increase due to MDR, and $I_{45}$ is the stellar moment of inertia in units of $10^{45}$g cm$^{-2}$ \citep{1988ApJ...331..784D, 1992ApJ...390..541W}.
The initial spin-down timescale, $\tau$, in general, can be identified as the observed break time, $t_{\rm b}$,
\begin{equation}
\tau=t_{b}.
\label{eq:tb}
\end{equation}
and the characteristic spin-down luminosity $L_{0}$, in general, should include two terms,
\begin{equation}
L_{0}=L_{\rm X}+L_{\rm K}=(L_{\rm X,iso}+L_{\rm K,iso}),
\label{eq:L0obs}
\end{equation}
where $L_{\rm X,iso}$ is the X-ray isotropic luminosity due to the internal dissipation of the magnetar wind, which is the observed X-ray luminosity of the internal plateau and measured at the observed break time (one can only derive an upper limit for external plateaus), and
\begin{equation}
L_{\rm K,iso}=E_{\rm K,iso}(1+z)/t_{b},
\label{eq:LK}
\end{equation}
is the kinetic isotropic luminosity which is injected into the blast wave during the energy injection phase.
It depends on the isotropic kinetic energy $E_{\rm K,iso}$ after the injection phase is over and the observed break time $t_{\rm b}$.
The isotropic kinetic energy $E_{\rm K,iso}$ can be derived from afterglow modeling, as discussed above.

Since it is not possible to completely rule out the magnetar progenitor for the Silver sample, we investigate the magnetar parameters for both the Silver and Bronze samples.

\begin{figure*}
\centering
\includegraphics[angle=0,scale=0.80]{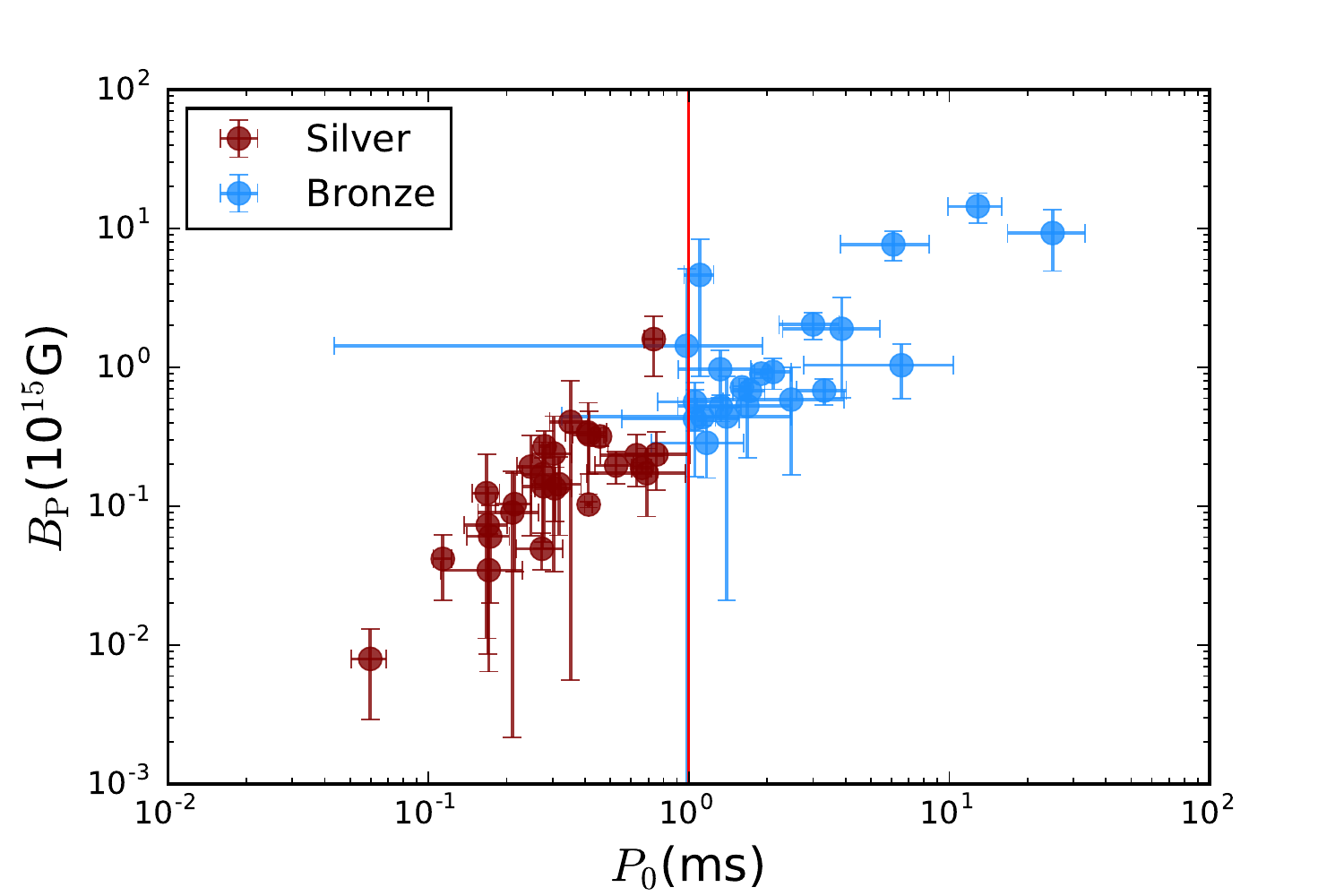}
\caption{Inferred magnetar parameters, the initial spin period $P_{\rm 0}$ vs. the surface polar cap magnetic field strength $B_{\rm p}$, derived from the Silver and Bronze samples. The vertical solid line is the breakup spin period for a neutron star \citep{2004Sci...304..536L}.}\label{magnetar}
\end{figure*}

The derived values of $P_{\rm 0}$ and $B_{\rm P}$ are presented in Table 3 and Fig.\ref{magnetar}.
We found that all bursts in the Bronze sample fall in a reasonable range in the ($P_{\rm 0}-B_{\rm P}$) plot (Fig. \ref{magnetar}). $B_{\rm P}$ has a typical value of $\sim$ $10^{15}$G, and is in the range [$10^{14}$G,$10^{16}$G]; no case have $B_{\rm P}$ below $10^{14}$G, and only one case is greater than $10^{16}$G. We find that most derived $P_{\rm 0}$ are close to 1 ms; this is consistent with the breakup limit of a neutron star, which is about 0.96 ms \citep{2004Sci...304..536L}.\footnote{We note that four of the five short GRBs are obtained in the Bronze sample, while the rest of the 19 GRBs are long GRBs.}.

However, we also found that approximately half of the Silver sample bursts have $B_{\rm P} < 10^{14}$G. Moreover, most of the bursts have $P_{\rm 0}$ that is far away below to 1 ms. This result implies that the Silver sample bursts are more likely to have a BH central engine.

Using the derived value of $P_0$, the total rotation energy of the millisecond magnetar within this scenario is
\begin{equation}
E_{\rm rot}=\frac{1}{2}I\Omega^{2}\simeq 2\times10^{52} {\rm erg} M_{1.4}R_{6}^{2}P_{0,-3}^{-2},
\end{equation}
where $I$ is the moment of inertia, $\Omega=2\Pi/P_{0}$ is the initial angular frequency of the neutron star, $M_{1.4}=M/1.4M_{\bigodot}$, and $E_{\rm rot}$ is the maximal total energy available and can be compared to the observed and inferred energies (see \S \ref{sec:energy}).

In conclusion, the properties of the bursts in the Bronze sample are consistent with the predictions of the magnetar central engine model.

\subsection{Total Energy and Correlations} \label{sec:energy}

\begin{figure*}
\centering
\includegraphics[angle=0, scale=0.8]{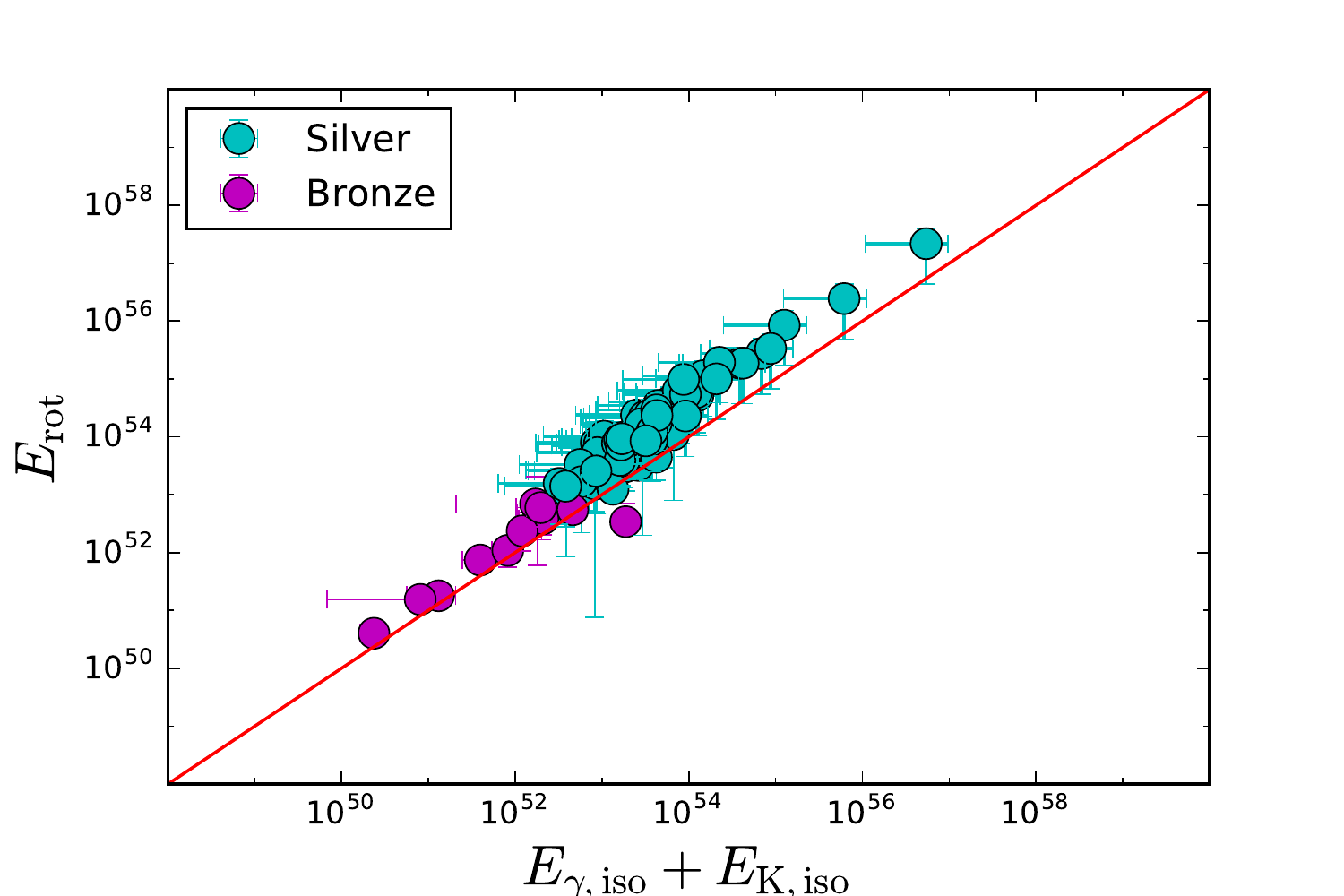}
\caption{Comparison between ($E_{\gamma, \rm iso}$+$E_{\rm K,iso}$) and $E_{\rm rot}$. The equal line is represented by the continuous line.}\label{ErotEkEx}
\end{figure*}

In Fig.\ref{ErotEkEx}, we compare the inferred $E_{\gamma,\rm iso}$+$E_{\rm K,iso}$ with the total rotation energy $E_{\rm rot}$ of the BH (Silver; \S \ref{sec:BH}) and the millisecond magnetar (Bronze; \S \ref{sec:magnetar}).
It is found that the GRBs are generally above and slightly above the $E_{\rm rot}=E_{ \gamma,\rm iso}+E_{\rm K,iso}$ line. This is consistent with the hypothesis, namely, that all of the emission energy ultimately comes from the spin energy of the BH or magnetar.

\begin{figure*}
\centering
\includegraphics[angle=0,scale=0.35]{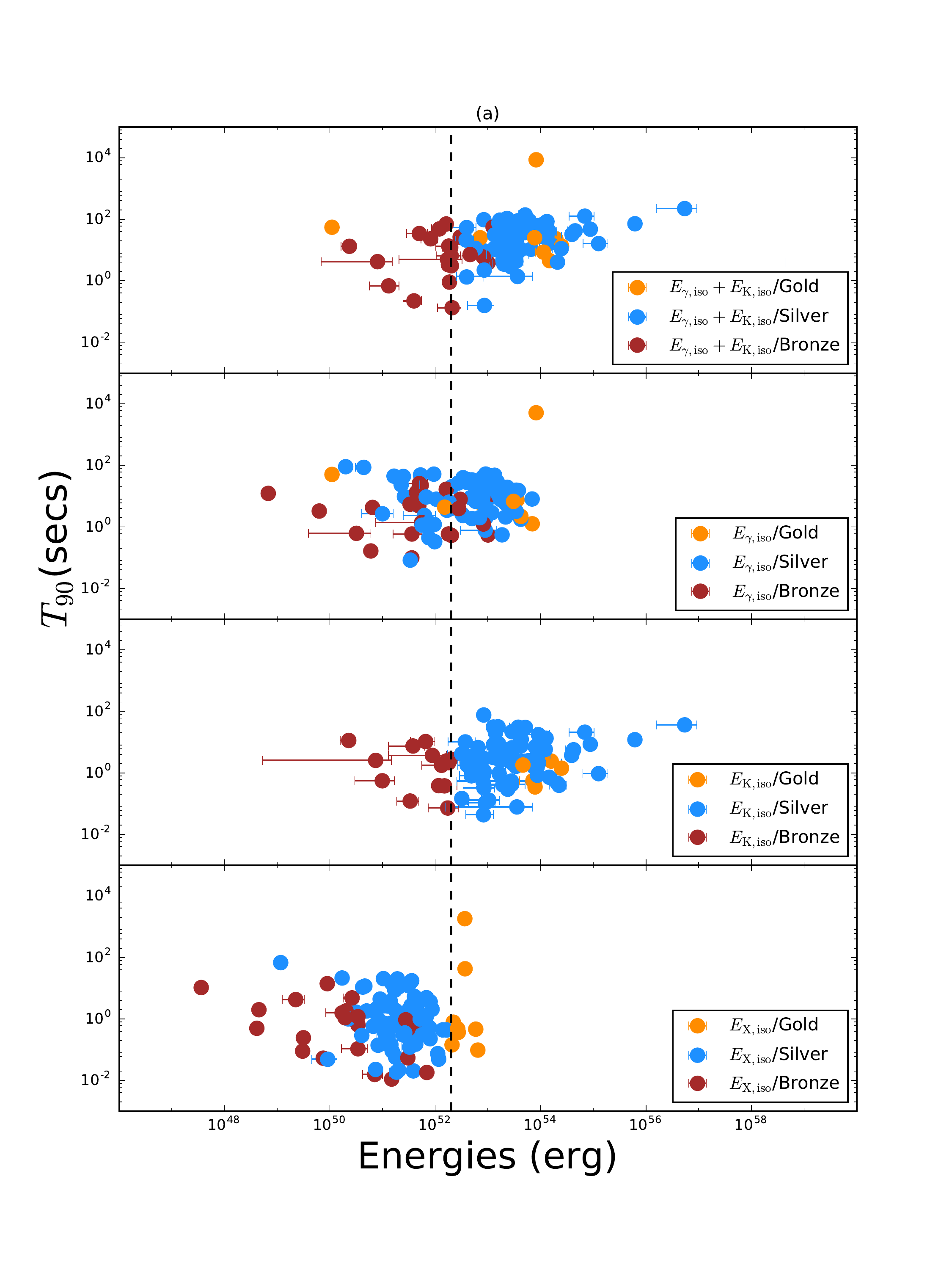}
\includegraphics[angle=0,scale=0.35]{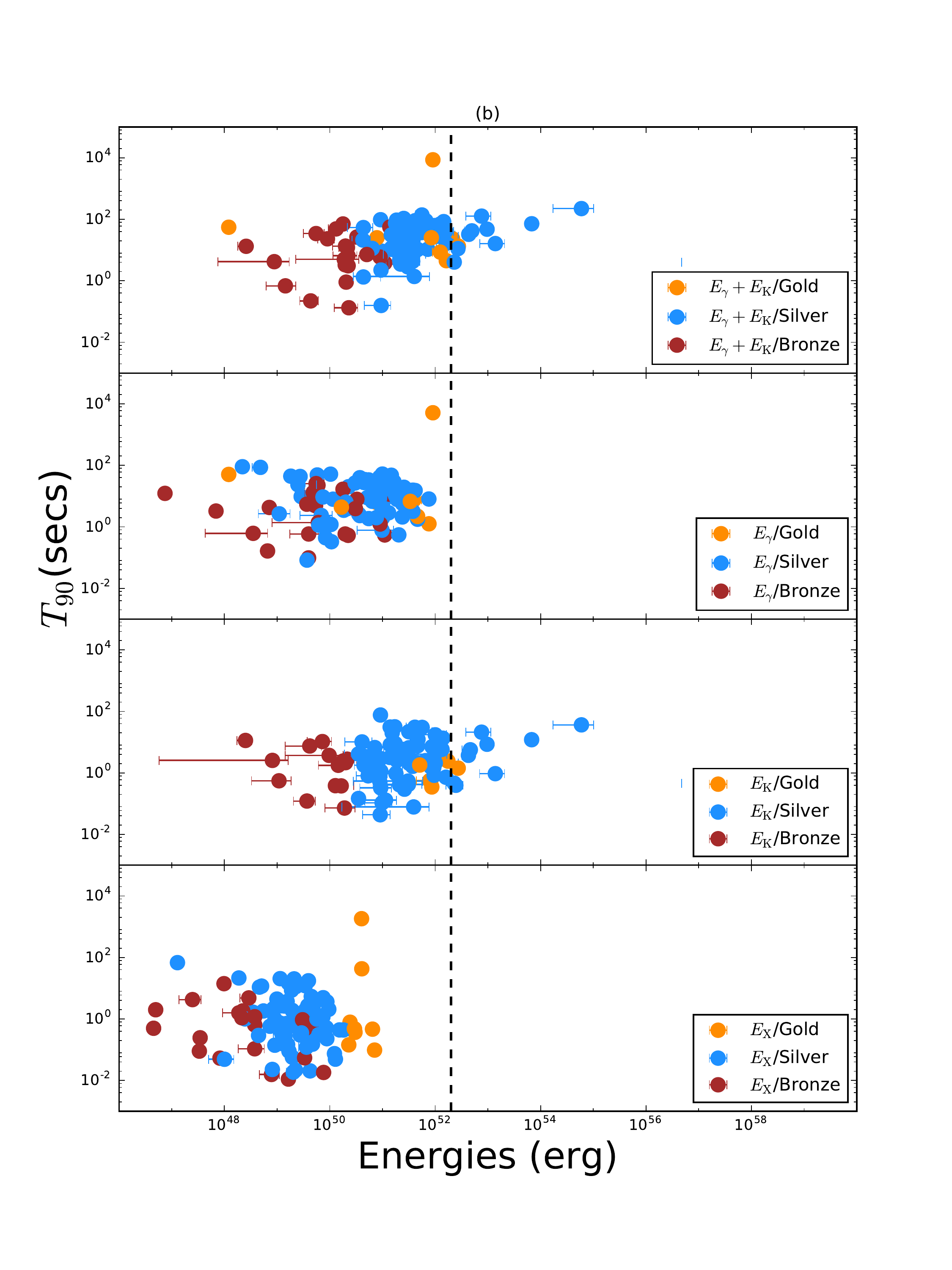}
\caption{$T_{90}$ (rest frame) as a function of energy for the Gold (orange dots), the Silver (blue dots) and the Bronze (brown dots) samples. The vertical dashed line represents energy equal to 2$\times$10$^{52}$ erg.
Left panel: all energies without a beam correction; right panel: the same as the left panel but for all energies with a beam correction.}\label{T90Energetics}
\end{figure*}

In Fig.\ref{T90Energetics}, we show the $T_{90}$ (rest frame) as a function of various energies for both the BH and magnetar candidate samples.
All isotropic energies for most cases are above 2$\times$10$^{52}$ erg while all other energies for most cases with a beam correction are below 2$\times$10$^{52}$ erg.

\begin{figure*}
\includegraphics[angle=0,scale=0.60]{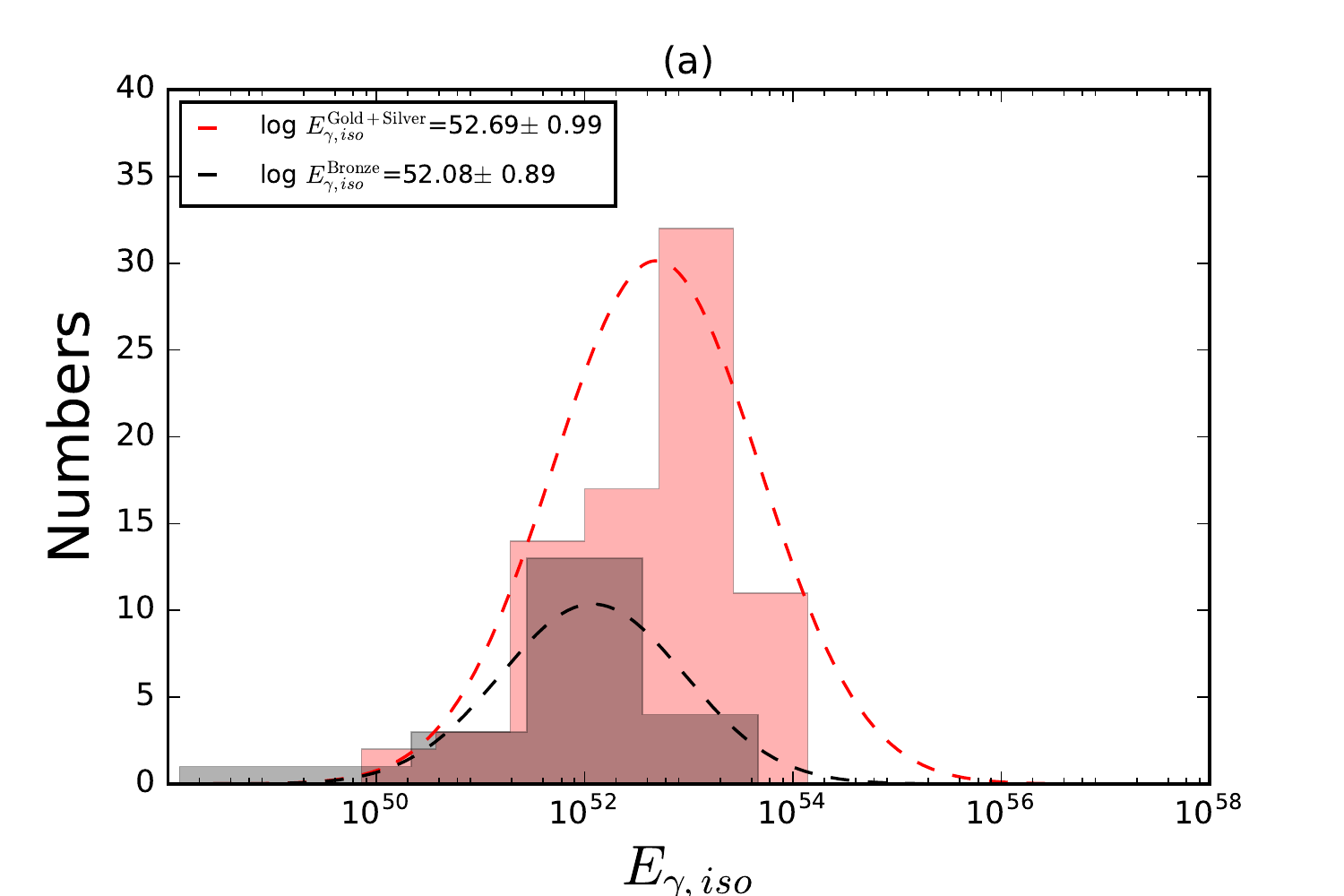}
\includegraphics[angle=0,scale=0.60]{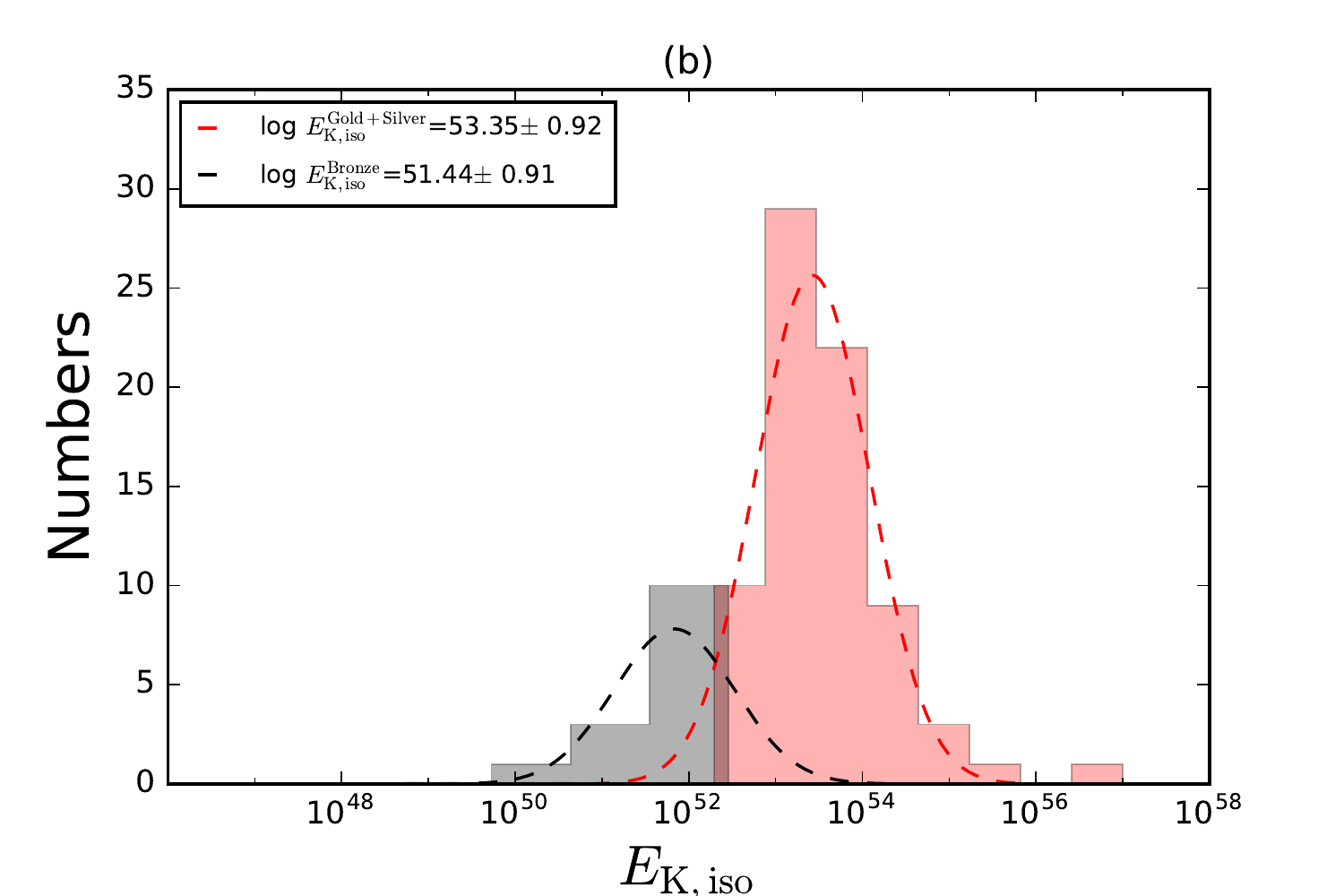}
\includegraphics[angle=0,scale=0.60]{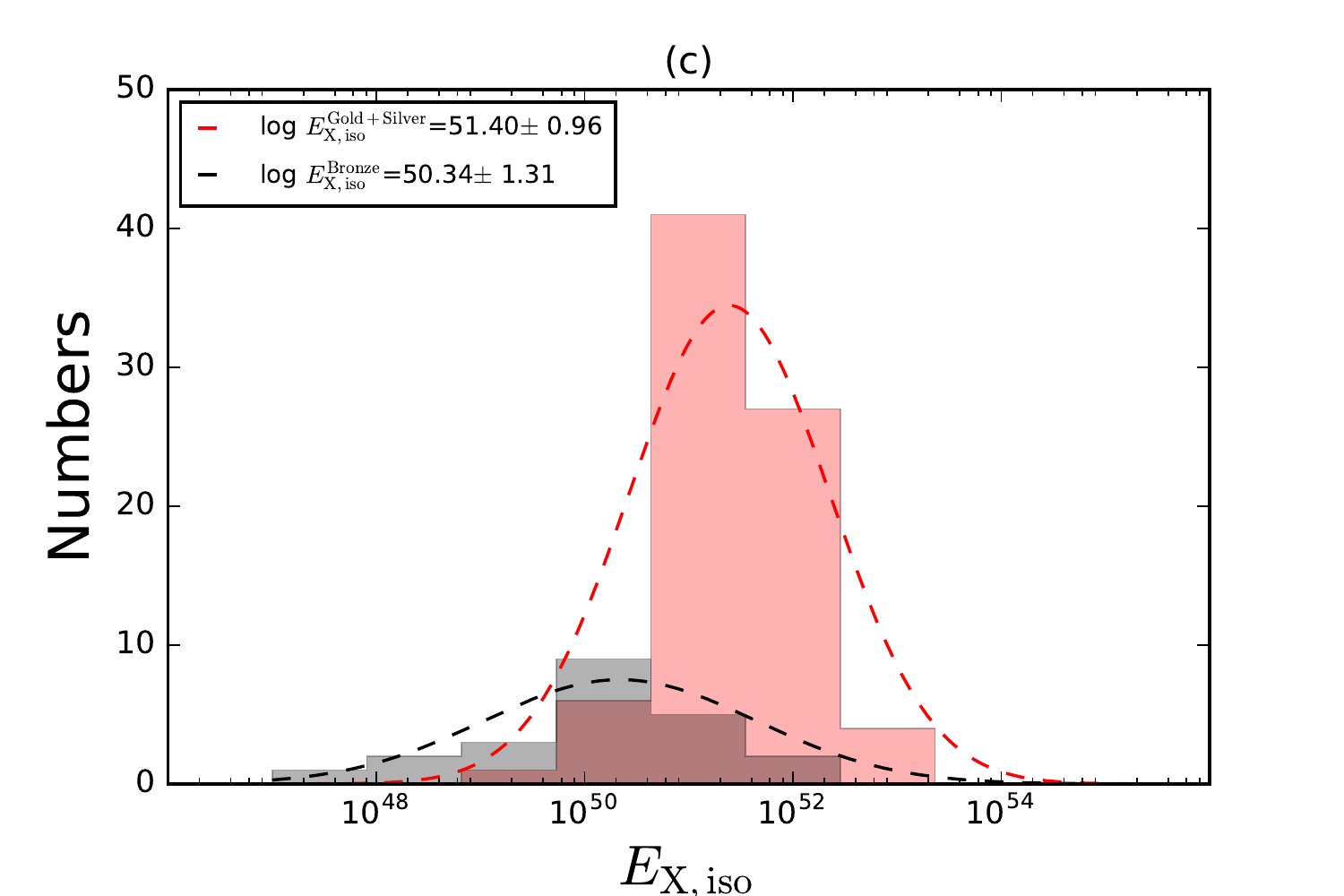}
\includegraphics[angle=0,scale=0.60]{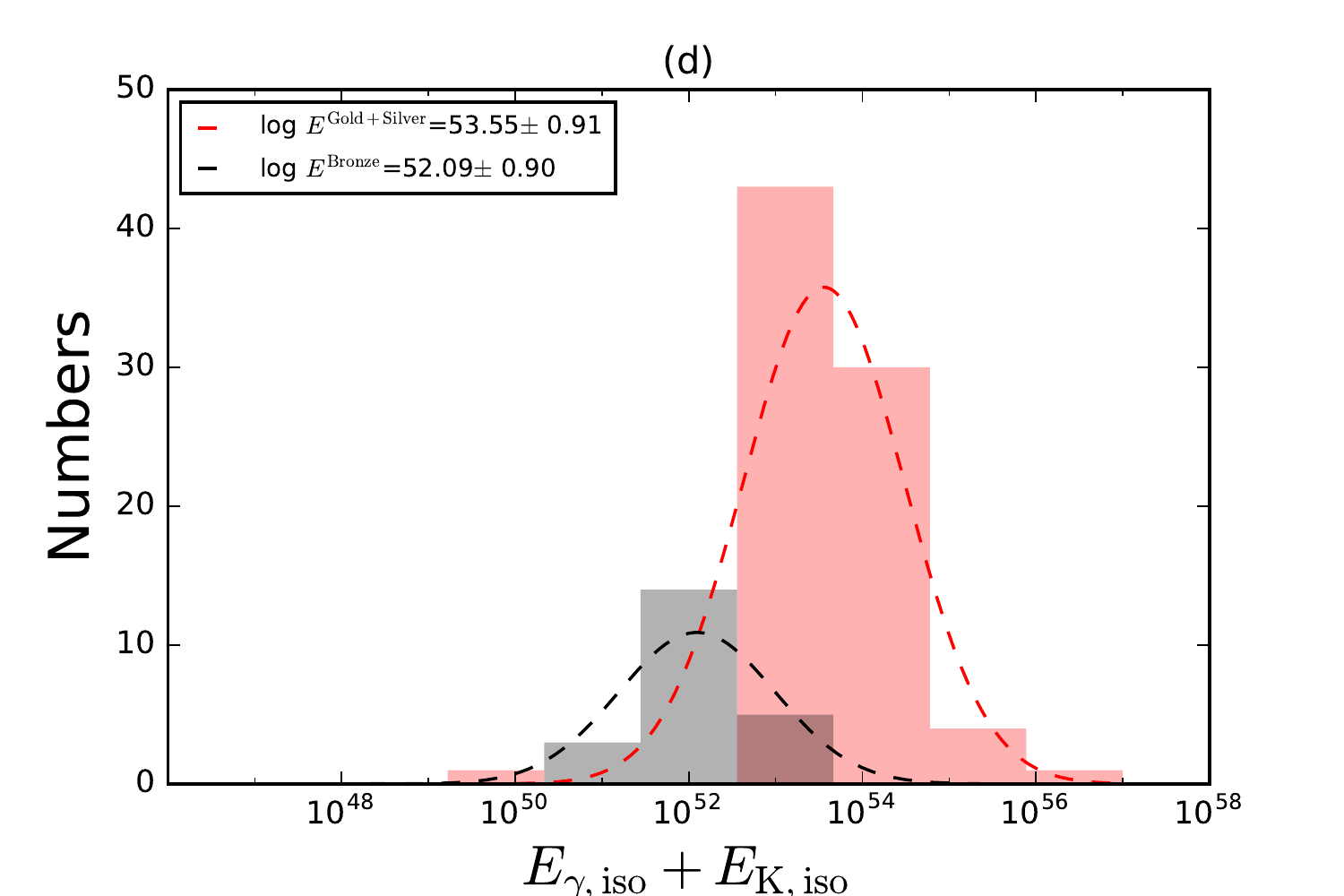}
\caption{Comparisons of the energy histograms of the BH candidate samples  (Gold+Silver) and the magnetar candidate sample (Bronze). The dashed lines represent its best Gaussian fits.} \label{EnergyDis}
\end{figure*}

In Fig.\ref{EnergyDis}, we compare the histograms of the isotropic energies ($E_{\gamma,iso}$, $E_{\rm K,iso}$, $E_{\rm X,iso}$) of the BH and magnetar candidate samples. The typical values of all kinds of energies ($E_{\gamma,\rm iso}$, $E_{K,\rm iso}$, $E_{X,\rm iso}$) for the BH candidate samples (Gold+Silver) are systemically larger than those for the magnetar sample (Bronze).

\begin{figure*}
\centering
\includegraphics[angle=0,scale=1.0]{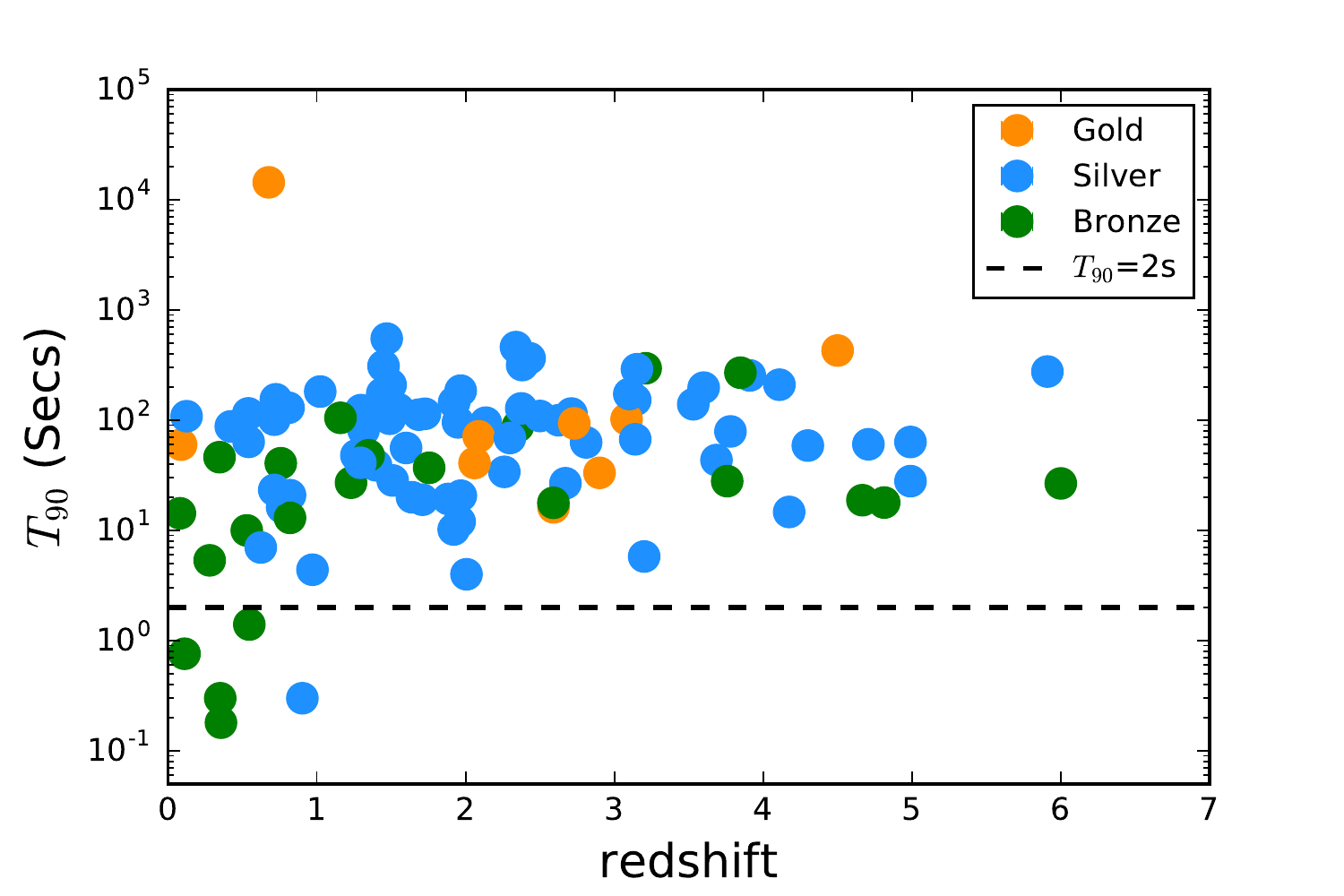}
\caption{$T_{90}$ (rest frame) as a function of redshift for the Gold (red dots), Silver (black dots) and Bronze (blue dots) samples. The vertical dashed line represents $T_{90}$ equal to 2 s.} \label{T90z}
\end{figure*}

Fig.\ref{T90z} displays the $T_{90}$ (rest frame) as a function of redshift for the samples. Two interesting findings: first, four out of the five short GRBs are in the magnetar candidate sample (Bronze); second, the redshifts of all five short bursts are below 1.

\begin{figure*}
\includegraphics[angle=0,scale=0.60]{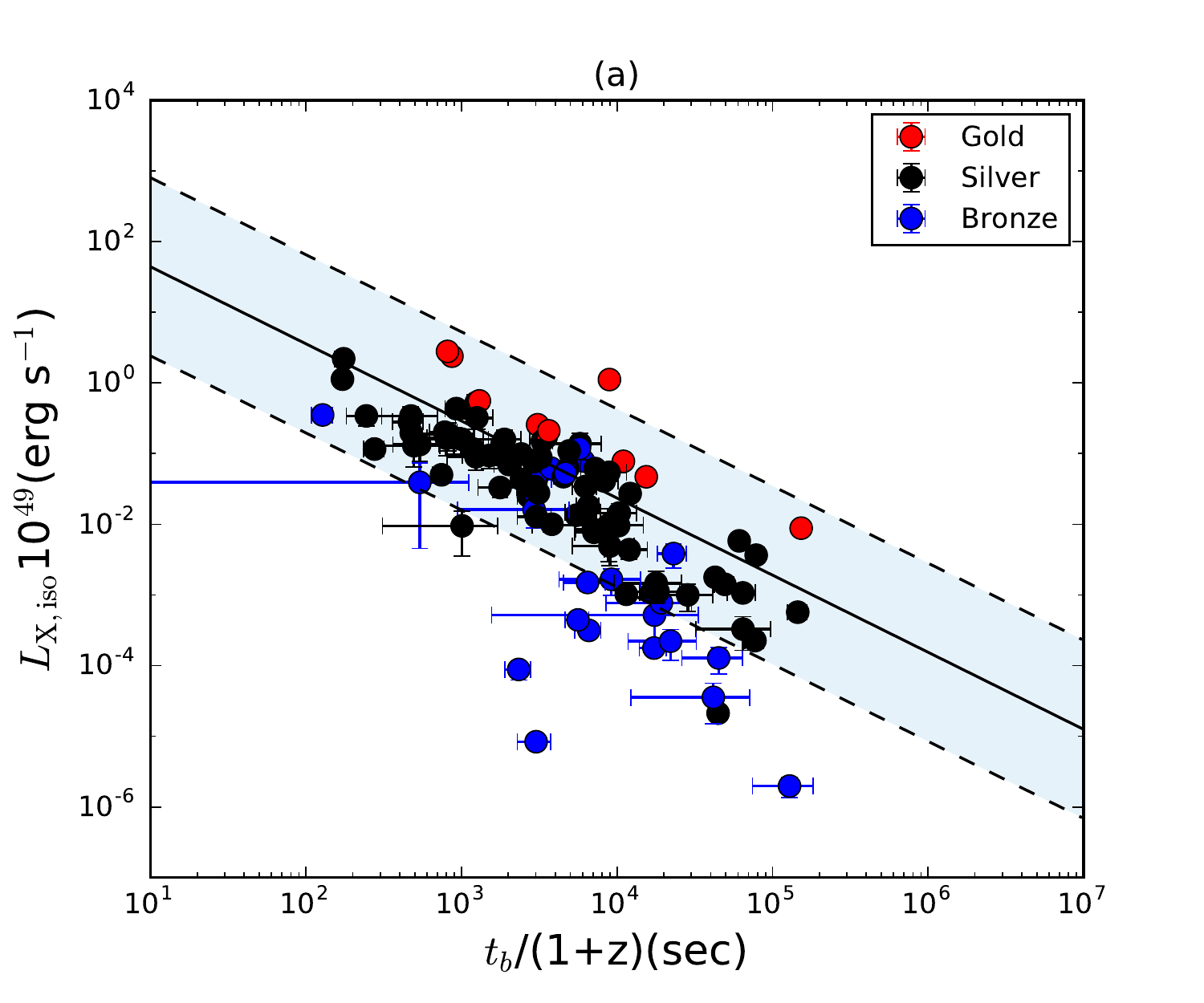}
\includegraphics[angle=0,scale=0.60]{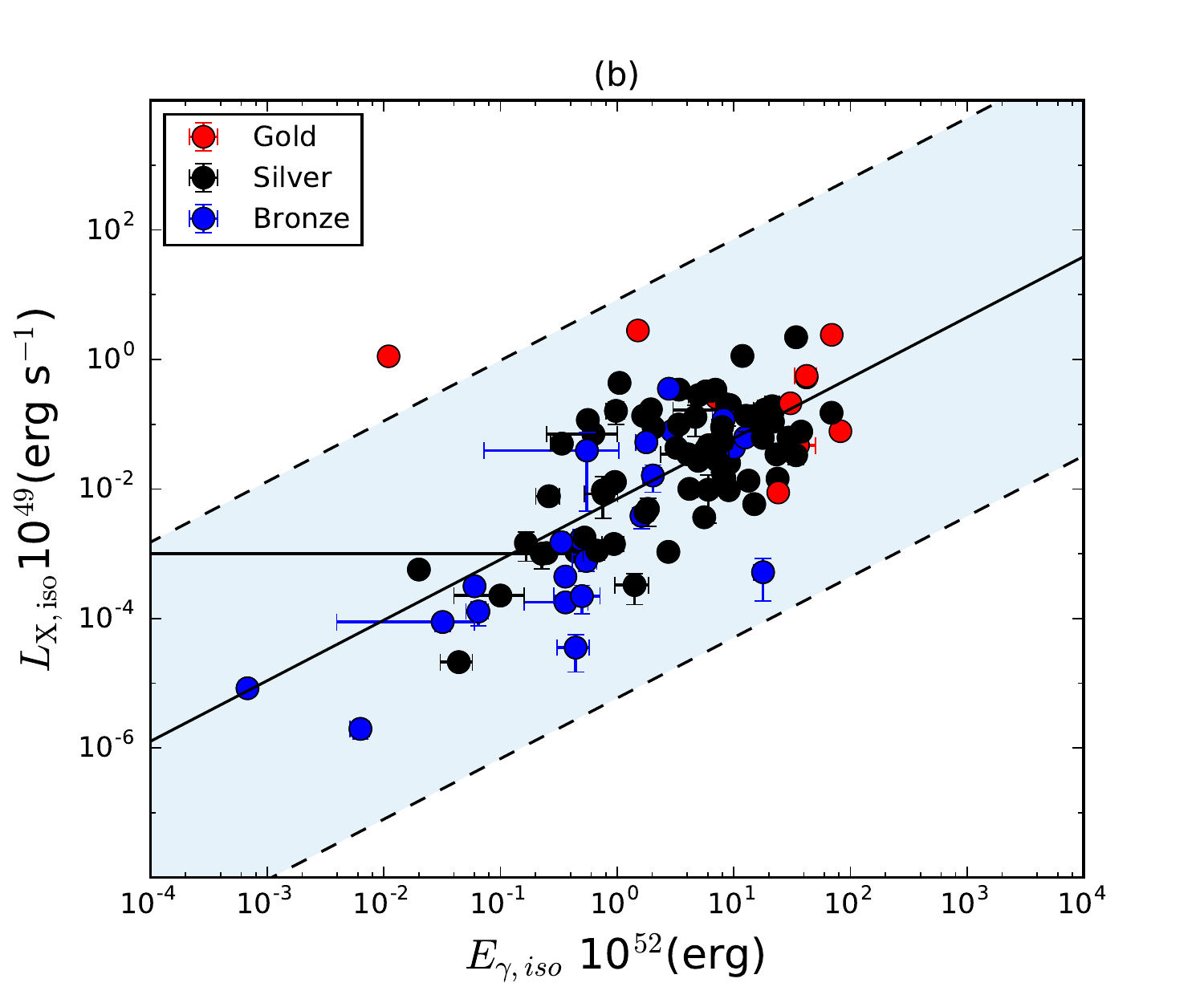}
\caption{The $L_{\rm X,iso}-t_{\rm b}/(1+z)$ and $L_{\rm X,iso}-E_{\gamma,\rm iso}$ correlations for the GRBs in various BH/magnetar samples.
The solid line is a power-law fitting to the Gold and Silver sample GRBs, and the two dashed lines denote the $2\sigma$ region of the fits. Color conventions are the same as in Fig. \ref{efficiency}.}\label{EisoLbtb}
\end{figure*}

Fig.\ref{EisoLbtb} shows the correlations of $L_{\rm X,iso}-E_{\gamma,\rm iso}$ and $L_{\rm X,iso}-t_{\rm b}$/(1+z) for the entire sample (BH candidate Gold+Silver samples and magnetar candidate Bronze sample).
We find that a higher isotropic $\gamma$-ray energy implies a higher isotropic X-ray break luminosity, and the bursts of the Gold sample cluster into the region of high isotropic $\gamma$-ray energy and isotropic X-ray break luminosity. The best fit to the correlation is $\log L_{\rm X, iso, 49}=(-1.33\pm 0.10)+(1.06\pm0.09)\log E_{\gamma, \rm iso,52}$, with a Spearman correlation coefficient of 0.78 and a chance probability of $p<10^{-4}$.
For the $L_{\rm X,iso}$-$t_{\rm b}$/(1+z) correlations, the best fit is $\log L_{\rm X, iso, 49}=(2.74\pm 0.40)+(-1.09\pm0.09)\log t_{\rm b}$/(1+z), with a Spearman correlation coefficient of 0.79 and a chance probability of $p<10^{-4}$. The anticorrelation between the luminosity and the rest-frame duration of the plateau phase is consistent with the previous findings \citep{2008MNRAS.391L..79D, 2010ApJ...722L.215D, 2013ApJ...774..157D, 2015MNRAS.451.3898D,2017A&A...600A..98D, 2011A&A...526A.121D, 2013MNRAS.430.1061R,2014MNRAS.443.1779R, 2017MNRAS.472.1152R}.

\begin{figure*}
\includegraphics[angle=0, scale=0.6]{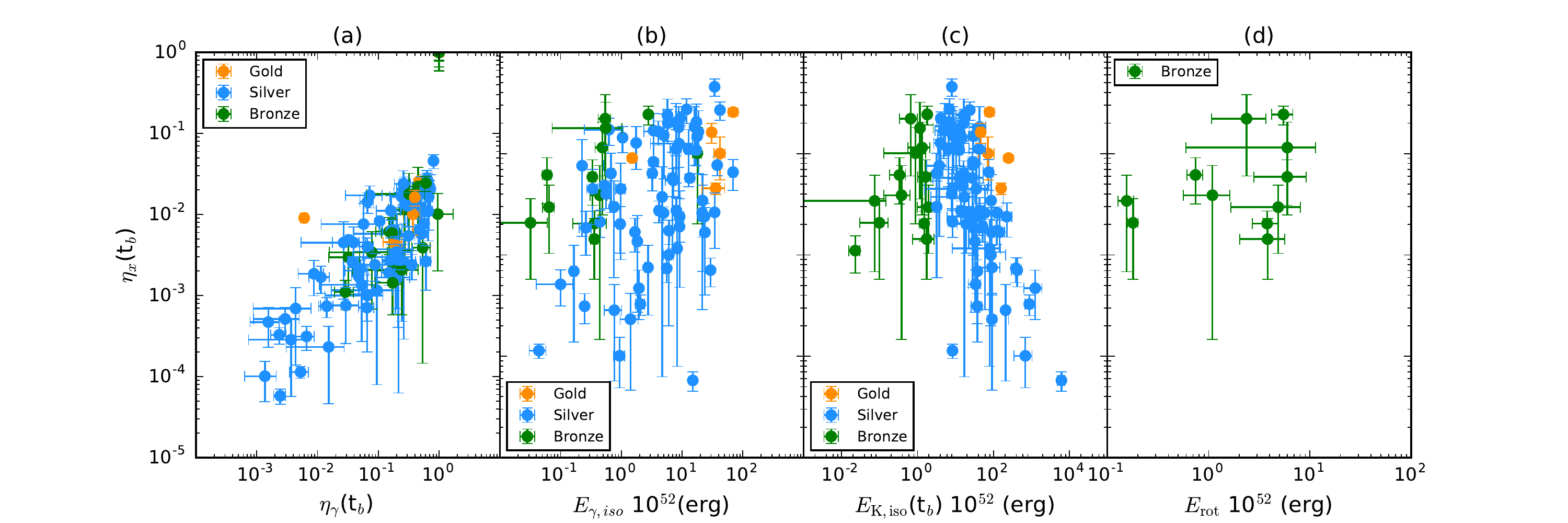}
\caption{Scatter plots of the X-ray efficiency $\eta_{\rm X}$ vs. several parameters: $\eta_{\gamma}(t_{\rm b})$, $E_{\gamma,\rm iso}$, $E_{\rm K,iso}$, and $E_{\rm rot}$. Different colors represent different samples.}\label{pairrelation}
\end{figure*}

In Fig.\ref{pairrelation}, we show the scatter plots of $\eta_{\rm X}$ versus $\eta_{\gamma}(t_{\rm b})$, $E_{\gamma, \rm iso}$, $E_{\rm K,iso}(t_{\rm b})$, and $E_{\rm rot}$ (Bronze). One interesting finding is the $\eta_{\rm X}$ and $E_{\rm K,iso}(t_{\rm b})$ presenting an anticorrelation for the BH candidate samples (Gold+Silver) while presenting a positive correlation for the magnetar candidate sample (Bronze).

\section{Comparison of Statistical Characters}

Since most of the bursts in our sample have an external plateau, this indicates that our GRB samples conform well with the external shock models, and they offer an excellent sample to study external shock model parameters. Here we investigate whether there are noticeable differences between the BH candidate (Gold+Silver) and magnetar candidate (Bronze) bursts for the observation properties and model parameters.

In addition, we also investigate the external plateau and the internal plateaus in our bursts.

\subsection{Temporal indices $\alpha$}

\begin{figure*}
\includegraphics[angle=0,scale=0.50]{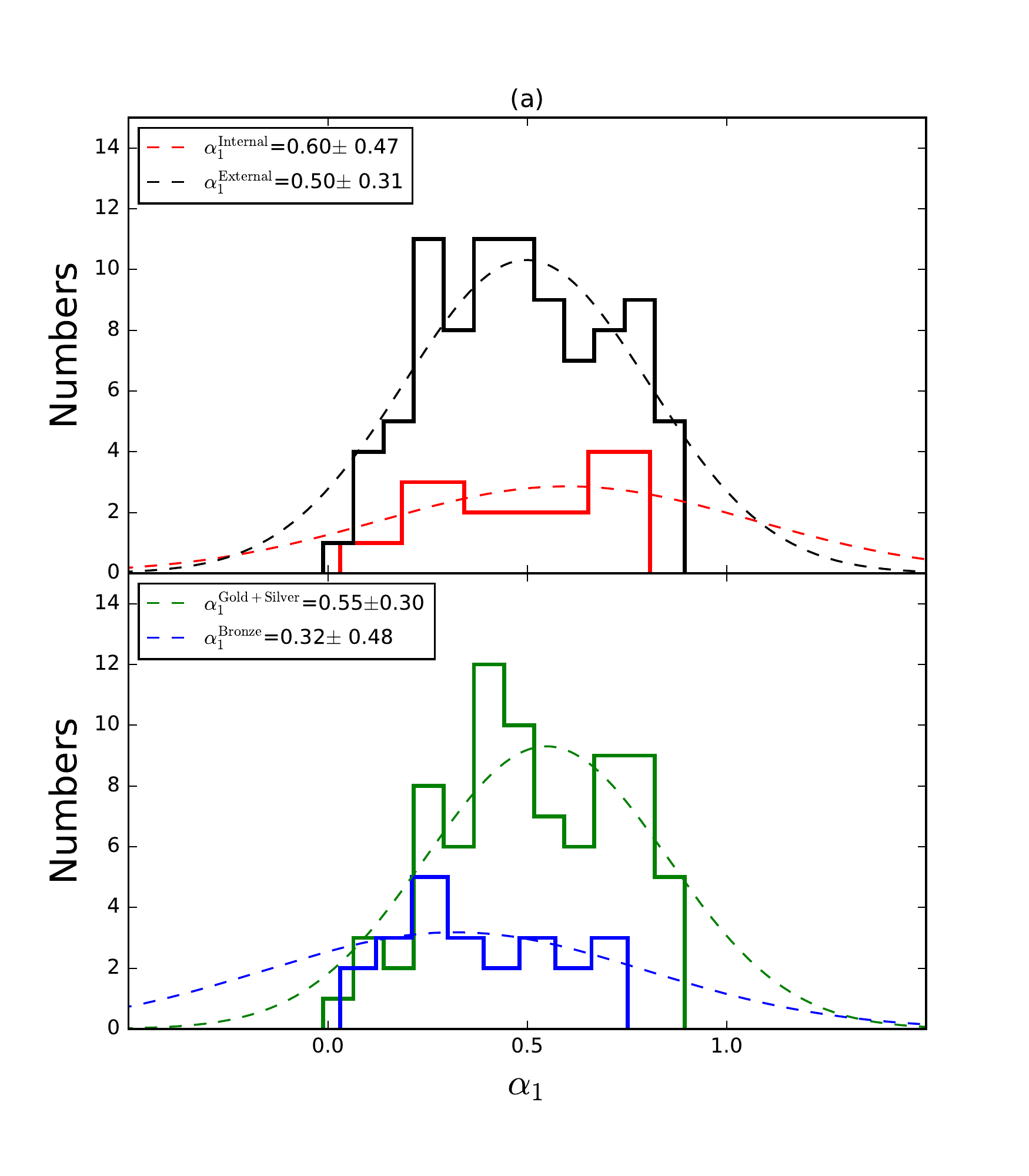}
\includegraphics[angle=0,scale=0.50]{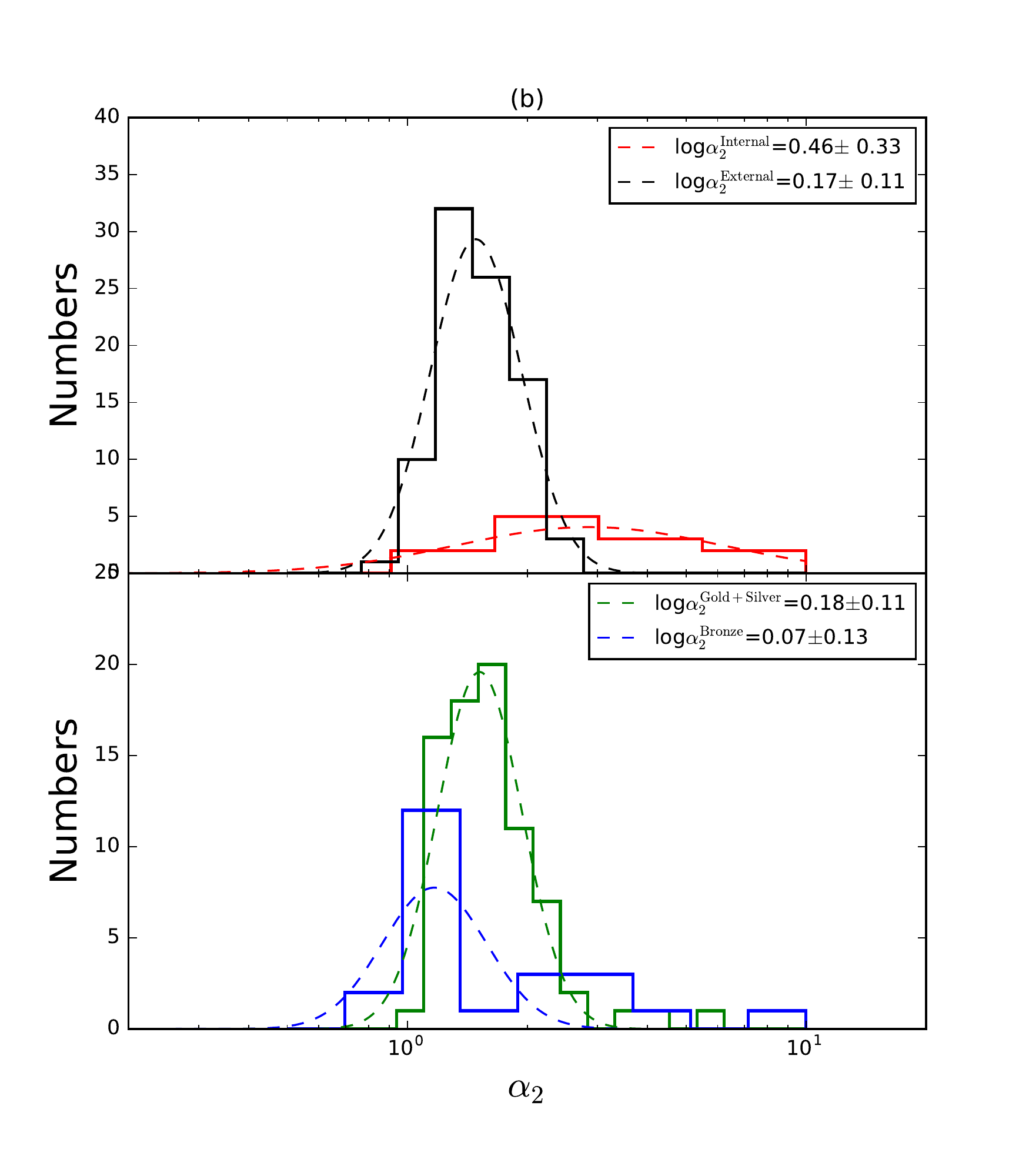}
\includegraphics[angle=0,scale=0.50]{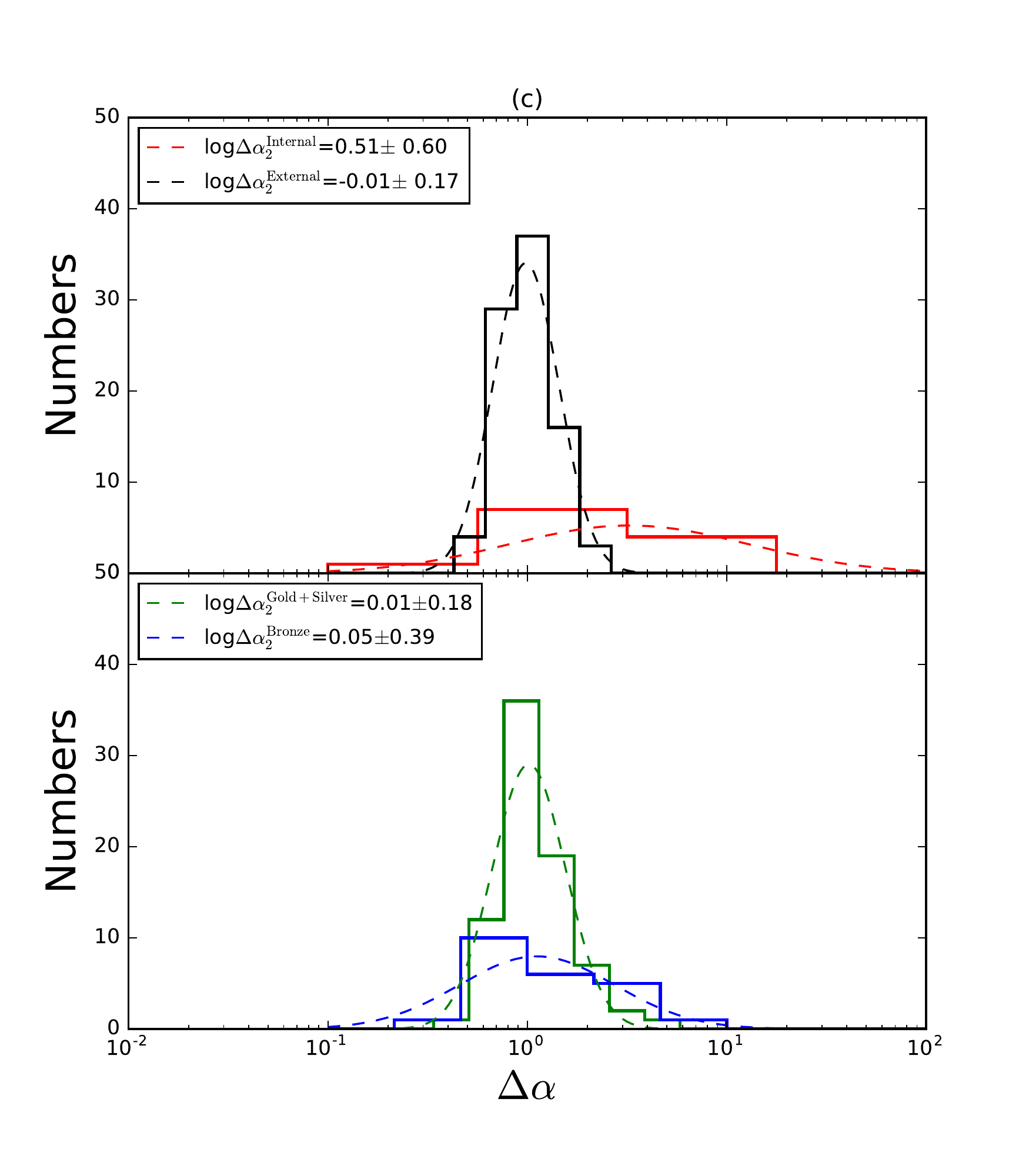}
\caption{Distributions (solid lines) of temporal indices for the subsamples and their best Gaussian fits (dashed lines). For each figure, the top panel compares the internal and external plateau samples, and the bottom panel compares the BH candidate samples (Gold+Silver) and magnetar candidate sample (Bronze). Different colors represent different subsamples. (a) $\alpha_{1}$ distributions, (b) $\alpha_{2}$ distributions, and (c) $\Delta\alpha$ distributions.}\label{alpha}
\end{figure*}

Fig.\ref{alpha} displays the distributions of the temporal indices $\alpha$ in different subsamples and different temporal segments. They are all well fitted with Gaussian distributions for each sample.
For the $\alpha_{\rm x,1}$ distributions, we have $\alpha_{\rm x,1}$=0.55$\pm$0.30 for the BH candidate samples and $\alpha_{\rm x,1}$=0.32$\pm$0.48 for the magnetar candidate sample.
The same analysis based on the type of plateau gives $\alpha_{\rm x,1}$=0.60$\pm$0.47 for the internal plateau sample, $\alpha_{\rm x,1}$=0.50$\pm$0.31 for the external plateau ones (Fig.\ref{alpha}a).
For the $\alpha_{\rm x,2}$ distributions, we have log$\alpha_{\rm x,2}$=0.18$\pm$0.11 for the BH candidate samples and log$\alpha_{\rm x,2}$=0.07$\pm$0.13 for the magnetar candidate samples.
For the internal/external plateau samples, one has log$\alpha_{\rm x,2}$=0.46$\pm$0.33 for the internal plateau sample and log$\alpha_{\rm x,2}$=0.17$\pm$0.11 for the external plateau sample (Fig.\ref{alpha}b).

Another self-consistency check is to compare the observed change of the decay slope ($\Delta\alpha_{\rm x}=\alpha_{\rm x,2}-\alpha_{\rm x,1}$).
We find log $\Delta\alpha_{\rm x}$=0.01$\pm$0.18 for the BH candidate samples and log $\Delta\alpha_{\rm x}$=0.05$\pm$0.39 for the magnetar candidate, and log$\Delta\alpha_{\rm x}$=0.51$\pm$0.60 for the internal plateau sample and log$\Delta\alpha_{\rm x}$=-0.01$\pm$0.17 for the external plateau sample (Fig.\ref{alpha}c).

The temporal indices (both $\alpha_{1}$ and $\alpha_{2}$) for the BH candidate samples (Gold+Silver), in general, are steeper than those for the magnetar sample (Bronze). For the internal and external examples, the prebreak temporal indices $\alpha_{1}$ are consistent with each other, but the postbreak temporal indices $\alpha_{2}$ for the internal sample is much steeper than those for the external sample, consistent with an internal origin. The degrees of the break $\Delta \alpha$ for the BH candidate samples (Gold+Silver) is significantly less than those of the Silver and the magnetar samples (Bronze).

\subsection{Photon spectral indices $\Gamma$}

\begin{figure*}
\includegraphics[angle=0,scale=0.50]{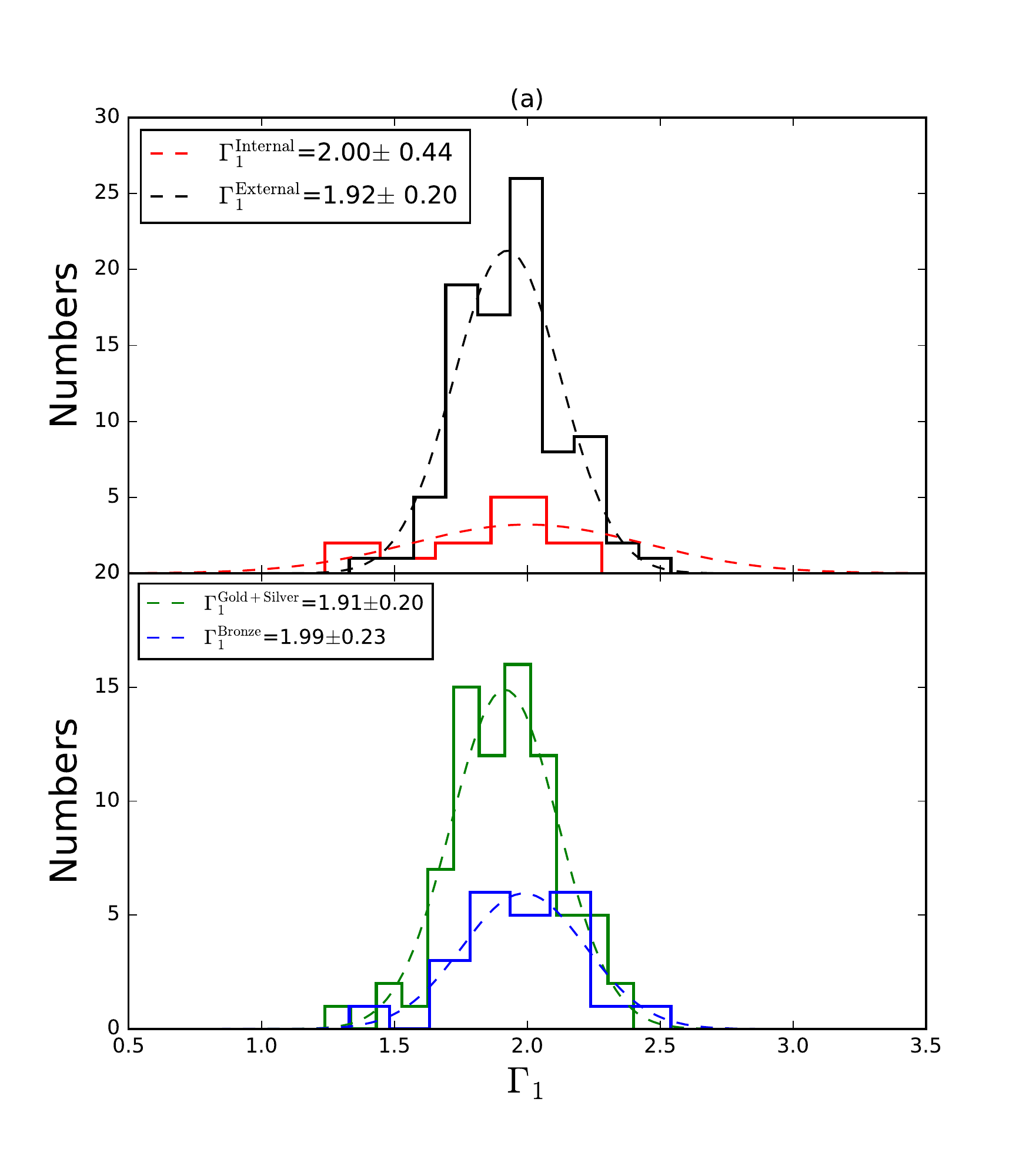}
\includegraphics[angle=0,scale=0.50]{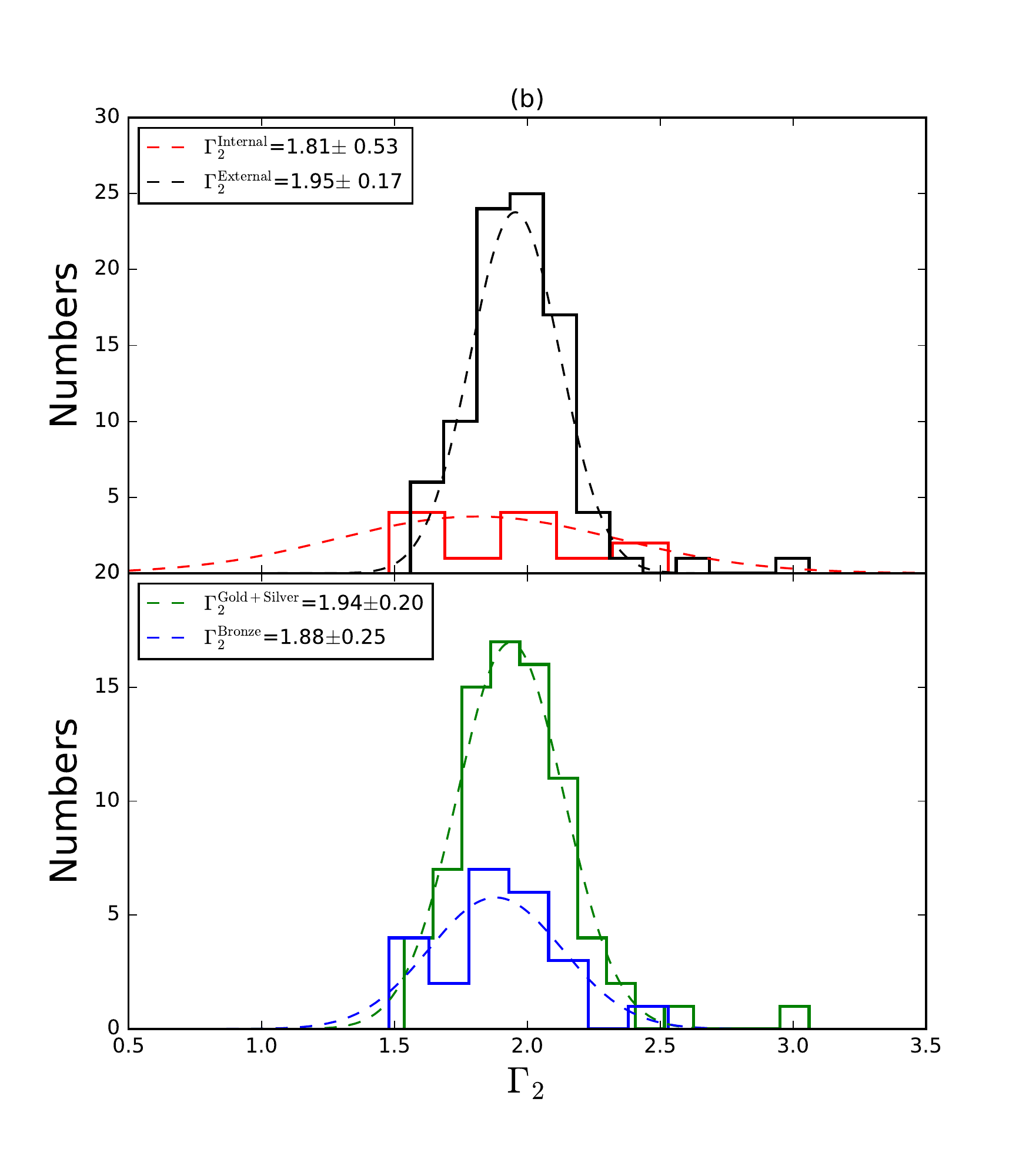}
\includegraphics[angle=0,scale=0.50]{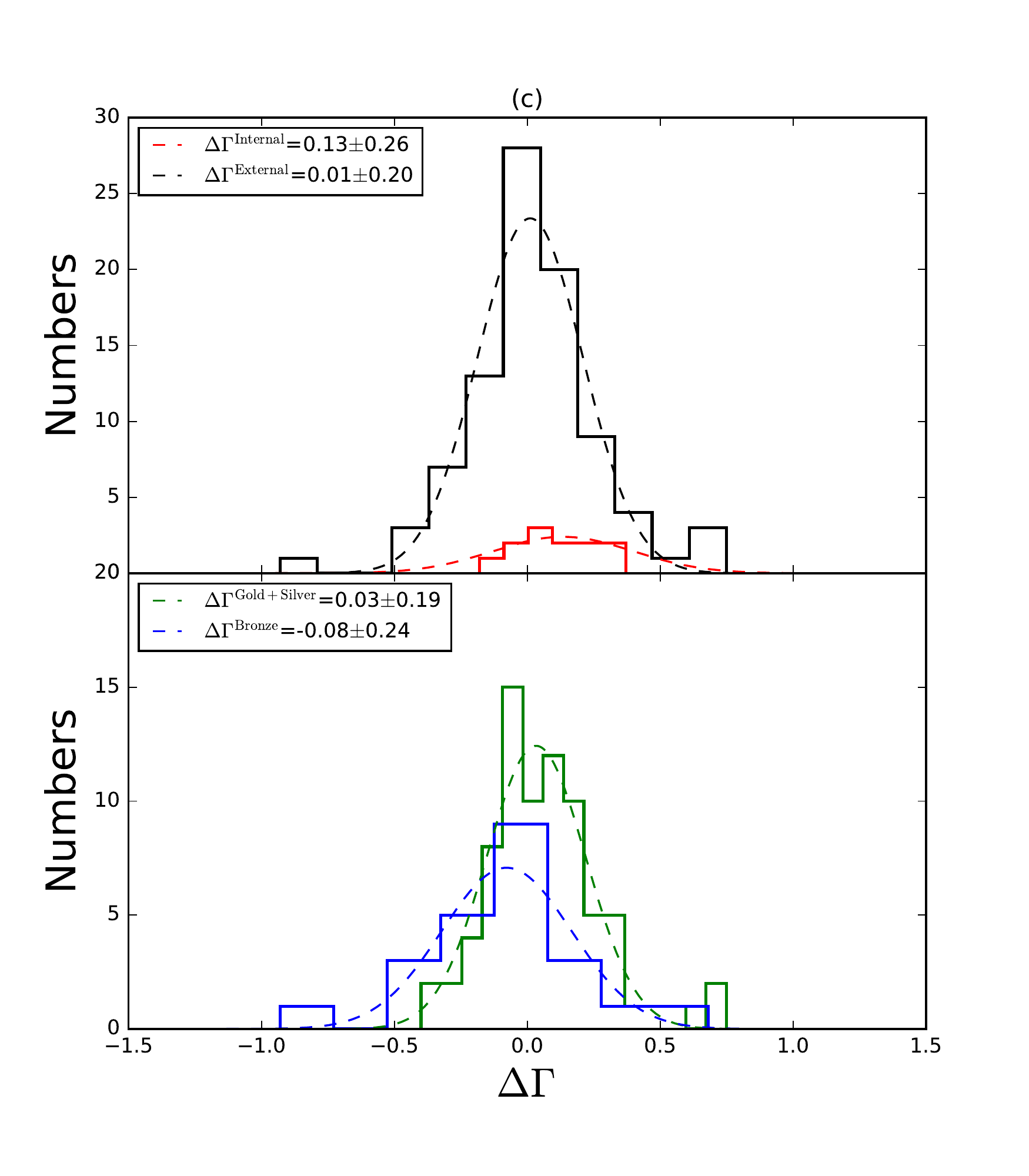}
\caption{Distributions (solid lines) of the photon spectral indices for the subsamples and their best Gaussian fits (dashed lines). The same symbol as Fig.\ref{alpha} but for the photon spectral index ($\Gamma$) distributions. (a) $\Gamma_{1}$ distributions, (b) $\Gamma_{2}$ distributions, and (c) $\Delta\Gamma$ distributions.}\label{beta}
\end{figure*}

Fig.\ref{beta} shows the distributions of the photon spectral indices $\Gamma$ in different subsamples and different plateau types. They are all well-fitted with Gaussian distributions for each sample.

For the $\Gamma_{\rm x,1}$ distributions, we have $\Gamma_{\rm x,1}=1.91\pm0.20$ for the BH candidate samples and $\Gamma_{\rm x,1}=1.99\pm0.23$ for the magnetar candidate sample (Fig.\ref{beta}a).
For the $\Gamma_{\rm x,2}$ distributions, one has $\Gamma_{\rm x,2}=1.94\pm0.20$ for the BH candidate samples and $\Gamma_{\rm x,2}=1.88\pm0.25$ for the magnetar candidate sample (Fig.\ref{beta}b).
For the $\Delta\Gamma_{\rm x}$ distributions, one has $\Delta\Gamma_{\rm x}=0.03\pm0.19$ for the BH candidate samples and log $\Delta\Gamma_{\rm x}=-0.08\pm0.24$ for the magnetar candidate sample (Fig.\ref{beta}c).

We also investigate the $\Gamma$ distributions in different plateau types.
For the internal plateau samples, one has $\Gamma_{\rm x,1}$=2.00$\pm$0.44, $\Gamma_{\rm x,2}$=1.81$\pm$0.53, and $\Delta\Gamma_{\rm x}$=0.13$\pm$0.26, respectively.
For the external plateau samples, one has $\Gamma_{\rm x,1}$=1.92$\pm$0.20, $\Gamma_{\rm x,2}$=1.95$\pm$0.17, and $\Delta\Gamma_{\rm x}$=0.01$\pm$0.20, respectively.

The spectral indices (both $\Gamma_{1}$ and $\Gamma_{2}$), and the degrees of the break $\Delta \Gamma$ for the different subsamples (the BH candidate samples and the magnetar candidate sample), in general, present the same distributions.
For the internal and external samples, a change is not found in the external sample, while a significant change in the internal sample is found.

The results indicate that there is no spectral evolution for the subsamples and the external plateau sample, while the internal plateau sample has a significant spectral change between a plateau phase and the following steeper decay phase. This is consistent with the internal plateau having an internal physical origin.

\subsection{Break time $t_{\rm b}$}

\begin{figure*}
\centering
\includegraphics[angle=0,scale=0.60]{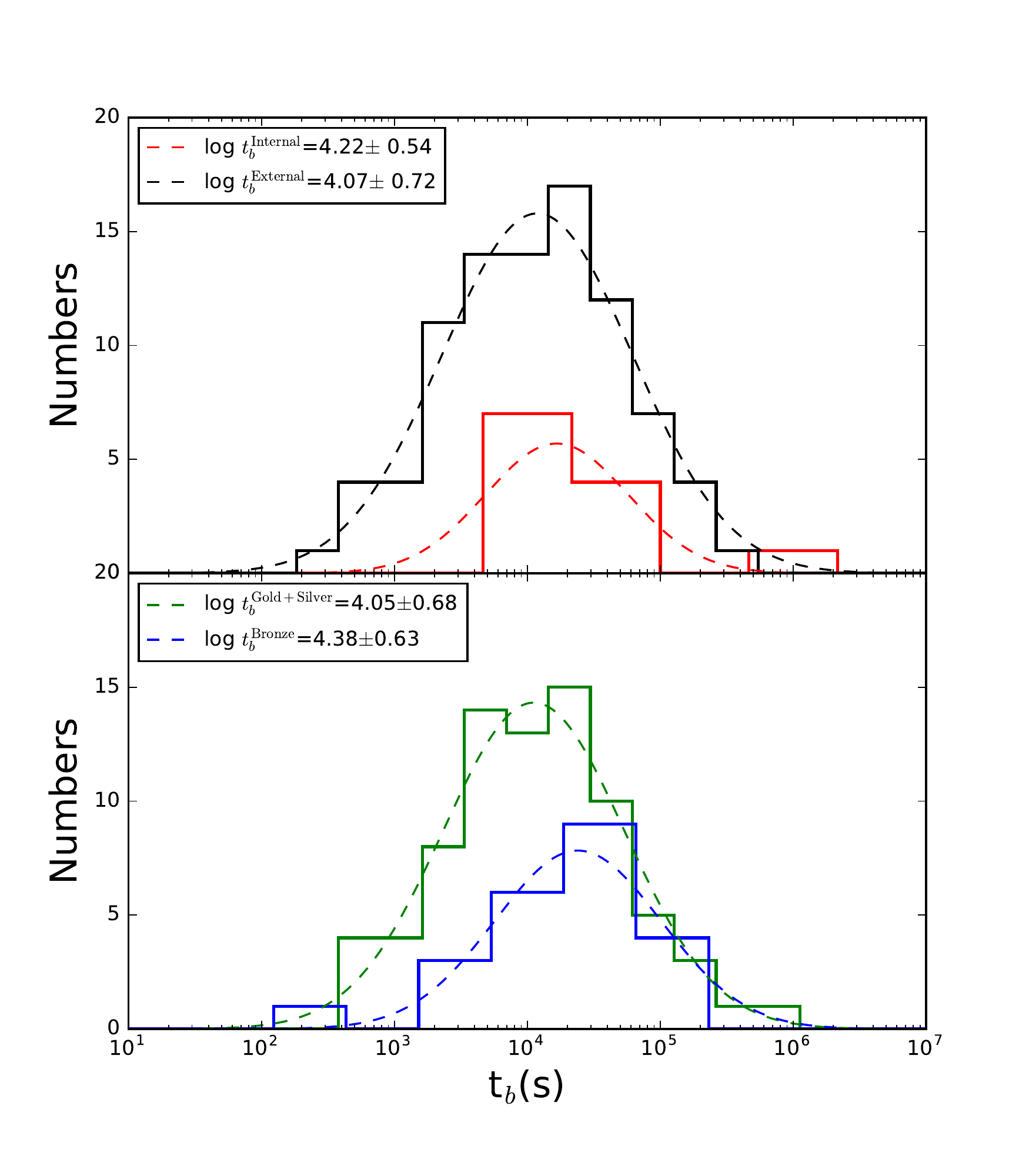}
\caption{Distributions (solid lines) of the break time ($t_{\rm b}$) for the subsamples and their best Gaussian fits (dashed lines). The same symbols as in Figure \ref{alpha} are used but for the break time ($t_{\rm b}$) distributions.}\label{breaktime}
\end{figure*}

Fig.\ref{breaktime} shows the distributions of the observed break times ($t_{\rm b}$). A Gaussian fit to the distributions gives log($t_{\rm b}$/s)=(4.05$\pm$0.68) and log($t_{\rm b}$/s)=(4.38$\pm$0.63) for the BH candidate samples and the magnetar candidate sample, respectively. Separating the "internal" plateau and "external" plateau samples, we have log($t_{\rm b}$/s)=(4.22$\pm$0.54) for the "internal" plateau sample and log($t_{\rm b}$/s)=(4.07$\pm$0.72) for the "external" plateau sample. So, both external and internal plateau samples are consistent in terms of $t_{\rm b}$.

The typical break time for the  BH candidate samples (Gold+Silver) are statistically earlier than that of the magnetar candidate sample (Bronze). Similar results can be found between the internal (earlier) and external (later) samples. The results indicate that the end energy injection for the "external plateau" is in general earlier than that for the "internal plateau" on average.

\subsection{Electron spectral index $p$}

We derive the electron spectral indices {\it p}, derived using the temporal indices:
\begin{eqnarray}
\label{n} p=\left\{ \begin{array}{ll}
\frac{4\alpha_{2}+2}{3}, &  \nu>max{(\nu_{m},\nu_{c})}  \textrm{(ISM and wind)}\\
\frac{4\alpha_{2}+3}{3},   &  \nu_{m}<\nu<\nu_{c} \textrm{(ISM)}\\
\frac{4\alpha_{2}+1}{3}, &  \nu_{m}<\nu<\nu_{c} \textrm{(wind)}\\
\end{array} \right.
\end{eqnarray}

\begin{figure*}
\centering
\includegraphics[angle=0,scale=0.8]{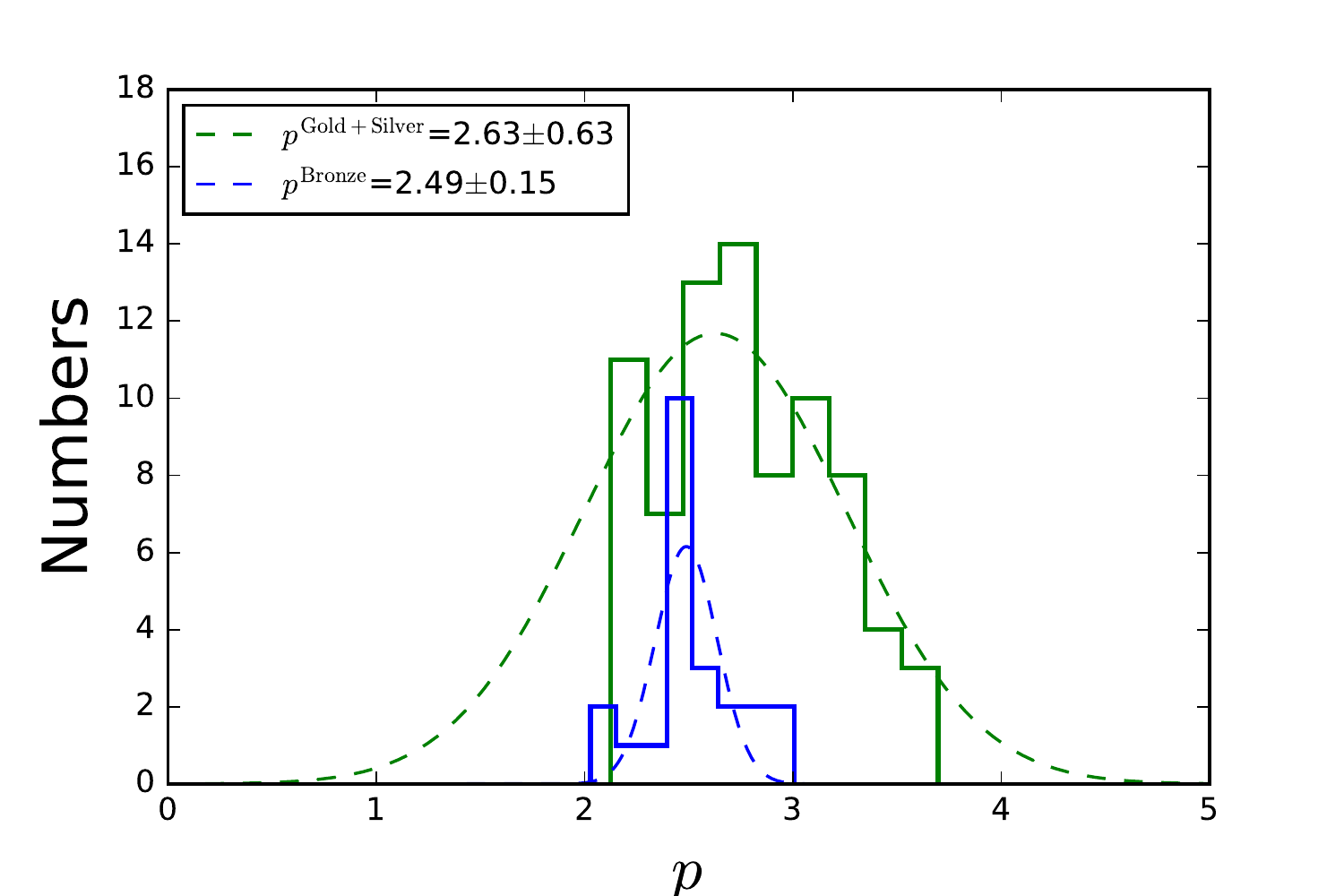}
\caption{Distributions (solid lines) of the electron spectral index ($p$) for the subsamples and their best Gaussian fits (dashed lines). The $p$ values are derived from the "external" plateau in the normal decay phase. The same symbols as in Figure \ref{alpha} are used but for the electron spectral index ($p$) distributions.}\label{p}
\end{figure*}

The {\it p} distributions for our subsamples are $p=2.63\pm0.63$
for the BH candidate samples and $p=2.49\pm0.15$ for the magnetar candidate sample (Fig.\ref{p}).
The global {\it p} value is within the range [1.89, 3.70], and it has a Gaussian distribution with a centre value $p=2.69\pm0.05$. Except for 11 GRBs that have internal plateaus, all the rest have $p>2.0$, except for one case (GRB 100302A) which had $p<2.0$ (Fig.\ref{p}).

The results are consistent with the typical value of {\it p} for relativistic shocks due to first-order Fermi acceleration \citep[e.g.,][]{2001MNRAS.328..393A,2002APh....18..213E}.

\subsection{Energy Injection Parameter {\it q}}

Similarly, the luminosity injection index \textit{q} (see Equation. \ref{eq:q}) can be determined from the temporal index $\alpha$ and the spectral index $\beta$ for different afterglow models; here, $\alpha$ and $\beta$ are measured in the shallow decay segment preceding the break:
\begin{eqnarray}
\label{n} q=\left\{ \begin{array}{ll}
\frac{2\alpha_{1}-2\beta_{1}+2}{1+\beta_{1}}, &  \nu>max{(\nu_{m},\nu_{c})}  \textrm{(ISM and wind)}\\
\frac{2\alpha_{1}-2\beta_{1}+2}{2+\beta_{1}}, &  \nu_{m}<\nu<\nu_{c} \textrm{(ISM)}\\
\frac{2\alpha_{1}-2\beta_{1}}{1+\beta_{1}},   &  \nu_{m}<\nu<\nu_{c}  \textrm{(wind)}\\
\end{array} \right.
\end{eqnarray}

\begin{figure*}
\centering
\includegraphics[angle=0,scale=0.6]{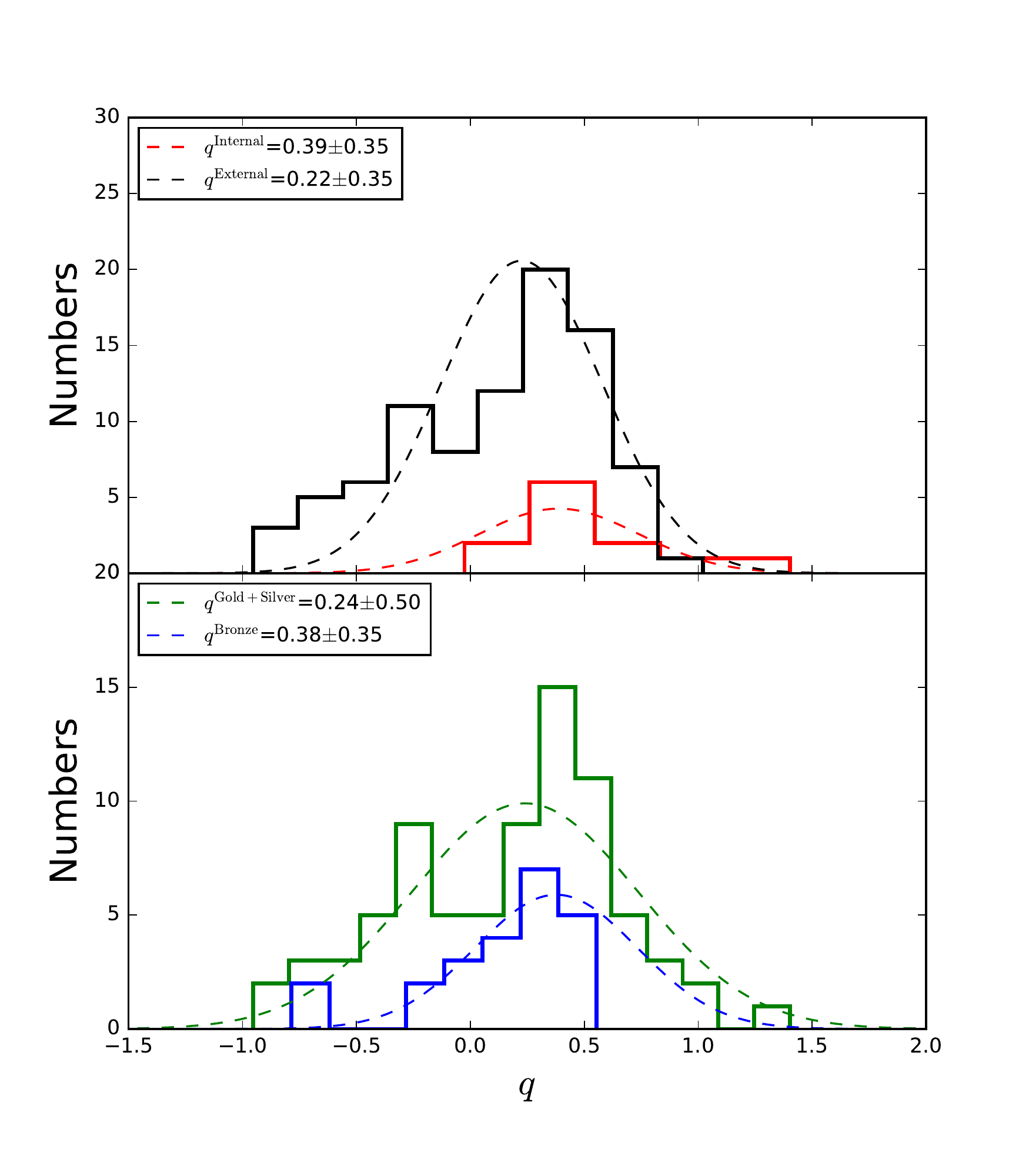}
\caption{Distributions (solid lines) of luminosity injection index $q$ for the sub-samples and their best Gaussian fits (dashed lines). $q$ values are derived from in a shallow decay segment. The same symbol as in Figure \ref{alpha} are used but for the luminosity injection index $q$ distributions.}\label{q}
\end{figure*}

The $q$ distributions for our subsamples are $q=0.24\pm0.50$ for the BH candidate samples, and $q=0.38\pm0.35$ for the magnetar candidate sample (Fig.\ref{q}).
Also, for the internal plateau, we have $q=0.39\pm0.35$, and for the external plateau, we have $q=0.22\pm0.35$.

The different subsamples present different central values; the typical value for the BH candidate samples (Gold+Silver) is slightly less than for the magnetar candidate sample (Bronze). The typical {\it q} value for the internal sample is greater than the external sample. Totally, the central {\it q} value for subsamples is around 0.3, which is consistent with the predictions of an energy injection model \citep{2006ApJ...642..354Z}.

\section{Discussions and Conclusions}

In the analysis above, several assumptions have been made, which, if relaxed, make the magnetar energy budget even more strained. For instance, the radiative efficiency is unknown but could indeed be low (e.g. \citealt{2017MNRAS.472.3058B}). Therefore, the limiting radiative energy for a magnetar burst could be only a fraction of the maximum allowed rotational energy. Moreover, the energy estimated from the X-ray afterglows could underestimate the true energy if they are strongly suppressed by Compton cooling \citep{2015MNRAS.454.1073B}.
On the other hand, the total energy from a magnetar GRB should come from the rotational energy of the magnetar itself. Except for the kinetic energy $E_{\rm K,iso}$ ($t_{\rm b}$), which is measured at the end of the plateau phase that is considered in this paper, the energy of the GRB itself ($E_{\rm \gamma,iso}$) including GeV emission should also be considered and should be added to the total energy; therefore, it could cause some GRBs in our magnetar sample to exceed the maximum energy allowed by the magnetar.
The main assumption made is, however, on the sphericity of the magnetar wind, which we discuss in the next section.

Furthermore, the related detailed studies have been done by many authors \citep[e.g.,][]{2011A&A...526A.121D, 2013MNRAS.430.1061R,2014MNRAS.443.1779R, 2017MNRAS.472.1152R, 2015ApJ...813...92R}. For instance, \cite{2011A&A...526A.121D} considered long GRBs to be formed by rapidly spinning, newly born magnetars, and claimed that the high spin-down luminosity caused by MRD losses represents a natural mechanism for prolonged energy injection in the external shock in the first hours after the formation of the magnetars.
The individual light curves of the different cases can be fitted well by their model, and the initial spin period $P_{0}$ and magnetic dipole field $B_{\rm p}$, which are derived from their fitting, also match well the values expected from the magnetar model. The results are consistent with our findings.
The simulation results in \cite{2015ApJ...813...92R} suggest that the if a rapidly spinning magnetar powered the GRB X-ray plateau, then magnetars are required to have two different progenitors and formation paths and different magnetic field formation efficiencies.
\cite{2013MNRAS.430.1061R} show evidence of energy injection for many short GRBs. In addition, they suggest that the remnant of neutron-star-neutron-star mergers could produce an unstable magnetar rather than collapse directly and immediately to a BH.
The unstable magnetar powers the X-ray plateau.
The results show that nearly half of events forming magnetars could collapse to a BH after a few hundred seconds.
Moreover, in \cite{2014MNRAS.443.1779R}, they studied GRB magnetar central engines using the observed plateau luminosity and duration correlation based on both long and short GRBs, and they found that the magnetar emission most likely is narrowly beamed ($<20^{o}$) and has a lower efficiency ($\lesssim 20\%$) in converting rotational energy into the observed X-ray plateau emission.

\subsection {Sphericity of the magnetar wind}

One important assumption in the analysis above is whether or not the emitted energy is isotropic, since this affects the derived energetics of the bursts. In the framework of a BH central engine, the rotational energy can naturally be assumed to be extracted through the jet and is deposited into the X-ray plateau. In this case, a beaming correction needs to be made: $E_{\rm K} = f_{\rm b} E_{\rm K, iso}$.
On the other hand, for a magnetar central engine, the rotational energy is initially emitted in an isotropic wind. Indeed, for short GRBs, the total energy is typically expected to be emitted quasi-spherically (e.g. \citealt{2014MNRAS.439.3916M, 2006MNRAS.372L..19F}). Here, we have also assumed the energy to be emitted quasi-isotropically even for long GRBs during their X-ray afterglow phase (see, e.g. \citealt{2010A&A...518A..27M, 2011A&A...526A.121D, 2014MNRAS.443...67M}). Then only a small part of the rotational energy of a magnetar is injected into the external shock of the jet. This means that for the magnetar sample, using this assumption, we should not use the beaming correction, so that $f_{\rm b}$ = 1.

We note, however, that it is still unclear whether the energy release for long GRBs during their X-ray afterglow phase is spherical or collimated \citep{2011MNRAS.413.2031M}, and there needs to be further theoretical and numerical modeling, as well as observational constraints, to understand the late-time energy release from a magnetar.
Indeed, there are arguments for a collimated outflow from a magnetar central engine. For instance, the progenitor star of long GRBs is expected to cause a collimation of the emitted energy, at least during the prompt phase \citep{2007MNRAS.380.1541B,2009MNRAS.396.2038B}, and this would change the derived total energetics. However, for very late times, such as during the X-ray plateau phase, which is a hundred seconds to a day following the trigger of the GRB, there should be less material surrounding the magnetar. Therefore, the magnetar emission could be expected to come out more isotropically.

Similarly, \cite{2014ApJ...785...74L} found, by using {\it Swift} data of GRB afterglows, that while an isotropic assumption for the emitted energy is reasonable for short GRBs, for most long GRBs, this assumption leads to unrealistic values of the derived rotation period of the neutron star. \cite{2014ApJ...785...74L} therefore argue that this means that the assumption of isotropic energy release during the afterglow phase is not correct. We note, however, that it could also point to the fact that the progenitor is not a magnetar.

By contrast, an indirect argument for quasi-isotropic magnetar winds is given in \cite{2014MNRAS.443...67M} who studied the energetics of GRBs associated with SNe. They noted that the SN energy is narrowly distributed around the characteristic magnetar energy of $\sim 10^{52}$ erg and is always larger than the GRB energy. On the other hand, the GRB energy varies a lot. They therefore suggest that the central engine for the SN-associated GRBs is a magnetar and that the SN is driven by a close-to-spherical magnetar wind. A consequence is that the jet does not drive the stellar explosion; indeed, it is not clear how the jet would transfer energy to the star after it has escaped.
\cite{2014MNRAS.443...67M} had to make an assumption, though, that the magnetar wind is not fully isotropic based on the fact that SNe, in general, are observed to have some degree of asphericity (e.g. \citealt{2002ApJ...565..405M, 2007ApJ...668L..19T}). The asphericity is, however, assumed to be much smaller than the highly collimated jetted flows from BH central engines, with only $f_{\rm b} < \sim 0.5-0.2$.
Moreover, \cite{2015Natur.523..189G} argued for a magnetar central engine in GRB 111209A, since the associated SN 2011kl did not show any evidence for Ni production. However, its properties are peculiar compared to other GRB/SN cases and might thus not be representative (\citealt{2017AdAst2017E...5C}). In the case of GRB 111209A, a correction for collimation of $f_b < 1/50$ was required for $E_{\gamma}$, which would suggest an initial jetted outflow.  On the other hand, considering the gamma-ray radiative efficiency found in \S \ref{Rfficiency} (Eq.\ref{Eq:efficiency}, Fig.\ref{efficiency} and Fig.\ref{efficiencyDis}) and the derived $E_{\rm X, iso}$ the total energy budget could be strongly strained for the simple estimates of a magnetar central engine, in particular if $E_{\rm X}$ is assumed to be  isotropic. Indeed, \cite{2016MNRAS.457.2761C} argues that additional energy sources, such as radioactive heating, have to be considered apart from the magnetar itself for a consistent scenario.

Following these lines of argument, the assumption made in this paper of $f_{\rm b} = 1$ might be too high. Instead, by employing a factor of $f_{\rm b} <∼ 0.5-0.2$, as argued by \cite{2014MNRAS.443...67M}, the energies become smaller, and more bursts are consistent with being magnetar-bound. For instance, using $f_{\rm b} =$ 0.5 (0.2), there are 2 (18) bursts in the Silver sample that would be classified as possibly harboring a magnetar central engine and thus belonging to the Bronze sample (see Table 5). We also calculate the beaming correction, $f_{\rm b}$, that individual bursts need in order to reach the magnetar limit (green line in Fig.\ref{angle}a). These $f_{\rm b}$ values are significantly larger than the $f_{\rm b}$ values obtained from the jet break. On the other hand, it shows that only a slight anisotropy is needed in order for the total energy to get below the magnetar limit. In such cases, the central engine could still be a magnetar.

This points toward the fact that there could be a continuum of degrees of asphericity (or collimation) of the magnetar wind. By assuming a fully isotropic wind as we did in this paper, then most bursts are inconsistent with the magnetar central engine (our Gold and Silver samples). Only a small fraction of bursts, both short GRBs and long GRBs in our Bronze sample, are consistent with a magnetar central engine assumption. By contrast, \cite{2014ApJ...785...74L} assume a high degree of collimation and find a large fraction of all bursts to be consistent with the magnetar central engine.
As argued above, the degree of asphericity could lie in between the assumption of \cite{2014ApJ...785...74L} (high degree of collimation) and this paper (isotropic wind) and could, for instance, be determined by varying the mass of the envelope surrounding the magnetar, which would have a different impact from the focusing of the wind.

\subsection {Conclusions}

In this paper, we systematically analyzed the \emph{Swift}/XRT light curves having a plateau phase in 101 GRBs which were detected before 2017 May. Assuming an isotropic magnetar wind, we draw the following conclusions:

\begin{itemize}

\item
A Gold sample, which includes 9 out of the 101 GRBs. Both $E_{\rm X,iso}$ and $E_{\rm K,iso}$ exceed the energy budget of a magnetar.
Therefore, these most likely harbor a BH central engine.

\item
A Silver sample, which includes 69 out of the 101 GRBs. Estimates of their kinetic energy $E_{\rm K,iso}$ are larger than the energy budget of a magnetar. Likewise, these most likely have a BH central engine.

\item
A Bronze sample, which includes 23 out of the 101 GRBs. These bursts have energy output lower than the maximal energy budget of magnetars, and as such can be considered candidates for the magnetar central engine.

\item
We further test the data with the BZ mechanism based on the BH model, and find that the observations of the "Gold" and "Silver" samples are consistent with expectations.
We also test the magnetar model for the Silver and Bronze samples and find that the magnetar surface magnetic filed ($B_{\rm p}$) and initial spin period ($P_{0}$) fall into a reasonable range for the Bronze sample but not for the Silver sample. This implies that the Bronze sample is consistent with having a magnetar central engine, but the Silver sample, in general, is inconsistent with the magnetar model.
\end{itemize}

We also compare the properties of the two central engine candidate samples, and the two types of plateau bursts.

\begin{itemize}

\item BH candidate (Gold+Silver) and Magnetar candidate (Bronze) samples.

 We found that the observational properties ($\alpha_{1}$, $\alpha_{2}$, $\Delta\alpha$, $\Gamma_{1}$, $\Gamma_{2}$, and $\Delta\Gamma$) and the model parameters ({\it q}), in general, do not differ significantly. This indicates that it could be insensitive to the central engine, or in the case where these parameters were to be sensitive to the central engine, it would argue for a common central engine type. However, one interesting result is that the {\it p} value distribution for the magnetar candidate sample is significantly narrower than that for the BH candidate sample.
A significant difference is also found for $t_{\rm b}$, {\it z}, $T_{90}$, $L_{\rm X,iso}$, energies ($E_{\rm X,iso}$, $E_{\rm K,iso}$, and $E_{\gamma,\rm iso}$), and {\it p}, which could imply a different physical reason for the different samples.

\item Internal plateau and External plateau sample.

For the internal and the external plateaus, we found that $\alpha_{2}$, $\Delta\alpha$, $\Gamma_{2}$, $\Delta\Gamma$, $t_{\rm b}$, and {\it q} in general, present different distributions.
The results are consistent with the expectation for different physical progenitors for the two types of plateaus, while the $\alpha_{1}$ and $\Gamma_{1}$ generally show similar distributions.

\end{itemize}

{\it We conclude that, under the assumption of isotropic energy release from a magnetar, most GRBs should harbor a BH central engine.}

\acknowledgments
We acknowledge the use of the public data from the {\it Swift} data archive and the UK {\it Swift} Science Data Center.
We thank Damien B\'egu\'e, Yu Wang, David Yu, H\"usne Dereli, Jin-Jun Geng, Yun-Feng Liang, and Christoffer Lundman for helpful discussions.
We appreciate the valuable comments by the referee and thank Bing Zhang for useful discussions that greatly improved the paper.
This work is supported by the National Basic Research Program (``973" Program) of China (grant No. 2014CB845800), the National Key Research and Development Program of China (grant No. 2017YFA0402600), the National Natural Science Foundation of China (grant Nos. 11725314, 11573014, 11673068, and 11773010), the Youth Innovation Promotion Association (2011231), the Key Research Program of Frontier Sciences (QYZDB-SSW-SYS005), the Strategic Priority Research Program “Multi-waveband gravitational wave Universe” (grant No. XDB23000000) of the Chinese Academy of Sciences, and the Swedish National Space Board, the Swedish Research Council.
L.L. acknowledges the support from the Erasmus Mundus Joint Doctorate Program through grant No. 2013-1471 from the EACEA of the European Commission.
F.R. is supported by the G\"oran Gustafsson Foundation for Research in Natural Sciences and Medicine. Part of this work made use of our private Python library.

\newpage
\clearpage
\startlongtable
\begin{deluxetable}{lccccccccccccccccccccc}
\tablewidth{0pt}
\tabletypesize{\scriptsize}
\tablecaption{The X-Ray Observation Properties and Fitting Results of Our Samples}
\tablenum{1}
\tablehead{
\colhead{GRB}
&\colhead{$T_{\rm s}-T_{\rm e}$\tablenotemark{a}}
&\colhead{$t_{\rm b}$\tablenotemark{a}}
&\colhead{$\alpha_1$}
&\colhead{$\alpha_2$}
&\colhead{$\Gamma_{\rm x,1}$\tablenotemark{b}}
&\colhead{$\Gamma_{\rm x,2}$\tablenotemark{b}}
&\colhead{$S_{\rm X}$\tablenotemark{c}}
&\colhead{$L_{\rm X,iso}$\tablenotemark{d}}
&\colhead{$E_{\rm X,iso}$\tablenotemark{d}}\\
&
\colhead{(ks)}&
\colhead{(ks)}&
&
&
&
&
\colhead{($10^{-7}$ erg cm$^{-2}$)}&
\colhead{($10^{47}$ erg s$^{-1}$)}&
\colhead{($10^{50}$ erg)}}
\startdata
\hline
&Gold\\
\hline
050401&0.13-548&5.1$\pm$0.6&0.55$\pm$0.01&1.47$\pm$0.06&1.78$\pm$0.21&1.79$\pm$0.15&10.88$\pm$0.57&55.73$\pm$5.10&208.70$\pm$10.91&\\
050730&4.01-408&9.7$\pm$0.5&0.72$\pm$0.12&2.69$\pm$0.04&1.43$\pm$0.06&1.63$\pm$0.05&11.57$\pm$0.64&111.72$\pm$12.58&368.39$\pm$20.39&\\
060607A&0.46-171&12.6$\pm$0.3&0.44$\pm$0.02&3.60$\pm$0.09&1.51$\pm$0.10&1.56$\pm$0.08&12.91$\pm$0.17&25.64$\pm$1.27&273.82$\pm$3.52&\\
061222A&0.21-1443&47.6$\pm$3.1&0.70$\pm$0.01&1.74$\pm$0.03&1.83$\pm$0.06&2.07$\pm$0.09&20.18$\pm$0.39&4.70$\pm$0.34&220.53$\pm$4.25&\\
080721&0.11-1396&3.1$\pm$0.2&0.80$\pm$0.01&1.65$\pm$0.01&1.81$\pm$0.02&1.96$\pm$0.06&40.29$\pm$0.68&239.60$\pm$12.48&644.10$\pm$10.93&\\
100902A&1.19-1489&839.9$\pm$48.9&0.73$\pm$0.02&4.53$\pm$0.00&2.28$\pm$0.16&2.38$\pm$0.54&6.93$\pm$0.10&0.88$\pm$0.11&268.32$\pm$3.79&\\
111209A&0.43-49&18.4$\pm$0.1&0.81$\pm$0.00&5.64$\pm$0.05&1.24$\pm$0.01&1.54$\pm$0.05&288.62$\pm$0.33&7.79$\pm$0.05&364.77$\pm$0.41&\\
140206A&0.42-1316&13.6$\pm$1.9&0.73$\pm$0.02&1.41$\pm$0.02&1.68$\pm$0.09&1.81$\pm$0.06&15.34$\pm$0.72&21.01$\pm$3.23&265.97$\pm$12.53&\\
150403A&0.08-2834&2.5$\pm$0.1&0.50$\pm$0.01&1.45$\pm$0.01&1.67$\pm$0.01&1.75$\pm$0.04&55.11$\pm$0.58&279.55$\pm$5.79&588.12$\pm$6.20&\\
\hline
&Silver\\
\hline
050315&5.37-888&231.4$\pm$29.4&0.68$\pm$0.03&1.94$\pm$0.16&1.89$\pm$0.07&2.16$\pm$0.21&6.64$\pm$0.29&0.37$\pm$0.05&64.21$\pm$2.83&\\
050319&0.44-1389&36.8$\pm$10.8&0.49$\pm$0.04&1.56$\pm$0.20&1.92$\pm$0.08&2.17$\pm$0.23&3.40$\pm$0.41&5.43$\pm$0.88&78.17$\pm$9.38&\\
050505&2.89-1127&26.1$\pm$3.2&0.77$\pm$0.05&1.80$\pm$0.06&1.98$\pm$0.08&2.00$\pm$0.10&3.09$\pm$0.17&10.91$\pm$1.72&111.58$\pm$6.13&\\
050802&0.32-830&6.6$\pm$0.7&0.66$\pm$0.03&1.66$\pm$0.04&1.76$\pm$0.09&1.90$\pm$0.09&3.46$\pm$0.16&4.32$\pm$0.60&26.44$\pm$1.21&\\
050803&0.42-1329&16.3$\pm$1.3&0.39$\pm$0.03&1.82$\pm$0.06&1.81$\pm$0.10&2.56$\pm$0.24&3.58$\pm$0.16&0.10$\pm$0.01&1.71$\pm$0.08&\\
060115&5.37-418&39.9$\pm$27.0&0.60$\pm$0.18&1.46$\pm$0.22&2.16$\pm$0.19&2.15$\pm$0.29&0.42$\pm$0.13&0.97$\pm$0.68&11.19$\pm$3.42&\\
060210&3.83-1581&28.4$\pm$10.4&0.83$\pm$0.07&1.41$\pm$0.05&2.02$\pm$0.06&1.97$\pm$0.09&5.31$\pm$0.82&13.82$\pm$5.64&165.07$\pm$25.61&\\
060418&0.36-537&1.3$\pm$0.4&0.87$\pm$0.13&1.57$\pm$0.05&1.94$\pm$0.21&1.91$\pm$0.23&1.85$\pm$0.36&13.46$\pm$5.73&10.94$\pm$2.15&\\
060502A&0.33-1581&30.0$\pm$9.2&0.55$\pm$0.04&1.19$\pm$0.06&1.91$\pm$0.14&1.84$\pm$0.17&2.29$\pm$0.27&0.44$\pm$0.11&13.88$\pm$1.65&\\
060604&1.24-767&22.3$\pm$5.5&0.40$\pm$0.09&1.27$\pm$0.06&2.07$\pm$0.16&2.07$\pm$0.17&0.73$\pm$0.10&0.77$\pm$0.17&8.28$\pm$1.13&\\
060605&0.20-127&8.7$\pm$0.6&0.46$\pm$0.03&2.19$\pm$0.08&1.97$\pm$0.12&2.08$\pm$0.15&1.45$\pm$0.05&13.47$\pm$1.18&42.73$\pm$1.43&\\
060614&1.00-2770&50.2$\pm$2.6&0.12$\pm$0.04&1.97$\pm$0.04&1.73$\pm$0.10&1.83$\pm$0.12&3.02$\pm$0.12&(2.12$\pm$0.14)e-3&(11.71$\pm$0.48)e-2&\\
060714&0.30-1245&4.6$\pm$1.6&0.43$\pm$0.11&1.31$\pm$0.05&1.81$\pm$0.15&2.00$\pm$0.18&1.03$\pm$0.20&9.04$\pm$3.17&17.58$\pm$3.36&\\
060729&0.40-11845&65.8$\pm$2.2&0.15$\pm$0.02&1.42$\pm$0.01&2.04$\pm$0.04&2.03$\pm$0.05&13.08$\pm$0.31&0.18$\pm$0.01&10.41$\pm$0.25&\\
060814&1.00-1325&13.2$\pm$1.8&0.51$\pm$0.05&1.43$\pm$0.03&2.00$\pm$0.11&2.15$\pm$0.09&3.35$\pm$0.24&0.57$\pm$0.08&6.55$\pm$0.47&\\
060906&0.33-258&12.6$\pm$1.9&0.29$\pm$0.06&1.82$\pm$0.17&2.07$\pm$0.21&1.77$\pm$0.25&0.54$\pm$0.05&2.52$\pm$0.38&15.27$\pm$1.33&\\
060908&0.08-450&0.7$\pm$0.2&0.44$\pm$0.10&1.56$\pm$0.07&2.03$\pm$0.29&1.97$\pm$0.22&1.41$\pm$0.21&34.19$\pm$9.41&12.86$\pm$1.95&\\
061121&0.30-2097&6.7$\pm$0.5&0.41$\pm$0.03&1.48$\pm$0.02&1.95$\pm$0.12&1.82$\pm$0.06&9.55$\pm$0.43&7.68$\pm$0.72&44.70$\pm$2.02&\\
070129&1.15-1490&21.9$\pm$3.6&0.24$\pm$0.06&1.20$\pm$0.04&2.28$\pm$0.19&2.13$\pm$0.18&1.06$\pm$0.12&1.80$\pm$0.23&14.11$\pm$1.55&\\
070306&0.50-1036&30.3$\pm$1.7&0.13$\pm$0.03&1.88$\pm$0.05&1.79$\pm$0.09&2.11$\pm$0.18&6.77$\pm$0.27&2.72$\pm$0.14&40.47$\pm$1.64&\\
070508&0.08-646&0.9$\pm$0.1&0.48$\pm$0.02&1.43$\pm$0.01&1.77$\pm$0.04&1.74$\pm$0.07&9.47$\pm$0.33&19.66$\pm$1.08&17.62$\pm$0.62&\\
070529&0.17-445&1.7$\pm$0.8&0.64$\pm$0.13&1.29$\pm$0.05&1.75$\pm$0.21&2.10$\pm$0.30&0.70$\pm$0.14&12.95$\pm$6.48&10.52$\pm$2.16&\\
080310&1.35-378&10.7$\pm$0.9&0.13$\pm$0.09&1.63$\pm$0.06&1.96$\pm$0.17&1.82$\pm$0.14&0.87$\pm$0.06&2.77$\pm$0.26&12.39$\pm$0.91&\\
080430&0.30-2879&29.4$\pm$4.7&0.39$\pm$0.03&1.15$\pm$0.03&2.05$\pm$0.11&2.06$\pm$0.13&1.94$\pm$0.16&0.11$\pm$0.01&3.16$\pm$0.26&\\
080516&0.14-47&4.2$\pm$2.1&0.32$\pm$0.10&1.10$\pm$0.12&2.40$\pm$0.57&2.21$\pm$0.40&0.82$\pm$0.22&16.04$\pm$6.10&18.51$\pm$4.97&\\
080605&0.10-230&0.5$\pm$0.0&0.52$\pm$0.04&1.37$\pm$0.02&1.67$\pm$0.04&1.71$\pm$0.07&4.61$\pm$0.37&113.13$\pm$11.75&32.62$\pm$2.64&\\
080905B&0.20-895&4.2$\pm$1.1&0.28$\pm$0.11&1.46$\pm$0.04&1.53$\pm$0.16&2.11$\pm$0.12&3.80$\pm$0.62&32.05$\pm$10.09&51.92$\pm$8.47&\\
081008&0.52-261&18.5$\pm$4.7&0.86$\pm$0.04&1.85$\pm$0.14&2.05$\pm$0.13&1.83$\pm$0.32&3.88$\pm$0.24&1.48$\pm$0.46&38.19$\pm$2.39&\\
081221&0.26-506&0.6$\pm$0.1&0.25$\pm$0.12&1.32$\pm$0.02&1.99$\pm$0.07&2.04$\pm$0.11&1.98$\pm$0.35&219.92$\pm$22.88&24.91$\pm$4.36&\\
090407&5.47-978&120.7$\pm$17.4&0.59$\pm$0.04&1.95$\pm$0.14&2.24$\pm$0.13&1.84$\pm$0.37&3.05$\pm$0.19&0.14$\pm$0.02&17.12$\pm$1.06&\\
090418A&0.14-216&3.1$\pm$0.5&0.49$\pm$0.06&1.57$\pm$0.04&2.02$\pm$0.22&1.96$\pm$0.10&3.92$\pm$0.31&11.59$\pm$2.50&26.74$\pm$2.14&\\
090510&0.10-67&1.4$\pm$0.2&0.62$\pm$0.04&2.15$\pm$0.09&1.67$\pm$0.10&2.05$\pm$0.20&3.28$\pm$0.19&5.01$\pm$0.77&7.39$\pm$0.42&\\
090516&3.89-521&16.9$\pm$2.8&0.77$\pm$0.09&1.86$\pm$0.08&2.05$\pm$0.09&2.16$\pm$0.12&1.68$\pm$0.17&14.99$\pm$3.36&56.37$\pm$5.73&\\
090529&6.85-774&234.1$\pm$118.0&0.51$\pm$0.10&1.74$\pm$0.54&1.90$\pm$0.44&1.66$\pm$0.58&0.41$\pm$0.08&0.03$\pm$0.02&6.67$\pm$1.27&\\
090530&0.38-791&41.6$\pm$15.6&0.54$\pm$0.04&1.30$\pm$0.13&2.03$\pm$0.15&1.93$\pm$0.32&1.15$\pm$0.16&0.11$\pm$0.03&5.02$\pm$0.69&\\
090618&0.40-2841&8.4$\pm$0.5&0.72$\pm$0.01&1.51$\pm$0.01&1.93$\pm$0.04&1.86$\pm$0.05&24.04$\pm$0.53&1.34$\pm$0.08&19.13$\pm$0.42&\\
091018&0.07-520&0.5$\pm$0.1&0.31$\pm$0.06&1.24$\pm$0.02&1.84$\pm$0.15&2.04$\pm$0.09&1.55$\pm$0.17&11.60$\pm$1.57&4.05$\pm$0.45&\\
100418A&0.40-2091&125.0$\pm$17.1&-0.01$\pm$0.05&1.66$\pm$0.10&1.89$\pm$0.29&1.84$\pm$0.17&2.09$\pm$0.23&(2.26$\pm$0.31)e-2&2.24$\pm$0.24&\\
100615A&0.20-163&19.8$\pm$4.5&0.39$\pm$0.03&1.26$\pm$0.10&2.24$\pm$0.19&1.92$\pm$0.25&12.82$\pm$1.44&4.10$\pm$0.60&67.44$\pm$7.57&\\
100621A&0.40-1769&99.3$\pm$20.6&0.86$\pm$0.03&1.66$\pm$0.08&2.33$\pm$0.10&3.06$\pm$0.31&25.66$\pm$0.99&0.11$\pm$0.03&20.58$\pm$0.80&\\
100704A&0.50-1051&33.3$\pm$6.4&0.68$\pm$0.03&1.38$\pm$0.05&2.04$\pm$0.10&1.92$\pm$0.17&5.10$\pm$0.32&6.15$\pm$1.14&139.15$\pm$8.85&\\
100814A&1.00-4848&148.9$\pm$7.3&0.50$\pm$0.02&2.09$\pm$0.06&1.90$\pm$0.05&2.08$\pm$0.13&12.25$\pm$0.28&0.58$\pm$0.03&68.13$\pm$1.58&\\
100906A&0.30-202&13.2$\pm$1.3&0.76$\pm$0.03&2.12$\pm$0.80&1.96$\pm$0.10&2.05$\pm$0.15&11.20$\pm$0.35&3.73$\pm$0.53&59.73$\pm$1.86&\\
111008A&0.34-1014&7.4$\pm$0.9&0.29$\pm$0.04&1.34$\pm$0.03&1.92$\pm$0.09&1.88$\pm$0.13&2.57$\pm$0.22&52.83$\pm$6.48&116.40$\pm$9.81&\\
111123A&4.22-183&42.9$\pm$12.3&0.81$\pm$0.08&2.02$\pm$0.29&2.23$\pm$0.20&2.31$\pm$0.47&0.99$\pm$0.10&1.44$\pm$0.52&21.80$\pm$2.28&\\
111228A&0.50-2572&6.5$\pm$0.8&0.22$\pm$0.05&1.23$\pm$0.02&2.01$\pm$0.15&2.15$\pm$0.10&3.29$\pm$0.30&1.00$\pm$0.11&4.63$\pm$0.42&\\
120327A&0.30-146&3.0$\pm$0.6&0.60$\pm$0.06&1.51$\pm$0.06&1.69$\pm$0.14&1.82$\pm$0.17&1.97$\pm$0.20&20.09$\pm$4.75&35.78$\pm$3.65&\\
120712A&0.10-67&4.6$\pm$1.5&0.89$\pm$0.04&1.64$\pm$0.10&2.01$\pm$0.15&2.25$\pm$0.20&1.12$\pm$0.08&15.03$\pm$6.22&36.00$\pm$2.55&\\
120811C&0.30-81&1.7$\pm$0.8&0.36$\pm$0.15&1.16$\pm$0.09&1.78$\pm$0.15&2.24$\pm$0.21&0.86$\pm$0.28&34.08$\pm$11.77&14.41$\pm$4.75&\\
120922A&0.74-751&3.4$\pm$1.6&0.43$\pm$0.19&1.09$\pm$0.04&2.20$\pm$0.24&1.98$\pm$0.15&0.87$\pm$0.31&18.93$\pm$8.11&18.53$\pm$6.56&\\
121024A&3.90-236&29.7$\pm$12.7&0.75$\pm$0.11&1.67$\pm$0.18&1.93$\pm$0.14&1.58$\pm$0.27&1.07$\pm$0.19&0.49$\pm$0.23&13.87$\pm$2.48&\\
121128A&0.20-70&1.6$\pm$0.2&0.52$\pm$0.07&1.68$\pm$0.04&1.91$\pm$0.11&1.97$\pm$0.13&2.91$\pm$0.26&43.37$\pm$7.23&34.86$\pm$3.13&\\
121211A&3.94-362&36.1$\pm$16.6&0.71$\pm$0.09&1.51$\pm$0.19&1.94$\pm$0.12&2.27$\pm$0.44&1.46$\pm$0.27&0.15$\pm$0.07&4.21$\pm$0.79&\\
130420A&0.76-1295&65.4$\pm$29.2&0.72$\pm$0.03&1.27$\pm$0.09&2.13$\pm$0.10&1.96$\pm$0.23&1.98$\pm$0.24&0.10$\pm$0.04&9.03$\pm$1.08&\\
130606A&5.32-238&13.0$\pm$3.6&0.44$\pm$0.30&1.82$\pm$0.14&1.79$\pm$0.17&1.85$\pm$0.21&0.57$\pm$0.16&15.88$\pm$5.53&32.93$\pm$9.38&\\
130612A&0.76-44&3.0$\pm$2.1&0.38$\pm$0.36&1.24$\pm$0.16&2.19$\pm$0.40&2.11$\pm$0.43&0.09$\pm$0.04&0.95$\pm$0.60&0.91$\pm$0.46&\\
131030A&0.36-1661&3.4$\pm$1.1&0.81$\pm$0.05&1.28$\pm$0.02&1.71$\pm$0.04&1.95$\pm$0.09&6.88$\pm$0.91&9.39$\pm$3.26&31.22$\pm$4.13&\\
131105A&0.41-734&4.8$\pm$1.3&0.23$\pm$0.13&1.20$\pm$0.05&1.90$\pm$0.17&1.99$\pm$0.18&1.04$\pm$0.20&3.32$\pm$0.92&7.74$\pm$1.50&\\
140512A&0.23-264&17.7$\pm$1.7&0.79$\pm$0.01&1.68$\pm$0.05&1.76$\pm$0.07&1.90$\pm$0.09&24.66$\pm$0.64&0.96$\pm$0.10&35.81$\pm$0.94&\\
140518A&0.29-19&2.6$\pm$0.6&0.19$\pm$0.10&1.53$\pm$0.18&1.99$\pm$0.15&1.94$\pm$0.34&0.43$\pm$0.07&28.04$\pm$4.55&17.74$\pm$2.88&\\
140703A&3.81-79&14.5$\pm$1.6&0.60$\pm$0.10&2.28$\pm$0.10&1.78$\pm$0.10&1.94$\pm$0.15&3.66$\pm$0.30&16.93$\pm$2.78&80.03$\pm$6.66&\\
141121A&56.28-1393&358.5$\pm$51.9&0.44$\pm$0.12&2.47$\pm$0.26&1.79$\pm$0.23&1.72$\pm$0.21&2.58$\pm$0.24&0.06$\pm$0.01&14.93$\pm$1.39&\\
150910A&0.20-224&7.6$\pm$0.4&0.57$\pm$0.01&2.36$\pm$0.06&1.83$\pm$0.20&1.75$\pm$0.08&16.07$\pm$0.32&8.86$\pm$0.57&80.16$\pm$1.59&\\
151027A&0.40-908&4.4$\pm$0.2&0.22$\pm$0.02&1.68$\pm$0.02&2.03$\pm$0.04&2.01$\pm$0.06&17.07$\pm$0.50&9.88$\pm$0.47&30.99$\pm$0.91&\\
160121A&0.20-46&20.8$\pm$4.9&0.30$\pm$0.05&2.14$\pm$0.77&1.97$\pm$0.23&1.85$\pm$0.66&1.20$\pm$0.15&0.84$\pm$0.14&11.76$\pm$1.42&\\
160227A&1.00-1647&21.2$\pm$3.7&0.24$\pm$0.09&1.22$\pm$0.03&1.73$\pm$0.09&1.73$\pm$0.10&3.06$\pm$0.36&3.43$\pm$0.56&42.01$\pm$4.91&\\
160327A&0.30-42&4.8$\pm$1.9&0.40$\pm$0.15&1.60$\pm$0.14&1.71$\pm$0.22&2.03$\pm$0.23&0.45$\pm$0.09&16.51$\pm$7.14&20.33$\pm$4.04&\\
161117A&1.05-1365&7.5$\pm$1.4&0.36$\pm$0.11&1.17$\pm$0.03&2.11$\pm$0.86&1.95$\pm$0.08&2.17$\pm$0.29&3.42$\pm$0.57&13.82$\pm$1.86&\\
170113A&0.20-877&6.0$\pm$1.2&0.57$\pm$0.03&1.26$\pm$0.03&1.83$\pm$0.11&1.81$\pm$0.09&3.77$\pm$0.32&7.02$\pm$1.36&37.13$\pm$3.17&\\
\hline
&Bronze\\
\hline
051109B&0.18-158&3.3$\pm$0.8&0.26$\pm$0.10&1.31$\pm$0.08&2.22$\pm$0.33&1.86$\pm$0.34&0.23$\pm$0.04&(8.38$\pm$1.73)e-4&(3.63$\pm$0.57)e-3&\\
051221A&3.00-848&26.7$\pm$5.2&0.18$\pm$0.15&1.43$\pm$0.06&1.82$\pm$0.22&1.96$\pm$0.15&0.39$\pm$0.07&(1.77$\pm$0.34)e-2&0.32$\pm$0.06&\\
060526&1.00-314&97.1$\pm$20.7&0.63$\pm$0.60&2.89$\pm$0.48&1.91$\pm$0.16&1.91$\pm$0.45&1.22$\pm$0.08&0.38$\pm$0.14&27.66$\pm$1.77&\\
060708&0.20-1228&8.8$\pm$2.2&0.59$\pm$0.04&1.32$\pm$0.05&2.10$\pm$0.18&2.08$\pm$0.16&0.88$\pm$0.08&(8.73$\pm$2.07)e-3&0.14$\pm$0.01&\\
061021&4.20-4236&29.7$\pm$13.9&0.75$\pm$0.07&1.19$\pm$0.03&1.94$\pm$0.08&1.87$\pm$0.08&2.81$\pm$0.61&0.02$\pm$0.01&0.89$\pm$0.20&\\
061110A&3.00-756&73.2$\pm$51.6&0.19$\pm$0.27&1.16$\pm$0.25&2.11$\pm$0.72&1.65$\pm$0.46&0.14$\pm$0.06&(3.56$\pm$2.06)e-3&0.23$\pm$0.10&\\
061201&0.11-120&2.6$\pm$0.5&0.53$\pm$0.09&2.01$\pm$0.11&1.33$\pm$0.18&2.01$\pm$0.40&1.36$\pm$0.11&(8.83$\pm$2.57)e-3&(4.13$\pm$0.34)e-2&\\
070110&4.08-29&20.4$\pm$0.2&0.03$\pm$0.05&9.39$\pm$0.69&2.07$\pm$0.08&2.09$\pm$0.19&3.23$\pm$0.04&2.75$\pm$0.11&20.55$\pm$0.24&\\
080707&0.11-279&14.4$\pm$4.3&0.36$\pm$0.05&1.29$\pm$0.10&2.05$\pm$0.30&1.83$\pm$0.23&0.47$\pm$0.07&0.15$\pm$0.04&1.95$\pm$0.29&\\
081007&0.30-1534&68.9$\pm$28.9&0.75$\pm$0.03&1.35$\pm$0.10&2.01$\pm$0.13&2.03$\pm$0.21&2.66$\pm$0.26&0.01$\pm$0.01&2.03$\pm$0.20&\\
081029&2.76-268&18.2$\pm$1.2&0.43$\pm$0.06&2.70$\pm$0.15&1.92$\pm$0.10&2.04$\pm$0.21&1.03$\pm$0.05&6.24$\pm$0.61&31.11$\pm$1.42&\\
100219A&5.00-121&26.4$\pm$3.6&0.24$\pm$0.31&2.96$\pm$0.83&1.66$\pm$0.30&1.48$\pm$0.39&1.70$\pm$0.18&5.26$\pm$1.28&69.43$\pm$7.38&\\
100302A&1.29-1062&17.0$\pm$11.5&0.30$\pm$0.14&0.92$\pm$0.08&1.88$\pm$0.35&1.95$\pm$0.22&0.17$\pm$0.07&1.61$\pm$0.72&7.12$\pm$2.92&\\
100425A&0.34-498&25.3$\pm$13.7&0.48$\pm$0.06&1.16$\pm$0.14&2.22$\pm$0.26&1.91$\pm$0.39&0.43$\pm$0.09&0.17$\pm$0.07&3.43$\pm$0.74&\\
110715A&0.10-847&0.2$\pm$0.0&0.20$\pm$0.12&1.00$\pm$0.01&1.87$\pm$0.10&1.83$\pm$0.09&1.84$\pm$0.40&35.25$\pm$3.79&3.42$\pm$0.74&\\
110808A&4.00-617&40.9$\pm$37.2&0.33$\pm$0.20&1.07$\pm$0.19&2.54$\pm$0.49&1.61$\pm$0.39&0.35$\pm$0.18&0.05$\pm$0.03&1.71$\pm$0.87&\\
120422A&0.30-1703&164.6$\pm$69.5&0.27$\pm$0.07&1.27$\pm$0.21&2.09$\pm$0.26&1.61$\pm$0.29&0.22$\pm$0.05&(1.98$\pm$0.61)e-4&0.05$\pm$0.01&\\
120521C&0.57-42&21.2$\pm$5.2&0.31$\pm$0.10&2.55$\pm$0.80&1.82$\pm$0.32&2.19$\pm$0.77&0.25$\pm$0.03&4.35$\pm$1.06&14.92$\pm$1.86&\\
130408A&9.55-79&27.2$\pm$2.7&0.55$\pm$0.21&3.87$\pm$0.42&2.26$\pm$0.19&2.53$\pm$0.25&1.04$\pm$0.09&11.83$\pm$2.66&30.26$\pm$2.63&\\
130603B&0.07-118&7.6$\pm$1.3&0.68$\pm$0.05&1.90$\pm$0.12&1.94$\pm$0.16&1.97$\pm$0.28&2.23$\pm$0.12&0.04$\pm$0.01&0.75$\pm$0.04&\\
140903A&0.21-111&8.9$\pm$1.7&0.10$\pm$0.07&1.28$\pm$0.09&1.70$\pm$0.21&1.56$\pm$0.20&0.93$\pm$0.13&(3.15$\pm$0.46)e-2&0.30$\pm$0.04&\\
151215A&0.30-175&1.9$\pm$2.0&0.54$\pm$0.20&1.17$\pm$0.14&2.23$\pm$0.22&1.91$\pm$0.55&0.22$\pm$0.11&3.92$\pm$3.47&3.42$\pm$1.76&\\
161108A&0.65-645&41.6$\pm$23.3&0.24$\pm$0.09&0.94$\pm$0.13&1.78$\pm$0.25&1.76$\pm$0.32&0.72$\pm$0.24&0.08$\pm$0.02&2.65$\pm$0.87&\\
\enddata
\tablenotetext{a}{The start, end, and the break times for the light curve fits.}
\tablenotetext{b}{The X-ray photo indices before and after the break time.}
\tablenotetext{c}{The fluence during the plateau phase, calculated by integrating the fitting light curve from start to break time.}
\tablenotetext{d}{The luminosity at the break time, in units of $10^{47} \rm erg s^{-1}$. The X-ray release energy during plateau phases, which is calculated by integrating the luminosity during the time interval of the plateau phase, in units of $10^{50}\rm erg$.}
\end{deluxetable}

\clearpage
\startlongtable
\begin{deluxetable}{lccccccccccccccccccccc}
\tablewidth{0pt}
\tabletypesize{\scriptsize}
\tablecaption{The Properties of the BH Candidate (Gold+Silver) Samples}
\tablenum{2}
\tablehead{
\colhead{GRB}
&\colhead{$T_{90}$}
&\colhead{$E_{\gamma\rm iso}$\tablenotemark{a}}
&\colhead{$E_{\rm K,iso}(t_{b})$\tablenotemark{b}}
&\colhead{$E_{\rm K,iso}(t_{0})$\tablenotemark{b}}
&\colhead{$a_{0}$\tablenotemark{c}}
&\colhead{$p$}
&\colhead{$q$}
&\colhead{$z$}\\
&
\colhead{(s)}&
\colhead{($10^{52}$ erg)}&
\colhead{($10^{52}$ erg)}&
\colhead{($10^{52}$ erg)}&
\colhead{($10^{-1}$)}&
&
&
\colhead{(redshift)}&
}
\startdata
\hline
&Gold\\
\hline
050401&33.3&42.23$\pm$8.95&71.53$\pm$29.27&0.75$\pm$0.50&1.39$\pm$0.29&2.29$\pm$0.06&-0.26$\pm$0.14&2.90&\\
050730&60.0&0.01$\pm$0.00&...\tablenotemark{d}&...&...&2.50$\pm$0.10&1.06$\pm$0.12&3.967&\\
060607A&102.2&7.14$\pm$0.34&...&...&...&2.50$\pm$0.10&0.74$\pm$0.12&3.0749&\\
061222A&71.4&35.64$\pm$14.53&158.72$\pm$17.52&19.66$\pm$6.52&2.06$\pm$0.12&3.33$\pm$0.03&0.61$\pm$0.06&2.088&\\
080721&16.2&69.31$\pm$4.40&78.49$\pm$6.24&30.10$\pm$2.92&1.45$\pm$0.06&3.20$\pm$0.01&0.71$\pm$0.02&2.591&\\
100902A&428.8&24.09$\pm$1.96&...&...&...&2.50$\pm$0.10&0.27$\pm$0.11&4.5&\\
111209A&14400.0&82.27$\pm$10.88&...&...&...&2.50$\pm$0.10&1.40$\pm$0.01&0.677&\\
140206A&93.6&30.68$\pm$1.12&46.28$\pm$8.34&21.92$\pm$8.46&1.12$\pm$0.10&2.88$\pm$0.02&0.79$\pm$0.10&2.73&\\
150403A&40.9&1.51$\pm$0.17&248.80$\pm$15.64&4.02$\pm$0.31&2.57$\pm$0.08&2.26$\pm$0.01&-0.21$\pm$0.01&2.06&\\
\hline
&Silver\\
\hline
050315&95.6&5.59$\pm$0.58&388.63$\pm$97.31&70.55$\pm$24.44&3.19$\pm$0.40&3.58$\pm$0.16&0.55$\pm$0.06&1.949&\\
050319&152.5&8.02$\pm$1.70&41.83$\pm$19.26&2.86$\pm$1.66&1.06$\pm$0.25&3.08$\pm$0.20&0.39$\pm$0.07&3.240&\\
050505&58.9&21.69$\pm$1.78&224.36$\pm$77.07&15.65$\pm$6.21&2.44$\pm$0.43&2.73$\pm$0.06&-0.21$\pm$0.06&4.27&\\
050802&19.0&3.23$\pm$0.47&16.33$\pm$4.93&0.55$\pm$0.21&0.67$\pm$0.11&2.55$\pm$0.04&-0.12$\pm$0.07&1.71&\\
050803&87.9&0.25$\pm$0.02&37.50$\pm$12.04&10.09$\pm$6.43&1.01$\pm$0.17&3.09$\pm$0.06&0.64$\pm$0.15&0.42&\\
060115&139.6&6.05$\pm$1.16&85.68$\pm$119.93&26.30$\pm$40.87&1.52$\pm$0.67&2.61$\pm$0.22&0.41$\pm$0.27&3.53&\\
060210&255.0&16.04$\pm$0.89&42.78$\pm$20.56&16.78$\pm$8.86&1.08$\pm$0.28&2.88$\pm$0.05&0.53$\pm$0.07&3.91&\\
060418&103.1&12.75$\pm$1.53&6.50$\pm$3.16&4.02$\pm$2.29&0.42$\pm$0.10&3.10$\pm$0.05&0.63$\pm$0.21&1.489&\\
060502A&28.4&1.76$\pm$0.10&4.05$\pm$1.39&0.32$\pm$0.21&0.33$\pm$0.06&2.59$\pm$0.06&0.44$\pm$0.12&1.51&\\
060604&95.0&0.26$\pm$0.06&29.57$\pm$11.70&4.19$\pm$3.01&0.89$\pm$0.19&2.36$\pm$0.06&0.32$\pm$0.20&2.1357&\\
060605&79.1&1.67$\pm$0.16&145.59$\pm$51.59&0.48$\pm$0.22&1.97$\pm$0.38&3.25$\pm$0.08&-0.52$\pm$0.07&3.78&\\
060614&108.7&0.04$\pm$0.01&8.33$\pm$1.26&0.14$\pm$0.03&0.48$\pm$0.04&2.96$\pm$0.04&-0.70$\pm$0.06&0.125&\\
060714&115.0&7.95$\pm$1.50&5.39$\pm$2.21&1.38$\pm$0.81&0.38$\pm$0.09&2.75$\pm$0.05&0.44$\pm$0.15&2.711&\\
060729&115.3&0.53$\pm$0.05&15.56$\pm$0.76&0.19$\pm$0.03&0.65$\pm$0.02&2.90$\pm$0.01&0.07$\pm$0.03&0.54&\\
060814&145.3&6.09$\pm$0.42&32.66$\pm$7.73&9.20$\pm$4.16&0.94$\pm$0.12&2.57$\pm$0.03&0.51$\pm$0.15&1.9229&\\
060906&43.5&9.13$\pm$1.42&28.23$\pm$24.05&0.05$\pm$0.05&0.87$\pm$0.31&2.76$\pm$0.17&-0.75$\pm$0.11&3.686&\\
060908&19.3&3.39$\pm$0.15&4.86$\pm$2.02&1.01$\pm$0.67&0.36$\pm$0.08&3.08$\pm$0.07&0.27$\pm$0.23&1.8836&\\
061121&81.3&37.89$\pm$3.43&28.81$\pm$5.64&0.23$\pm$0.07&0.88$\pm$0.09&2.31$\pm$0.02&-0.55$\pm$0.07&1.314&\\
070129&460.6&8.06$\pm$0.83&42.92$\pm$12.30&2.04$\pm$1.24&1.08$\pm$0.16&2.27$\pm$0.04&-0.03$\pm$0.17&2.3384&\\
070306&209.5&4.96$\pm$0.41&126.93$\pm$17.89&5.76$\pm$2.00&1.85$\pm$0.14&3.51$\pm$0.05&0.25$\pm$0.08&1.4959&\\
070508&20.9&9.29$\pm$0.69&4.09$\pm$0.63&0.18$\pm$0.03&0.33$\pm$0.02&2.24$\pm$0.01&-0.33$\pm$0.03&0.82&\\
070529&109.2&4.70$\pm$0.50&17.06$\pm$15.08&17.91$\pm$22.88&0.68$\pm$0.24&2.39$\pm$0.05&1.02$\pm$0.39&2.4996&\\
080310&365.0&7.20$\pm$1.86&15.82$\pm$4.29&0.34$\pm$0.13&0.65$\pm$0.09&2.50$\pm$0.06&-0.85$\pm$0.10&2.42&\\
080430&16.2&0.45$\pm$0.04&8.37$\pm$2.10&0.39$\pm$0.25&0.48$\pm$0.06&2.20$\pm$0.03&0.33$\pm$0.13&0.767&\\
080516&5.8&0.98$\pm$0.18&35.50$\pm$37.61&0.92$\pm$1.82&0.98$\pm$0.36&2.14$\pm$0.12&-0.07$\pm$0.47&3.2&\\
080605&20.0&11.88$\pm$1.01&6.90$\pm$1.77&1.12$\pm$0.33&0.43$\pm$0.06&2.16$\pm$0.02&-0.18$\pm$0.04&1.6398&\\
080905B&128.0&5.76$\pm$0.83&16.44$\pm$6.21&4.69$\pm$3.24&0.67$\pm$0.14&2.95$\pm$0.04&0.59$\pm$0.19&2.374&\\
081008&185.5&8.30$\pm$1.19&79.29$\pm$72.06&1.16$\pm$1.17&1.46$\pm$0.51&2.80$\pm$0.14&-0.18$\pm$0.08&1.9685&\\
081221&34.0&34.30$\pm$1.55&8.03$\pm$1.35&4.16$\pm$0.88&0.47$\pm$0.04&2.75$\pm$0.02&0.18$\pm$0.10&2.26&\\
090407&310.0&0.94$\pm$0.20&(6.90$\pm$3.49)e2&5.13$\pm$3.04&4.20$\pm$1.04&2.94$\pm$0.14&-0.58$\pm$0.07&1.4485&\\
090418A&56.0&17.06$\pm$2.31&12.57$\pm$2.99&1.45$\pm$0.87&0.58$\pm$0.07&3.10$\pm$0.04&0.31$\pm$0.17&1.608&\\
090510&0.3&0.34$\pm$0.07&8.24$\pm$4.58&0.51$\pm$0.31&0.47$\pm$0.13&3.20$\pm$0.09&-0.05$\pm$0.08&0.903&\\
090516&210.0&69.02$\pm$7.54&75.01$\pm$21.53&34.50$\pm$11.42&1.42$\pm$0.23&3.48$\pm$0.08&0.47$\pm$0.09&4.109&\\
090529&100.0&1.41$\pm$0.46&91.37$\pm$155.70&0.62$\pm$1.29&1.57$\pm$0.77&2.65$\pm$0.54&-0.41$\pm$0.27&2.625&\\
090530&48.0&0.68$\pm$0.08&3.19$\pm$1.40&0.14$\pm$0.11&0.29$\pm$0.07&2.73$\pm$0.13&0.34$\pm$0.12&1.266&\\
090618&113.2&13.39$\pm$1.09&12.72$\pm$0.98&3.08$\pm$0.43&0.59$\pm$0.02&3.01$\pm$0.01&0.53$\pm$0.04&0.54&\\
091018&4.4&0.56$\pm$0.06&7.96$\pm$1.96&2.91$\pm$1.49&0.46$\pm$0.06&2.33$\pm$0.02&0.51$\pm$0.22&0.971&\\
100418A&7.0&0.10$\pm$0.06&33.95$\pm$11.78&(4.53$\pm$4.21)e-4&0.96$\pm$0.18&2.55$\pm$0.10&-0.95$\pm$0.14&0.6235&\\
100615A&39.0&5.82$\pm$0.24&16.66$\pm$5.28&0.27$\pm$0.18&0.67$\pm$0.11&2.67$\pm$0.10&0.10$\pm$0.12&1.398&\\
100621A&63.6&2.75$\pm$0.29&91.35$\pm$62.62&4.43$\pm$4.08&1.57$\pm$0.48&2.88$\pm$0.08&0.45$\pm$0.11&0.542&\\
100704A&197.5&17.75$\pm$1.07&29.45$\pm$7.24&2.55$\pm$1.12&0.89$\pm$0.12&2.84$\pm$0.05&0.42$\pm$0.08&3.6&\\
100814A&174.5&15.01$\pm$0.97&(6.16$\pm$1.25)e3&35.06$\pm$8.51&9.68$\pm$0.41&3.12$\pm$0.06&-0.42$\pm$0.03&1.44&\\
100906A&114.4&29.48$\pm$0.98&(3.33$\pm$0.97)e2&3.49$\pm$1.40&2.96$\pm$0.42&3.16$\pm$0.80&-0.20$\pm$0.06&1.727&\\
111008A&63.46&42.11$\pm$3.66&23.77$\pm$3.92&3.15$\pm$0.88&0.80$\pm$0.07&2.78$\pm$0.03&0.25$\pm$0.08&4.9898&\\
111123A&290.0&23.78$\pm$1.47&89.24$\pm$71.83&20.07$\pm$18.06&1.55$\pm$0.53&3.70$\pm$0.29&0.36$\pm$0.15&3.1516&\\
111228A&101.2&4.17$\pm$0.45&13.89$\pm$2.91&1.82$\pm$0.90&0.61$\pm$0.07&2.31$\pm$0.02&0.21$\pm$0.17&0.714&\\
120327A&62.9&8.78$\pm$0.34&8.45$\pm$5.07&0.67$\pm$0.46&0.48$\pm$0.14&2.35$\pm$0.06&-0.10$\pm$0.11&2.813&\\
120712A&14.7&18.57$\pm$0.27&8.69$\pm$4.74&1.77$\pm$1.34&0.49$\pm$0.14&3.19$\pm$0.10&0.59$\pm$0.13&4.1745&\\
120811C&26.8&6.96$\pm$1.04&27.80$\pm$24.19&15.01$\pm$15.21&0.87$\pm$0.30&2.21$\pm$0.09&0.65$\pm$0.28&2.671&\\
120922A&173.0&21.38$\pm$3.36&59.62$\pm$41.06&18.00$\pm$16.35&1.27$\pm$0.39&2.12$\pm$0.04&0.21$\pm$0.30&3.1&\\
121024A&69.0&1.84$\pm$0.22&32.61$\pm$45.48&2.92$\pm$4.39&0.94$\pm$0.44&2.56$\pm$0.18&-0.19$\pm$0.12&2.298&\\
121128A&23.3&1.05$\pm$0.11&14.73$\pm$3.61&4.40$\pm$1.49&0.63$\pm$0.08&3.24$\pm$0.04&0.42$\pm$0.10&2.2&\\
121211A&182.0&0.17$\pm$0.02&37.76$\pm$37.17&24.19$\pm$26.08&1.01$\pm$0.37&2.68$\pm$0.19&0.80$\pm$0.19&1.023&\\
130420A&123.5&0.23$\pm$0.28&3.72$\pm$1.86&0.24$\pm$0.16&0.32$\pm$0.08&2.69$\pm$0.09&0.38$\pm$0.08&1.297&\\
130606A&276.58&21.67$\pm$2.04&86.86$\pm$92.64&25.01$\pm$28.81&1.53$\pm$0.59&2.76$\pm$0.14&-0.39$\pm$0.24&5.913&\\
130612A&4.0&0.76$\pm$0.11&3.19$\pm$4.66&1.02$\pm$1.75&0.29$\pm$0.13&2.33$\pm$0.16&0.17$\pm$0.51&2.006&\\
131030A&41.1&17.22$\pm$0.60&6.73$\pm$2.61&4.38$\pm$1.80&0.43$\pm$0.09&2.70$\pm$0.02&0.81$\pm$0.06&1.295&\\
131105A&112.3&34.31$\pm$2.63&22.24$\pm$10.29&4.48$\pm$3.53&0.78$\pm$0.19&2.26$\pm$0.05&0.35$\pm$0.25&1.686&\\
140512A&154.8&9.07$\pm$0.41&51.44$\pm$14.15&0.76$\pm$0.28&1.18$\pm$0.17&2.57$\pm$0.05&0.03$\pm$0.05&0.725&\\
140518A&60.5&4.99$\pm$0.92&8.41$\pm$5.03&1.25$\pm$0.86&0.48$\pm$0.14&3.04$\pm$0.18&0.14$\pm$0.13&4.707&\\
140703A&67.1&1.95$\pm$0.11&(1.26$\pm$0.58)e3&(2.53$\pm$1.26)e2&5.53$\pm$1.21&3.37$\pm$0.10&-0.20$\pm$0.10&3.14&\\
141121A&549.9&0.02$\pm$0.00&(5.42$\pm$3.90)e4&(4.10$\pm$3.31)e3&9.14$\pm$2.98&3.63$\pm$0.26&-0.39$\pm$0.17&1.47&\\
150910A&112.2&2.06$\pm$0.35&(8.75$\pm$2.03)e2&8.10$\pm$4.27&4.69$\pm$0.50&3.48$\pm$0.06&-0.29$\pm$0.13&1.359&\\
151027A&129.69&3.38$\pm$0.32&28.71$\pm$2.04&3.51$\pm$0.38&0.88$\pm$0.03&3.24$\pm$0.02&0.12$\pm$0.03&0.81&\\
160121A&12.0&0.77$\pm$0.25&(2.08$\pm$0.98)e2&8.52$\pm$4.08&0.24$\pm$0.07&3.18$\pm$0.77&-0.68$\pm$0.12&1.960&\\
160227A&316.5&4.06$\pm$1.70&12.76$\pm$2.62&1.92$\pm$0.77&0.59$\pm$0.06&2.63$\pm$0.03&0.38$\pm$0.11&2.38&\\
160327A&28.0&8.90$\pm$5.88&10.38$\pm$6.74&2.67$\pm$2.49&0.53$\pm$0.16&3.13$\pm$0.14&0.51$\pm$0.23&4.99&\\
161117A&125.7&23.29$\pm$0.76&42.15$\pm$11.73&9.31$\pm$17.07&1.07$\pm$0.15&2.23$\pm$0.03&0.23$\pm$0.92&1.549&\\
170113A&20.66&0.63$\pm$0.38&8.07$\pm$1.98&1.61$\pm$0.70&0.47$\pm$0.06&2.68$\pm$0.03&0.53$\pm$0.10&1.968&\\
\enddata
\tablenotetext{a}{We searched for $E_{\gamma, \rm iso}$ in published papers or is using the fluence and redshift extrapolated into 1-10$^{4}$keV (rest frame) with a spectral model and a \emph{k}-correction, in units of $10^{52}$ erg.}
\tablenotetext{b}{The kinetic energy, which is derived from the normal decay phase. The initial kinetic energy (upper limit) is estimated from the kinetic energy at break time ($t_{\rm b}$).}
\tablenotetext{c}{The BH spin parameter $a_\bullet$, in units of $10^{-1}$ s.}
\tablenotetext{d}{We did not calculate the kinetic energy for the bursts that have an internal plateau.}
\end{deluxetable}

\clearpage
\begin{deluxetable}{lccccccccccccccccccccc}
\tablewidth{0pt}
\tabletypesize{\scriptsize}
\tablecaption{The Properties of the Magnetar Candidate (Bronze) Sample}
\tablenum{3}
\tablehead{
\colhead{GRB}
&\colhead{$T_{90}$}
&\colhead{$E_{\gamma,iso}$}
&\colhead{$q$}
&\colhead{$E_{\rm K,iso}(t_{b})$\tablenotemark{a}}
&\colhead{$B_{p}$\tablenotemark{b}}
&\colhead{$P_{0}$\tablenotemark{b}}
&\colhead{$E_{\rm rot}$\tablenotemark{c}}
&\colhead{$z$}\\
&
\colhead{(s)}&
\colhead{($10^{52}$ erg)}&
&
\colhead{($10^{52}$ erg)}&
\colhead{($10^{15}$ G)}&
\colhead{($10^{-3}$ s)}&
\colhead{($10^{50}$ erg)}&
\colhead{(redshift)}&
}
\startdata
051109B&14.3&(6.79$\pm$1.06)e-4&0.02$\pm$0.22&0.02$\pm$0.01&14.44$\pm$3.54&12.87$\pm$3.05&(4.02$\pm$0.10)e-2&0.08\\
051221A&1.4&0.36$\pm$0.20&0.25$\pm$0.21&1.50$\pm$0.37&0.52$\pm$0.11&1.33$\pm$0.25&3.78$\pm$1.05&0.5465\\
060526&296.0&1.62$\pm$0.26&0.50$\pm$0.43&...\tablenotemark{d}&0.68$\pm$0.15&3.31$\pm$0.71&0.61$\pm$0.08&3.221\\
060708&10.2&0.96$\pm$0.10&0.47$\pm$0.22&0.52$\pm$0.22&1.71$\pm$0.47&2.51$\pm$0.67&1.06$\pm$0.23&1.92\\
061021&46.2&0.50$\pm$0.21&0.55$\pm$0.08&...&9.27$\pm$4.33&24.95$\pm$8.20&0.01$\pm$0.00&0.3463\\
061110A&40.7&0.44$\pm$0.14&0.05$\pm$0.51&0.38$\pm$0.26&0.59$\pm$0.42&2.47$\pm$1.49&1.09$\pm$0.53&0.758\\
061201&0.76&0.03$\pm$0.03&0.30$\pm$0.19&0.10$\pm$0.07&7.68$\pm$1.82&6.10$\pm$2.21&0.18$\pm$0.02&0.111\\
070110&88.4&2.97$\pm$0.26&-0.03$\pm$0.06&...&1.21$\pm$0.02&2.70$\pm$0.06&0.91$\pm$0.01&2.352\\
080707&27.1&0.33$\pm$0.05&0.20$\pm$0.22&1.64$\pm$0.62&0.56$\pm$0.22&1.06$\pm$0.30&5.97$\pm$3.18&1.23\\
081007&10.0&0.06$\pm$0.01&0.49$\pm$0.11&1.95$\pm$0.98&0.29$\pm$0.13&1.17$\pm$0.45&4.86$\pm$3.21&0.5295\\
081029&270.0&12.55$\pm$1.93&0.35$\pm$0.09&...&0.90$\pm$0.07&1.90$\pm$0.11&1.85$\pm$0.12&3.8479\\
100219A&18.8&1.78$\pm$0.35&0.44$\pm$0.36&...&0.68$\pm$0.10&1.72$\pm$0.24&2.26$\pm$0.37&4.6667\\
100302A&17.9&2.03$\pm$0.36&0.29$\pm$0.30&...&1.90$\pm$1.29&3.86$\pm$1.57&0.45$\pm$0.09&4.813\\
100425A&37.0&0.49$\pm$0.12&0.16$\pm$0.18&1.32$\pm$0.76&0.42$\pm$0.26&1.06$\pm$0.50&5.98$\pm$5.37&1.755\\
110715A&13.0&2.79$\pm$0.13&0.23$\pm$0.11&1.81$\pm$0.27&4.63$\pm$3.79&1.10$\pm$0.14&5.48$\pm$1.30&0.82\\
110808A&48.0&17.86$\pm$3.03&-0.12$\pm$0.28&0.88$\pm$0.69&0.44$\pm$0.42&1.40$\pm$1.05&3.41$\pm$3.65&1.35\\
120422A&5.35&(6.31$\pm$1.18)e-3&-0.79$\pm$0.13&0.07$\pm$0.08&1.04$\pm$0.44&6.56$\pm$4.00&0.15$\pm$0.03&0.283\\
120521C&26.7&10.05$\pm$1.30&0.35$\pm$0.28&...&0.93$\pm$0.23&2.11$\pm$0.37&1.50$\pm$0.25&6.0\\
130408A&28.0&8.21$\pm$1.63&0.18$\pm$0.18&...&0.44$\pm$0.06&1.13$\pm$0.14&5.24$\pm$1.14&3.757\\
130603B&0.18&0.36$\pm$0.05&-0.27$\pm$0.11&1.73$\pm$1.02&0.97$\pm$0.37&1.32$\pm$0.42&3.82$\pm$1.85&0.3564\\
140903A&0.3&0.06$\pm$0.01&-0.70$\pm$0.13&0.34$\pm$0.15&2.04$\pm$0.45&3.00$\pm$0.78&0.74$\pm$0.13&0.351\\
151215A&17.8&0.55$\pm$0.48&0.19$\pm$0.19&1.16$\pm$1.46&1.43$\pm$3.74&0.98$\pm$0.95&6.91$\pm$13.57&2.59\\
161108A&105.1&0.54$\pm$0.13&0.33$\pm$0.22&0.66$\pm$0.32&0.53$\pm$0.30&1.68$\pm$0.77&2.37$\pm$1.30&1.159\\
\enddata
\tablenotetext{a}{The kinetic energy, which is calculated from the normal decay segment; the kinetic energy of the afterglow. }
\tablenotetext{b}{Dipolar magnetic field strength at the polar cap in units of $10^{15}$ G, and the initial spin period of the magnetar in units of milliseconds, assuming an isotropic magnetar wind.}
\tablenotetext{c}{The total rotation energy of the millisecond magnetar.}
\tablenotetext{d}{We did not calculate the kinetic energy for the bursts that have an internal plateau.}
\end{deluxetable}

\clearpage
\startlongtable
\begin{deluxetable}{lccccccccccccccccccccc}
\tablewidth{0pt}
\tabletypesize{\scriptsize}
\tablecaption{The Results of the Estimation Jet-opening Angle}
\tablenum{4}
\tablehead{
\colhead{GRB}
&\colhead{$\theta_{j}$\tablenotemark{a}}
&\colhead{$f_{b}$\tablenotemark{a}}
&\colhead{$a_{j}$\tablenotemark{b}}
&\colhead{$\theta^{'}_{j}$\tablenotemark{c}}
&\colhead{$f^{'}_{b}$\tablenotemark{c}}\\
&
\colhead{($10^{-2}$)}&
\colhead{($10^{-2}$)}&
&
\colhead{($10^{-2}$)}&
\colhead{($10^{-2}$)}&
}
\startdata
\hline
&Gold\\
\hline
050401&15.09$\pm$0.62&1.14$\pm$0.09&10.79&6.99$\pm$0.39&0.24$\pm$0.03&\\
050730&...&...&...&59.76$\pm$1.62&17.36$\pm$0.91&\\
060607A&...&...&...&11.10$\pm$0.14&0.62$\pm$0.02&\\
061222A&21.42$\pm$0.22&2.29$\pm$0.05&14.64&7.31$\pm$0.77&0.27$\pm$0.11&\\
080721&21.85$\pm$0.16&2.38$\pm$0.03&17.75&6.15$\pm$0.10&0.19$\pm$0.01&\\
100902A&...&...&...&8.09$\pm$0.17&0.33$\pm$0.01&\\
111209A&...&...&...&5.88$\pm$0.20&0.17$\pm$0.01&\\
140206A&22.53$\pm$0.38&2.53$\pm$0.09&14.81&7.60$\pm$0.07&0.29$\pm$0.01&\\
150403A&19.21$\pm$0.11&1.84$\pm$0.02&5.77&16.62$\pm$0.49&1.38$\pm$0.09&\\
\hline
&Silver\\
\hline
050315&16.26$\pm$0.38&1.32$\pm$0.06&6.87&11.83$\pm$0.32&0.70$\pm$0.04&\\
050319&22.17$\pm$0.96&2.45$\pm$0.21&10.29&10.77$\pm$0.59&0.58$\pm$0.07&\\
050505&15.14$\pm$0.50&1.14$\pm$0.08&9.10&8.31$\pm$0.18&0.35$\pm$0.01&\\
050802&24.31$\pm$0.71&2.94$\pm$0.17&8.90&13.64$\pm$0.51&0.93$\pm$0.08&\\
050803&33.30$\pm$1.00&5.49$\pm$0.33&6.26&26.57$\pm$0.62&3.52$\pm$0.17&\\
060115&12.60$\pm$1.69&0.79$\pm$0.21&5.43&11.59$\pm$0.58&0.67$\pm$0.07&\\
060210&21.96$\pm$0.95&2.40$\pm$0.21&12.20&8.99$\pm$0.13&0.40$\pm$0.01&\\
060418&23.92$\pm$1.07&2.85$\pm$0.25&12.52&9.54$\pm$0.30&0.46$\pm$0.03&\\
060502A&37.92$\pm$1.21&7.10$\pm$0.45&11.85&15.98$\pm$0.24&1.28$\pm$0.04&\\
060604&20.74$\pm$0.76&2.14$\pm$0.16&3.95&26.25$\pm$1.56&3.43$\pm$0.50&\\
060605&7.39$\pm$0.25&0.27$\pm$0.02&2.28&16.18$\pm$0.40&1.31$\pm$0.07&\\
060614&57.78$\pm$0.84&16.23$\pm$0.46&6.93&41.67$\pm$3.32&8.57$\pm$2.38&\\
060714&28.93$\pm$1.14&4.16$\pm$0.33&13.39&10.79$\pm$0.53&0.58$\pm$0.06&\\
060729&81.80$\pm$0.38&31.63$\pm$0.28&18.69&21.86$\pm$0.51&2.38$\pm$0.12&\\
060814&30.50$\pm$0.66&4.61$\pm$0.20&13.17&11.57$\pm$0.21&0.67$\pm$0.02&\\
060906&11.94$\pm$0.88&0.71$\pm$0.10&5.73&10.41$\pm$0.42&0.54$\pm$0.05&\\
060908&21.96$\pm$0.83&2.40$\pm$0.18&8.14&13.47$\pm$0.16&0.91$\pm$0.02&\\
061121&34.03$\pm$0.64&5.73$\pm$0.21&23.64&7.19$\pm$0.17&0.26$\pm$0.01&\\
070129&24.81$\pm$0.69&3.06$\pm$0.17&11.52&10.75$\pm$0.29&0.58$\pm$0.03&\\
070306&21.11$\pm$0.28&2.22$\pm$0.06&8.65&12.20$\pm$0.26&0.74$\pm$0.04&\\
070508&30.55$\pm$0.42&4.63$\pm$0.13&14.73&10.36$\pm$0.20&0.54$\pm$0.02&\\
070529&17.39$\pm$1.41&1.51$\pm$0.24&7.02&12.37$\pm$0.34&0.77$\pm$0.04&\\
080310&16.64$\pm$0.47&1.38$\pm$0.08&7.51&11.07$\pm$0.75&0.61$\pm$0.11&\\
080430&49.60$\pm$1.17&12.05$\pm$0.56&10.86&22.81$\pm$0.50&2.60$\pm$0.12&\\
080516&6.36$\pm$0.57&0.20$\pm$0.04&1.71&18.60$\pm$0.87&1.73$\pm$0.17&\\
080605&16.90$\pm$0.41&1.43$\pm$0.07&8.69&9.72$\pm$0.21&0.47$\pm$0.02&\\
080905B&23.01$\pm$0.79&2.64$\pm$0.18&9.80&11.73$\pm$0.44&0.69$\pm$0.05&\\
081008&12.50$\pm$1.05&0.78$\pm$0.13&5.85&10.67$\pm$0.40&0.57$\pm$0.05&\\
081221&20.59$\pm$0.31&2.11$\pm$0.06&13.94&7.38$\pm$0.09&0.27$\pm$0.01&\\
090407&16.82$\pm$0.78&1.41$\pm$0.13&4.47&18.80$\pm$1.04&1.77$\pm$0.23&\\
090418A&15.38$\pm$0.37&1.18$\pm$0.06&8.68&8.85$\pm$0.31&0.39$\pm$0.03&\\
090510&11.78$\pm$0.60&0.69$\pm$0.07&2.40&24.56$\pm$1.28&3.01$\pm$0.35&\\
090516&13.30$\pm$0.37&0.88$\pm$0.05&10.80&6.15$\pm$0.17&0.19$\pm$0.01&\\
090529&17.12$\pm$2.75&1.46$\pm$0.47&5.06&16.90$\pm$1.42&1.43$\pm$0.32&\\
090530&31.32$\pm$1.35&4.86$\pm$0.41&7.64&20.48$\pm$0.64&2.09$\pm$0.14&\\
090618&49.17$\pm$0.34&11.85$\pm$0.16&26.06&9.42$\pm$0.20&0.44$\pm$0.02&\\
091018&25.15$\pm$0.57&3.14$\pm$0.14&5.84&21.49$\pm$0.64&2.31$\pm$0.14&\\
100418A&38.01$\pm$1.29&7.14$\pm$0.48&5.64&33.66$\pm$5.25&5.62$\pm$3.41&\\
100615A&13.80$\pm$0.38&0.95$\pm$0.05&5.89&11.70$\pm$0.13&0.69$\pm$0.02&\\
100621A&32.17$\pm$1.89&5.13$\pm$0.60&11.30&14.22$\pm$0.39&1.01$\pm$0.06&\\
100704A&20.22$\pm$0.48&2.04$\pm$0.10&11.54&8.76$\pm$0.14&0.38$\pm$0.01&\\
100814A&23.35$\pm$0.46&2.71$\pm$0.11&12.75&9.15$\pm$0.15&0.42$\pm$0.01&\\
100906A&10.25$\pm$0.29&0.53$\pm$0.03&6.67&7.68$\pm$0.07&0.29$\pm$0.00&\\
111008A&18.48$\pm$0.29&1.70$\pm$0.05&13.19&7.00$\pm$0.16&0.25$\pm$0.01&\\
111123A&9.50$\pm$0.62&0.45$\pm$0.06&5.84&8.12$\pm$0.13&0.33$\pm$0.01&\\
111228A&45.20$\pm$0.88&10.04$\pm$0.38&17.69&12.76$\pm$0.35&0.81$\pm$0.05&\\
120327A&12.11$\pm$0.65&0.73$\pm$0.08&5.75&10.52$\pm$0.11&0.55$\pm$0.01&\\
120712A&8.15$\pm$0.40&0.33$\pm$0.03&4.70&8.66$\pm$0.03&0.38$\pm$0.00&\\
120811C&8.48$\pm$0.72&0.36$\pm$0.06&3.79&11.17$\pm$0.44&0.62$\pm$0.05&\\
120922A&17.05$\pm$1.19&1.45$\pm$0.20&10.21&8.34$\pm$0.34&0.35$\pm$0.03&\\
121024A&12.93$\pm$1.65&0.83$\pm$0.21&4.09&15.79$\pm$0.49&1.25$\pm$0.08&\\
121128A&9.15$\pm$0.21&0.42$\pm$0.02&2.50&18.27$\pm$0.51&1.67$\pm$0.09&\\
121211A&17.91$\pm$1.84&1.60$\pm$0.33&3.03&29.53$\pm$0.89&4.34$\pm$0.26&\\
130420A&36.71$\pm$1.83&6.66$\pm$0.66&6.74&27.22$\pm$8.91&3.69$\pm$6.18&\\
130606A&8.70$\pm$0.88&0.38$\pm$0.08&5.23&8.31$\pm$0.20&0.35$\pm$0.02&\\
130612A&9.55$\pm$1.24&0.46$\pm$0.12&2.40&19.89$\pm$0.75&1.98$\pm$0.15&\\
131030A&37.50$\pm$1.36&6.95$\pm$0.50&21.22&8.83$\pm$0.08&0.39$\pm$0.01&\\
131105A&22.41$\pm$1.02&2.50$\pm$0.23&15.17&7.38$\pm$0.15&0.27$\pm$0.01&\\
140512A&16.23$\pm$0.43&1.31$\pm$0.07&7.78&10.43$\pm$0.12&0.54$\pm$0.01&\\
140518A&4.83$\pm$0.27&0.12$\pm$0.01&1.98&12.18$\pm$0.58&0.74$\pm$0.08&\\
140703A&4.98$\pm$0.23&0.12$\pm$0.01&1.60&15.55$\pm$0.23&1.21$\pm$0.04&\\
141121A&11.08$\pm$0.74&0.61$\pm$0.08&1.08&51.16$\pm$2.00&12.83$\pm$1.06&\\
150910A&9.53$\pm$0.20&0.45$\pm$0.02&3.10&15.33$\pm$0.67&1.18$\pm$0.11&\\
151027A&27.26$\pm$0.19&3.69$\pm$0.05&10.10&13.47$\pm$0.33&0.91$\pm$0.05&\\
160121A&5.78$\pm$1.13&0.17$\pm$0.06&1.46&19.82$\pm$1.65&1.96$\pm$0.53&\\
160227A&29.86$\pm$0.58&4.42$\pm$0.17&11.61&12.85$\pm$1.40&0.83$\pm$0.78&\\
160327A&6.23$\pm$0.35&0.19$\pm$0.02&2.97&10.48$\pm$1.80&0.55$\pm$0.65&\\
161117A&26.62$\pm$0.72&3.52$\pm$0.19&16.30&8.16$\pm$0.07&0.33$\pm$0.01&\\
170113A&26.20$\pm$0.58&3.41$\pm$0.15&6.26&20.89$\pm$3.28&2.18$\pm$2.39&\\
\hline
&Bronze\\
\hline
051109B&41.94$\pm$1.24&8.67$\pm$0.51&1.70&123.29$\pm$5.00&66.97$\pm$5.04&\\
051221A&40.77$\pm$1.01&8.19$\pm$0.40&8.45&24.10$\pm$3.51&2.90$\pm$3.66&\\
060526&...&...&...&16.32$\pm$0.67&1.33$\pm$0.13&\\
060708&57.94$\pm$2.16&16.32$\pm$1.18&15.47&18.71$\pm$0.53&1.75$\pm$0.10&\\
061021&...&...&...&22.15$\pm$2.45&2.45$\pm$1.30&\\
061110A&44.19$\pm$2.72&9.60$\pm$1.16&9.65&22.88$\pm$1.82&2.61$\pm$0.63&\\
061201&31.17$\pm$2.04&4.82$\pm$0.63&3.44&45.28$\pm$10.32&10.10$\pm$11.27&\\
070110&...&...&...&13.94$\pm$0.32&0.97$\pm$0.04&\\
080707&23.18$\pm$0.81&2.67$\pm$0.19&4.70&24.61$\pm$0.91&3.02$\pm$0.24&\\
081007&49.07$\pm$2.29&11.80$\pm$1.08&6.50&37.69$\pm$2.13&7.03$\pm$0.89&\\
081029&...&...&...&9.58$\pm$0.38&0.46$\pm$0.04&\\
100219A&...&...&...&15.92$\pm$0.80&1.27$\pm$0.14&\\
100302A&...&...&...&15.39$\pm$0.71&1.18$\pm$0.12&\\
100425A&27.32$\pm$1.49&3.71$\pm$0.40&6.12&22.29$\pm$1.41&2.48$\pm$0.39&\\
110715A&37.45$\pm$0.52&6.93$\pm$0.19&13.20&14.17$\pm$0.17&1.00$\pm$0.02&\\
110808A&33.05$\pm$2.64&5.41$\pm$0.86&18.88&8.74$\pm$0.39&0.38$\pm$0.04&\\
120422A&82.70$\pm$7.70&32.29$\pm$5.66&5.98&69.05$\pm$3.36&22.95$\pm$2.30&\\
120521C&...&...&...&10.15$\pm$0.34&0.52$\pm$0.04&\\
130408A&...&...&...&10.70$\pm$0.55&0.57$\pm$0.06&\\
130603B&20.06$\pm$1.08&2.00$\pm$0.21&4.15&24.12$\pm$0.81&2.90$\pm$0.20&\\
140903A&24.16$\pm$1.01&2.90$\pm$0.24&3.14&38.45$\pm$1.67&7.31$\pm$0.68&\\
151215A&16.97$\pm$1.96&1.44$\pm$0.33&3.92&21.59$\pm$4.87&2.33$\pm$2.59&\\
161108A&35.96$\pm$1.65&6.40$\pm$0.58&8.29&21.68$\pm$1.33&2.35$\pm$0.32&\\
\enddata
\tablenotetext{a}{Estimate of the jet-opening angle from the last data point of the \emph{Swift}/XRT observation using Eq.(\ref{eq:thetaj}).}
\tablenotetext{b}{$a_{j}$ value was estimated by setting the ratio of the $\theta_{j,0}$ equal to $\theta_{j}$.}
\tablenotetext{c}{Estimate of the jet-opening angle from the empirical relation of $\theta_{j,0}\approx5.0/\Gamma_{0}$ \citep{2012cosp...39.1354N}.}
\end{deluxetable}

\clearpage
\begin{deluxetable}{lccccccccccccccccccccc}
\tablewidth{0pt}
\tabletypesize{\scriptsize}
\tablecaption{The relationship between the $f_{\rm b}$ value and the sample size}
\tablenum{5}
\tablehead{
\colhead{$f_{\rm b}$}
&\colhead{Silver to Bronze\tablenotemark{a}}
&\colhead{GRB-SN Bursts\tablenotemark{b}}
&\colhead{Fractions}
&\colhead{Fractions}\\
&
\colhead{(Numbers)}&
\colhead{(Numbers)}&
\colhead{(for Silver to Bronze)}&
\colhead{(for GRB-SN Bursts)}&
}
\startdata
$f_{\rm b}$=0.7&0&0&0.0\%&0.0\%&\\
$f_{\rm b}$=0.6&1&0&1.6\%&0.0\%&\\
$f_{\rm b}$=0.5&2&0&3.2\%&0.0\%&\\
$f_{\rm b}$=0.4&5&0&8.1\%&0.0\%&\\
$f_{\rm b}$=0.3&6&0&11.3\%&0.0\%&\\
$f_{\rm b}$=0.2&18&0&29.0\%&0.0\%&\\
$f_{\rm b}$=0.1&62&3&100.0\%&100.0\%&\\
\enddata
\tablenotetext{a}{Number of Silver bursts that become Bronze bursts after applying a beaming correction ($E_{\rm K,iso}$).}
\tablenotetext{b}{Number of GRB-SN bursts from the Silver burst sample that become Bronze bursts after applying a beaming correction ($E_{\rm K,iso}$).}
\end{deluxetable}

\begin{figure}
\gridline{\fig{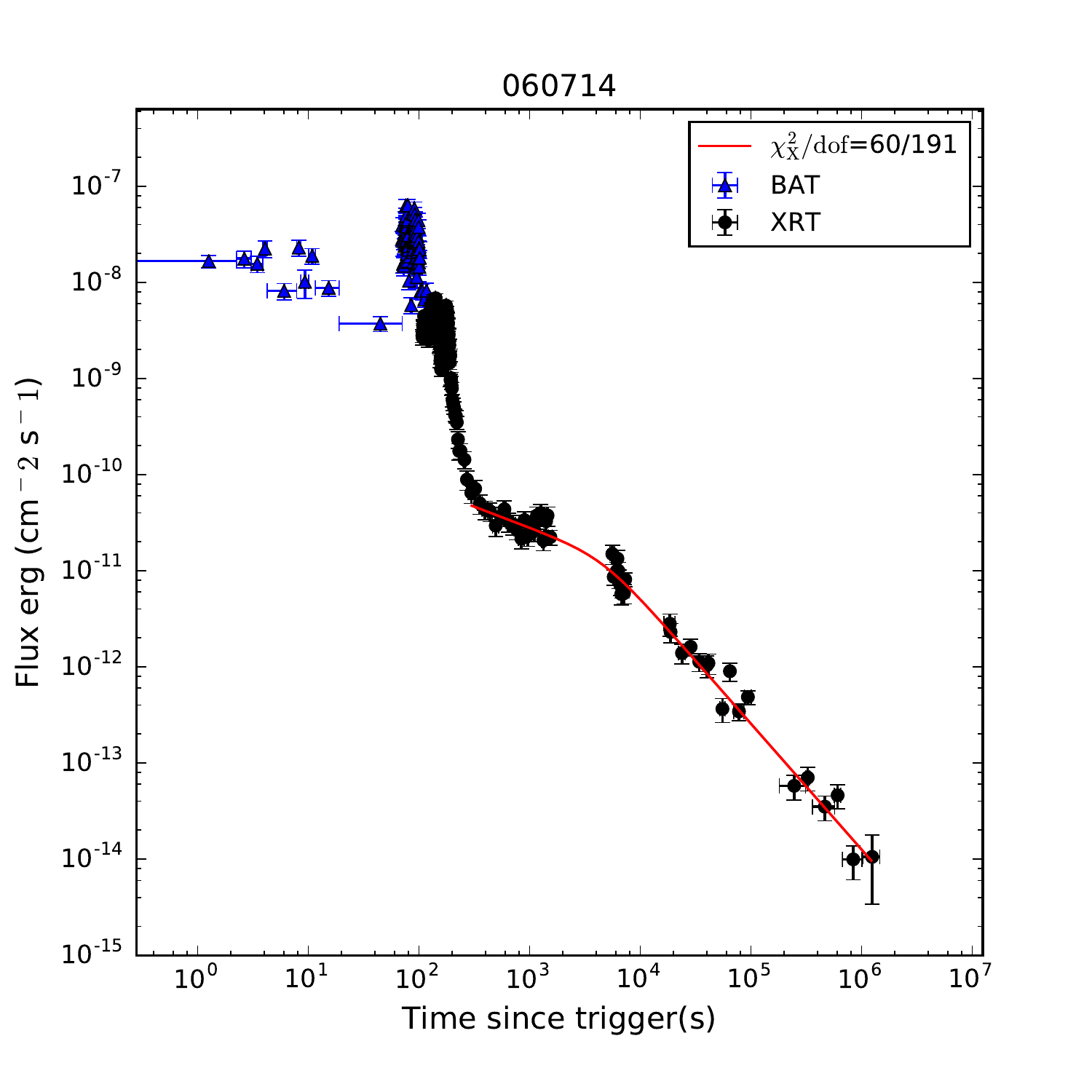}{0.28\textwidth}{}
          \fig{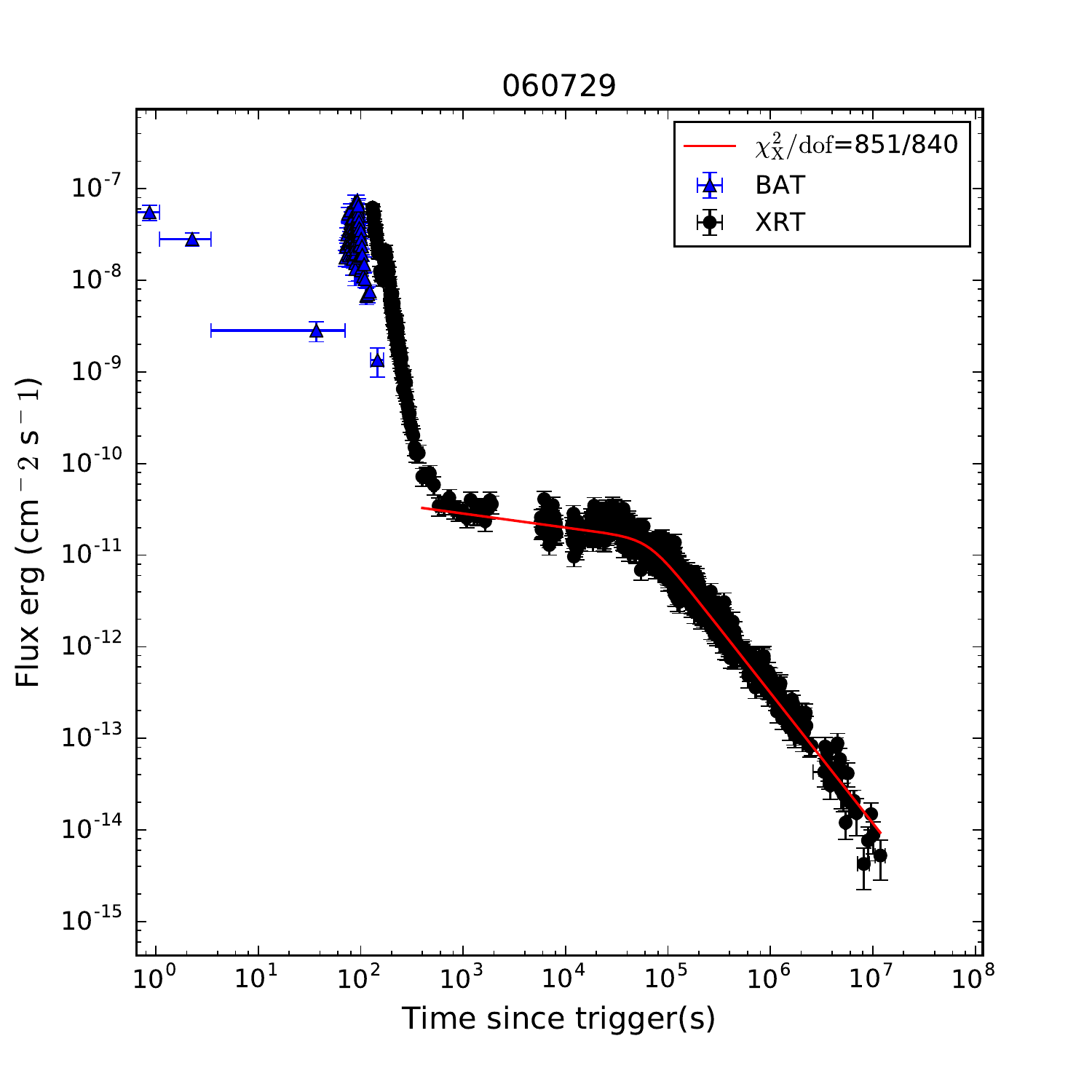}{0.28\textwidth}{}
          \fig{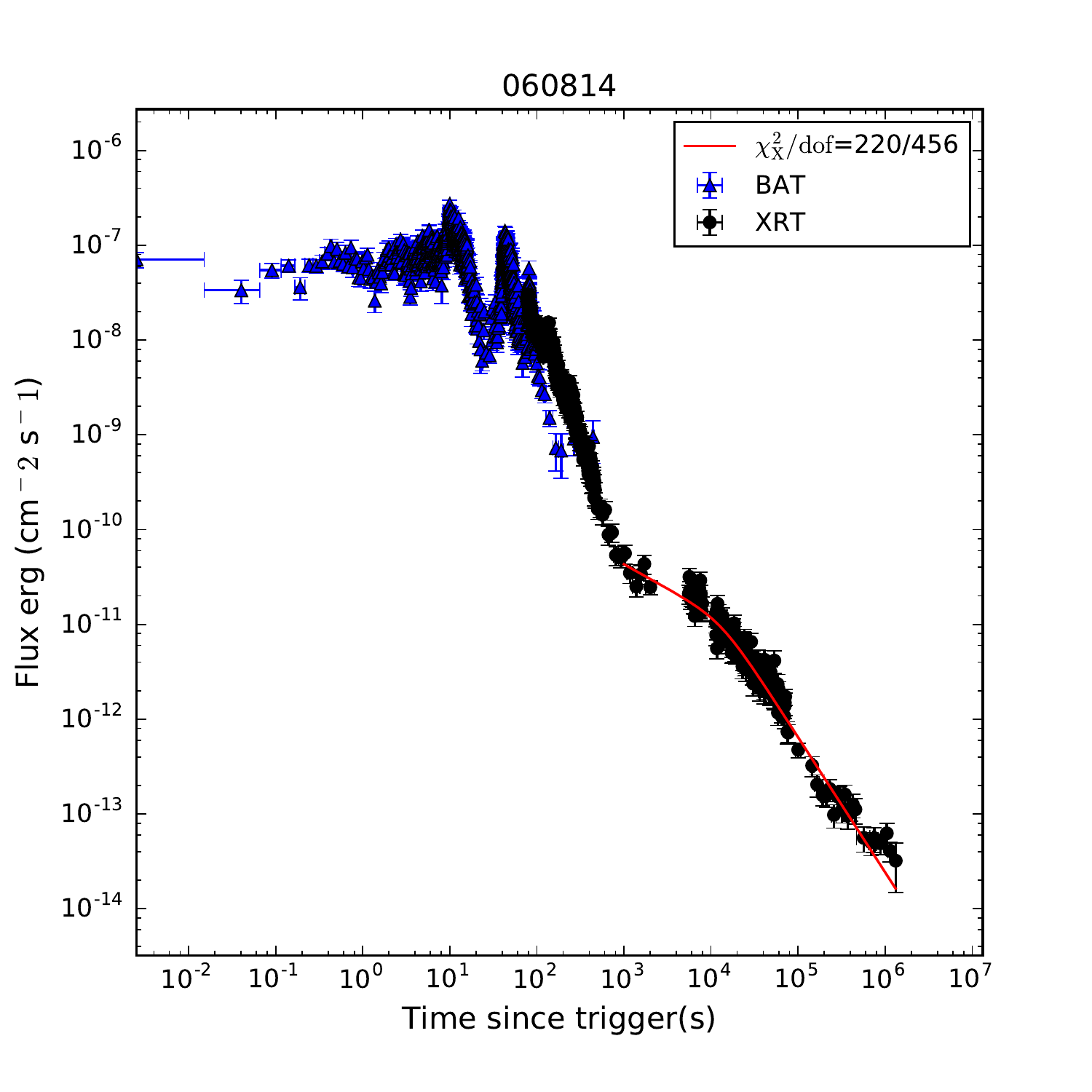}{0.28\textwidth}{}
          }
\gridline{\fig{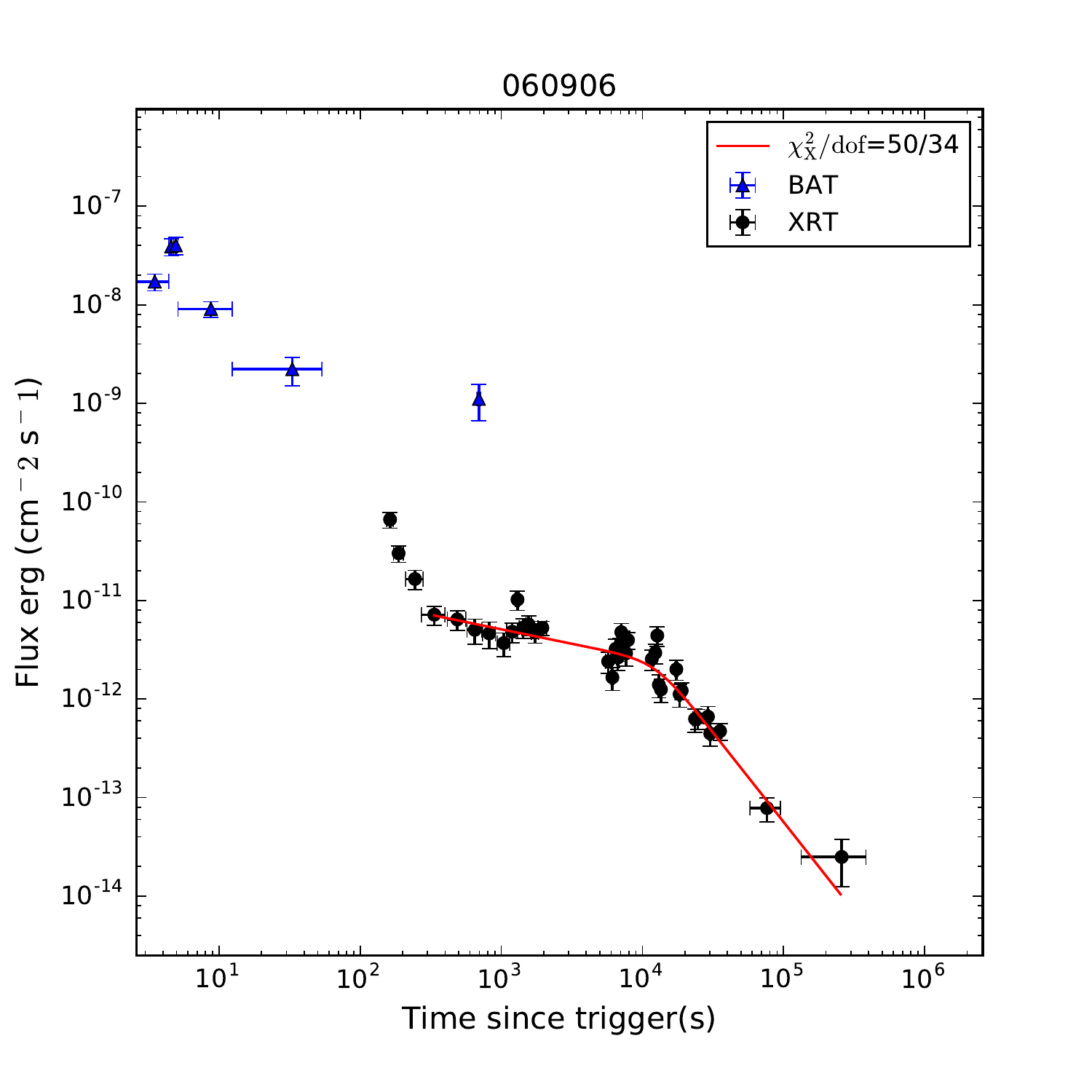}{0.28\textwidth}{}
          \fig{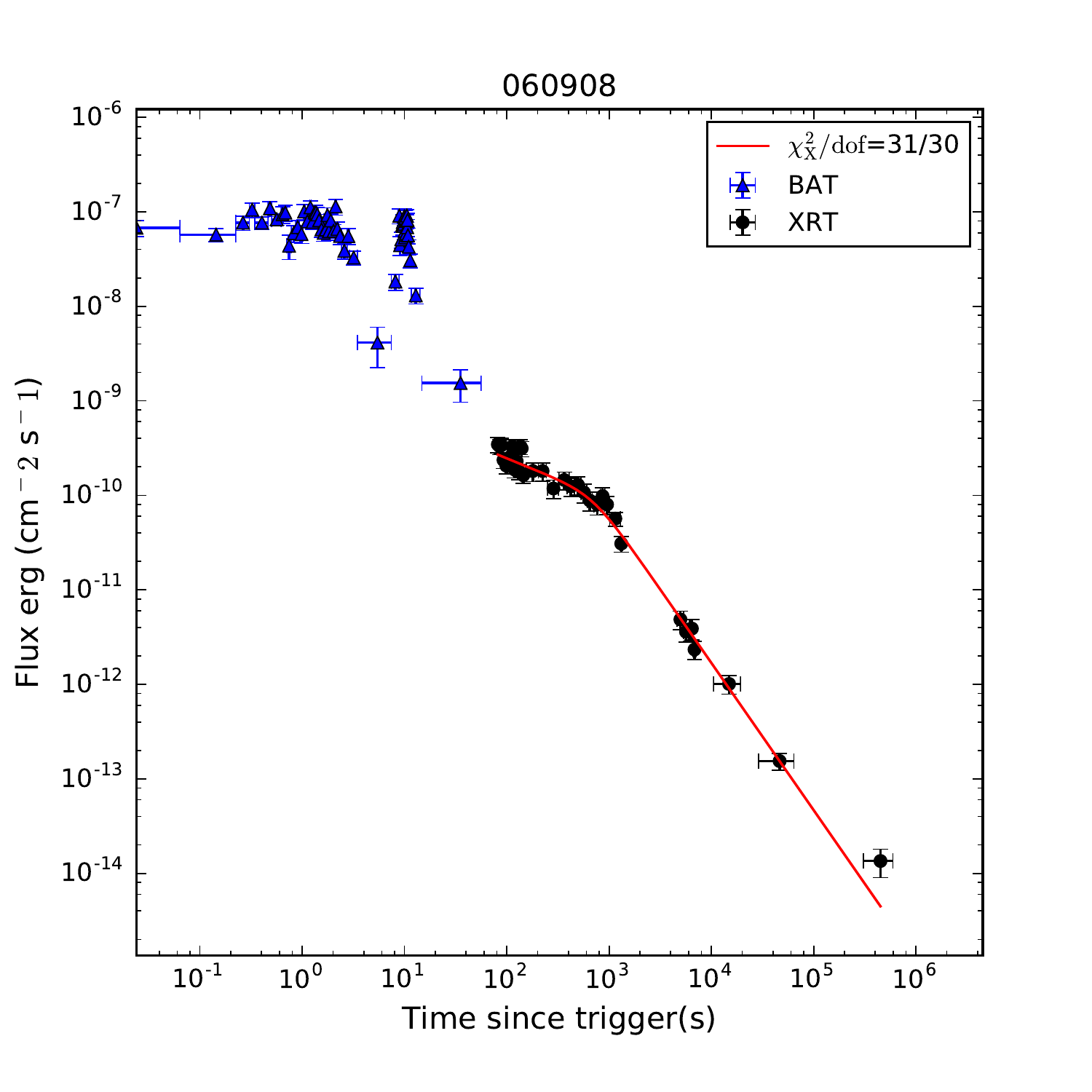}{0.28\textwidth}{}
          \fig{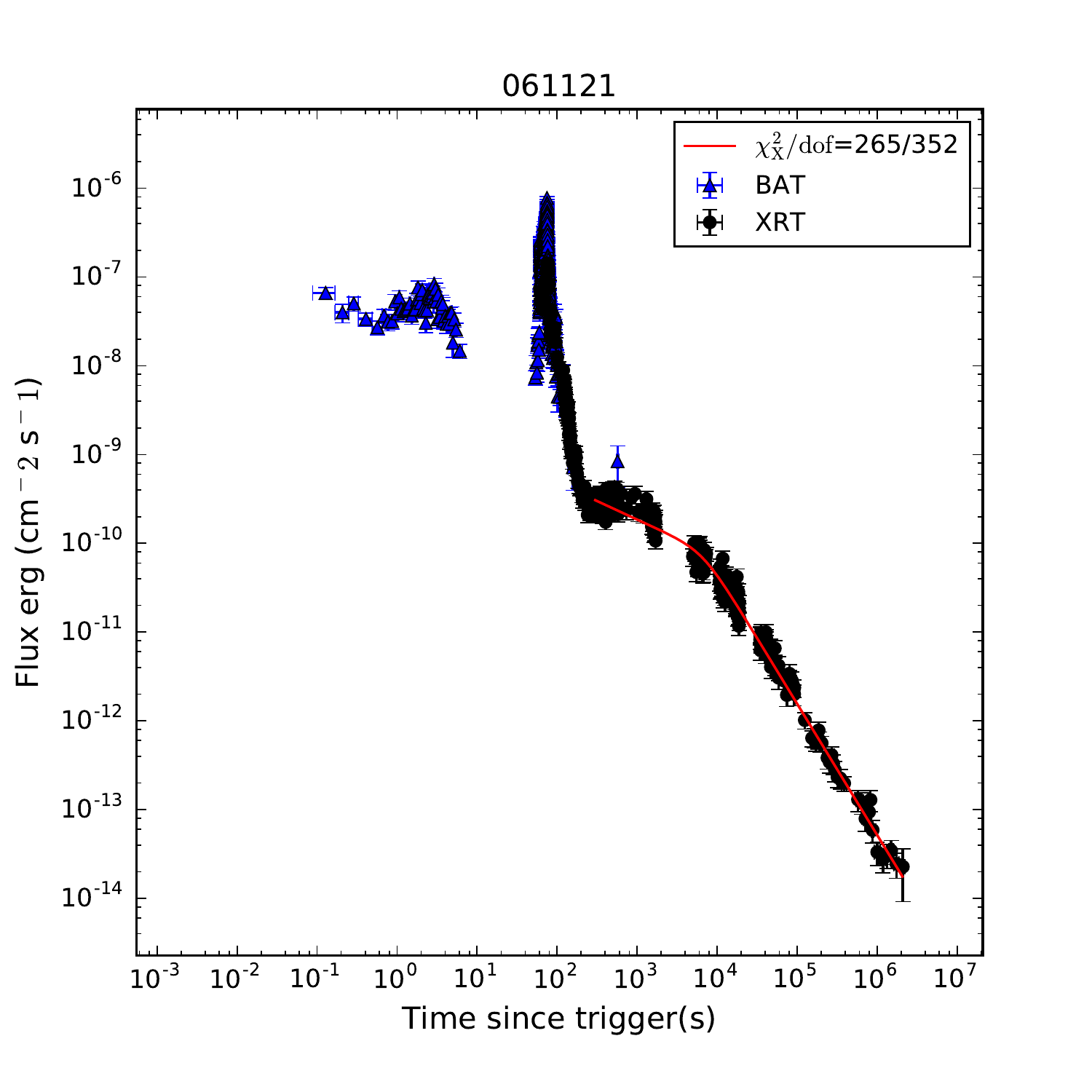}{0.28\textwidth}{}
          }
\gridline{\fig{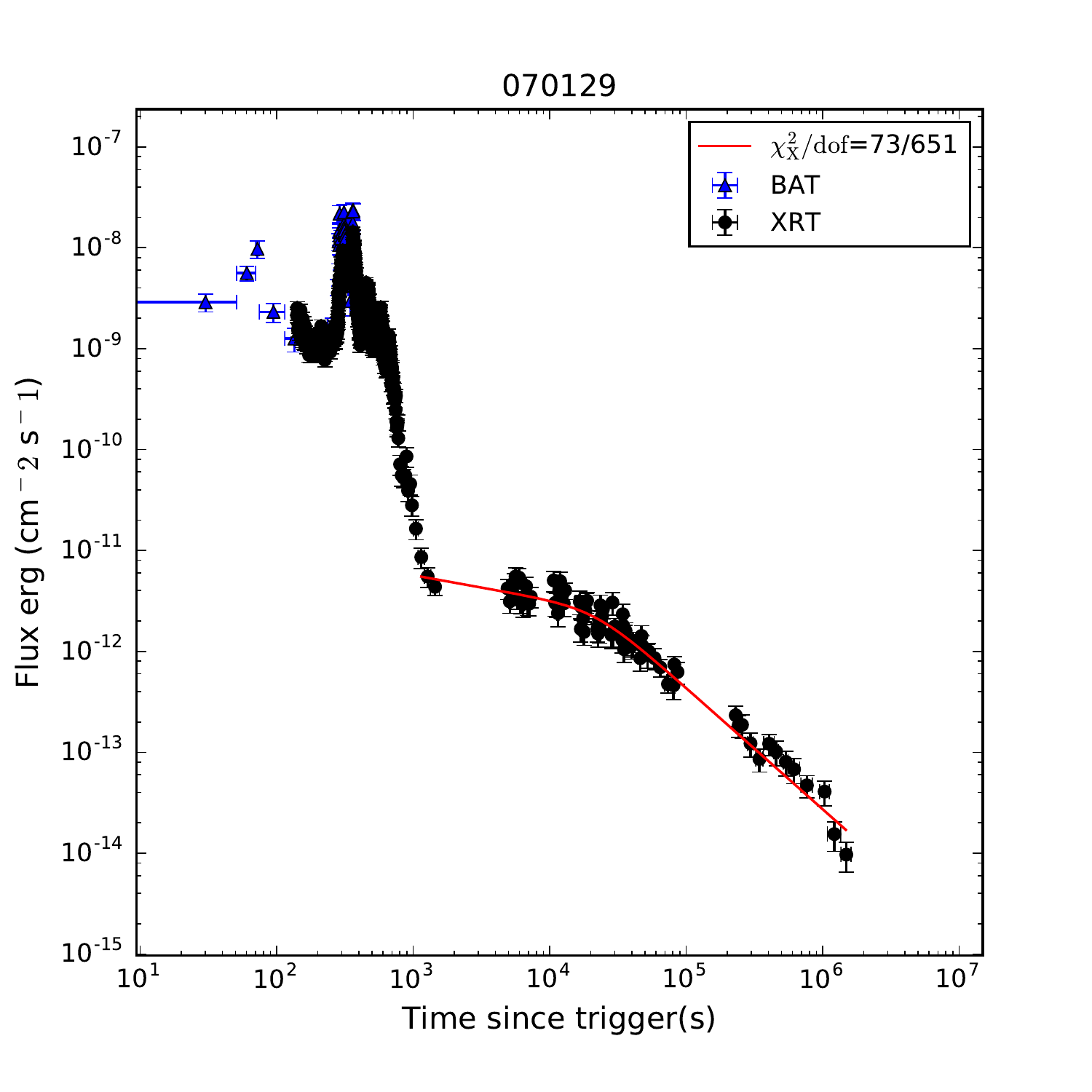}{0.28\textwidth}{}
          \fig{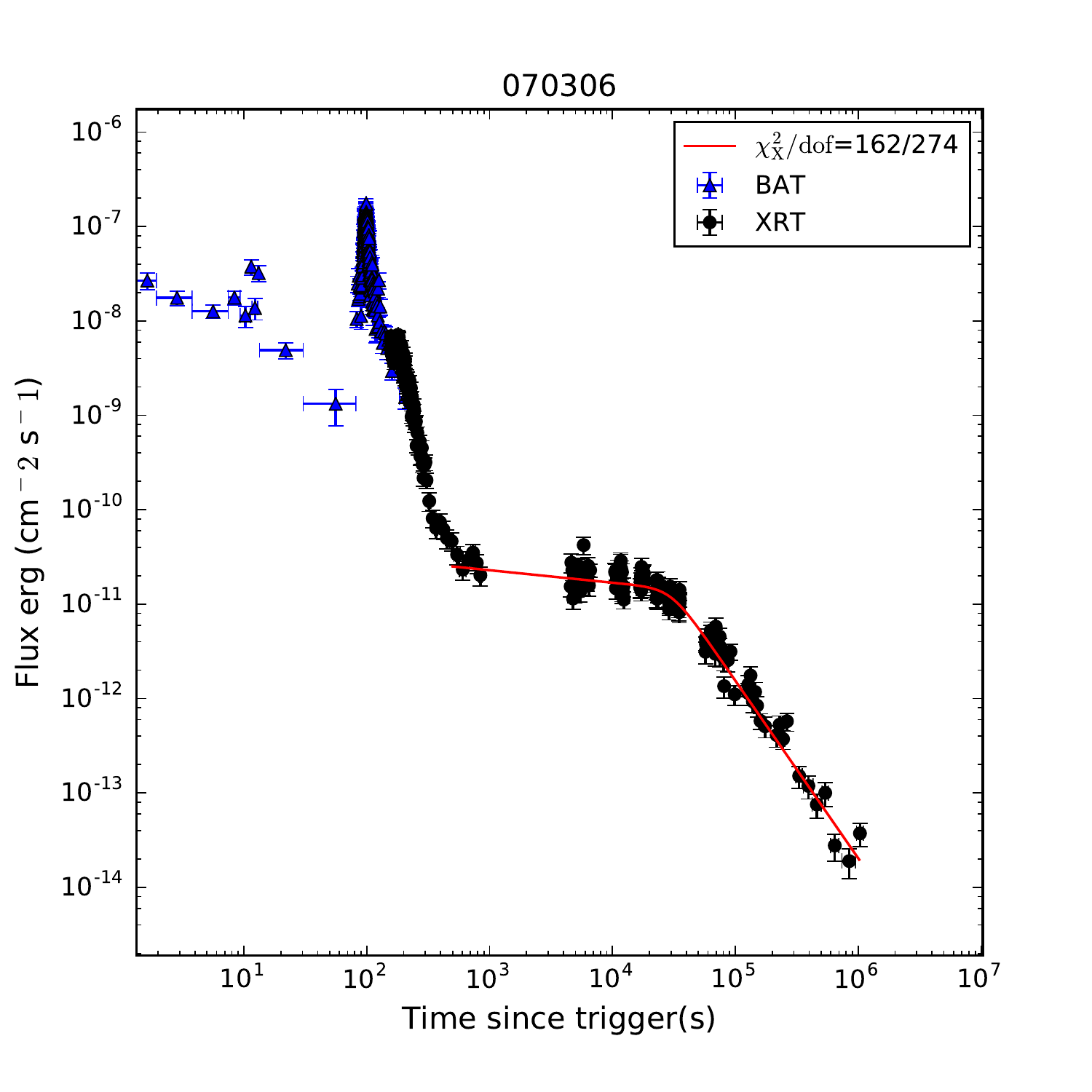}{0.28\textwidth}{}
          \fig{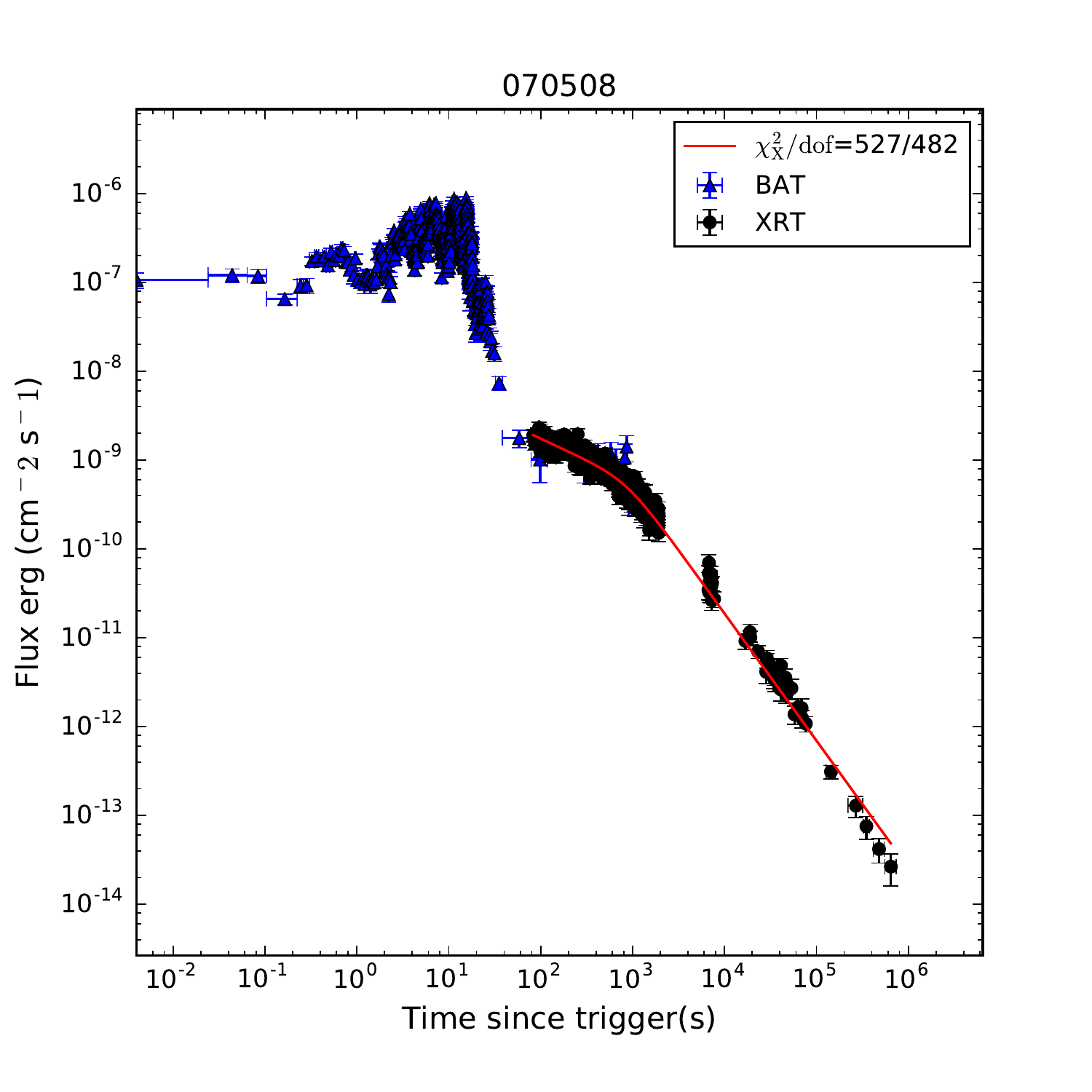}{0.28\textwidth}{}
          }
\gridline{\fig{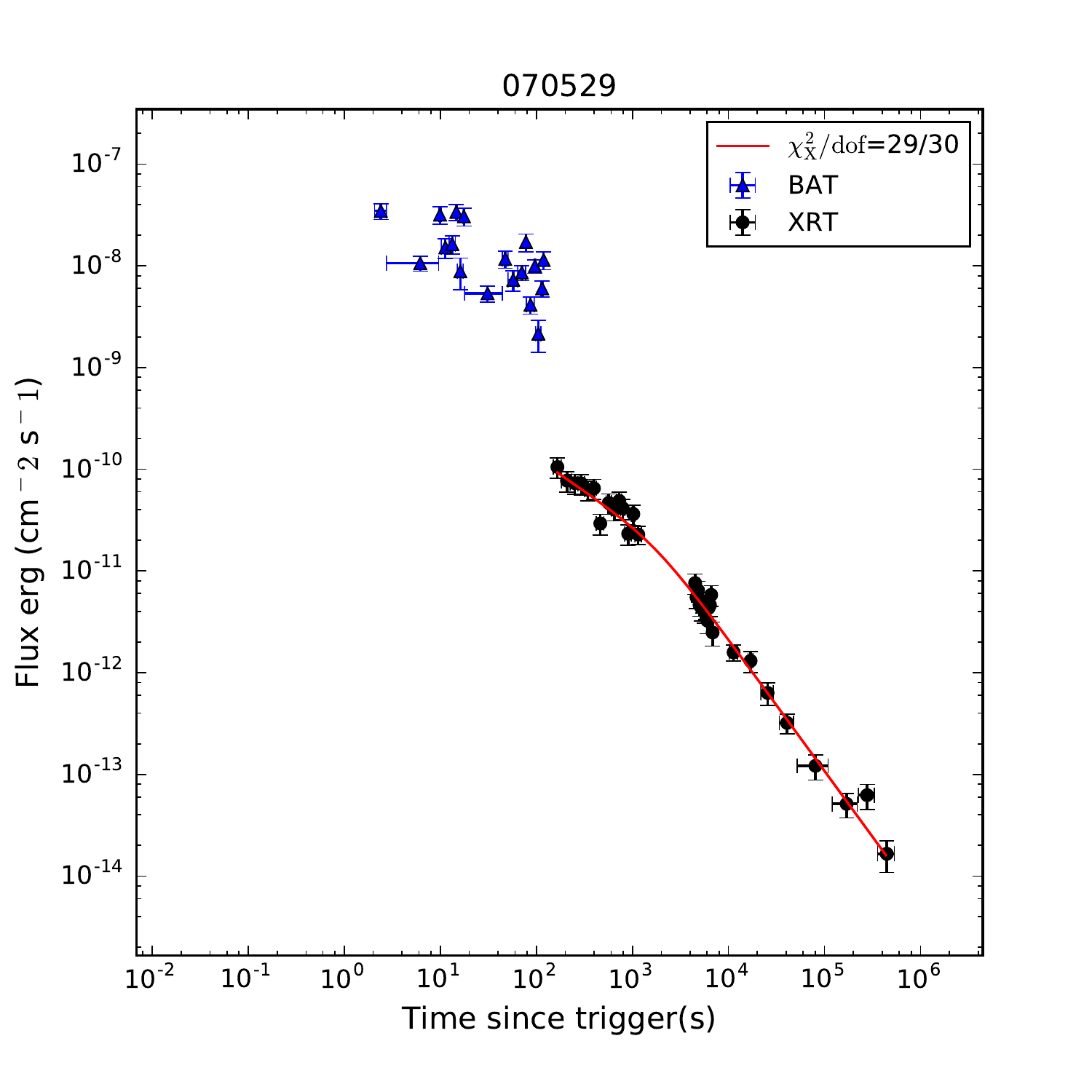}{0.28\textwidth}{}
          \fig{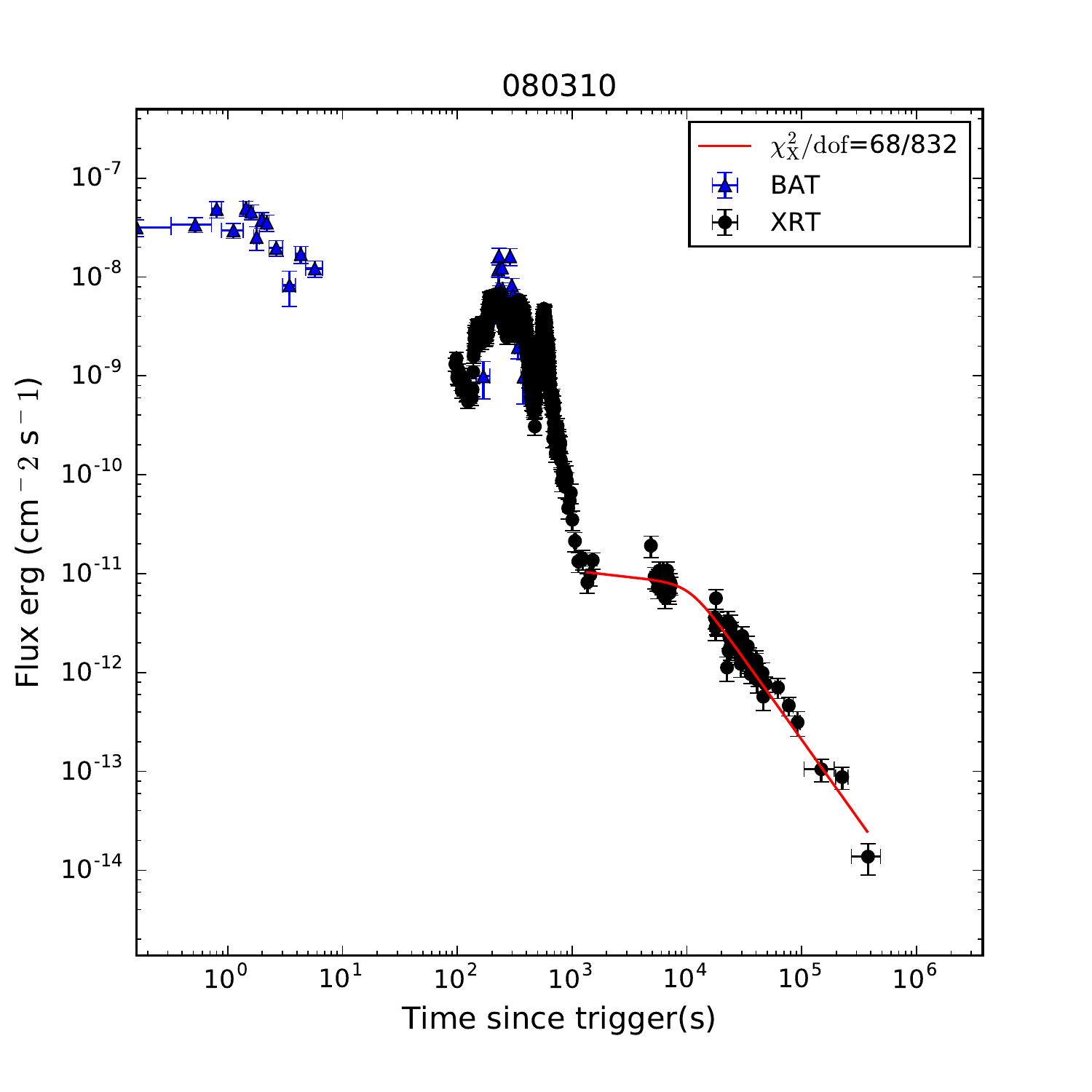}{0.28\textwidth}{}
          \fig{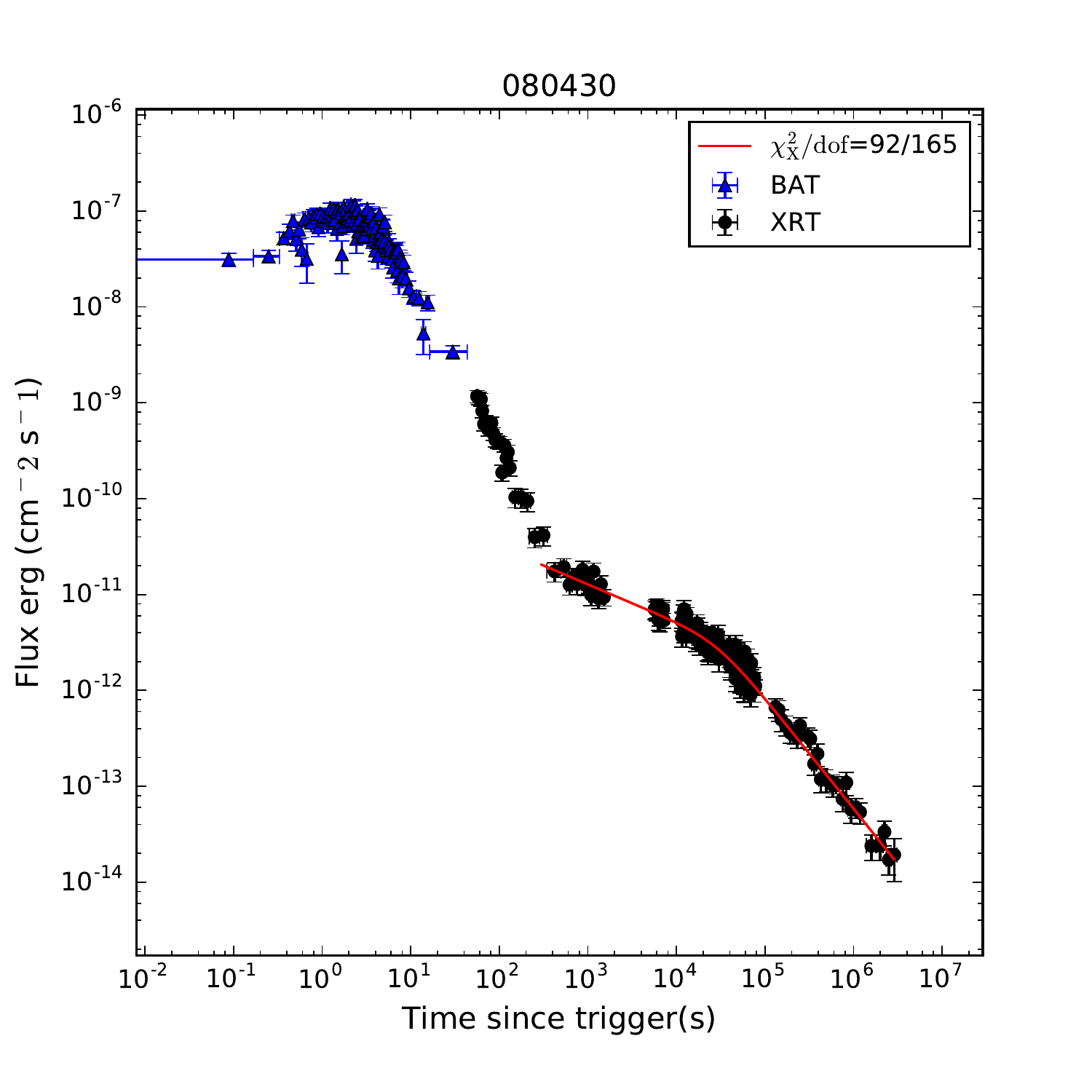}{0.28\textwidth}{}
          }
\center{Fig. \ref{Silver}--- Continued}
\end{figure}
\begin{figure}   
\gridline{\fig{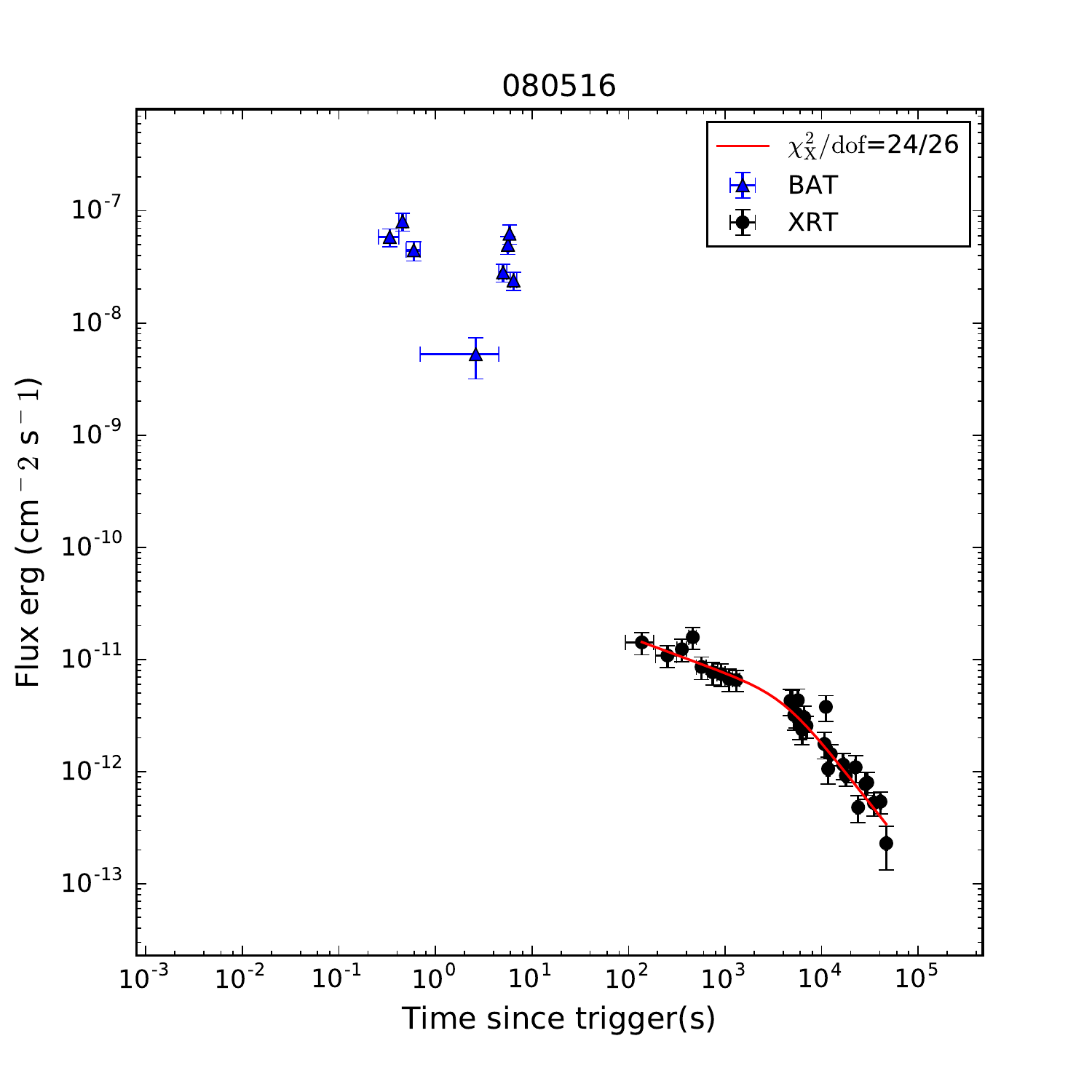}{0.28\textwidth}{}
          \fig{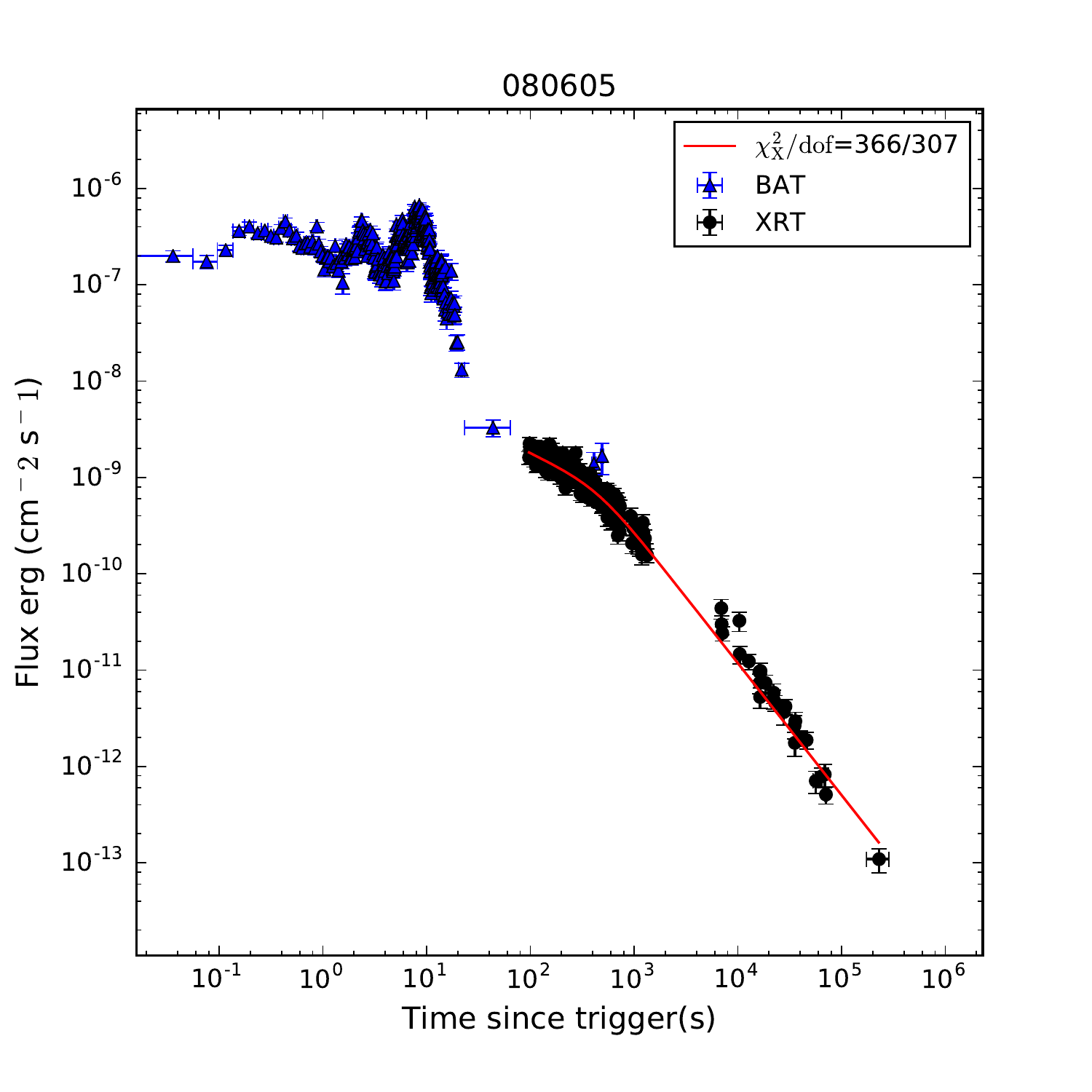}{0.28\textwidth}{}
          \fig{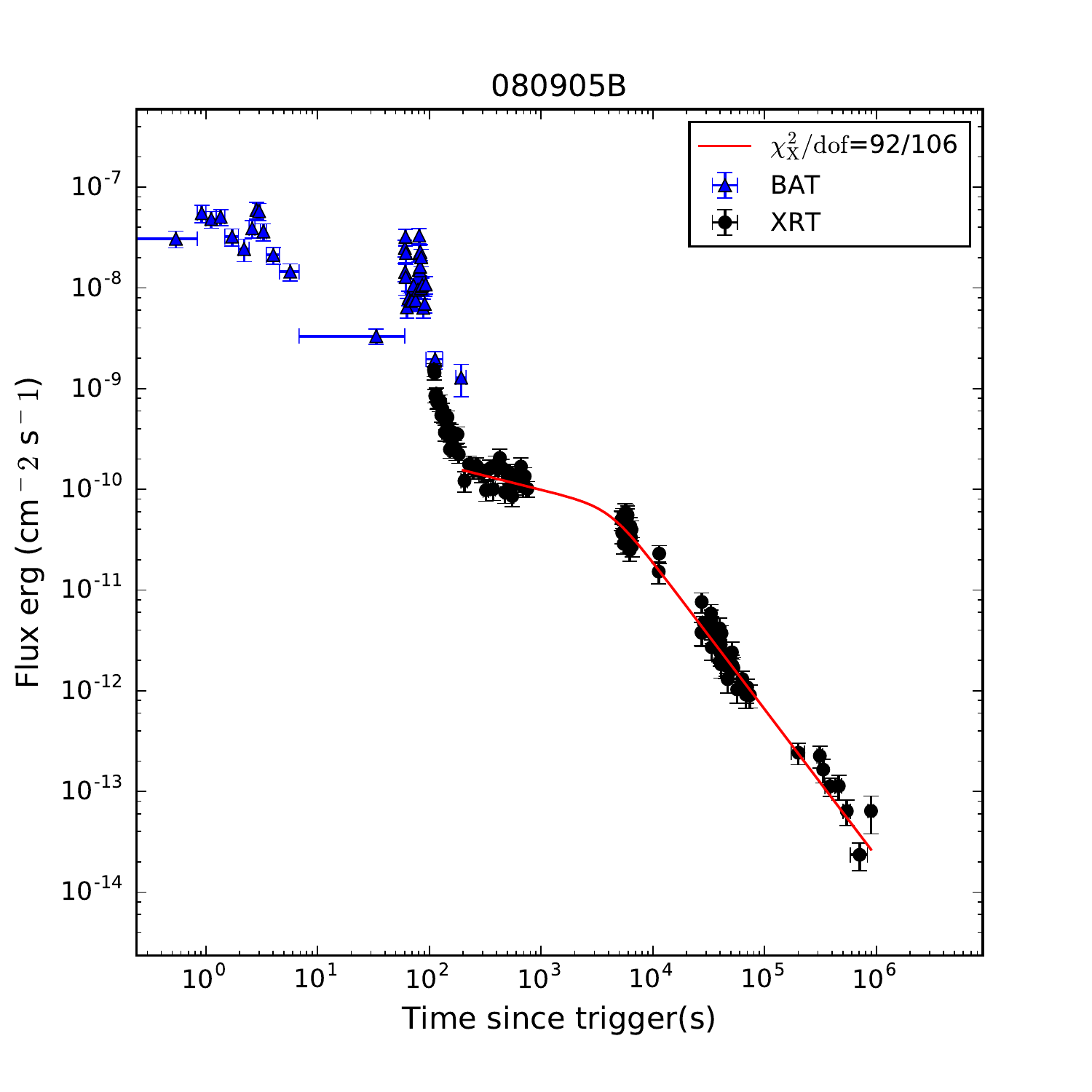}{0.28\textwidth}{}
          } 
\gridline{\fig{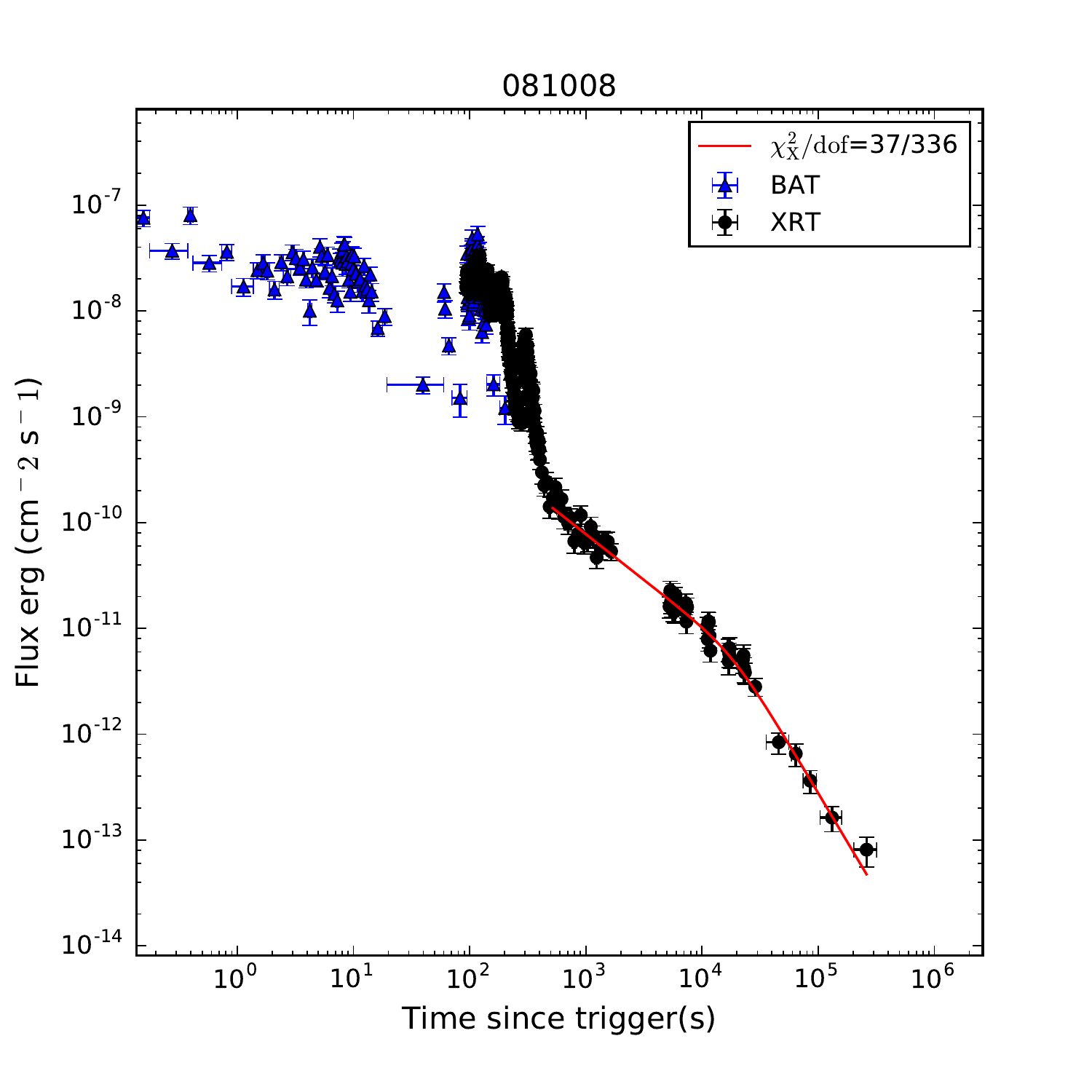}{0.28\textwidth}{}
          \fig{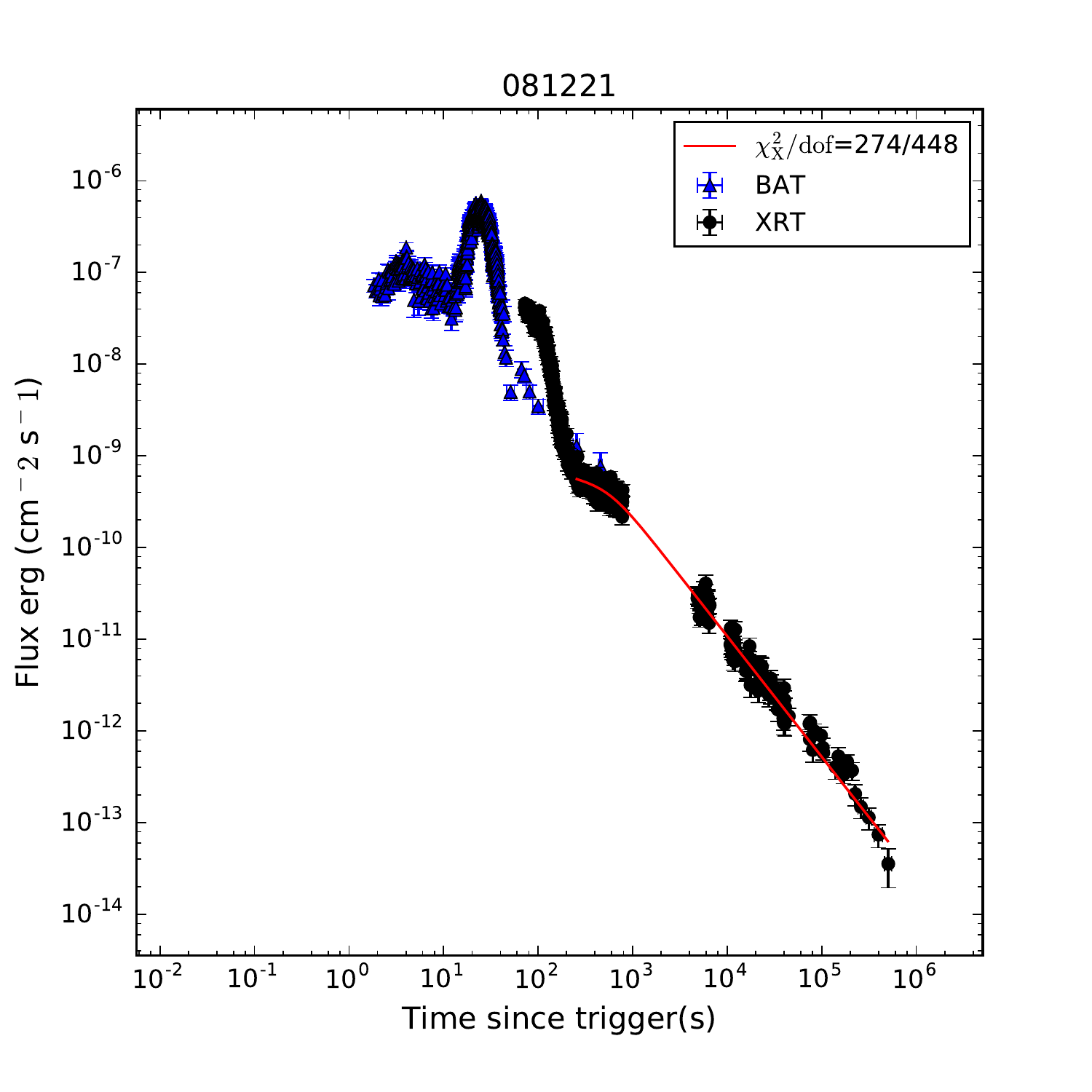}{0.28\textwidth}{}
          \fig{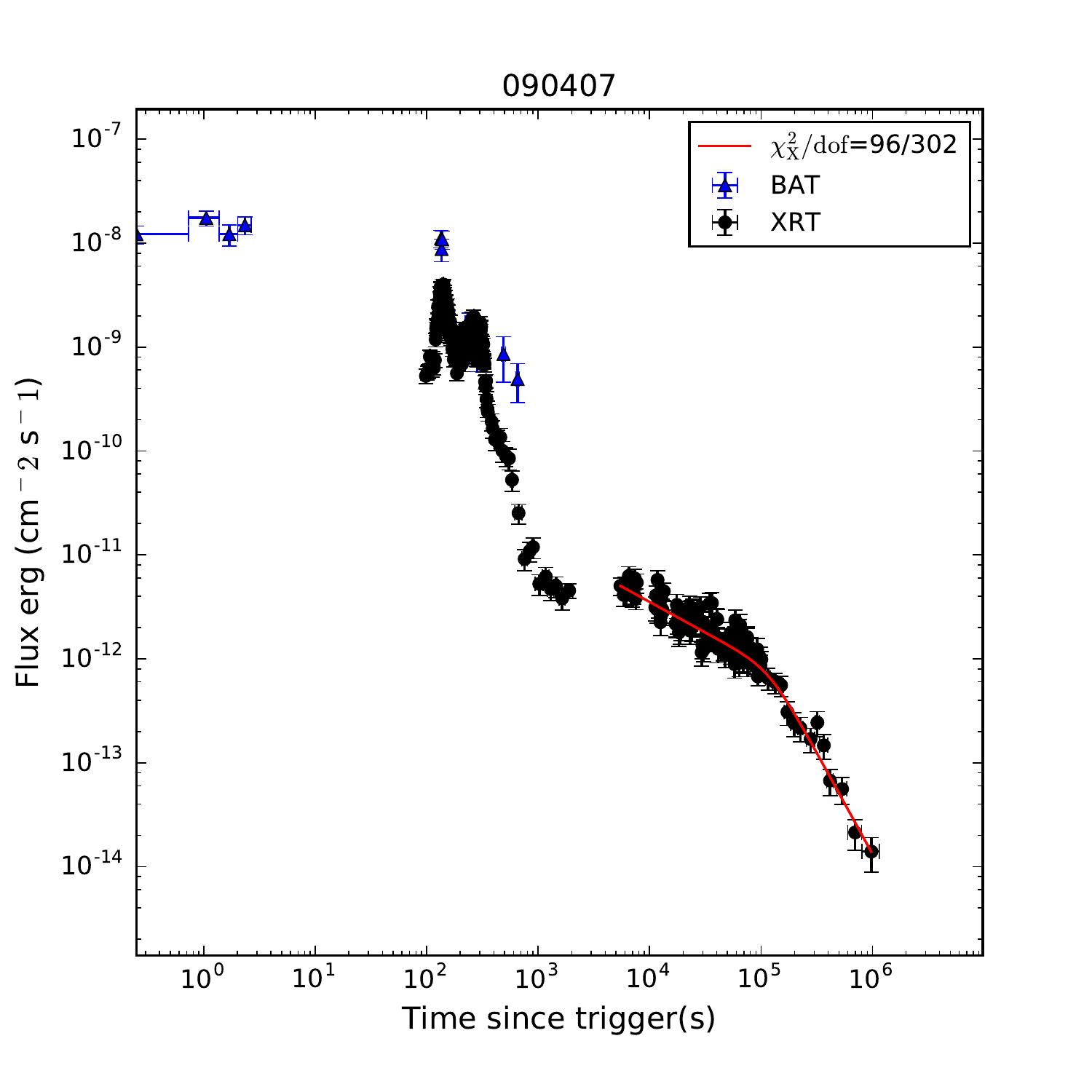}{0.28\textwidth}{}
          }
\gridline{\fig{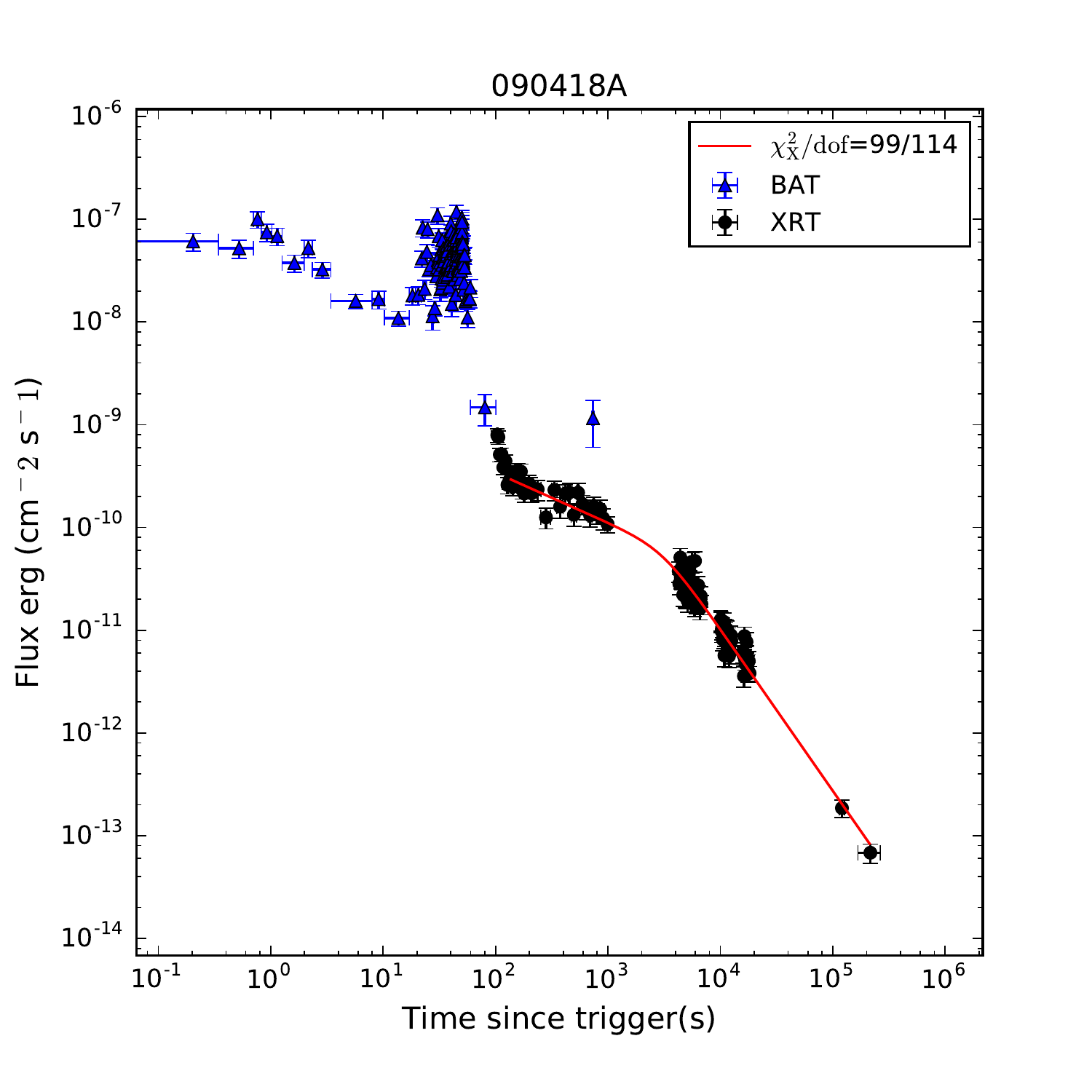}{0.28\textwidth}{}
          \fig{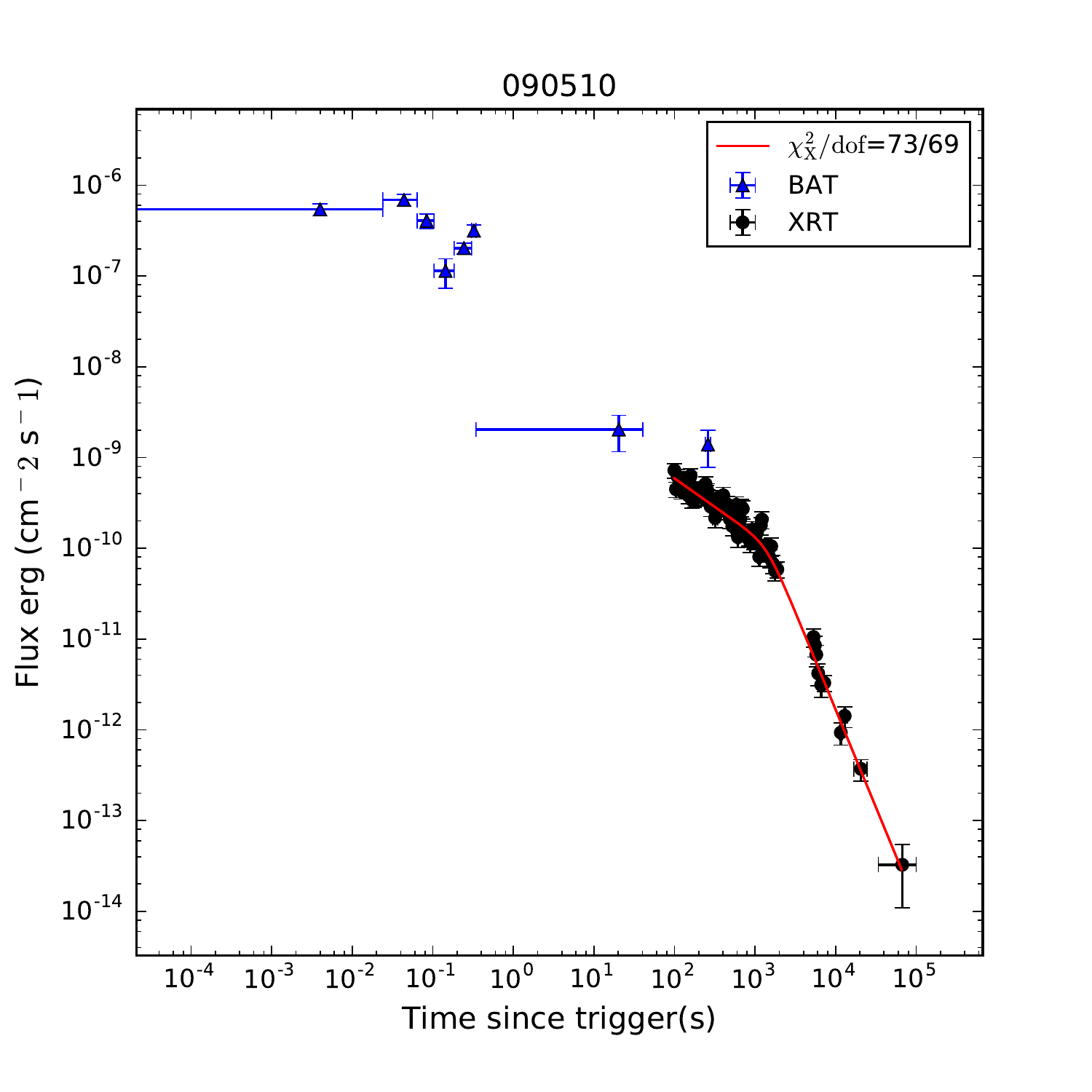}{0.28\textwidth}{}
          \fig{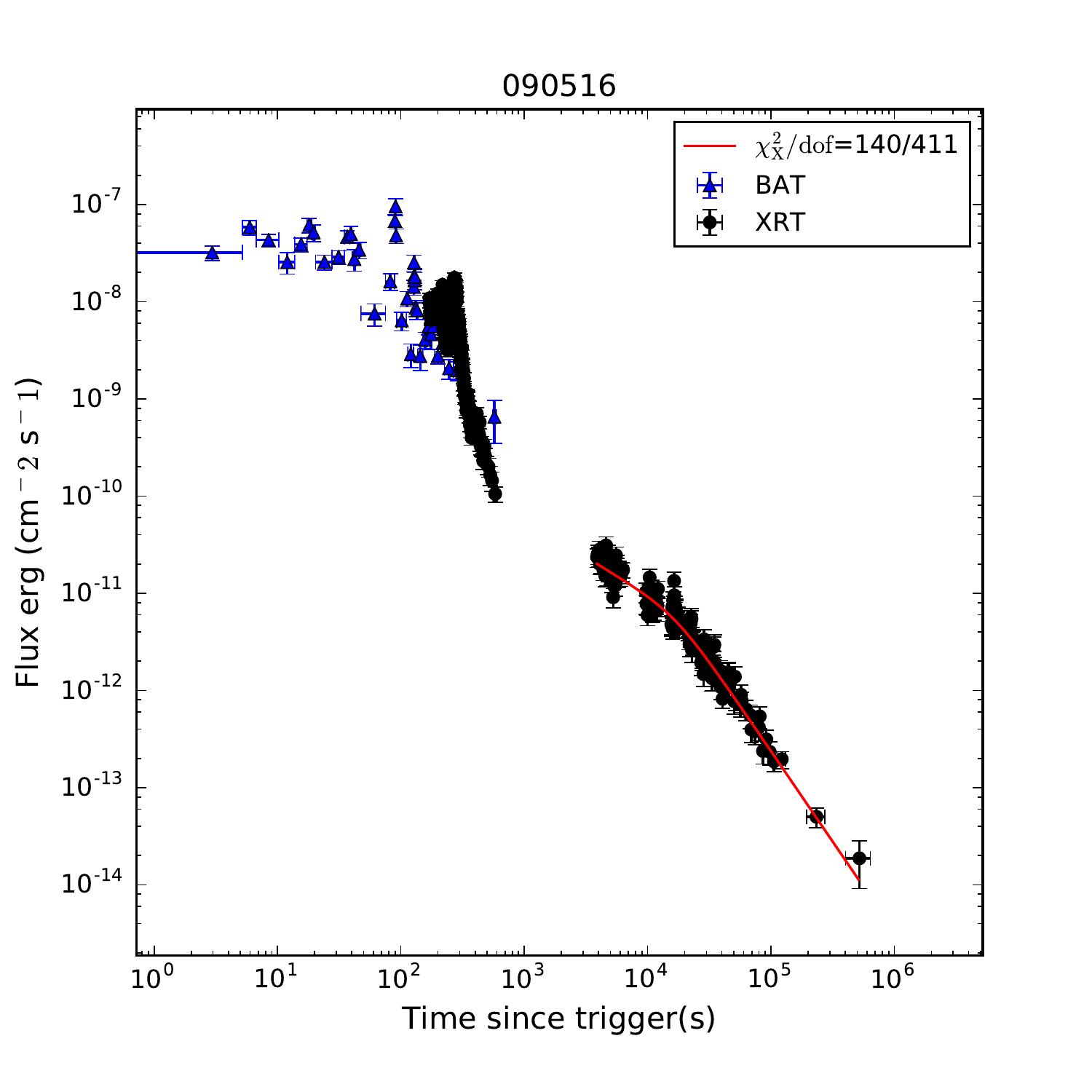}{0.28\textwidth}{}
          }
\gridline{\fig{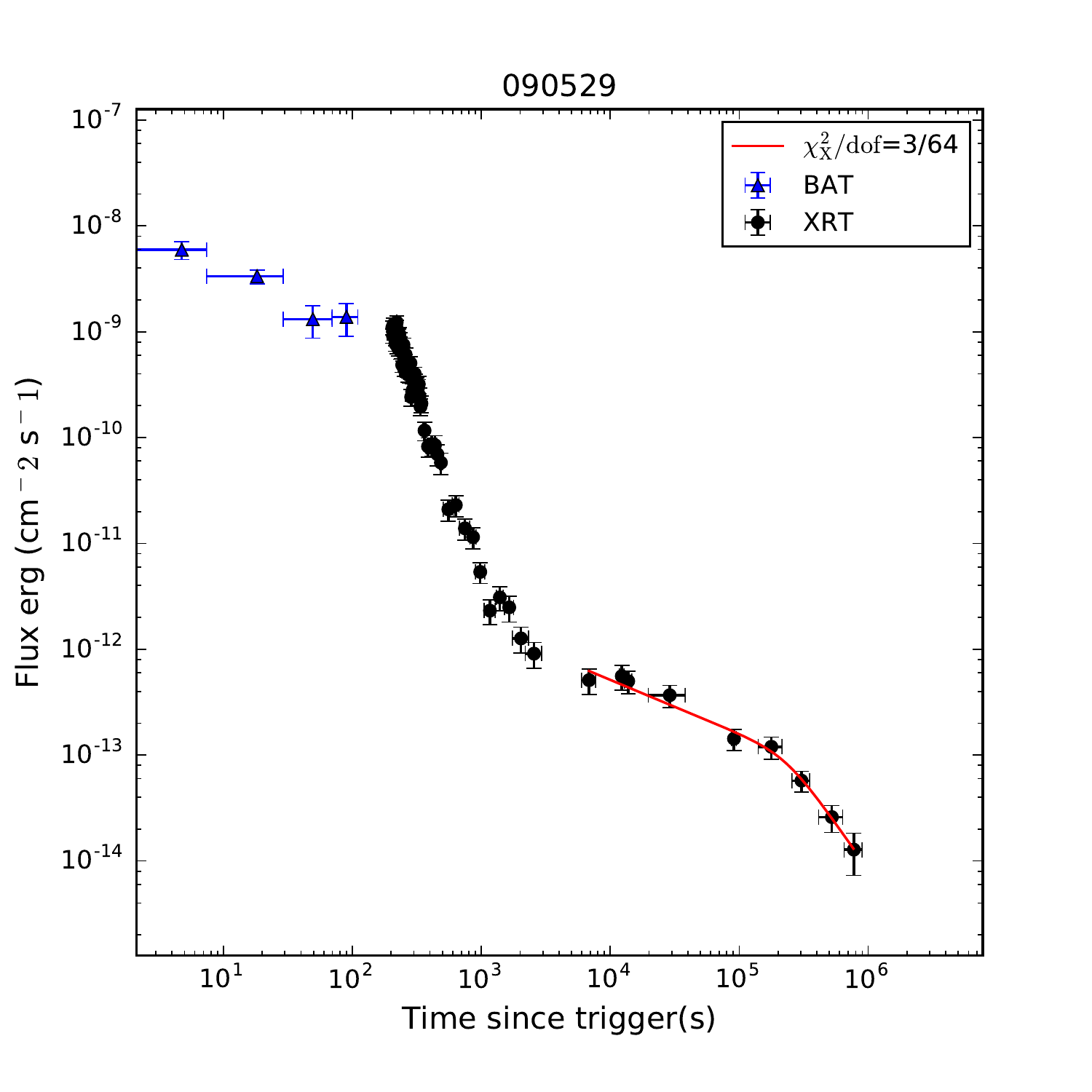}{0.28\textwidth}{}
          \fig{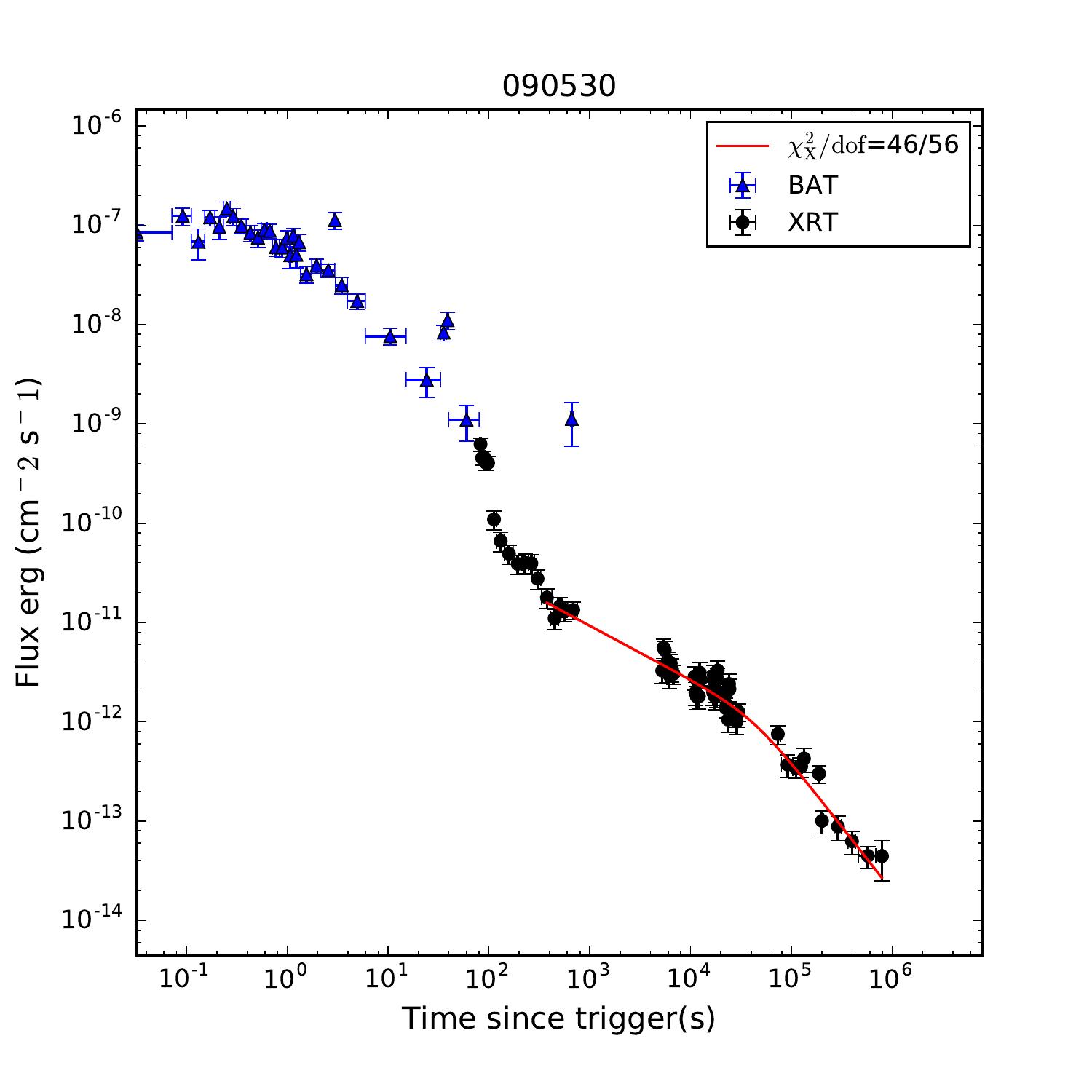}{0.28\textwidth}{}
          \fig{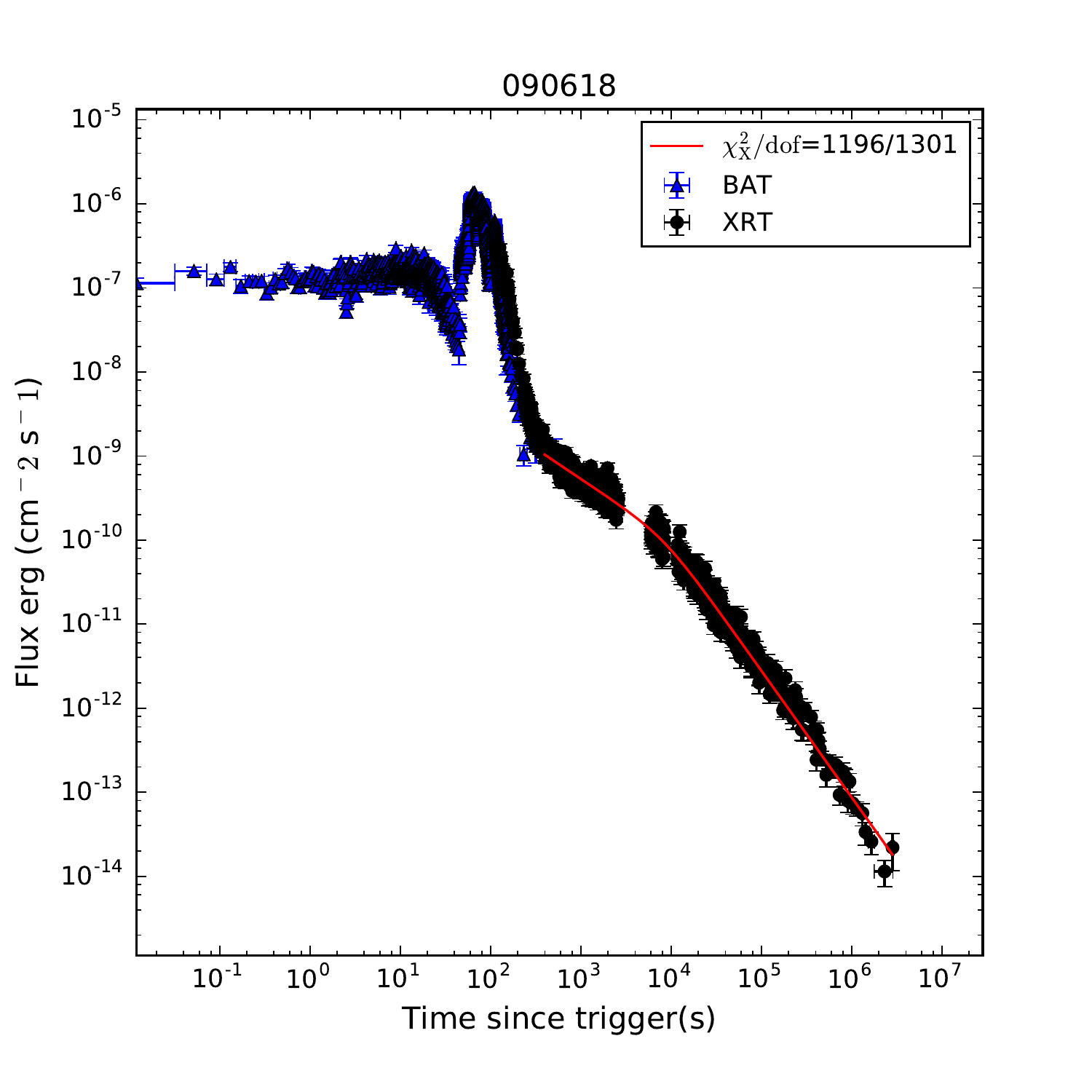}{0.28\textwidth}{}
          }
\center{Fig. \ref{Silver}--- Continued}
\end{figure}
\begin{figure}  
\gridline{\fig{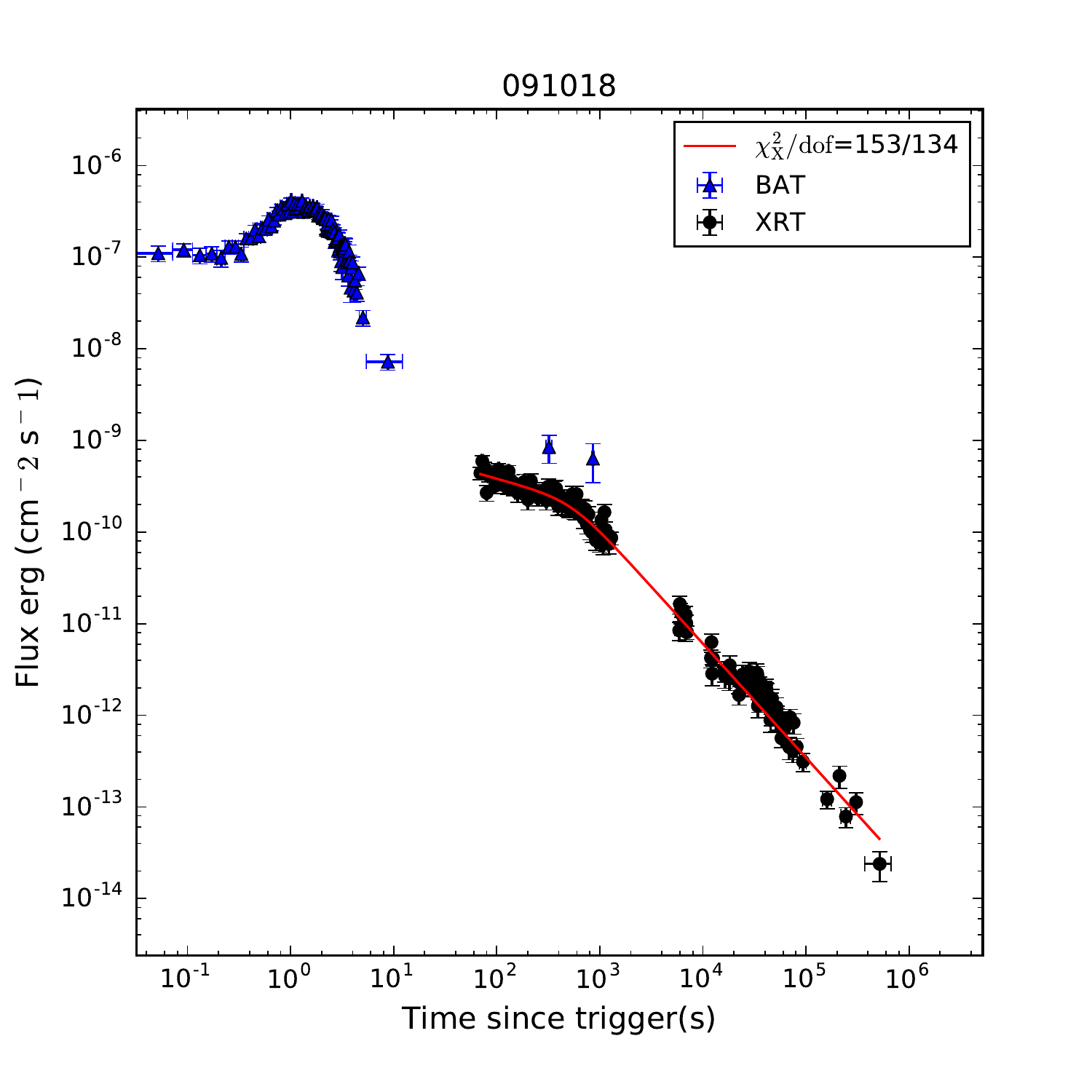}{0.28\textwidth}{}
          \fig{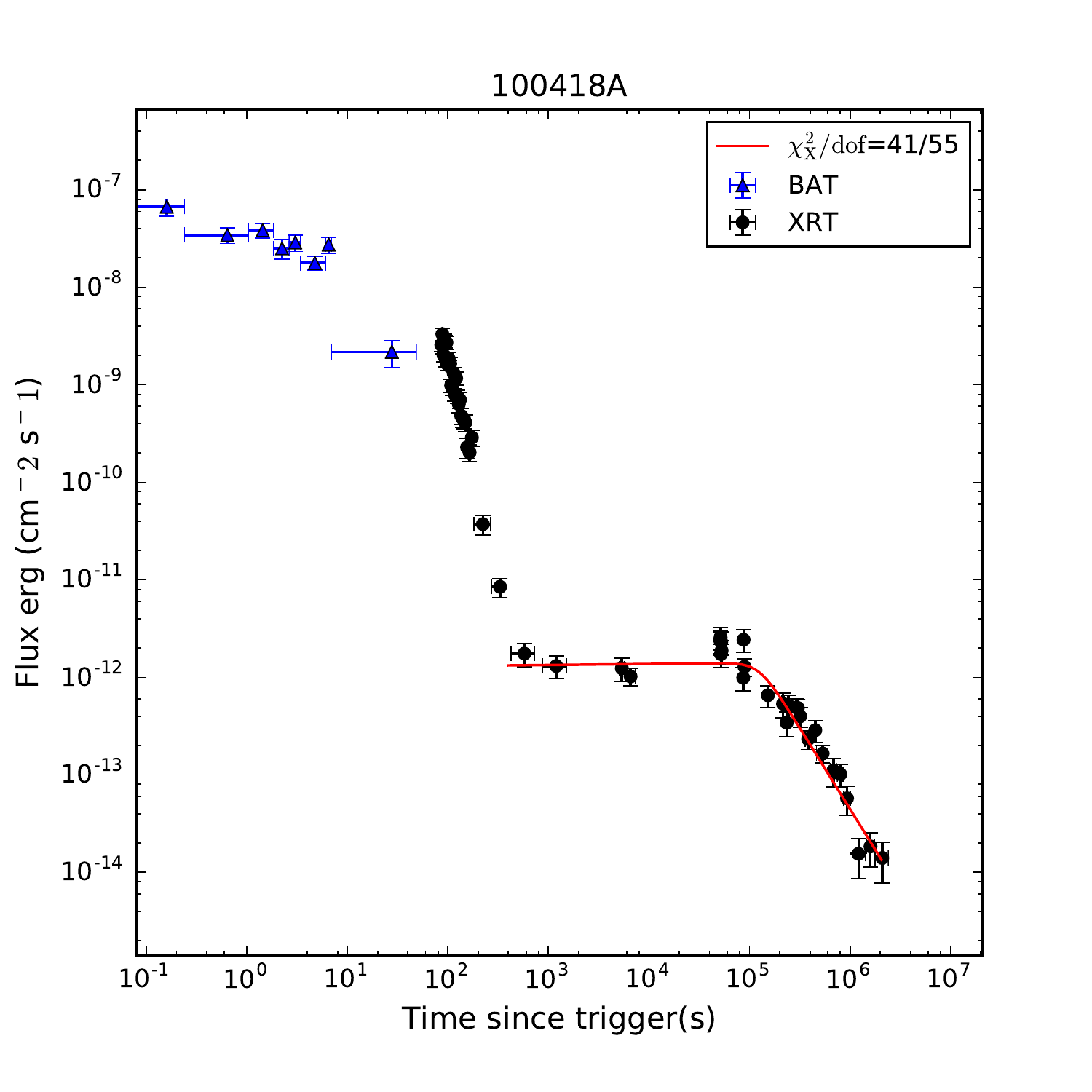}{0.28\textwidth}{}
          \fig{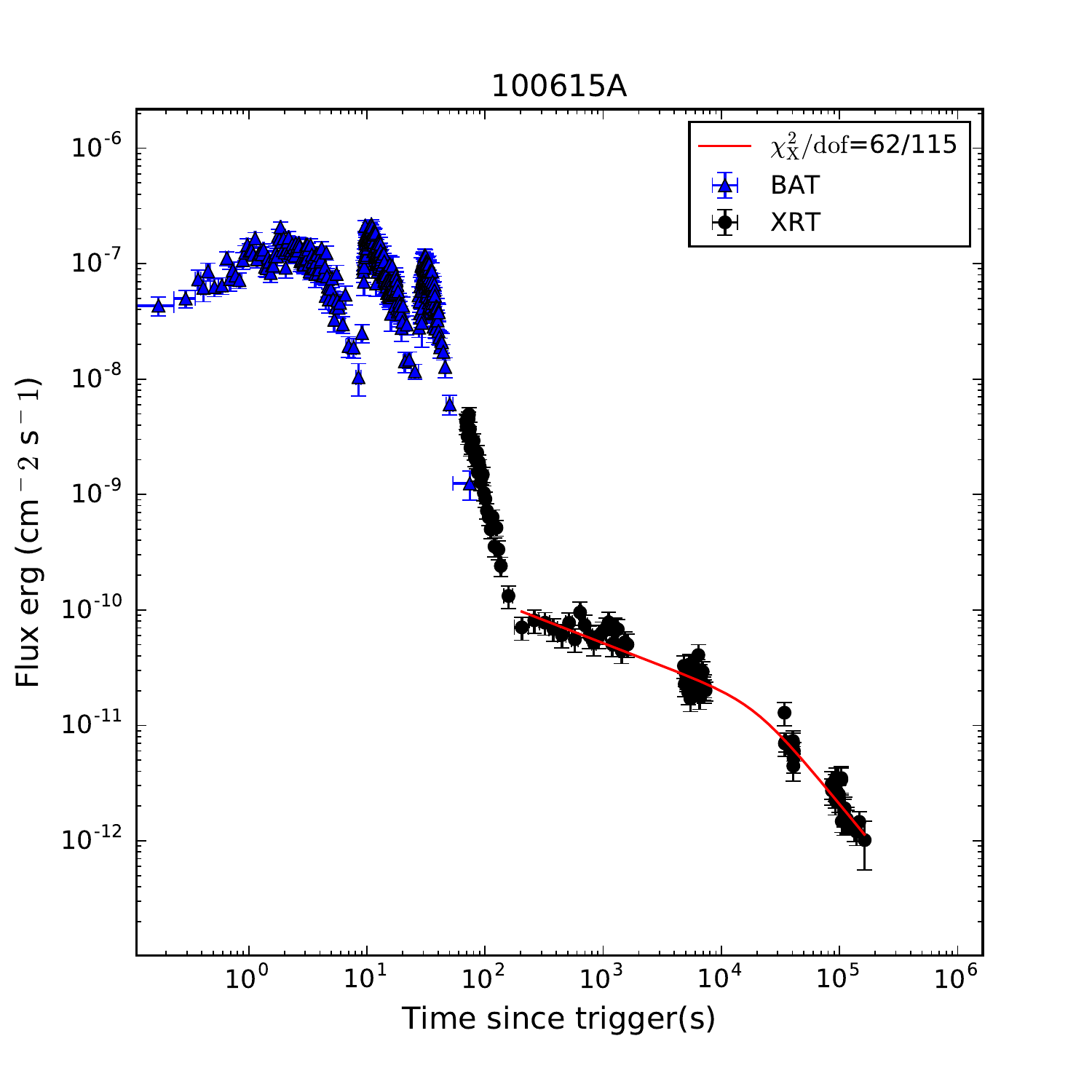}{0.28\textwidth}{}
          }
\gridline{\fig{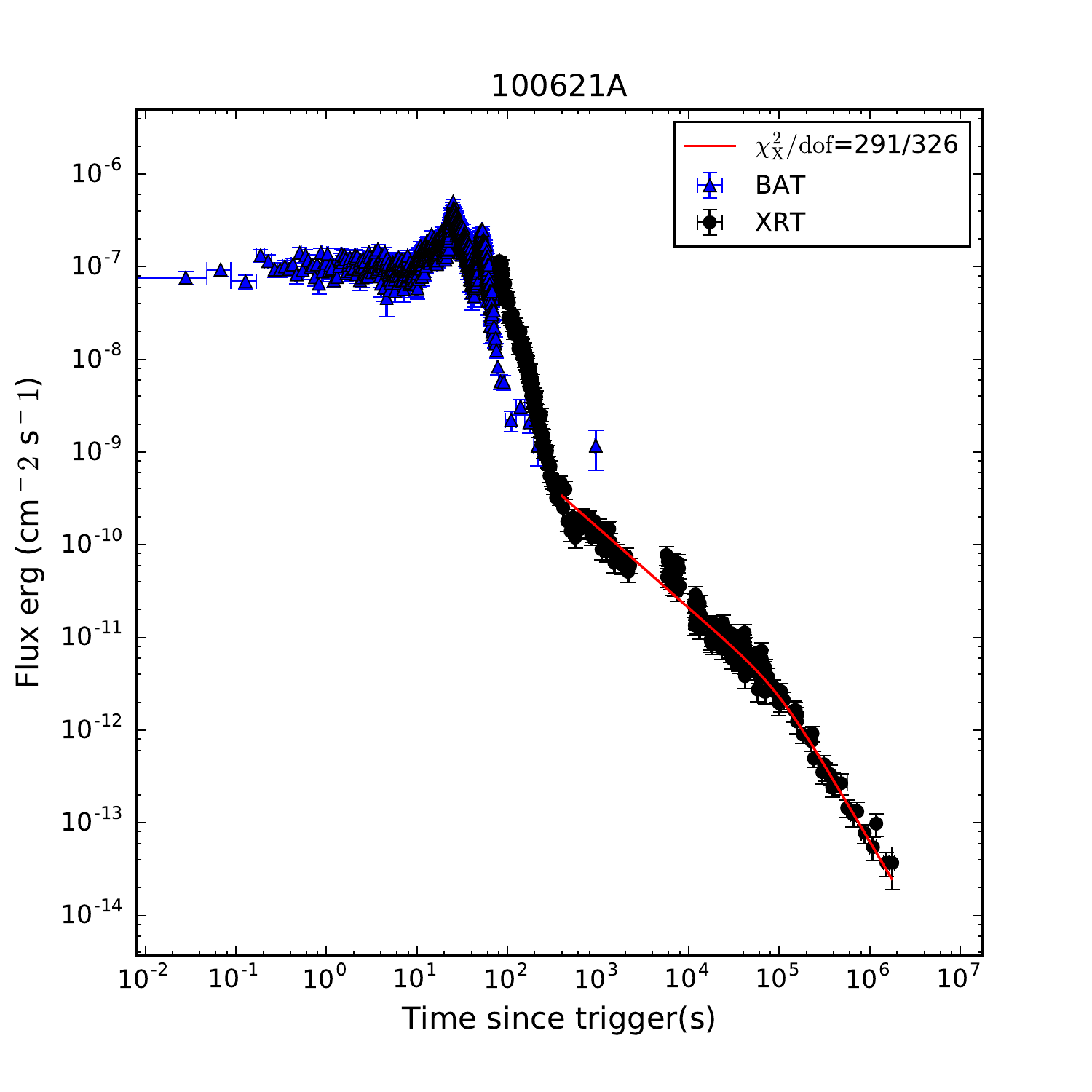}{0.28\textwidth}{}
          \fig{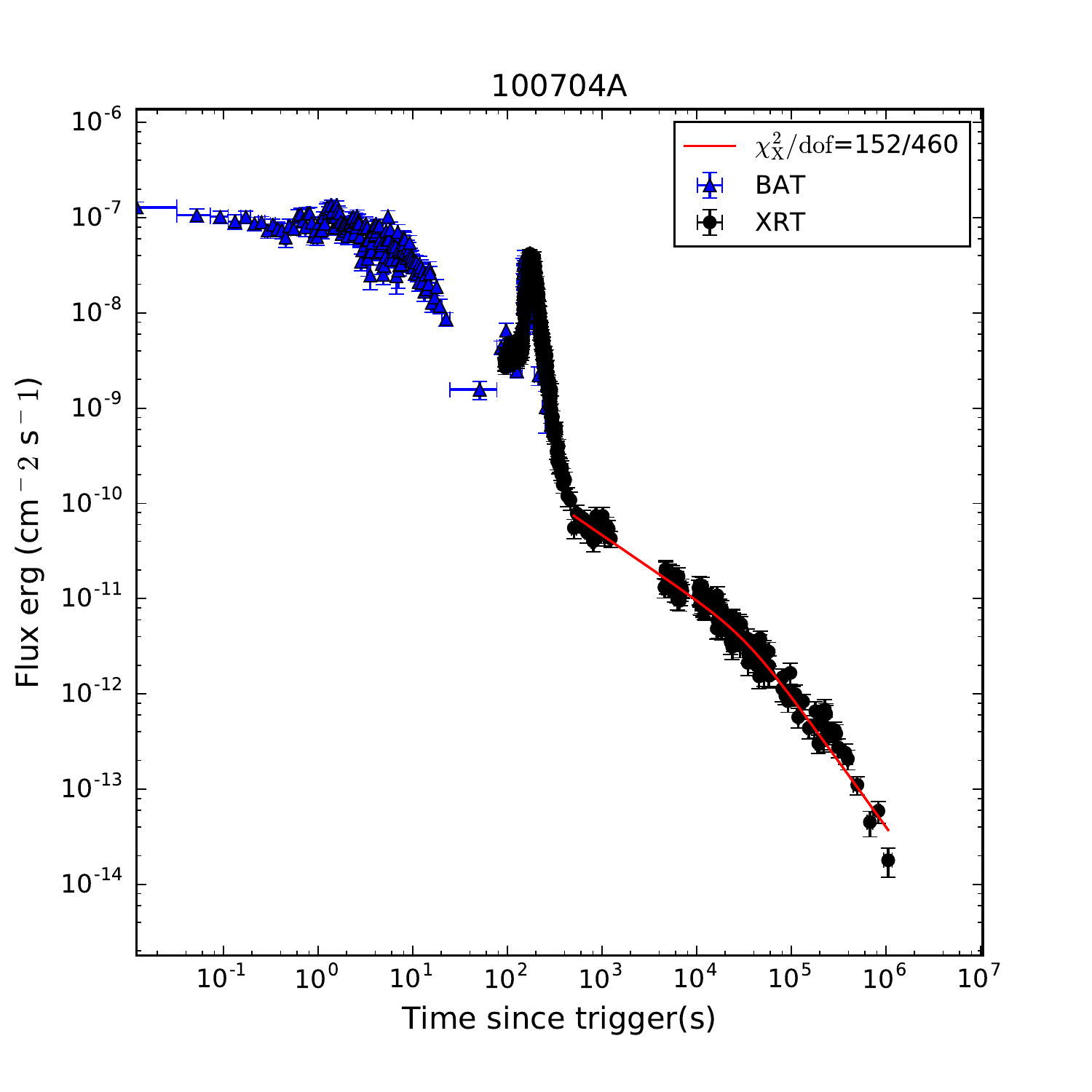}{0.28\textwidth}{}
          \fig{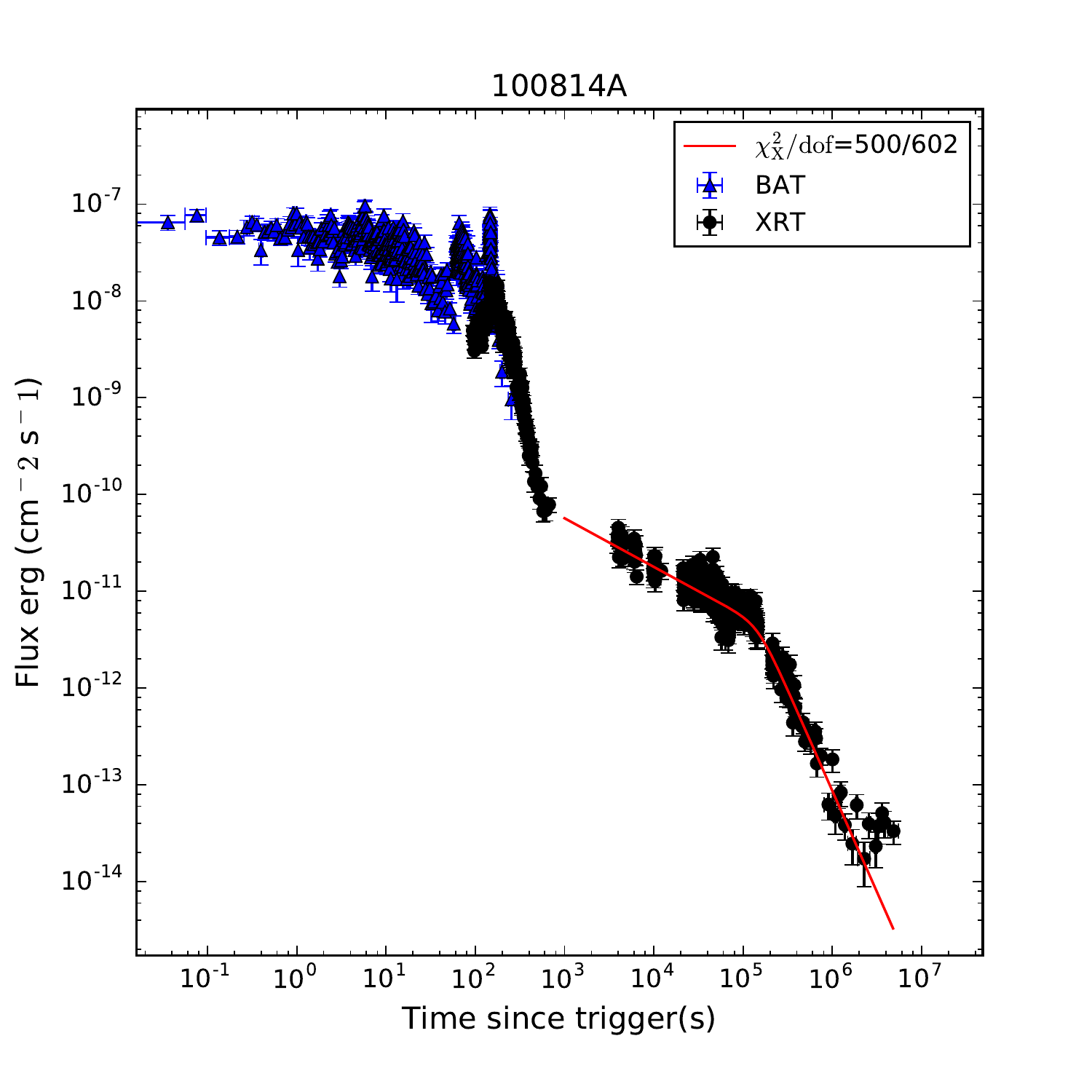}{0.28\textwidth}{}
          }
\gridline{\fig{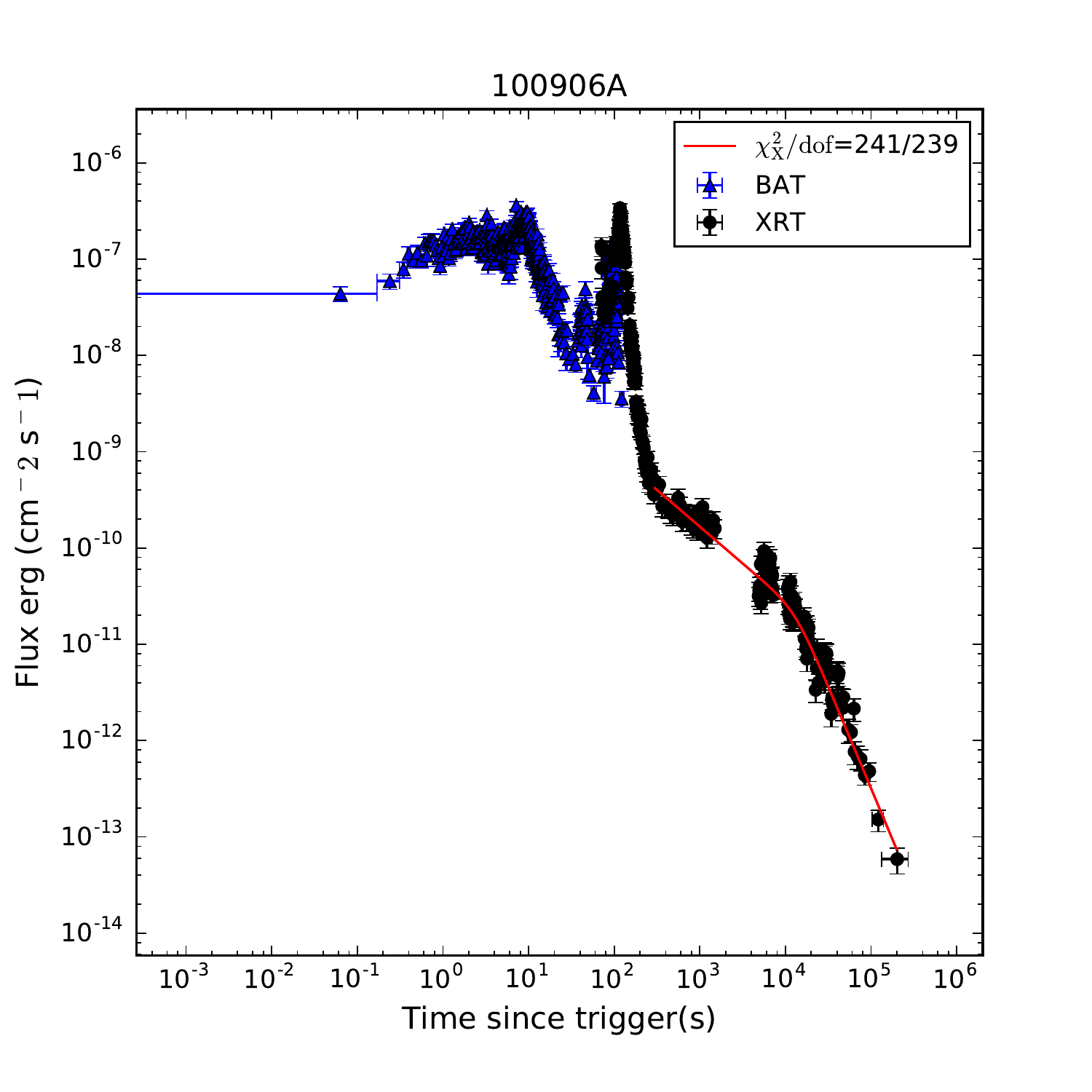}{0.28\textwidth}{}
          \fig{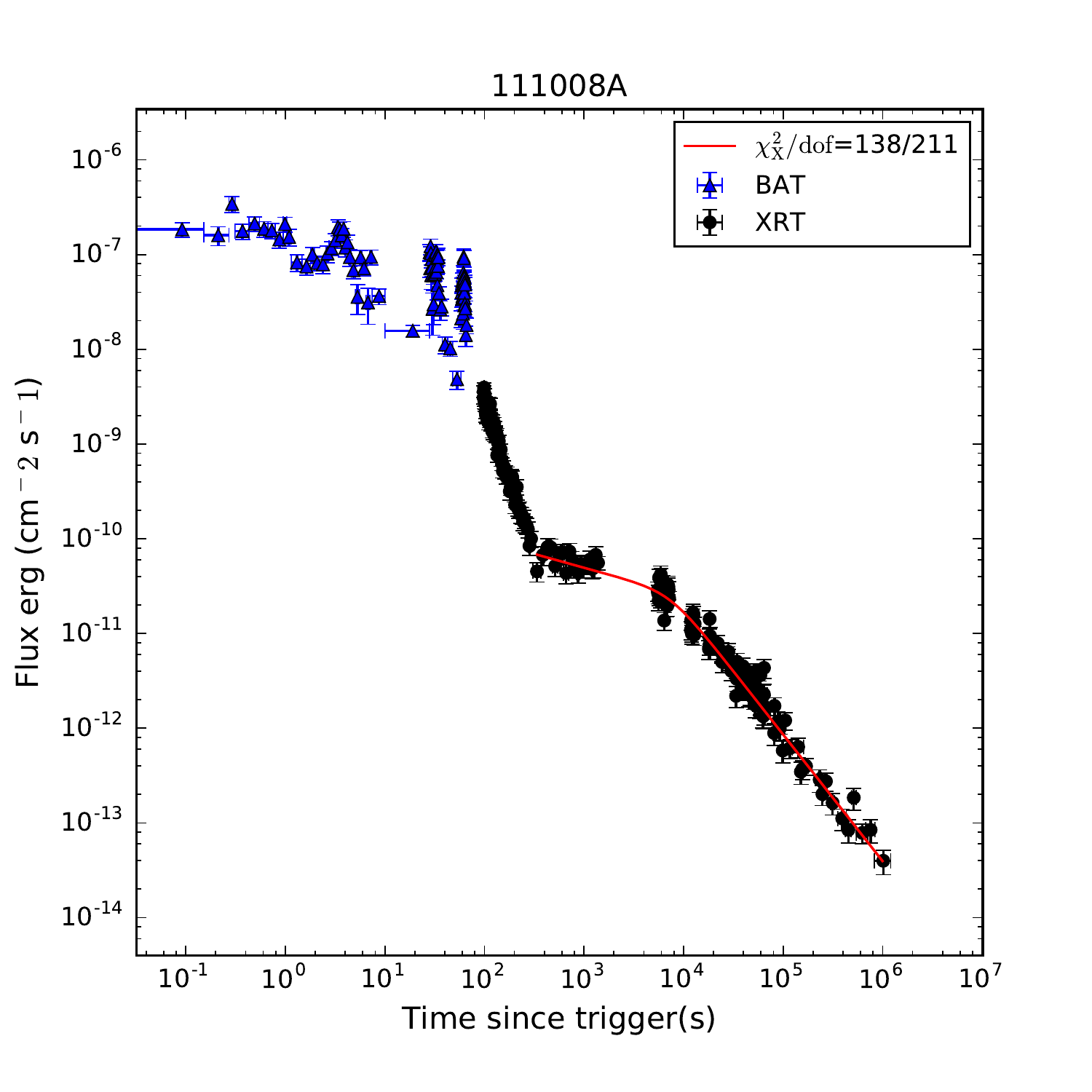}{0.28\textwidth}{}
          \fig{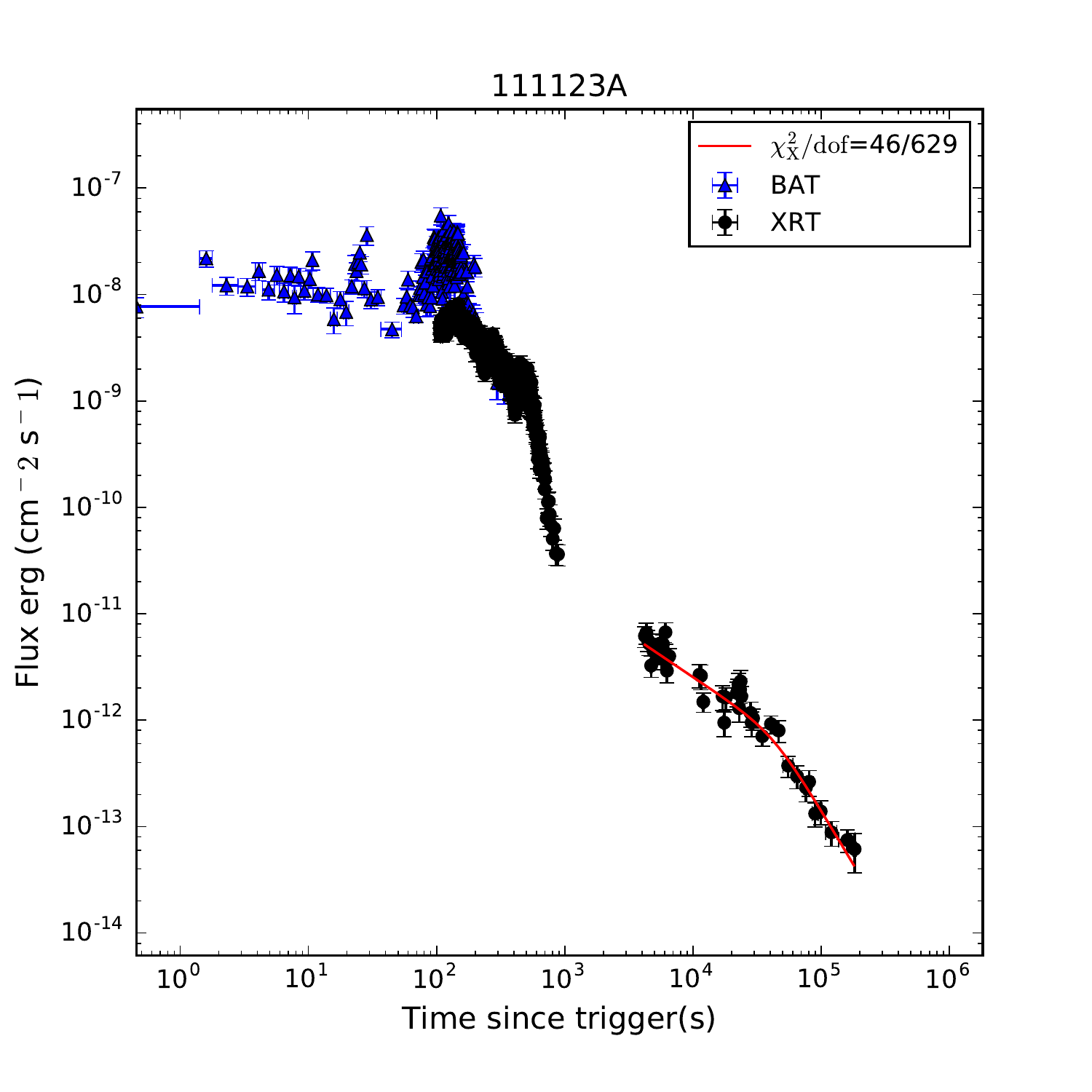}{0.28\textwidth}{}
          }
\gridline{\fig{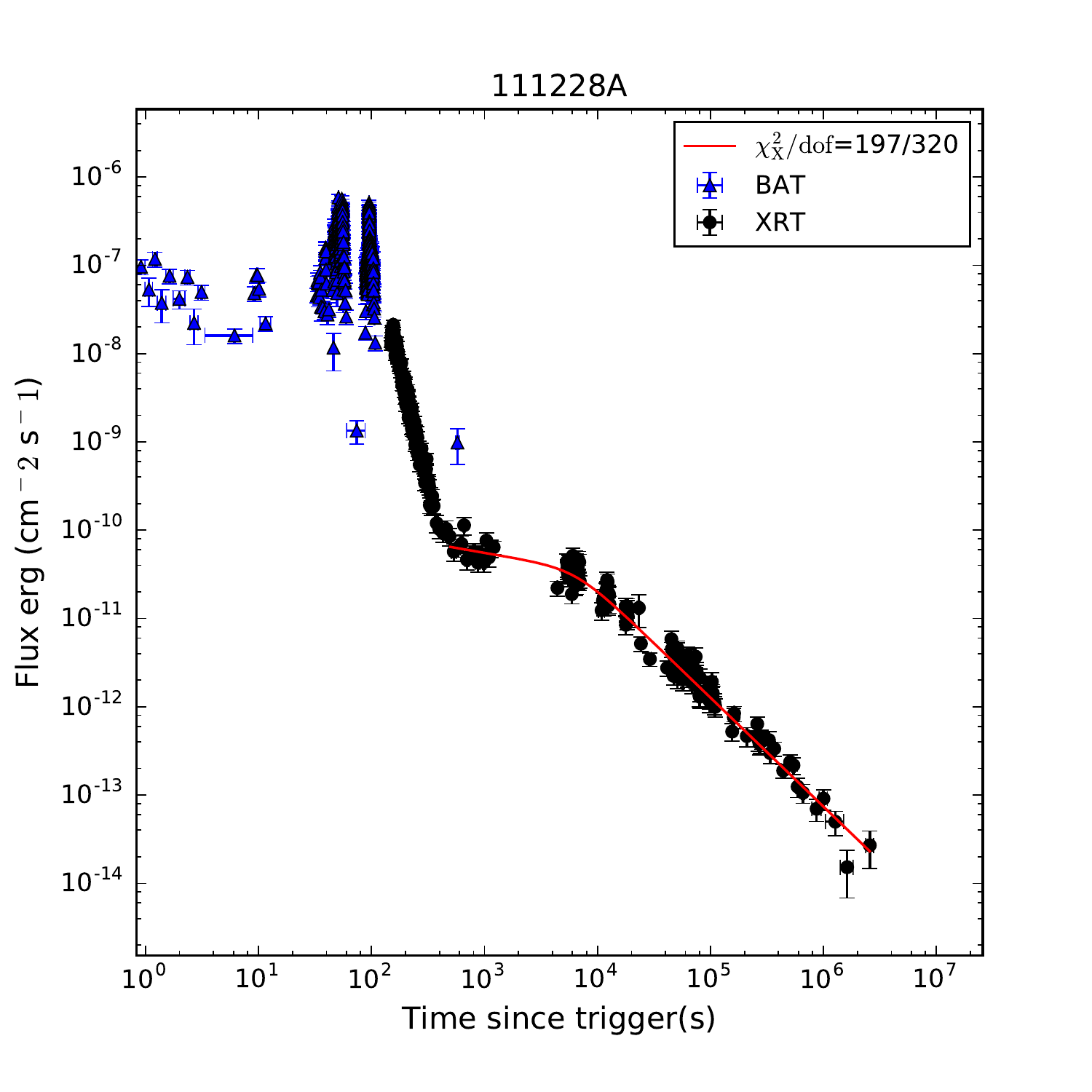}{0.28\textwidth}{}
          \fig{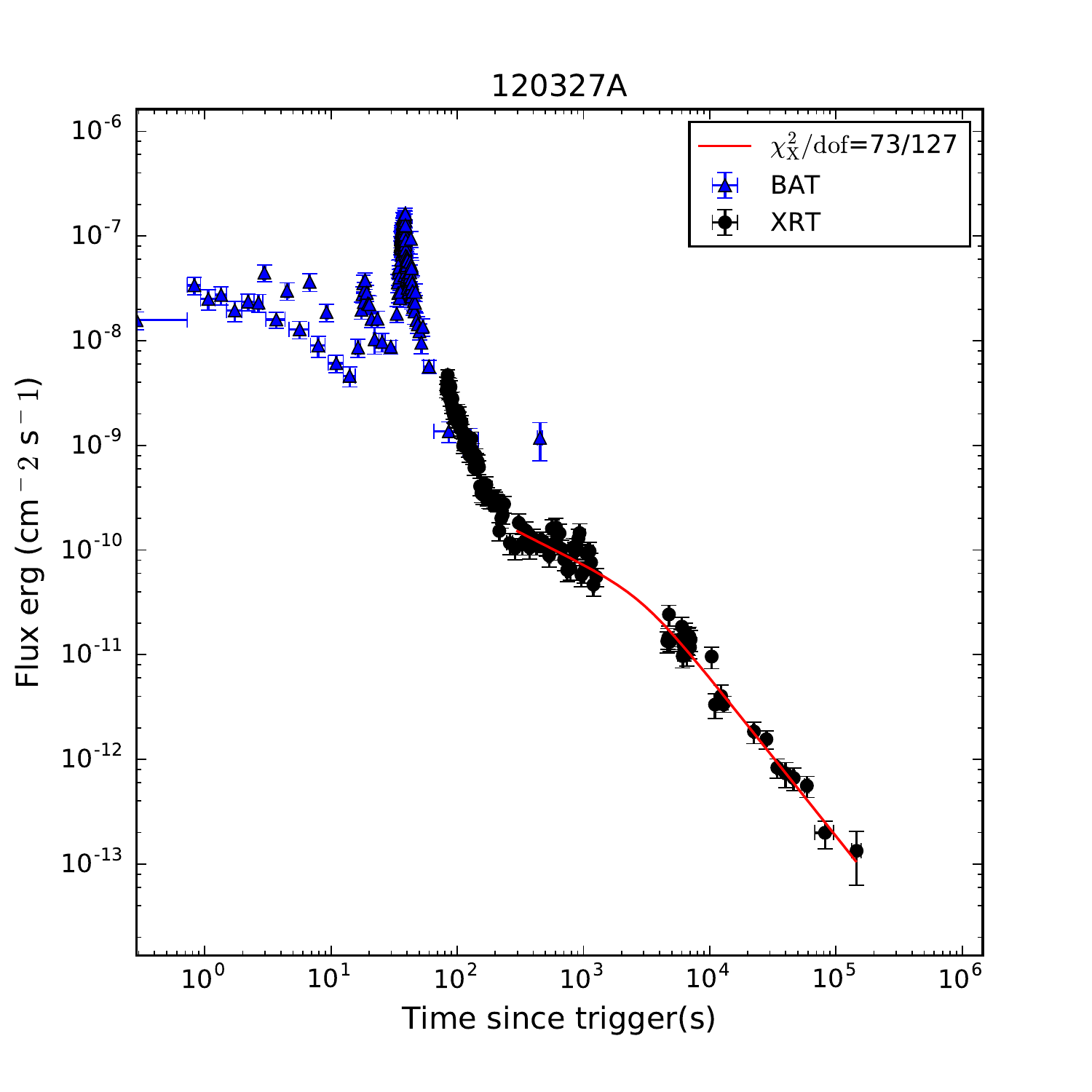}{0.28\textwidth}{}
          \fig{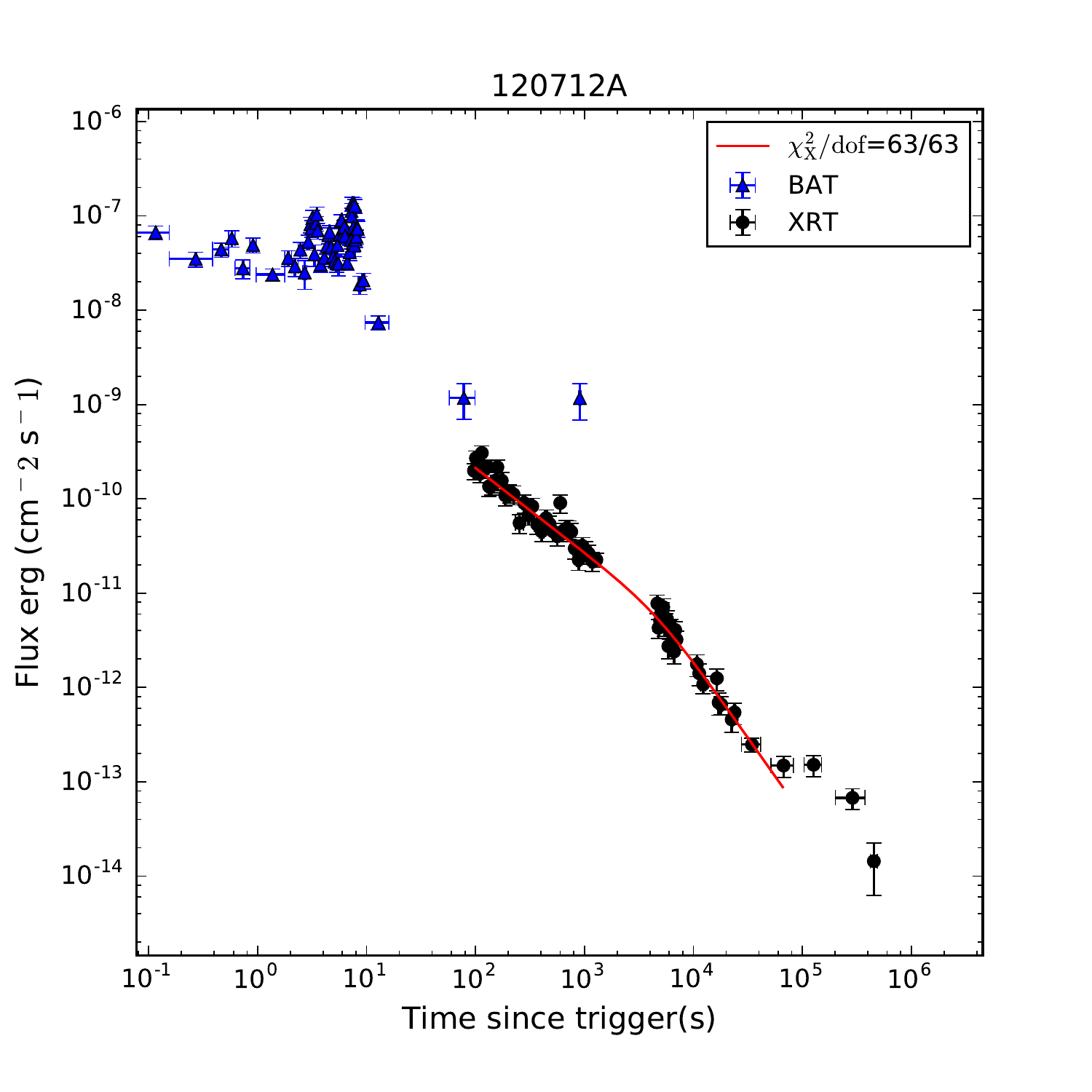}{0.28\textwidth}{}
          }
\center{Fig. \ref{Silver}--- Continued}
\end{figure}
\begin{figure}
\gridline{\fig{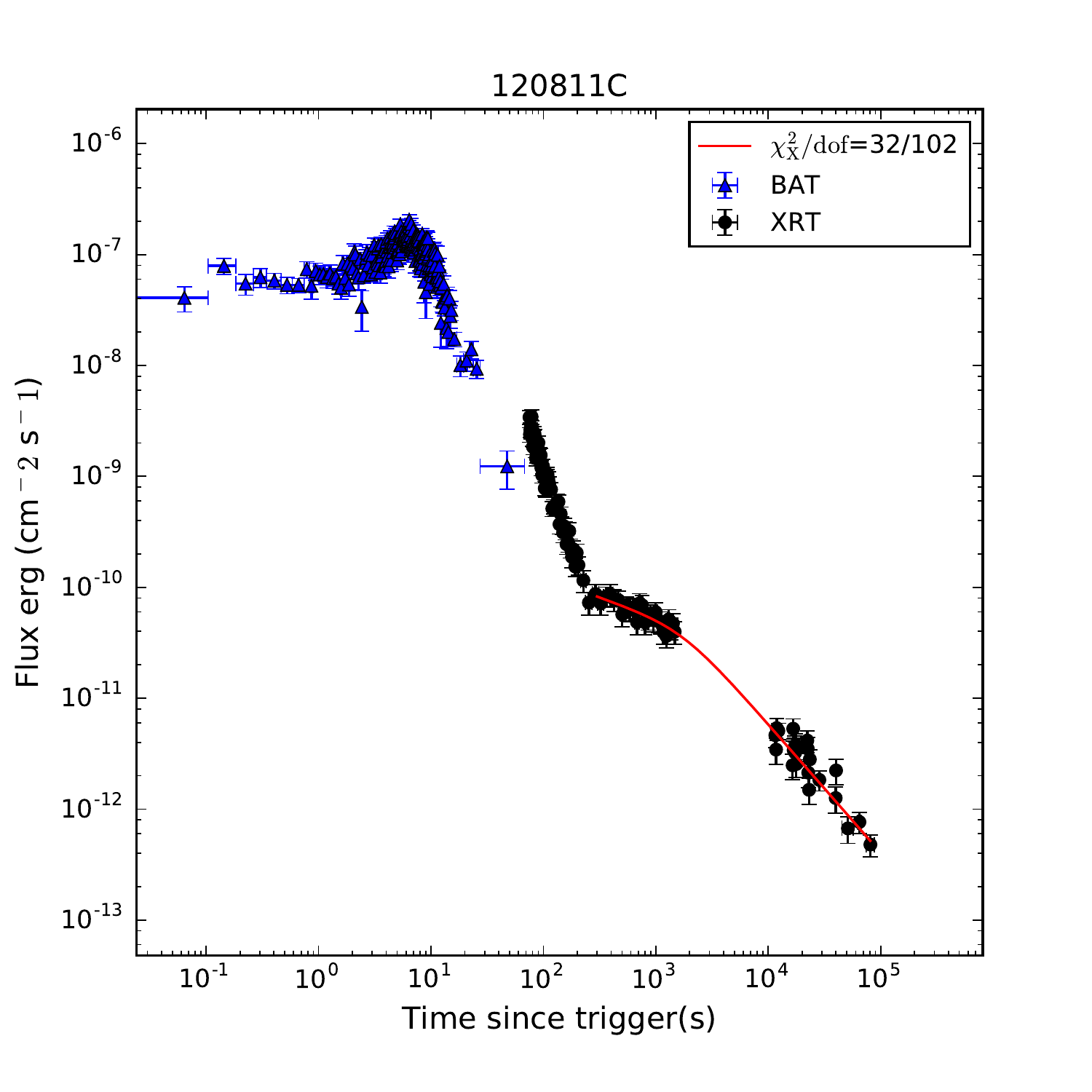}{0.28\textwidth}{}
          \fig{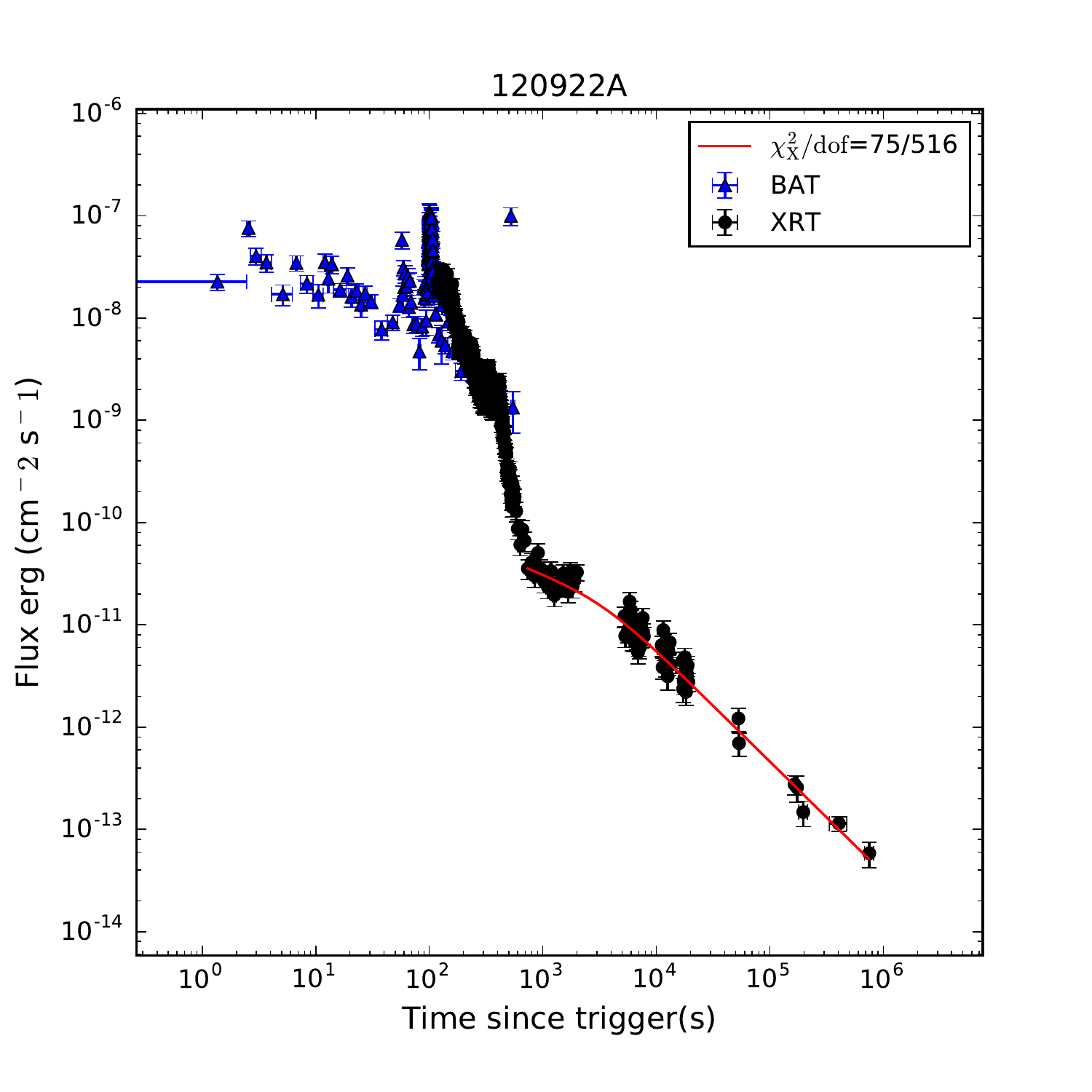}{0.28\textwidth}{}
          \fig{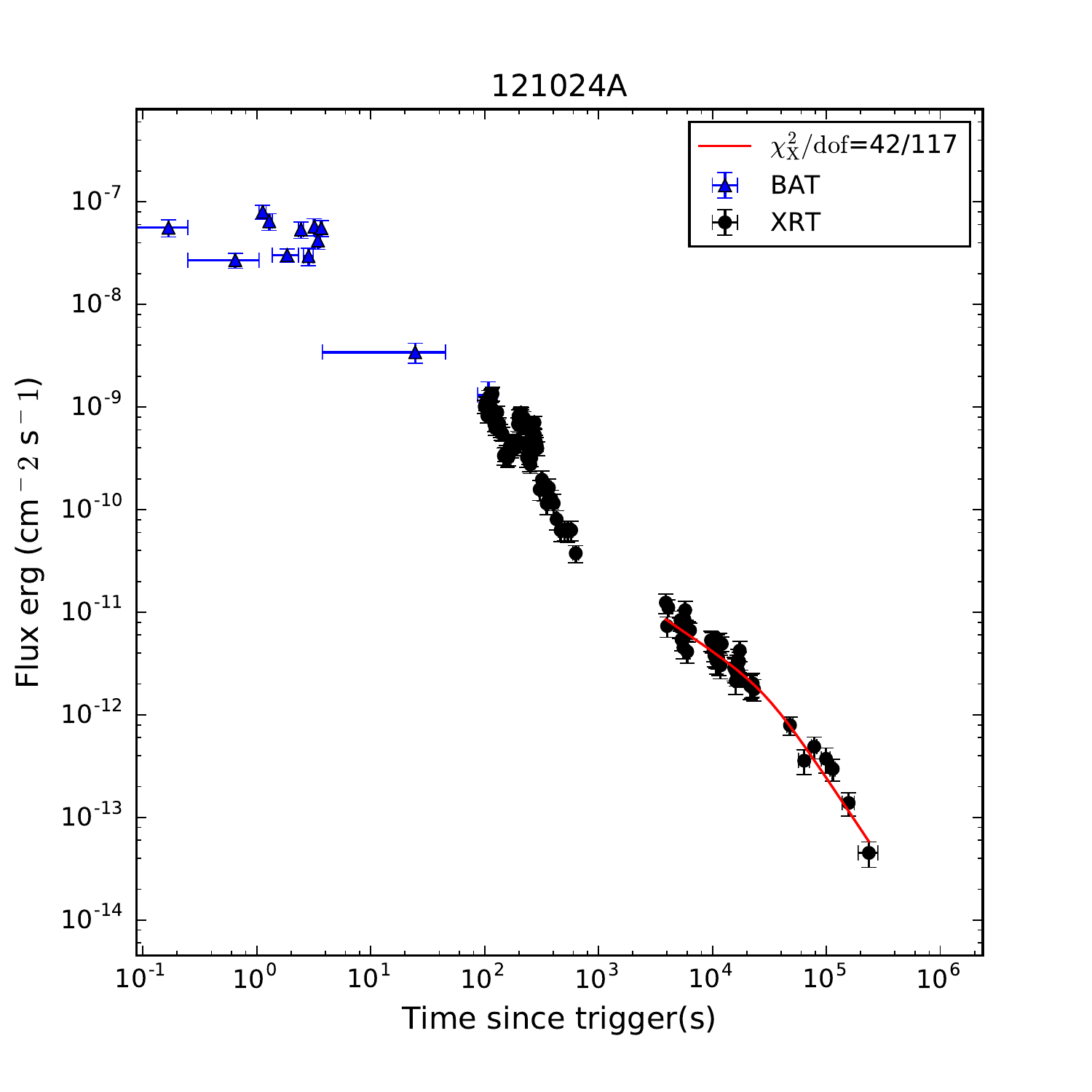}{0.28\textwidth}{}
          }
\gridline{\fig{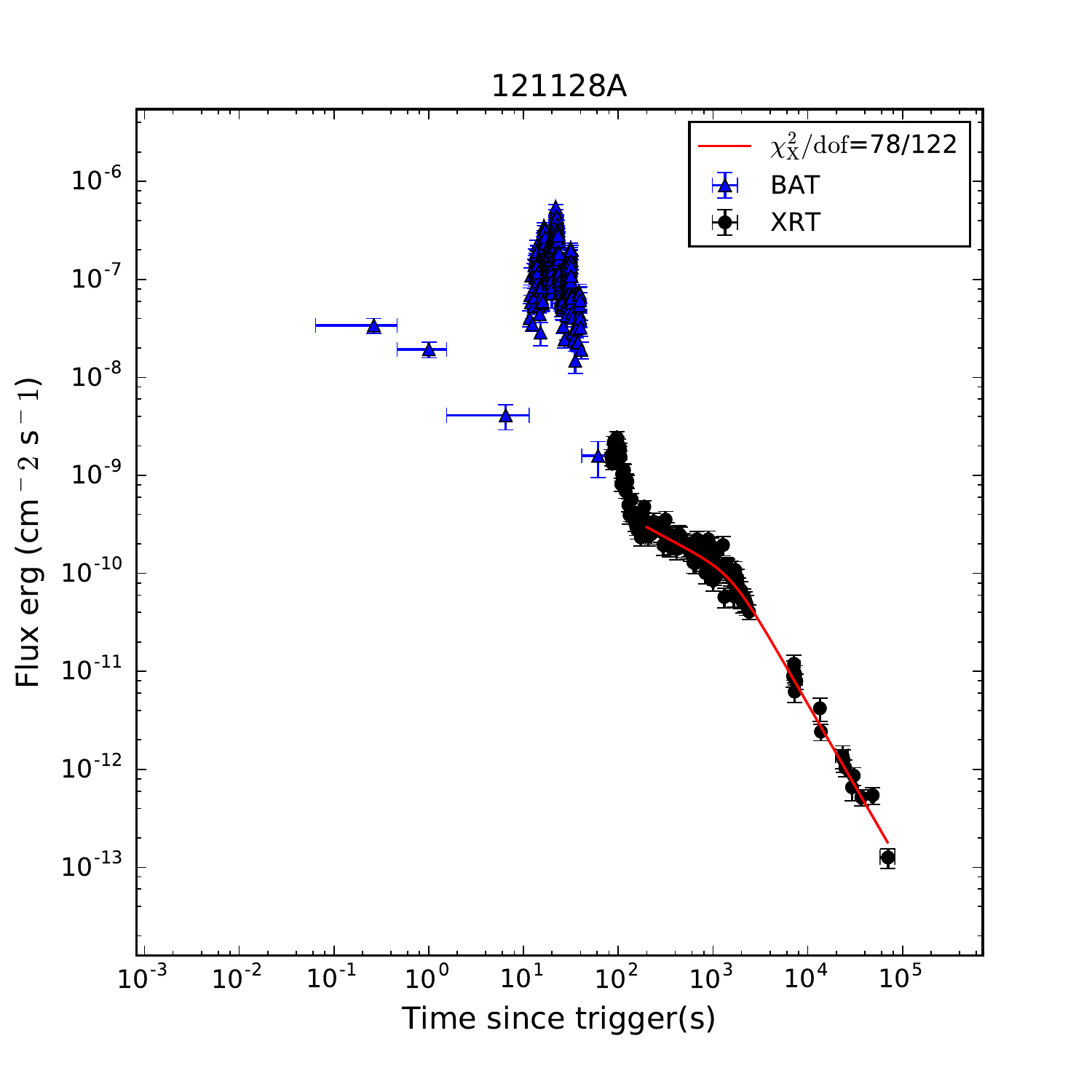}{0.28\textwidth}{}
          \fig{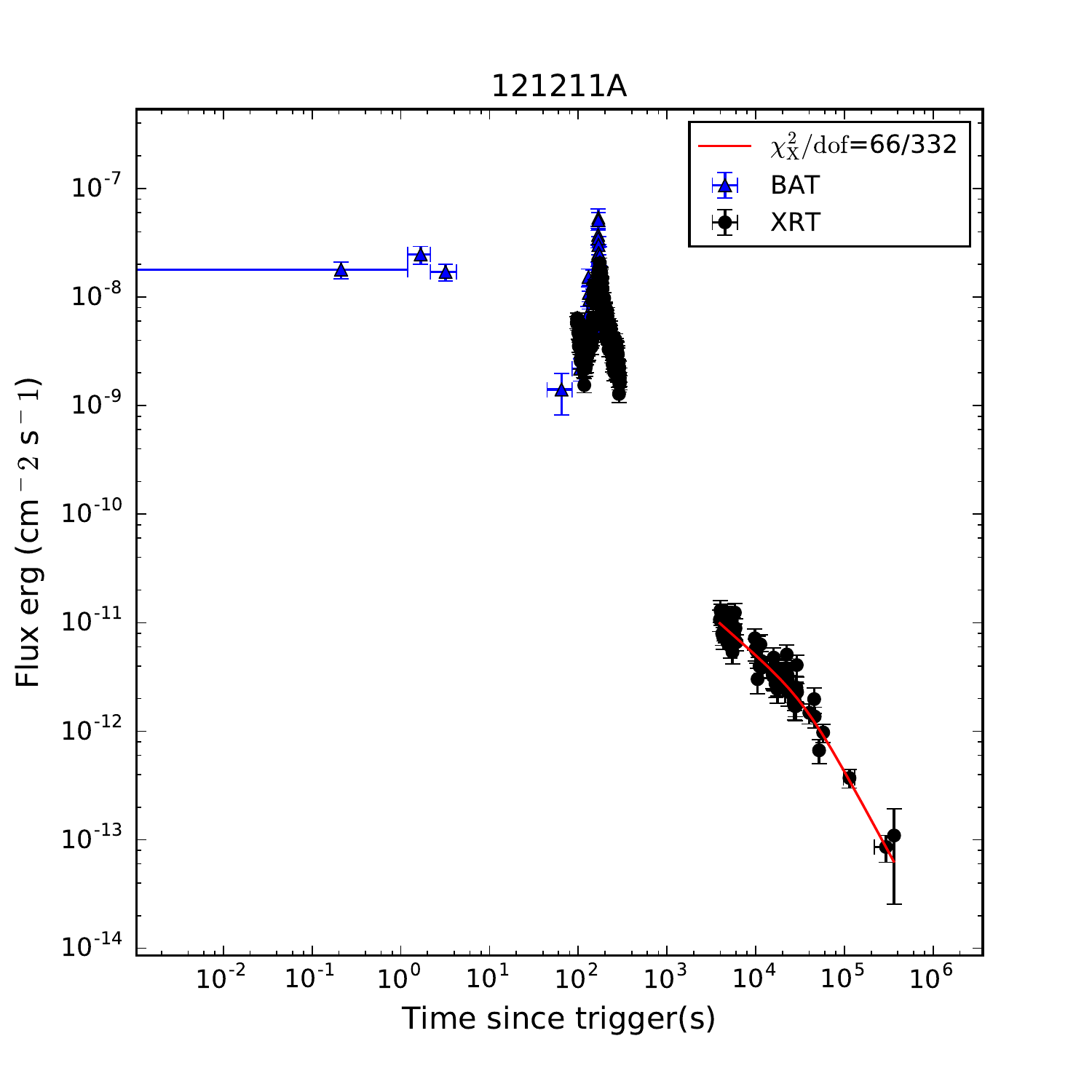}{0.28\textwidth}{}
          \fig{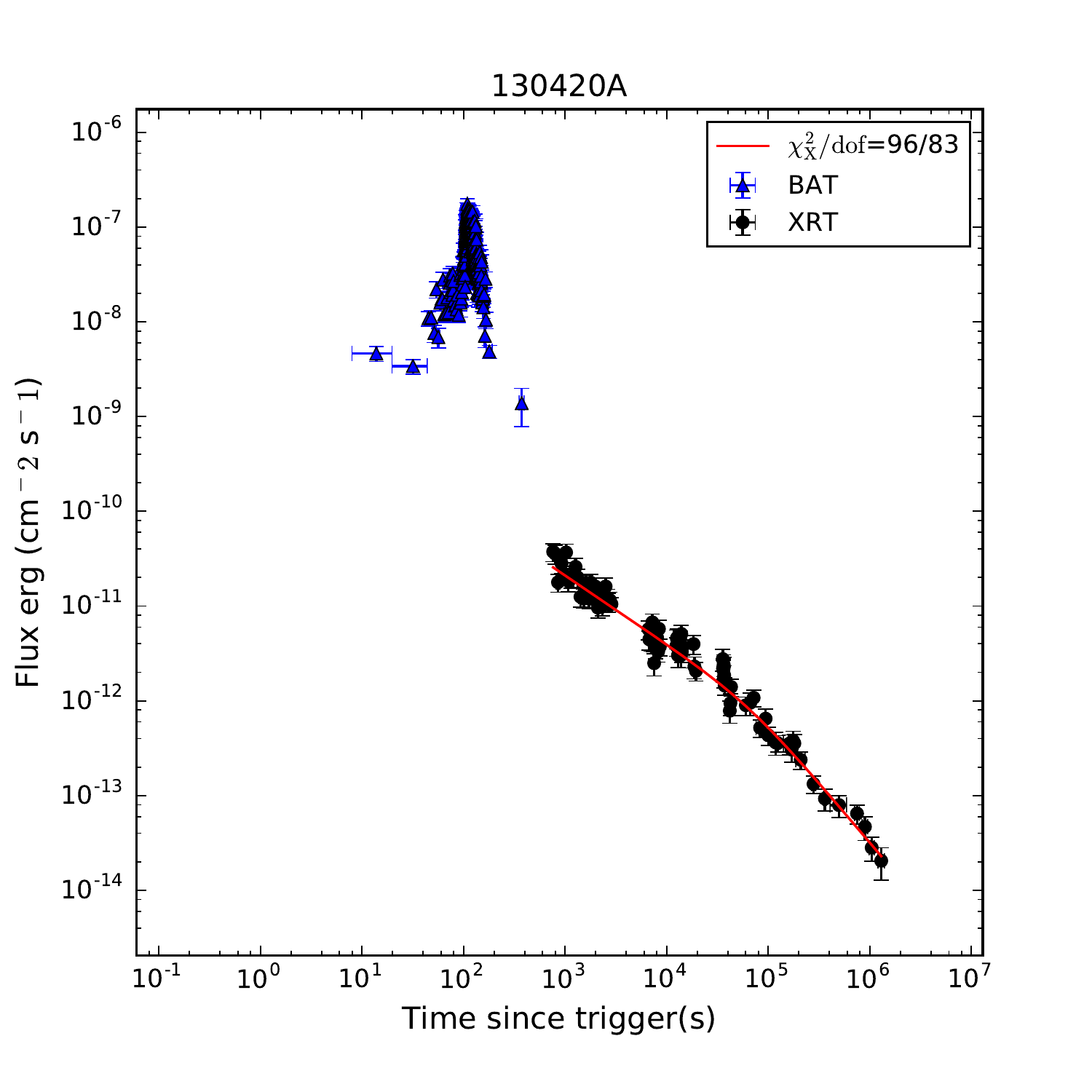}{0.28\textwidth}{}
          }    
\gridline{\fig{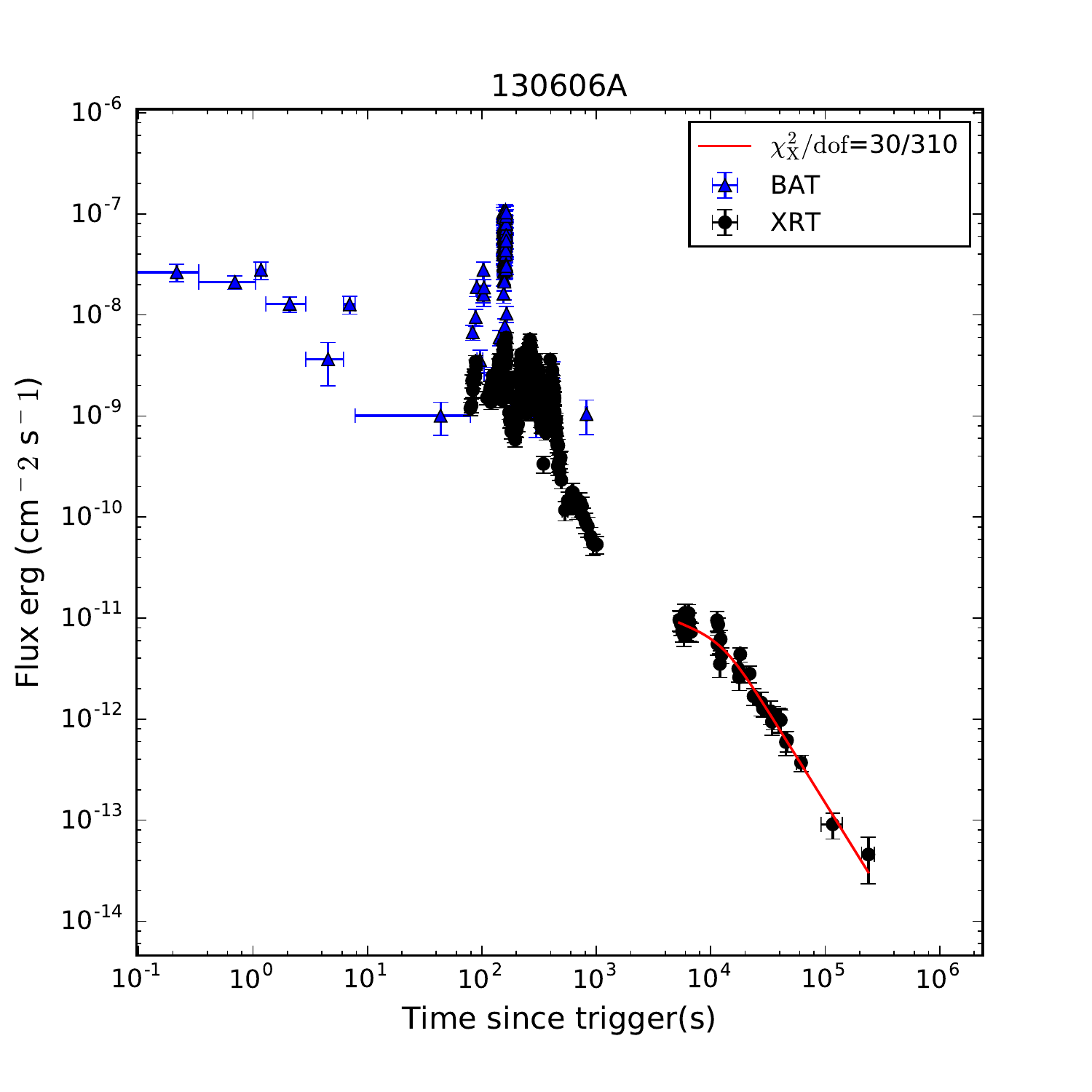}{0.28\textwidth}{}
          \fig{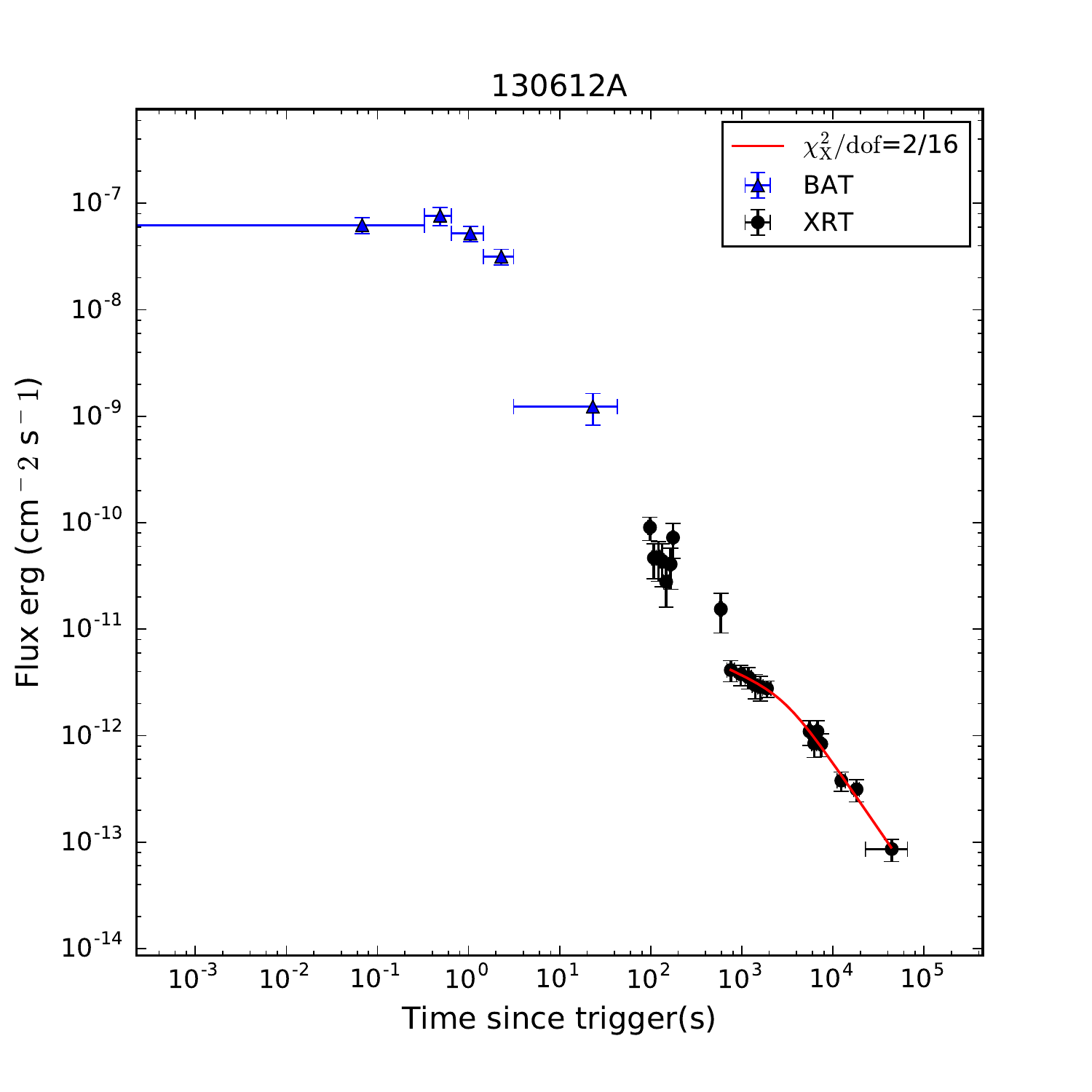}{0.28\textwidth}{}
          \fig{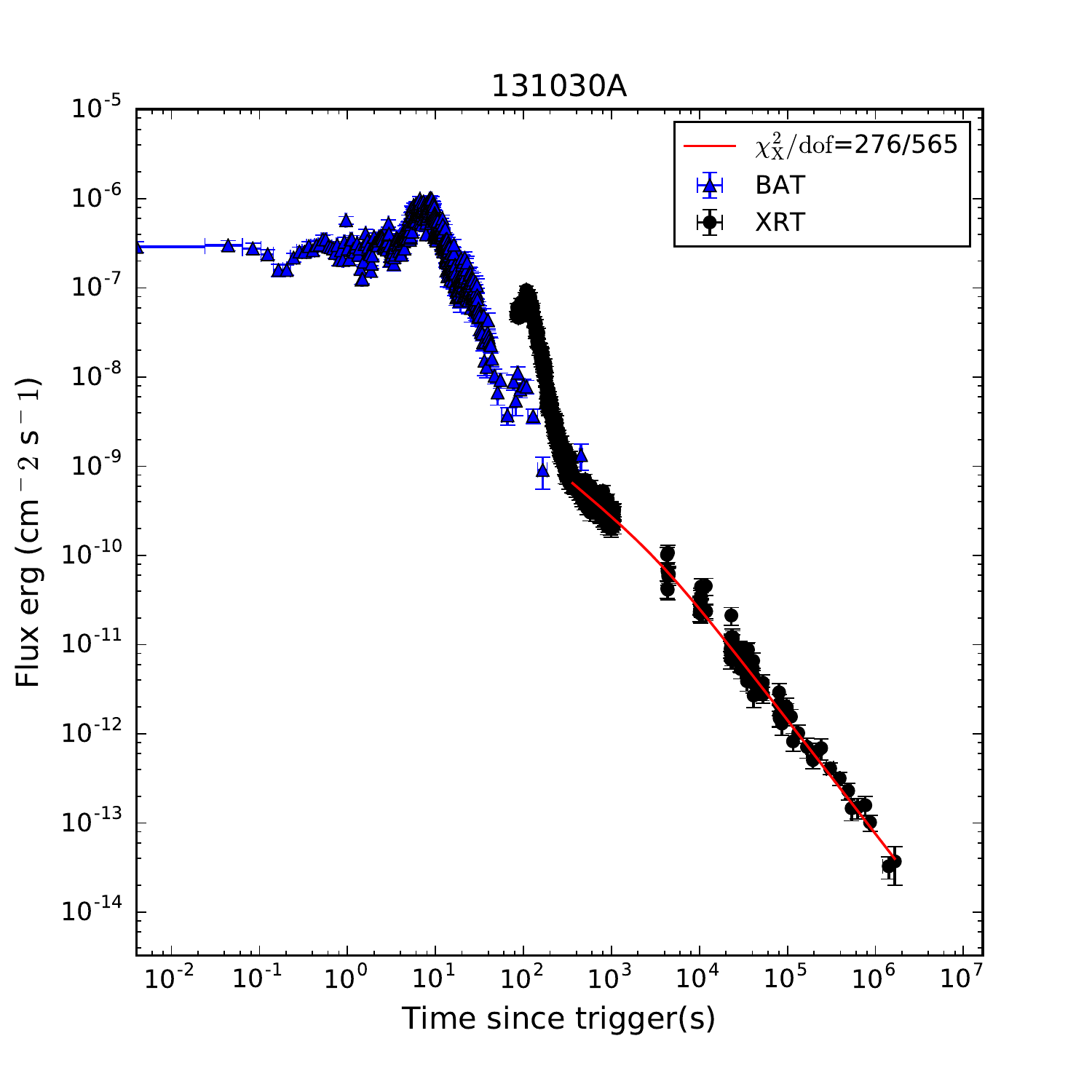}{0.28\textwidth}{}
          }
\gridline{\fig{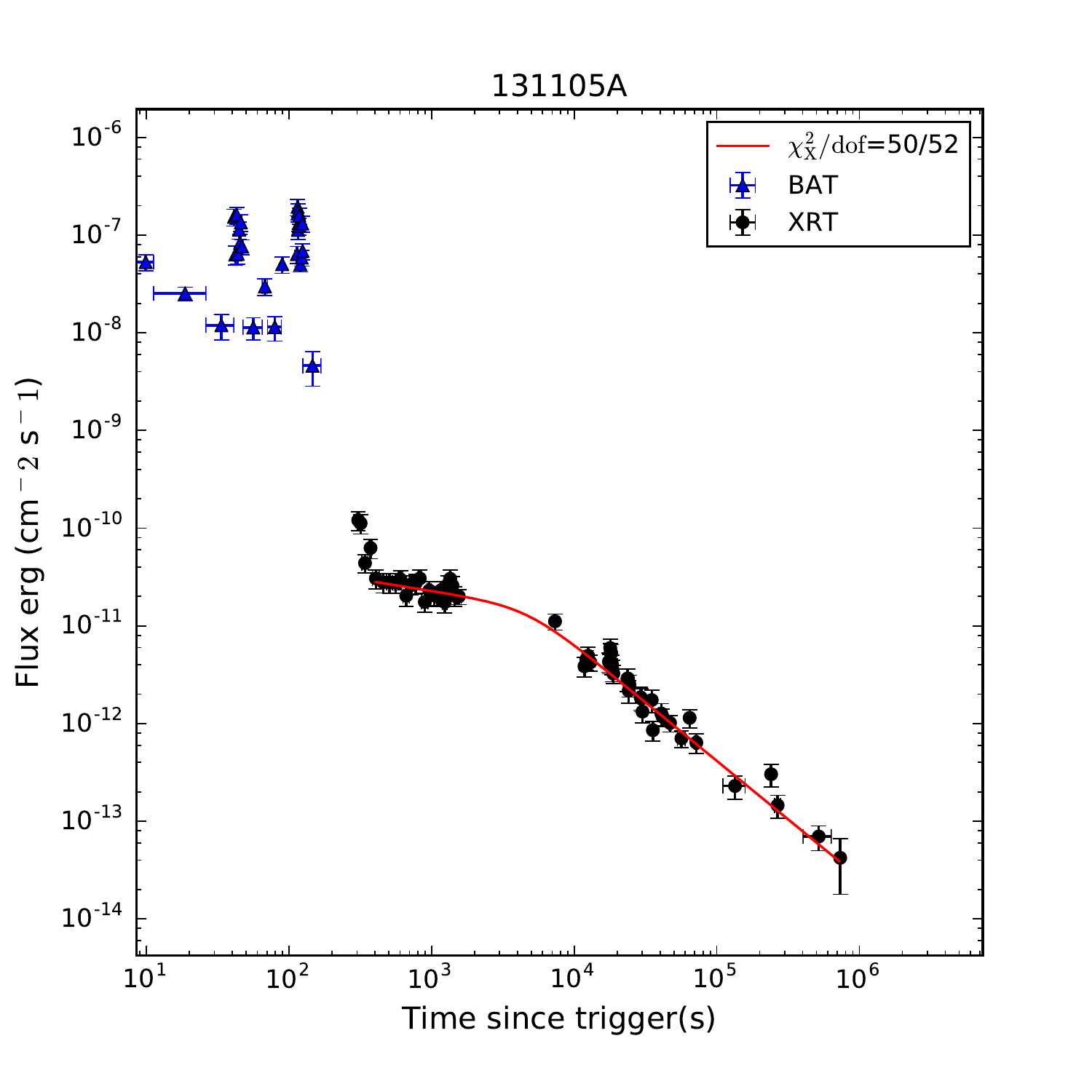}{0.28\textwidth}{}
          \fig{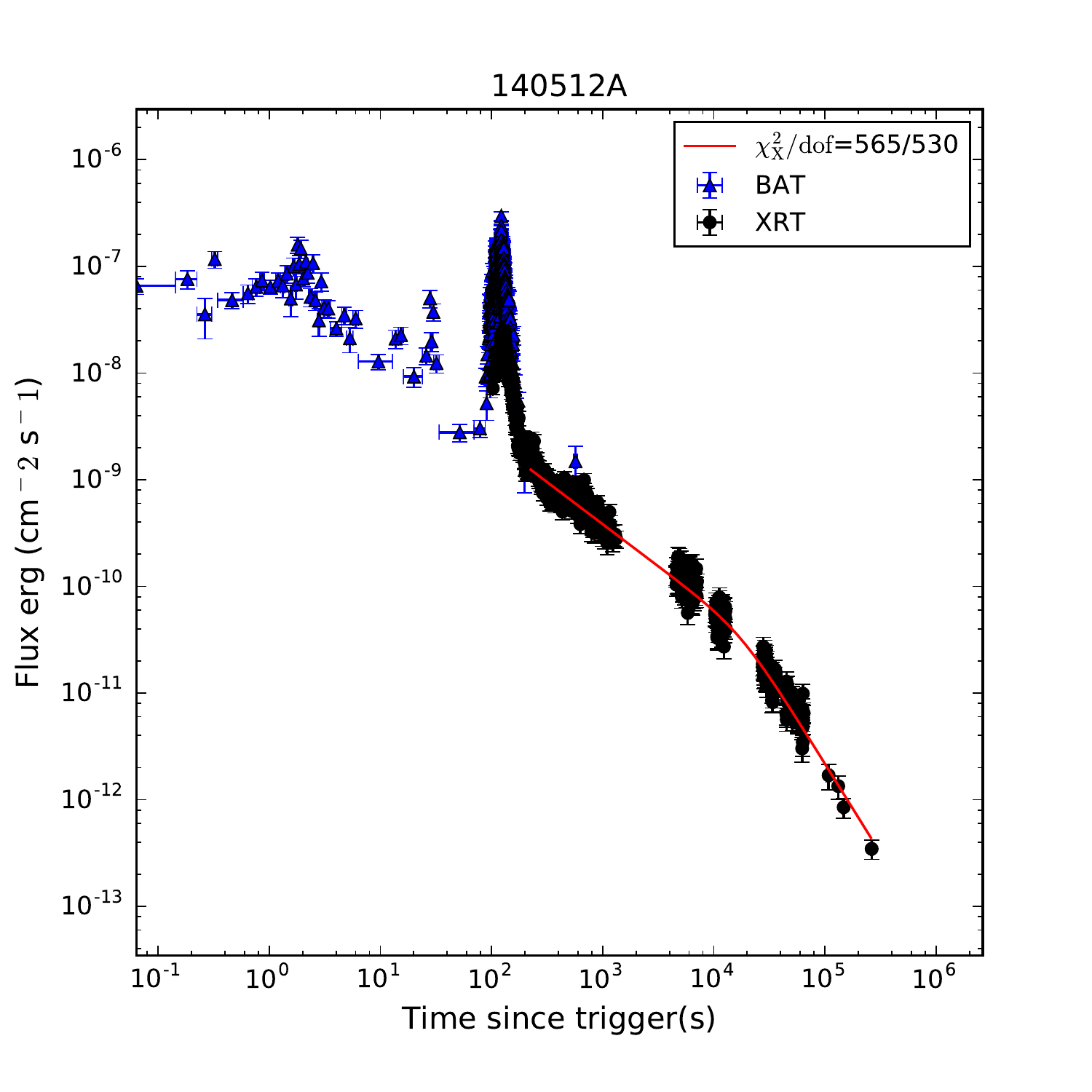}{0.28\textwidth}{}
          \fig{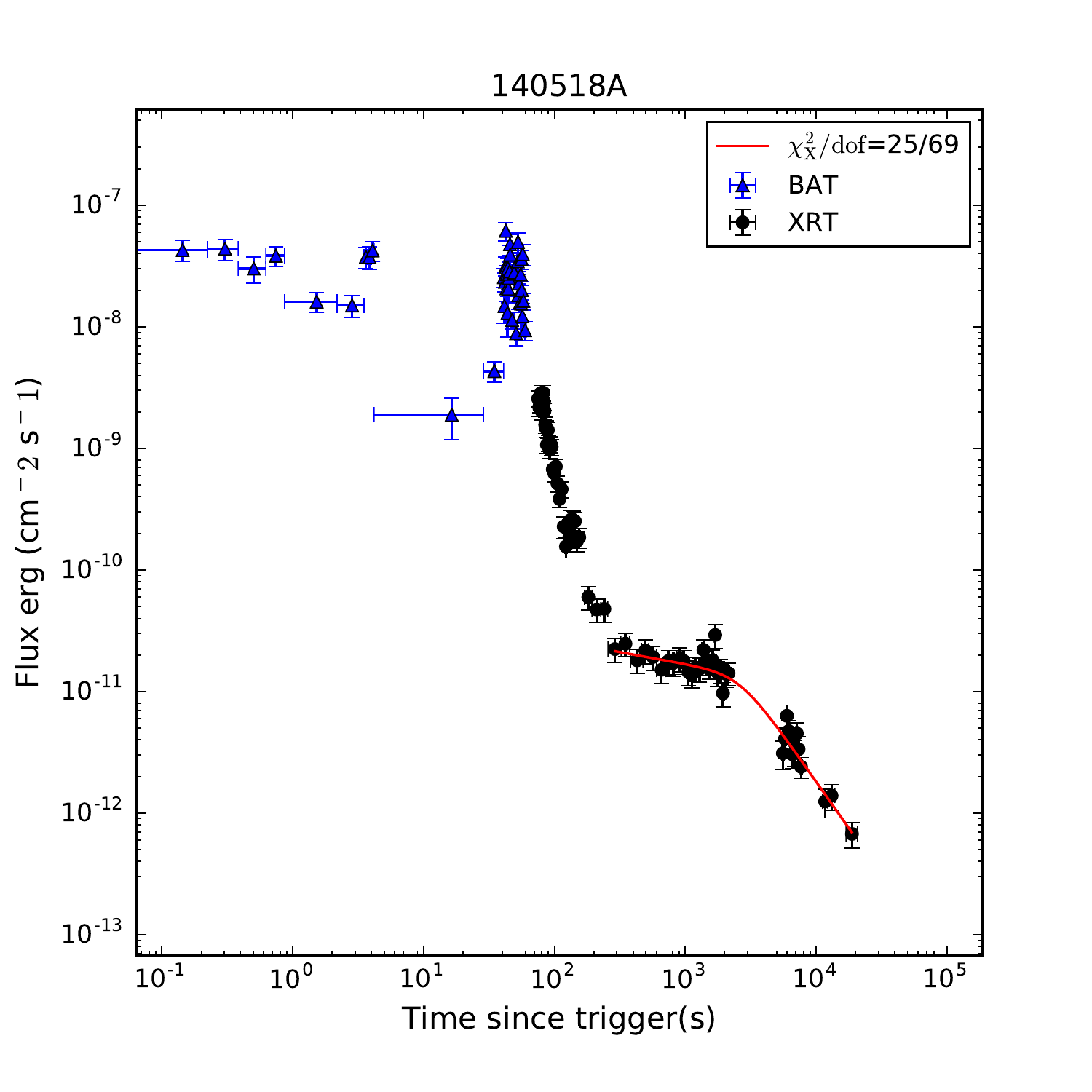}{0.28\textwidth}{}
          }        
\center{Fig. \ref{Silver}--- Continued}
\end{figure}
\begin{figure}
\gridline{\fig{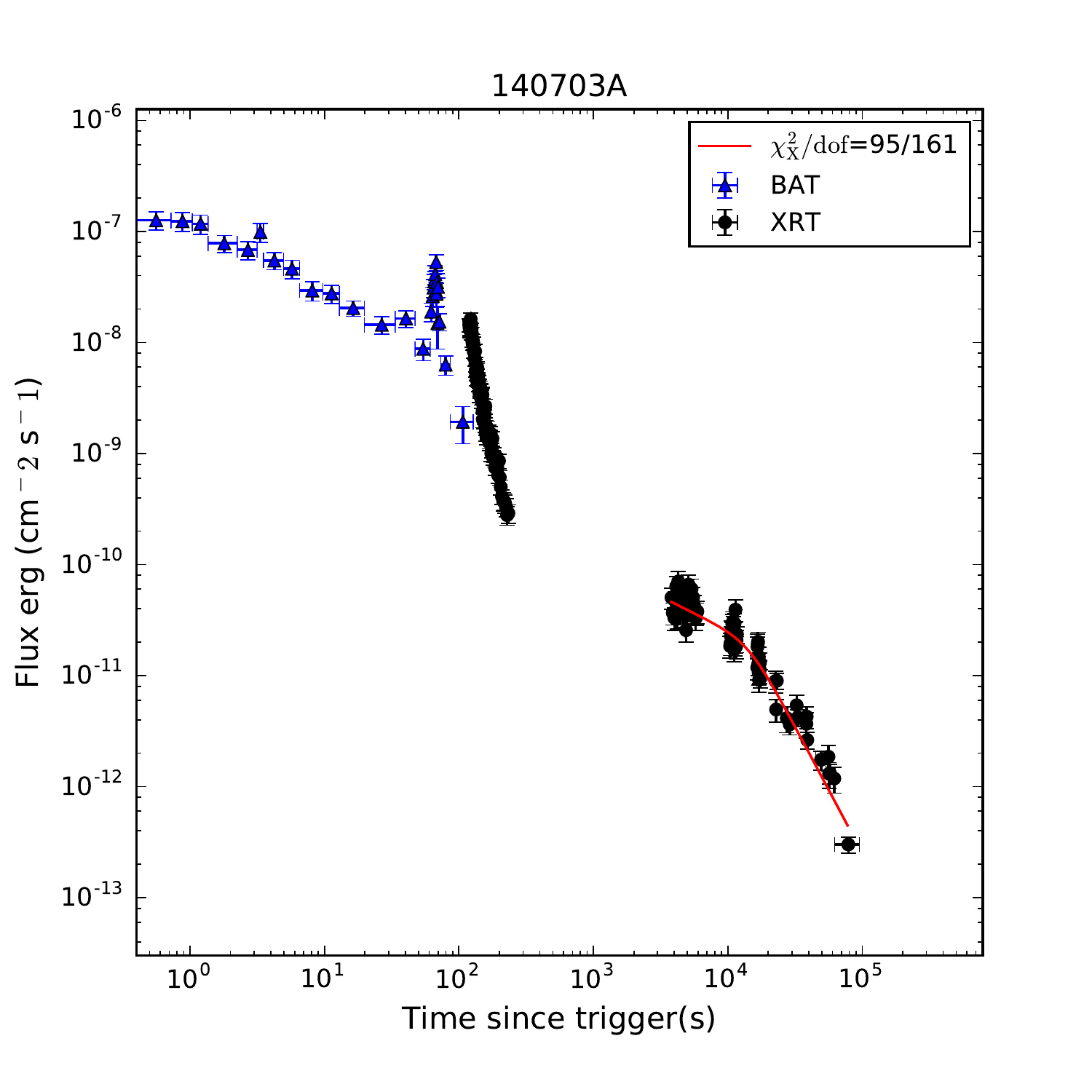}{0.28\textwidth}{}
          \fig{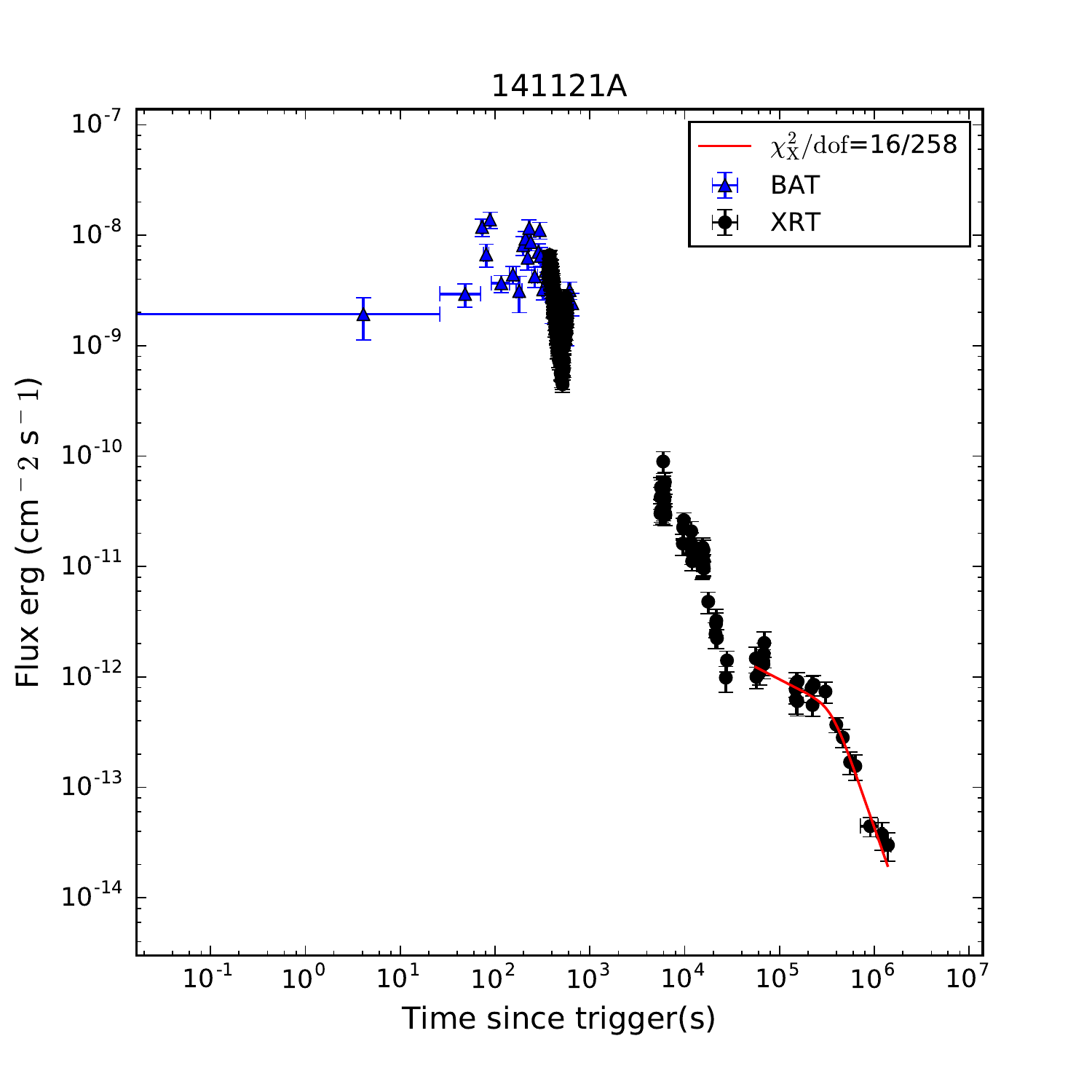}{0.28\textwidth}{}
          \fig{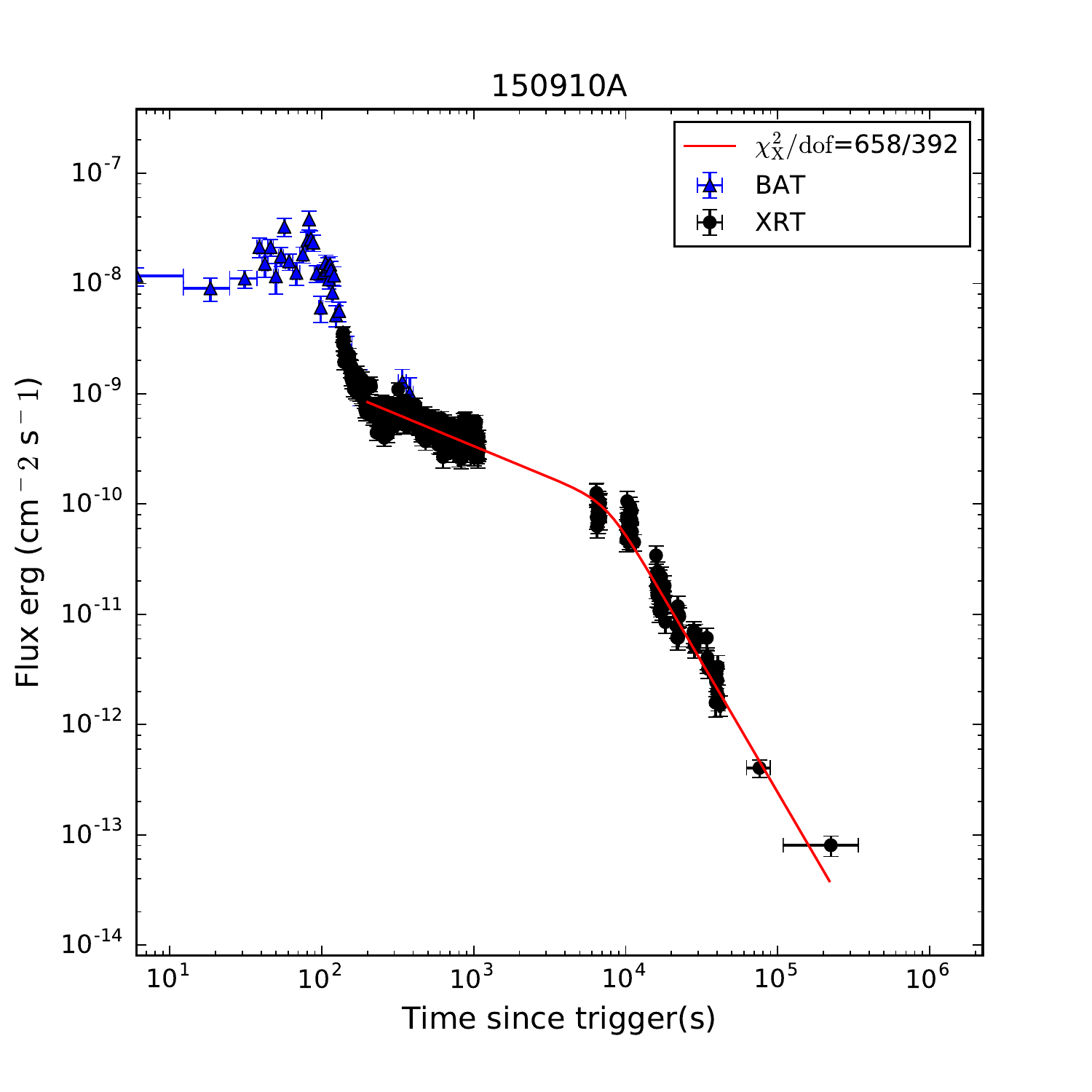}{0.28\textwidth}{}
          }
\gridline{\fig{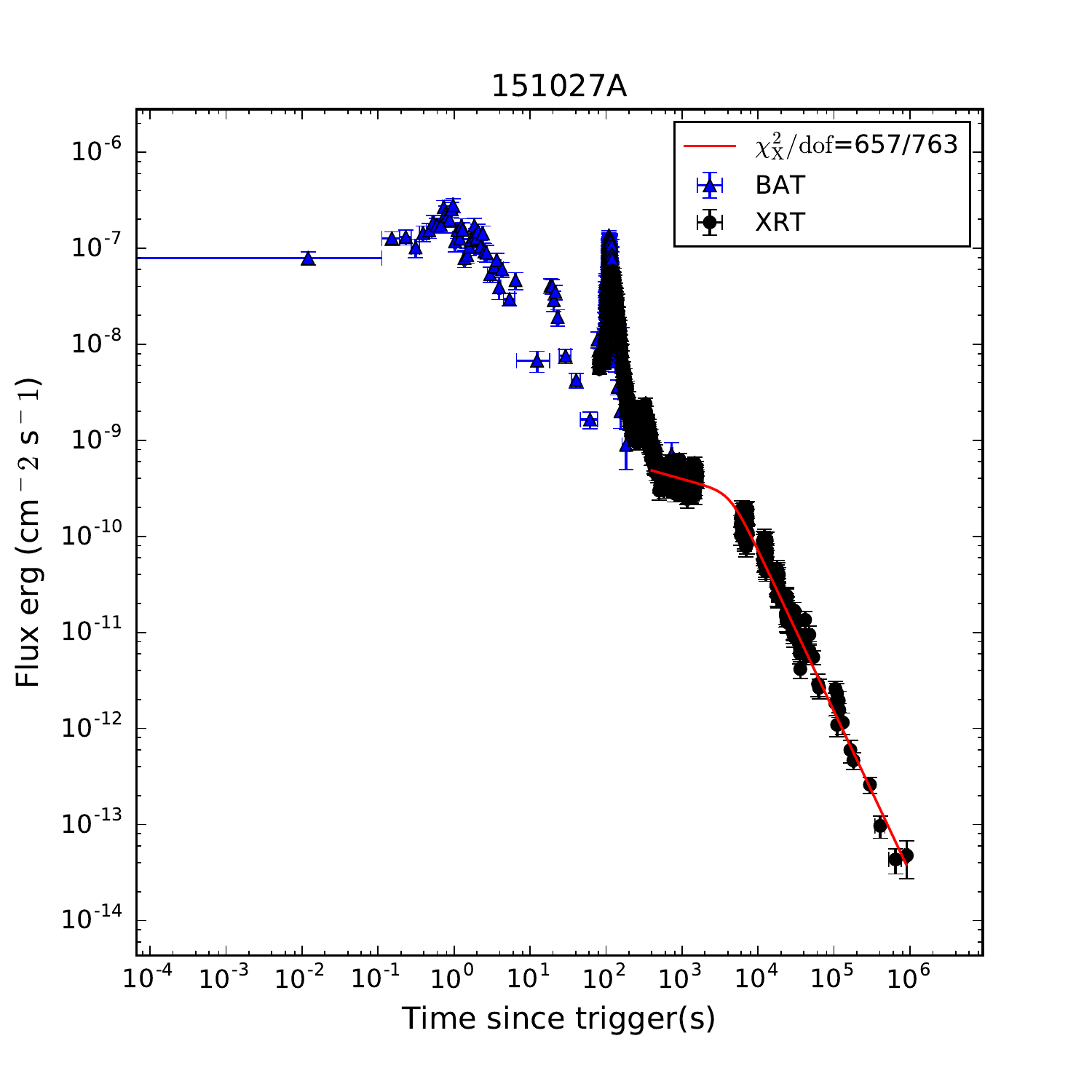}{0.28\textwidth}{}
          \fig{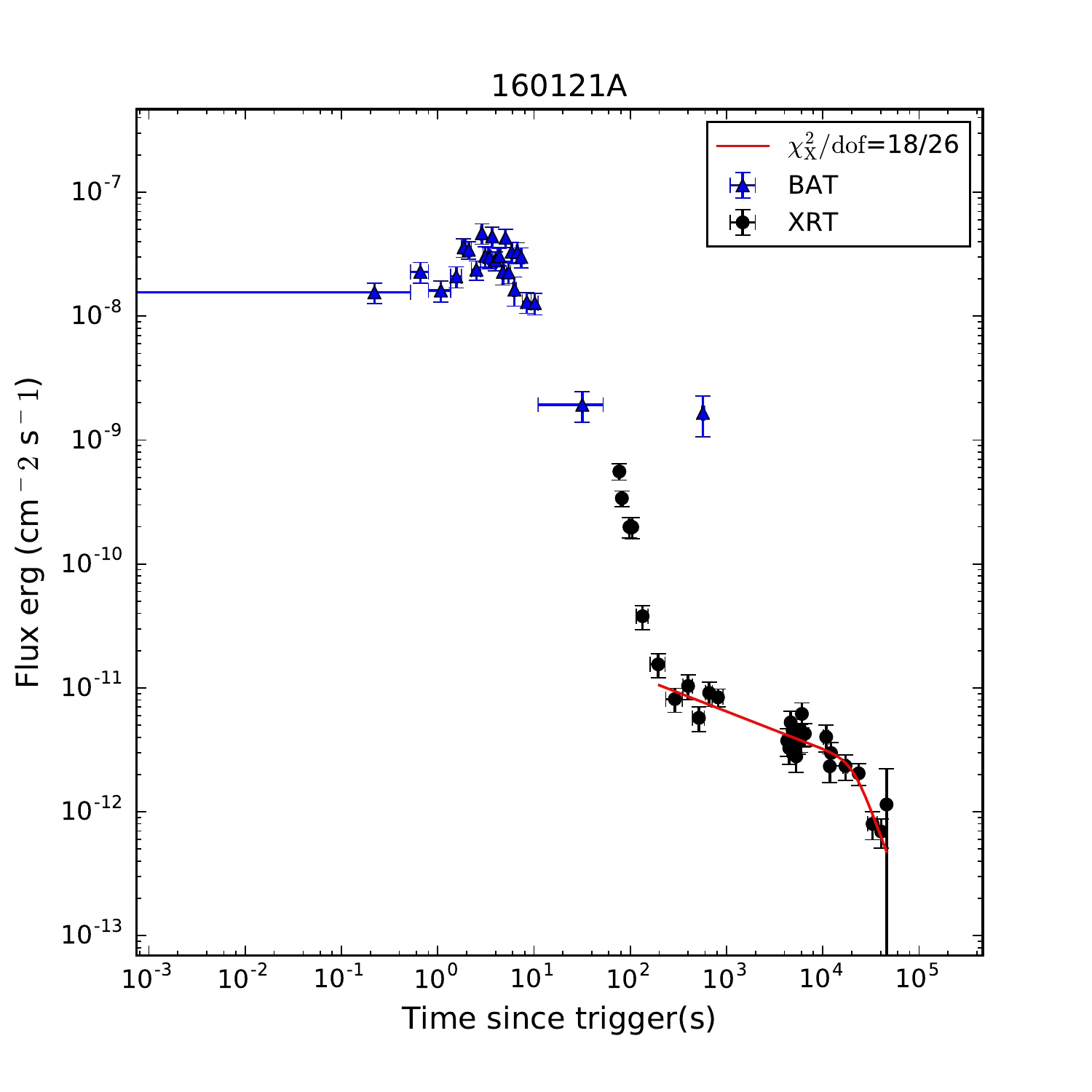}{0.28\textwidth}{}
          \fig{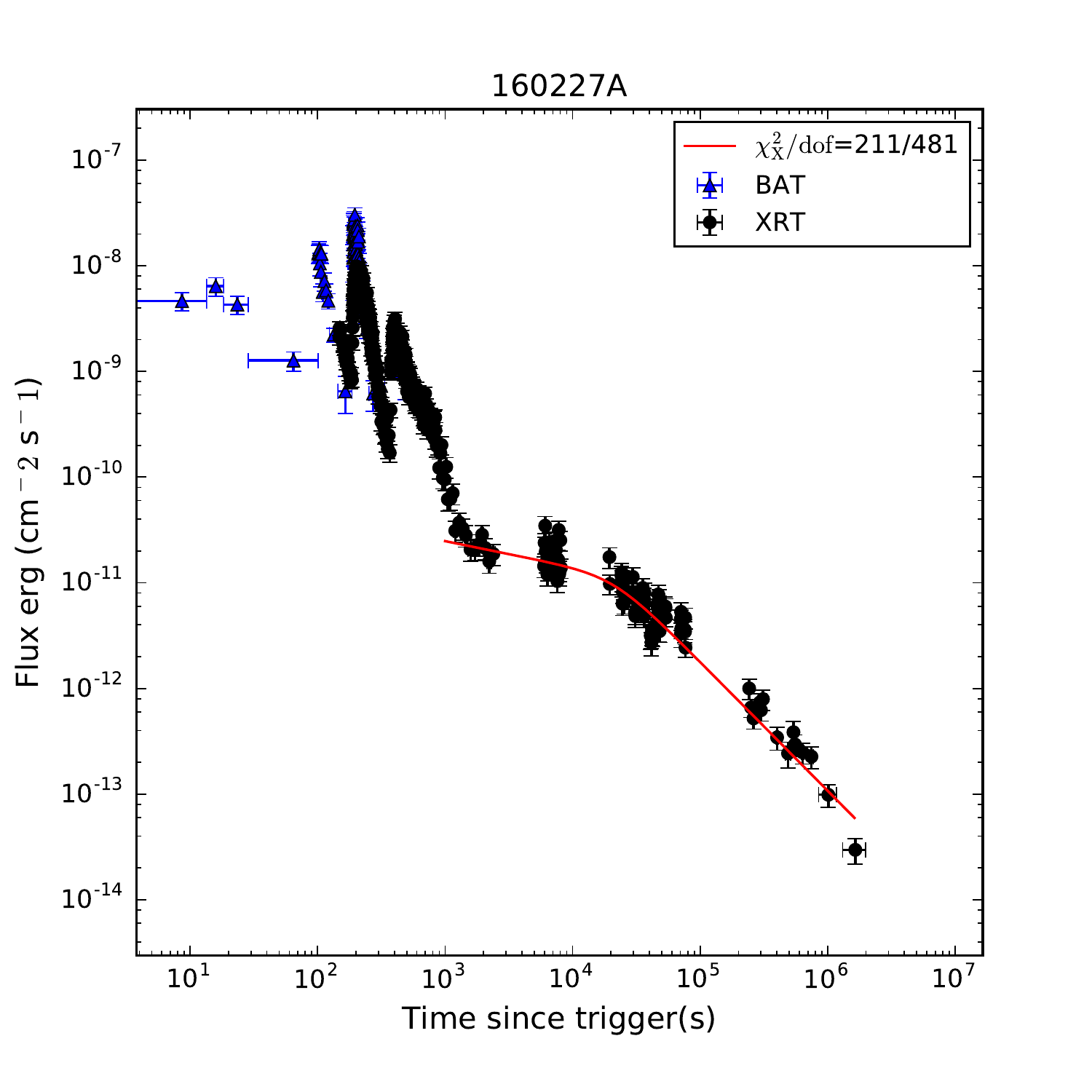}{0.28\textwidth}{}
          }    
\gridline{\fig{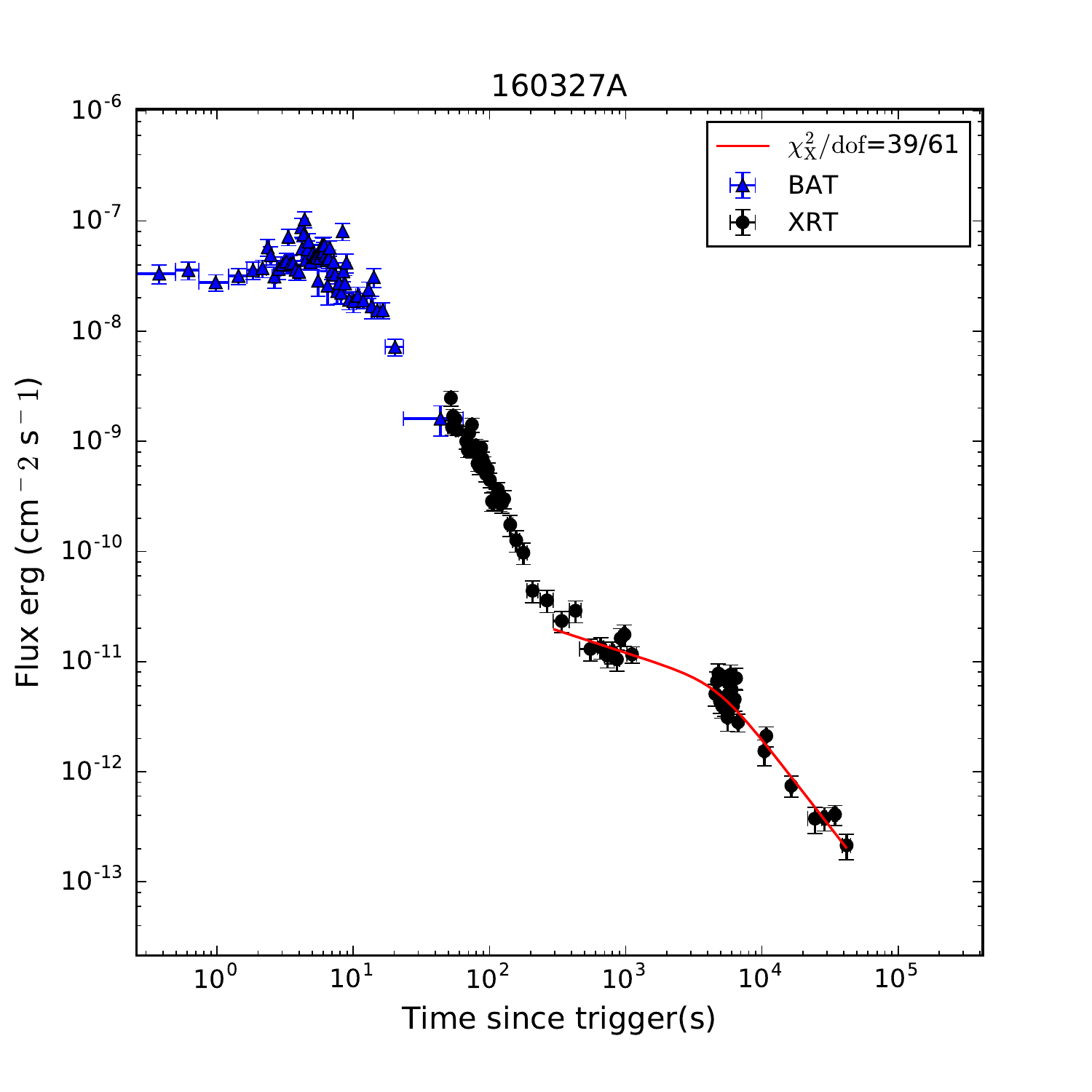}{0.28\textwidth}{}
          \fig{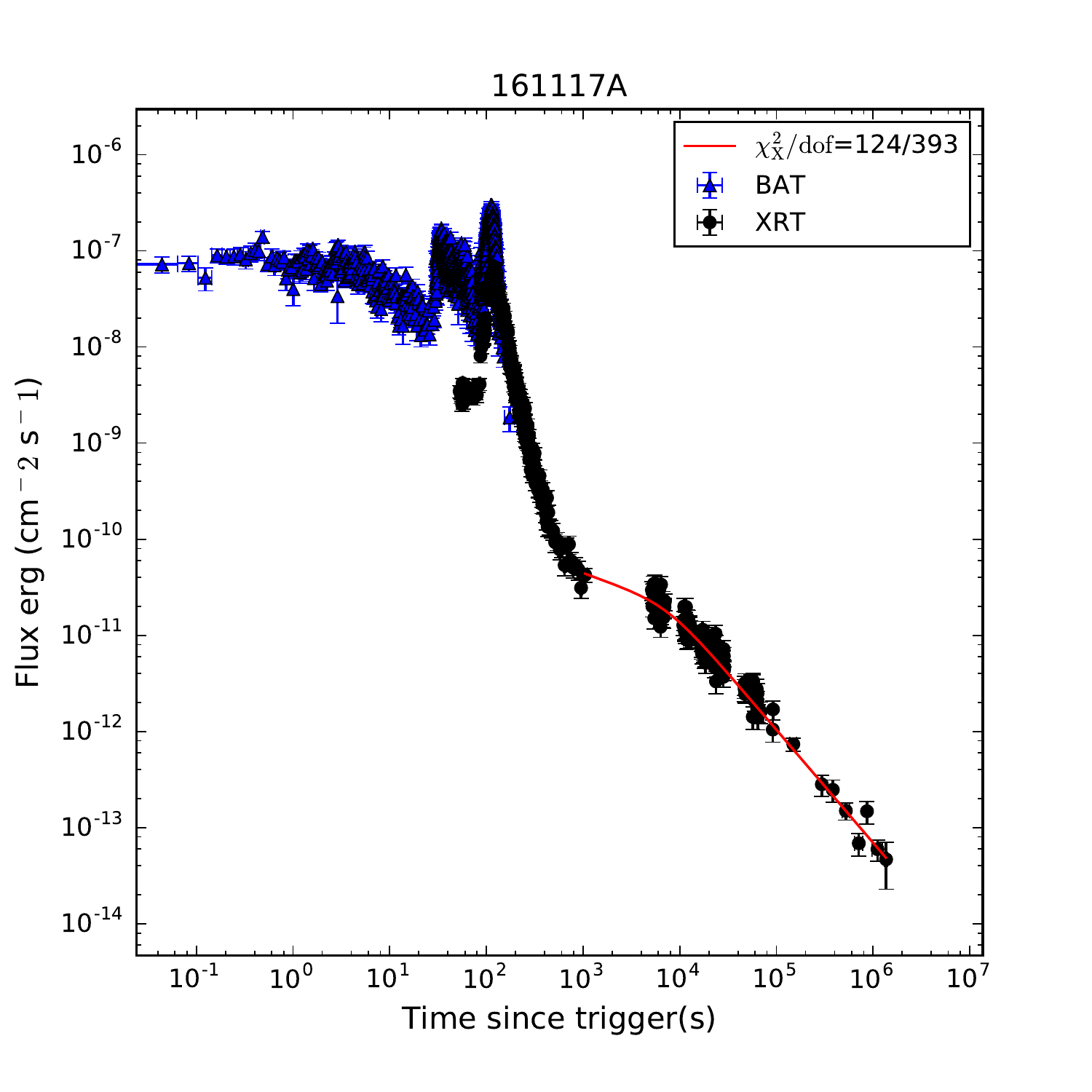}{0.28\textwidth}{}
          \fig{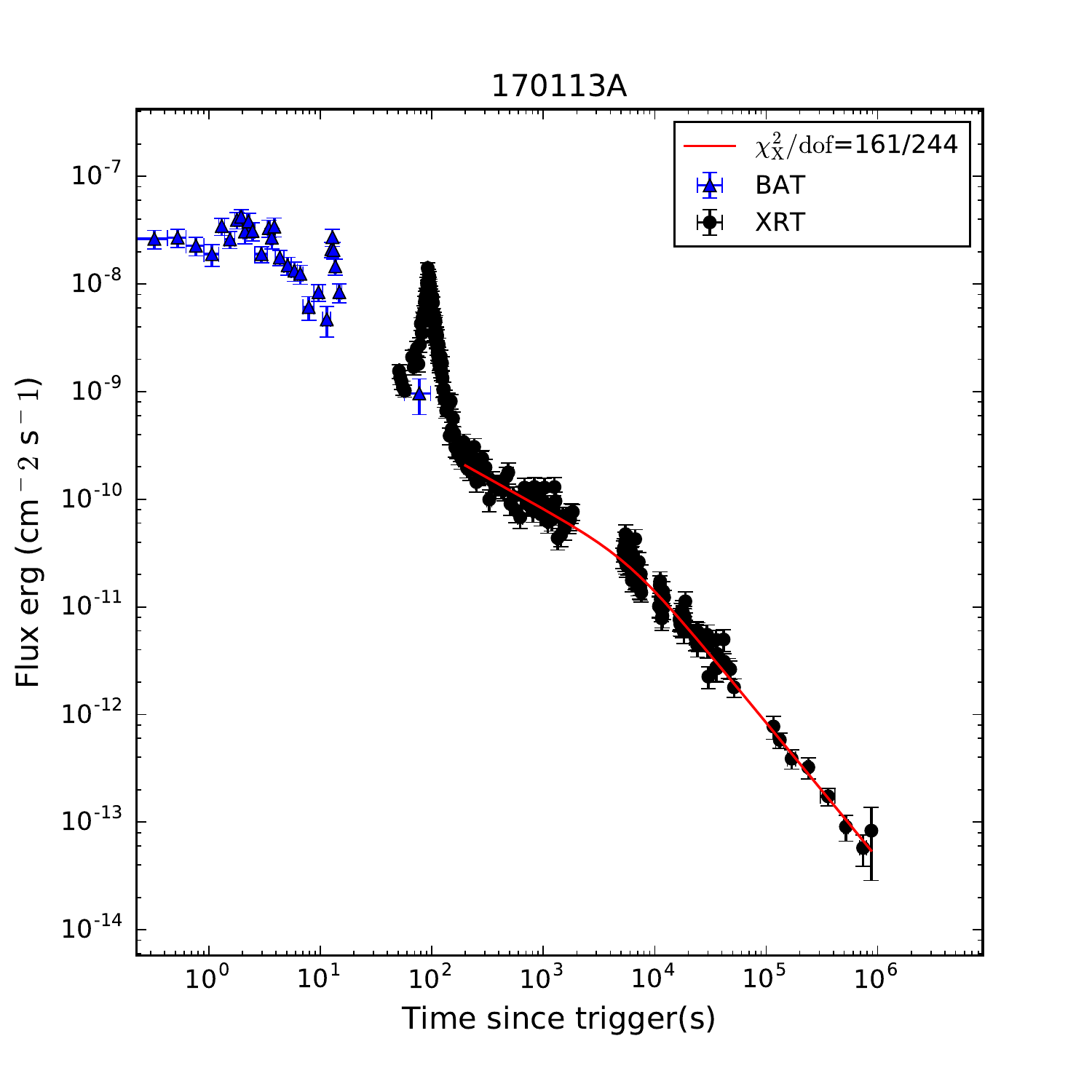}{0.28\textwidth}{}
          }
\center{Fig. \ref{Silver}--- Continued}
\end{figure}

\begin{figure}
\gridline{\fig{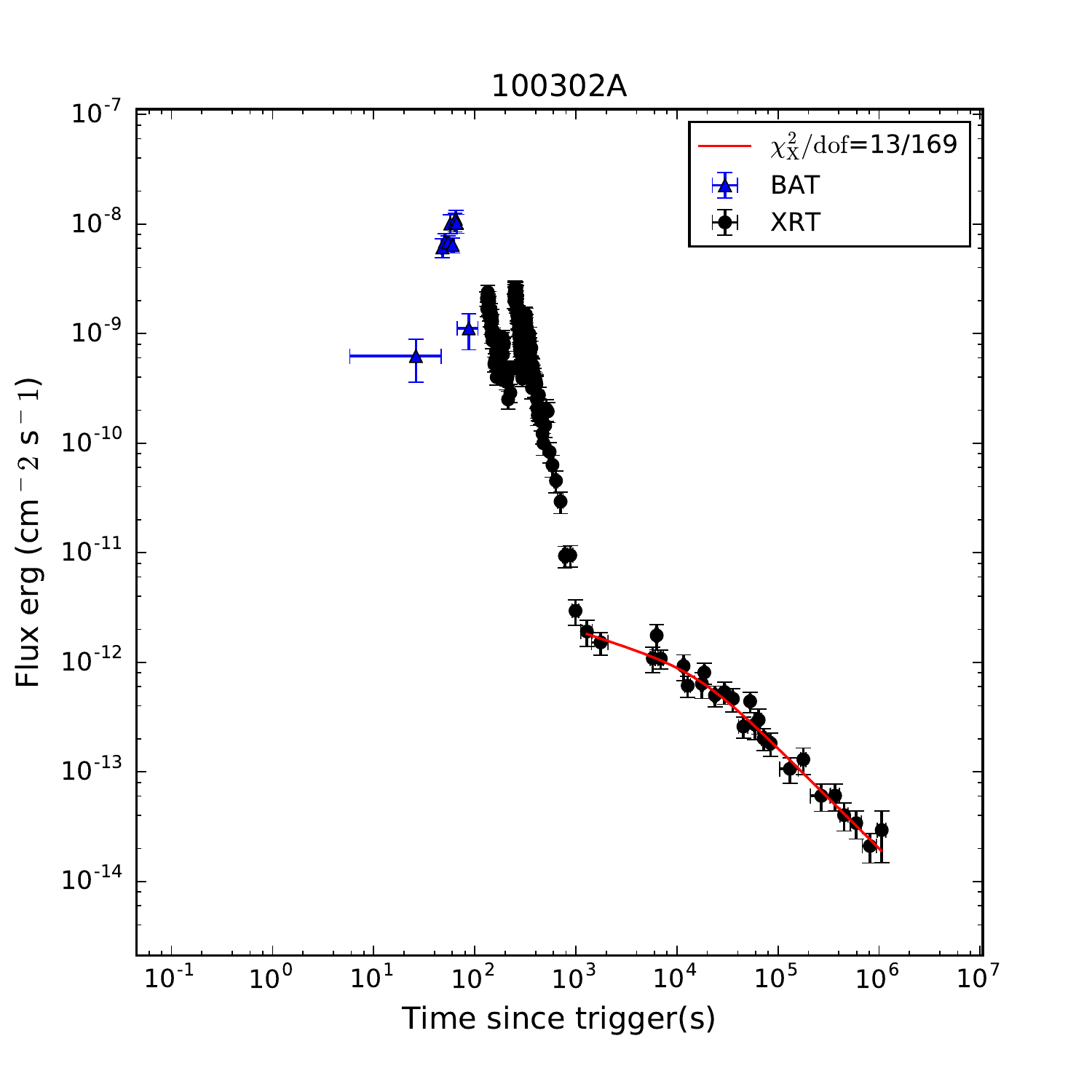}{0.28\textwidth}{}
          \fig{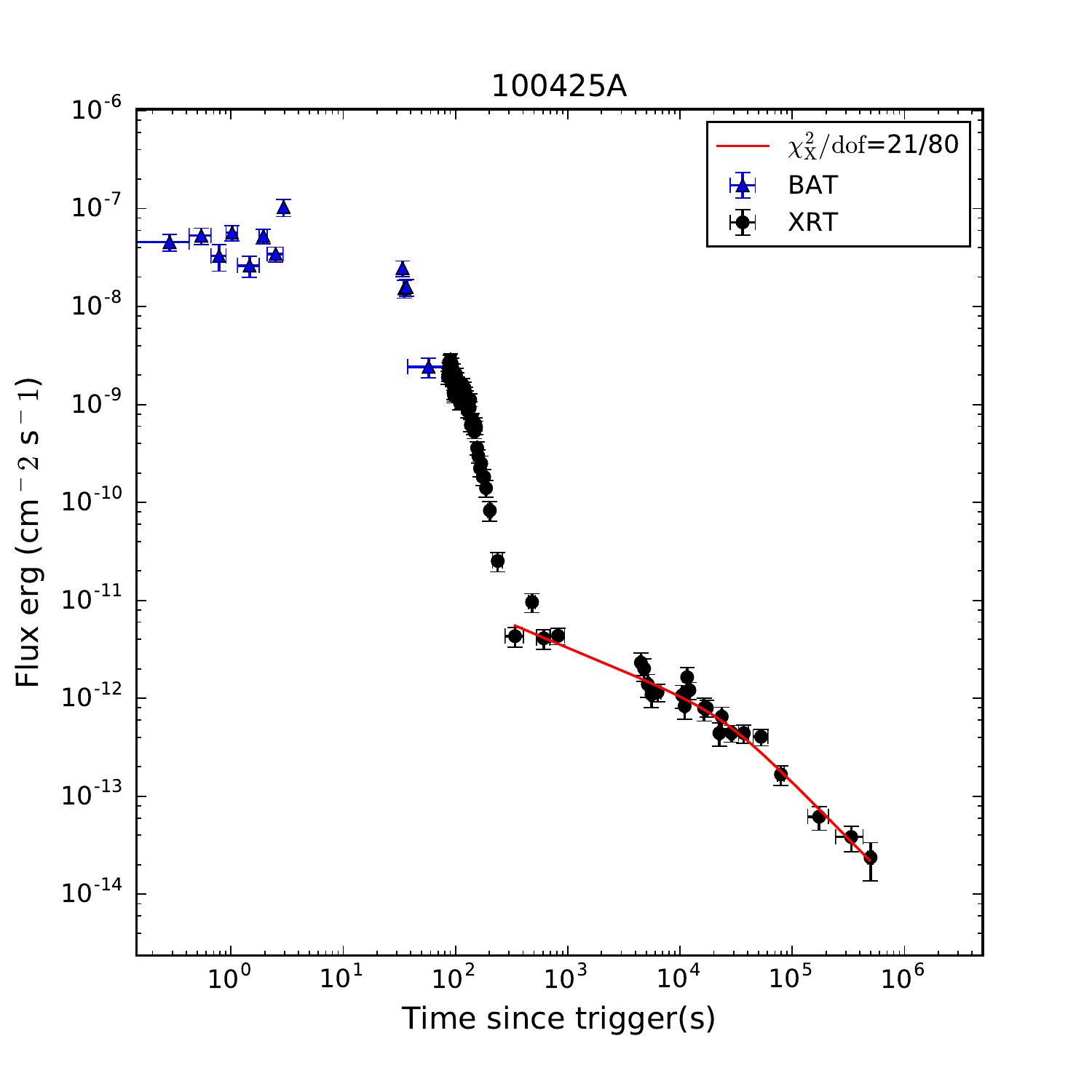}{0.28\textwidth}{}
          \fig{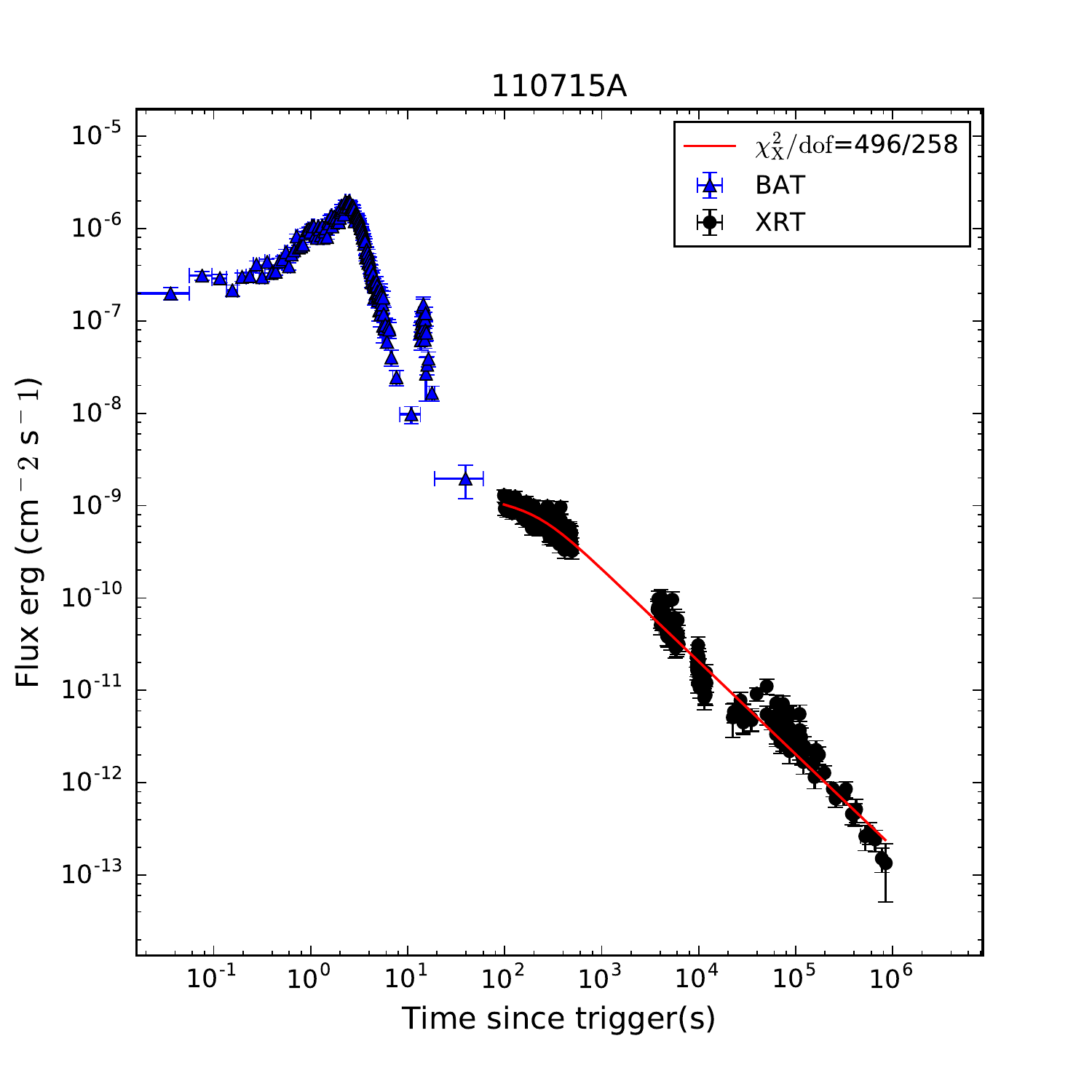}{0.28\textwidth}{}
          }
\gridline{\fig{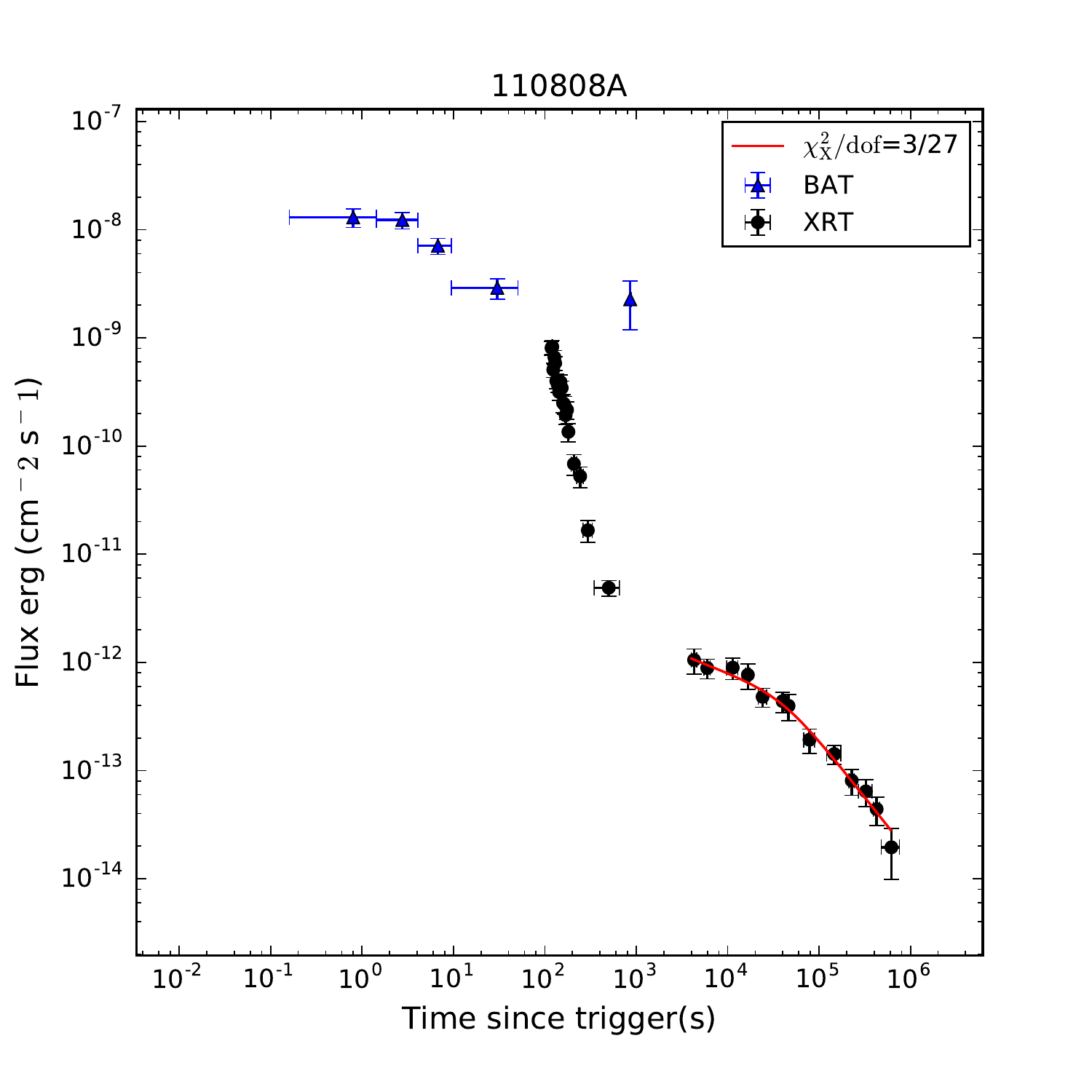}{0.28\textwidth}{}
          \fig{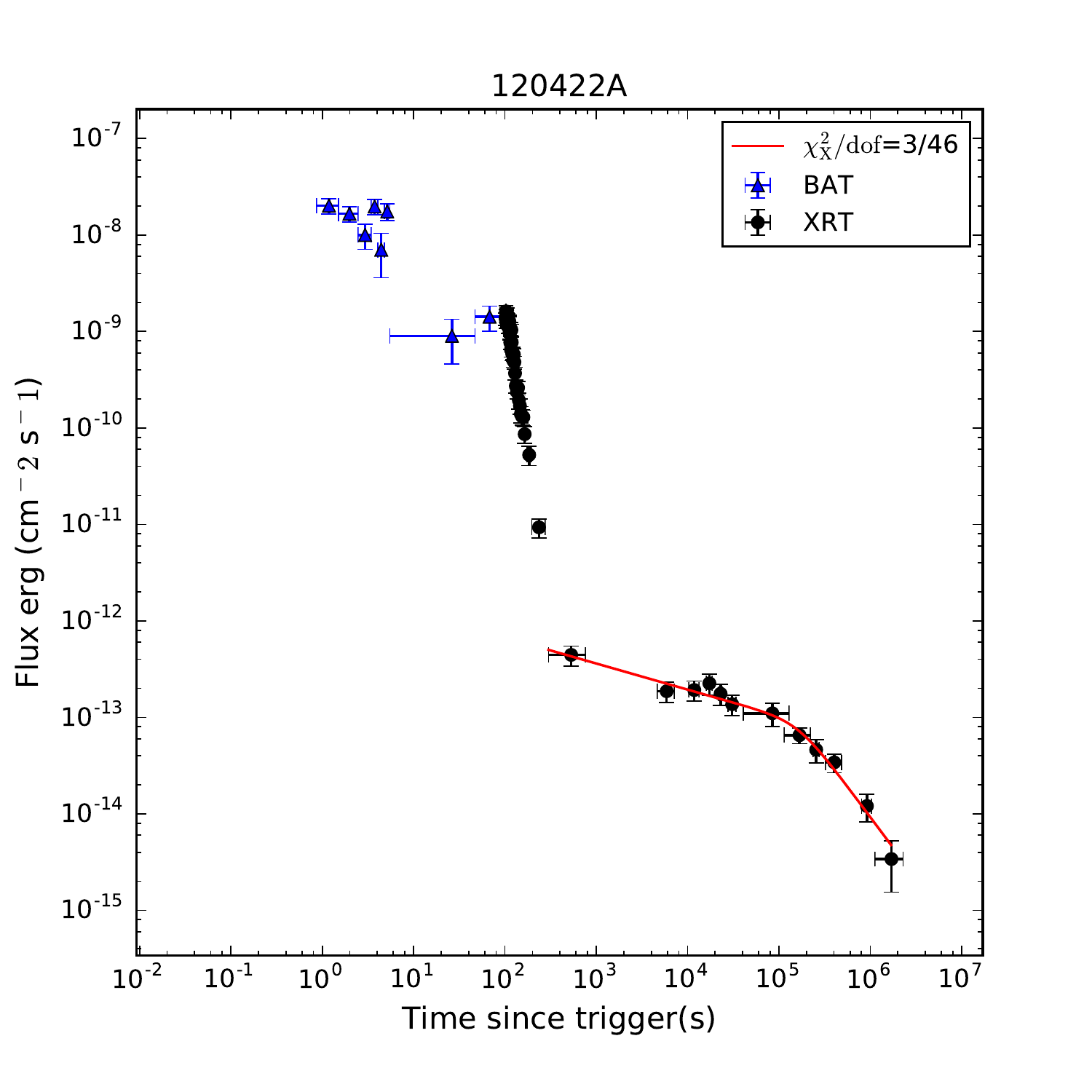}{0.28\textwidth}{}
          \fig{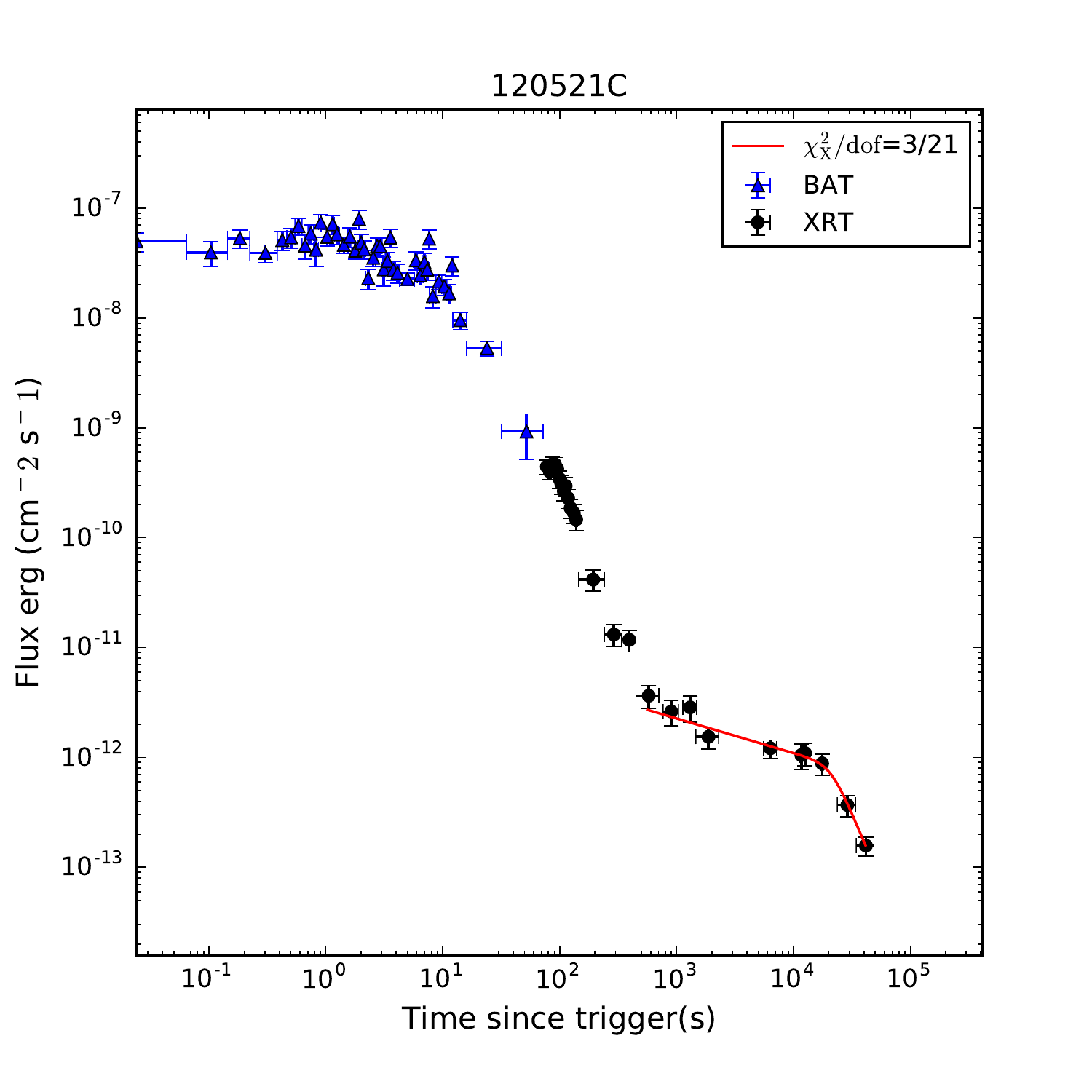}{0.28\textwidth}{}
          }
\gridline{\fig{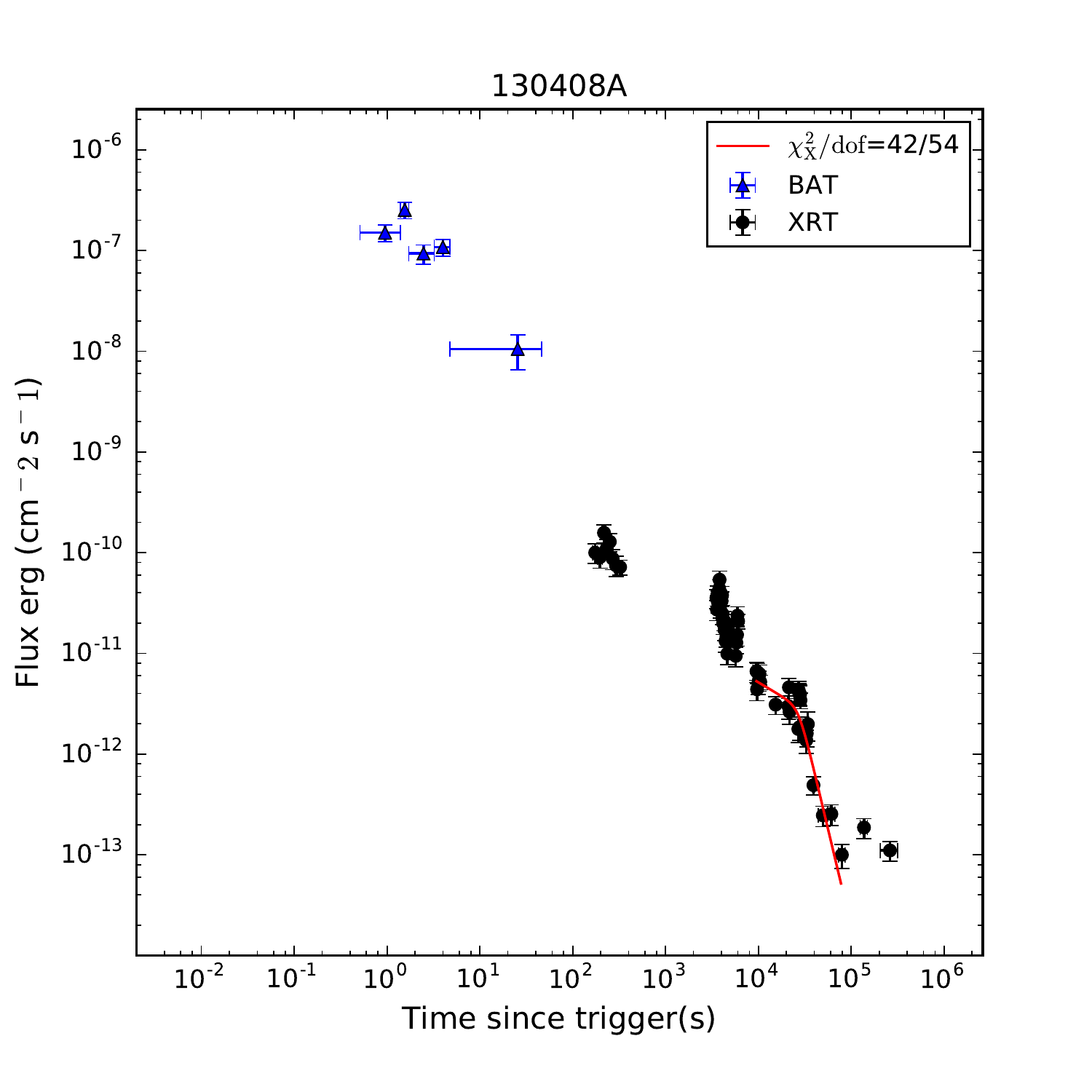}{0.28\textwidth}{}
          \fig{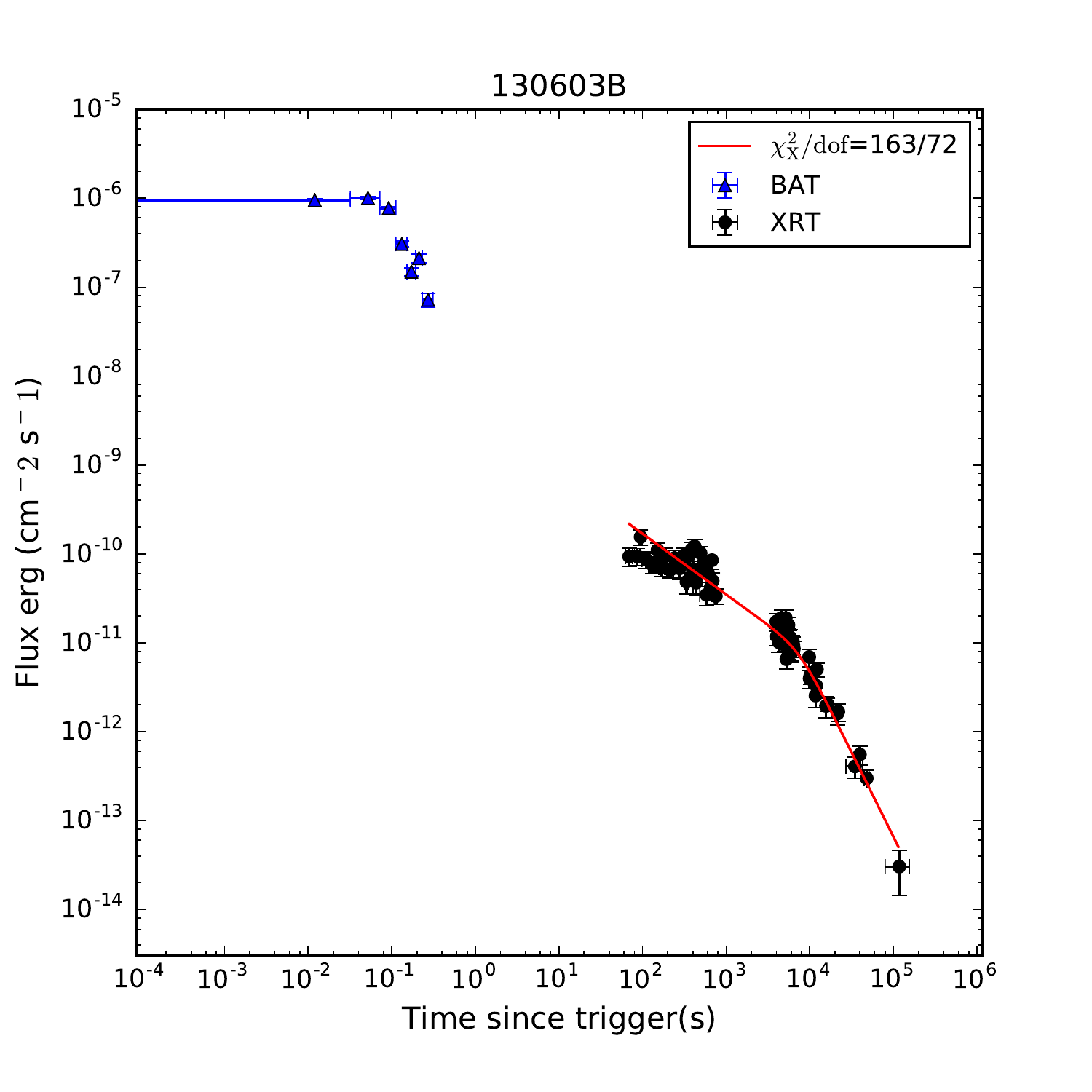}{0.28\textwidth}{}
          \fig{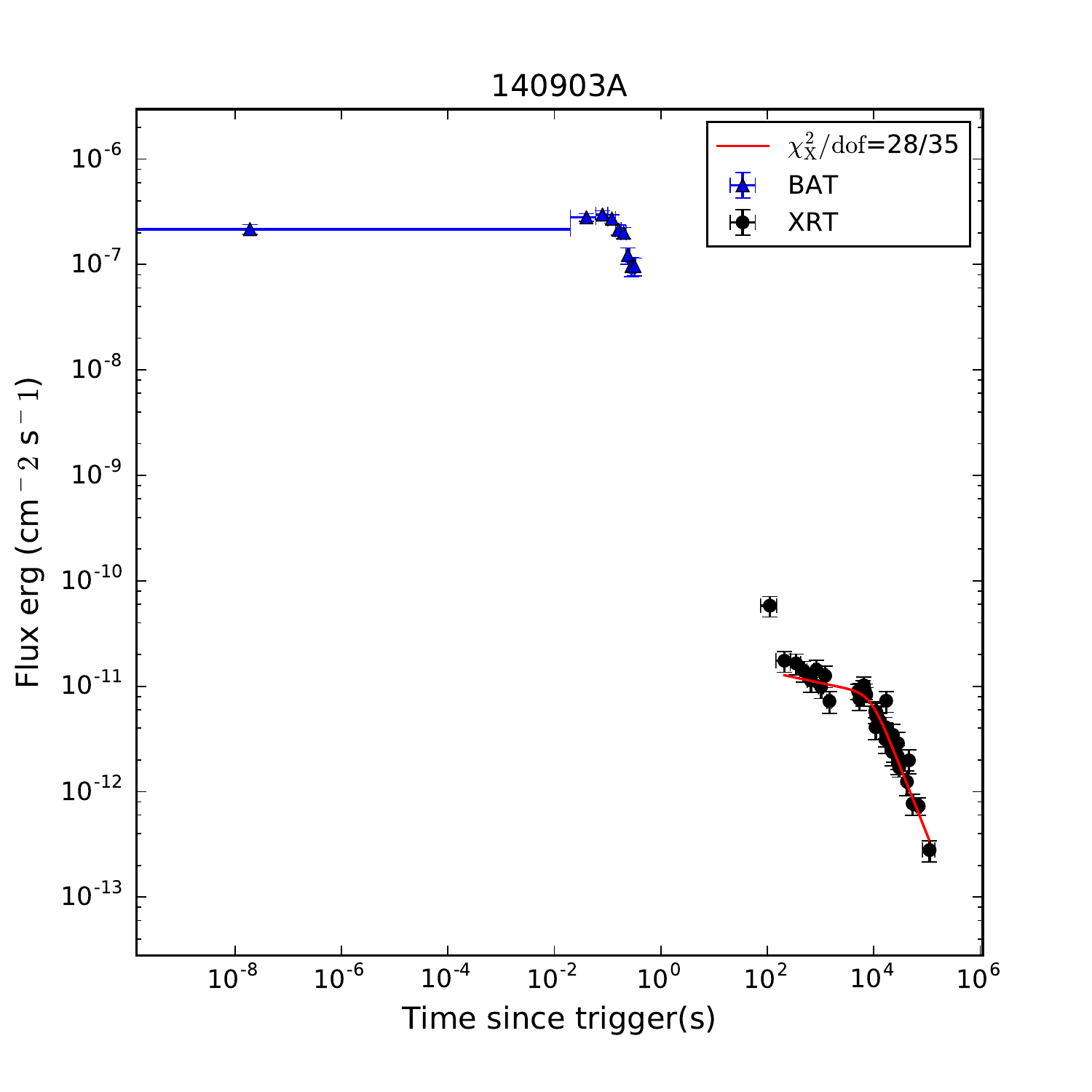}{0.28\textwidth}{}
          }
\gridline{\fig{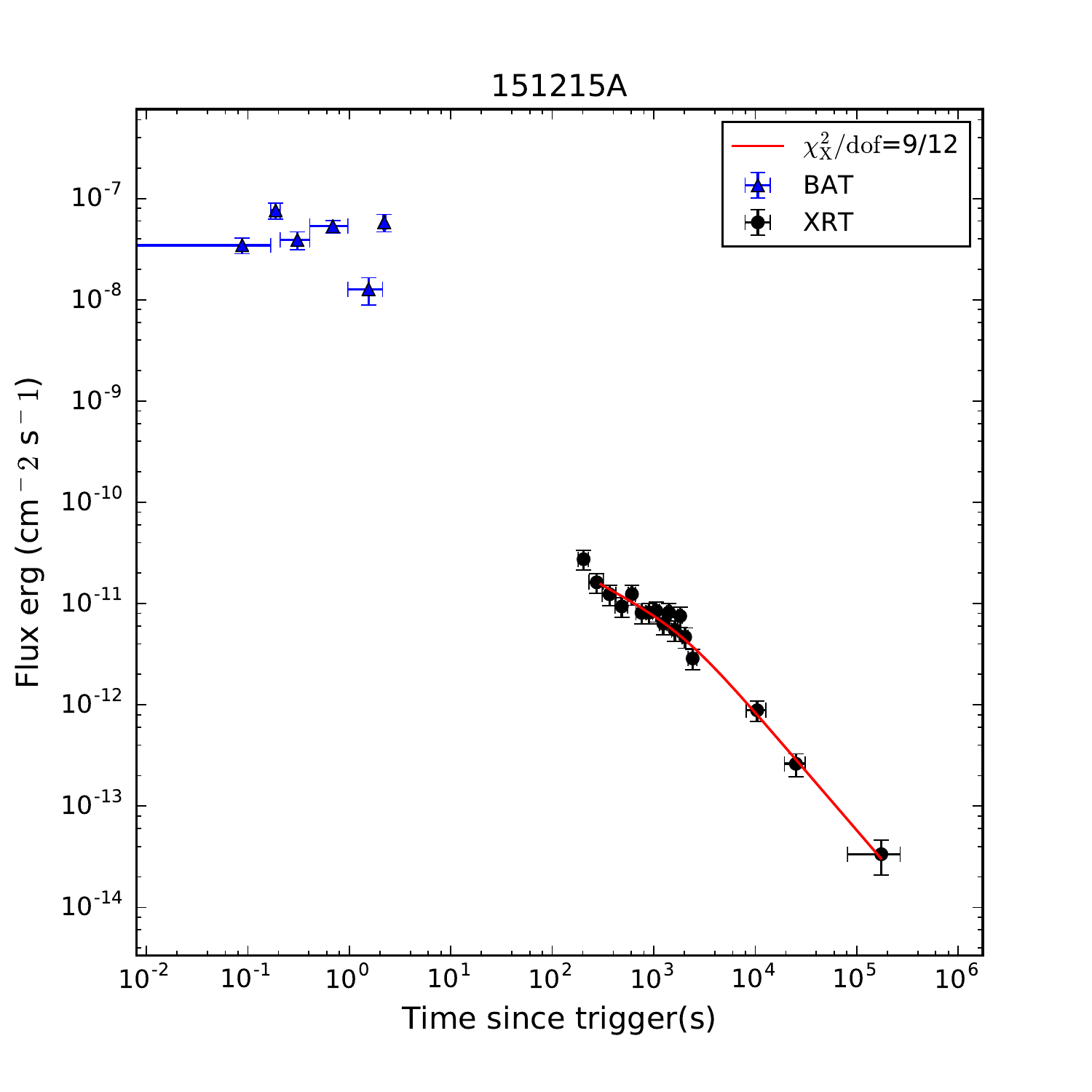}{0.28\textwidth}{}
          \fig{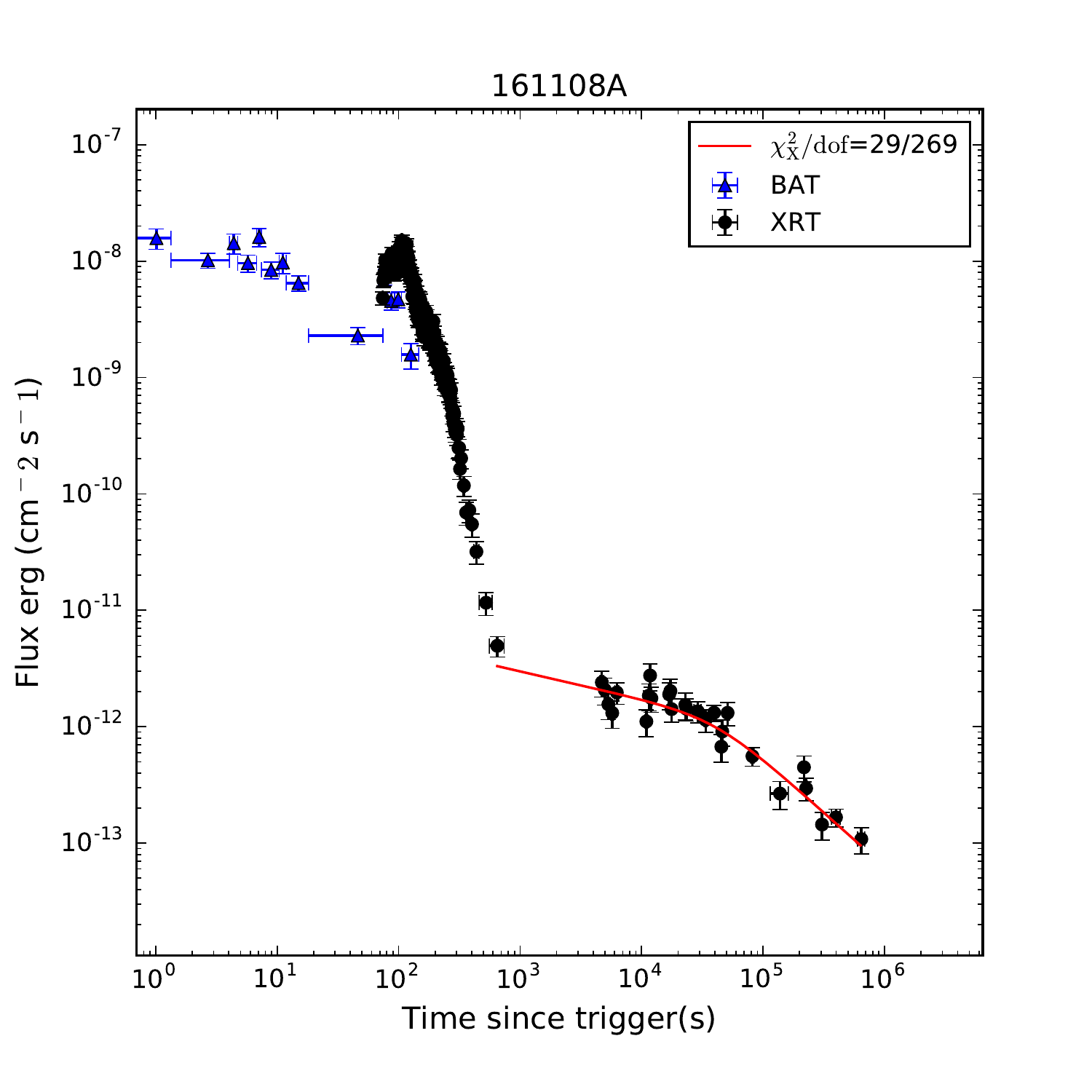}{0.28\textwidth}{}
          }
\center{Fig. \ref{Bronze}--- Continued}
\end{figure}

\clearpage

\clearpage

\clearpage

\clearpage

\clearpage

\clearpage

\clearpage

\clearpage

\clearpage

\clearpage

\clearpage

\clearpage

\clearpage

\clearpage

\clearpage

\clearpage

\clearpage

\clearpage

\clearpage

\clearpage

\clearpage
\vspace{5mm}
\facilities{{\it Swift}/XRT}
\software{Python}
\bibliography{./my_citation_D.bib}

\end{document}